\documentclass [11pt, a4paper, openright] {book}
\usepackage {a4, amsmath, alltt}

\usepackage[scriptsize, sf, bf]{subfigure}

\usepackage[figuresright]{rotating}

\usepackage{longtable}
\usepackage{lscape}

\usepackage[caption2]{ccaption}

\usepackage[scriptsize]{caption2}

\usepackage[innercaption]{sidecap}

\usepackage[sort]{natbib}
\usepackage{aas_macros}

\usepackage{sectsty}
\allsectionsfont{\sffamily}

\usepackage{fancyhdr}
\usepackage[Uwe]{fncychap}
\newcommand{\corr}{\mathrel{\widehat{=}}}

\newcommand{\mas}{{\rm mas}}

\usepackage{setspace}

\usepackage{url}

\usepackage{mathrsfs}

\usepackage{multirow}

\usepackage{german}

\usepackage{amssymb}

\usepackage{fix2col}

\usepackage{dcolumn}
\newcolumntype{.}{D{.}{.}{5}}
\newcolumntype{,}{D{~}{\,}{-1}}

\usepackage[T1]{fontenc}
\usepackage{times}
\usepackage{mathptmx}

\usepackage{graphics}

\usepackage{fancybox}

\begin{document}


\thispagestyle{empty}

\selectlanguage{german}

\doublespacing
\begin{center}
Gas Around Active Galactic Nuclei\\
and New Phase Calibration Strategies for High-Frequency VLBI\\
\vspace{2.1cm}
{\bf Dissertation}\\

zur\\
Erlangung des Doktorgrades (Dr. rer. nat.)\\
der\\
Mathematisch-Naturwissenschaftlichen Fakult\"at\\
der\\
Rheinischen Friedrich-Wilhelms-Universit\"at Bonn\\

\vspace{2.52cm}

vorgelegt von\\
Enno Middelberg\\
aus\\
Haren (Ems)\\

\vspace{1.68cm}

Bonn, April 2004
\end{center}
\singlespacing

%
%
%

\thispagestyle{empty}
\begin{figure}[htpb]
\begin{minipage}[b]{12cm}
Information bedeutet Horizonterweiterung, und zwar immer auf Kosten
der Nestw\"arme. Sie ist eine N\"otigung, anderem, fernem und fremdem
Geschick und Geschehen Aufmerksamkeit zu widmen.\\

\begin{flushright}
{\it Hans-J\"urgen Schultz}
\end{flushright}

\vspace{8cm}
\end{minipage}
\end{figure}

\selectlanguage{english}

\newpage
\tableofcontents
\newpage
\listoffigures
\newpage
\listoftables

\chapter{Introduction}

Active galactic nuclei, or AGN, have received considerable attention
during the last 40~years. When Maarten Schmidt recognized in 1963 that
the quasi-stellar object 3C\,273 has a redshift of almost 0.16
(\citealt{Schmidt1963}), it became immediately clear that there were
objects outside our galaxy with tremendous
luminosities. \cite{Zeldovich1964}, \cite{Salpeter1964} and
\cite{Lynden-Bell1969} suggested that extragalactic radio sources were
mainly driven by gas accreted into a disc around super-massive black
holes, and \cite{Blandford1979} suggested that the radio emission from
AGN is produced by a relativistic outflow of plasma along magnetic
field lines. Their model, with numerous modifications, is still
thought to be valid, and hence the basic foundations of what we know
about AGN today are almost 25~years old.\\

Like in all branches of astronomy, the progress in AGN science was
tightly correlated with technical improvements. Radio astronomy
started out with very low angular resolution due to relatively small
dishes and long wavelengths. This made $\lambda/D$, governing the
resolution of any observing instrument, very poor compared to optical
instruments. Almost simultaneously, however, astronomers started to
experiment with radio interferometers with progressively longer
baselines and finally, to combine interferometric measurements to
simulate a larger dish (\citealt{Ryle1960}). This technique is called
``aperture synthesis'' and was transferred from directly linked,
locally distributed antennas to spatially widely separated radio
telescopes in the 1970s. What is known as VLBI, or Very Long Baseline
Interferometry, today, is a combination of single-dish radio
astronomy, interferometry and aperture synthesis, and therefore
certainly one of the most advanced technical achievements of the 20th
century. Consequently, in 1974, Sir Martin Ryle was awarded the Nobel
prize in physics for his contribution.\\

AGN science and VLBI are tightly interrelated: no other instrument
yields the angular resolution necessary to spatially resolve the
innermost regions of AGN and details in the jets, and AGN are almost
the only objects that are observable with VLBI, with a few exceptions
like masers or extremely hot gases. Thus, one cannot live without the
other, but the connection has turned out to be very fruitful.\\

\section{The AGN Zoo}

The first AGN-related phenomena, although not recognized as such,
were reported as early as 1908 (\citealt{Fath1908}) and in the
following decades (\citealt{Slipher1917}, \citealt{Humason1932},
\citealt{Mayall1934}), when observers noticed bright emission lines
from nuclei of several galaxies, and the first extragalactic jet in
M87 was observed (\citealt{Curtis1918}). The first systematic study of
these objects was carried out by Carl Seyfert
(\citealt{Seyfert1943}). He noticed that emission lines from the
nuclei of six galaxies were unusually broadened (several $1000\,{\rm
km\,s^{-1}}$), but simply stated that this phenomenon was ``probably
correlated with the physical properties of the nucleus'', without any
further interpretation. In the 1950s and 1960s, the first radio sky
surveys were completed and catalogues, like the Third Cambridge
Catalogue, were published. When interferometric techniques were
further developed, all kinds of differences between the AGN turned up,
and following ancient habits, astronomers started to classify what
they observed. Meanwhile, they have created a colourful collection of
mostly phenotypical classes and an equally colourful bunch of acronyms
to describe these classes.

\subsection{The General Picture}

\begin{figure}[htpb!]
\centering
\includegraphics[width=\linewidth]{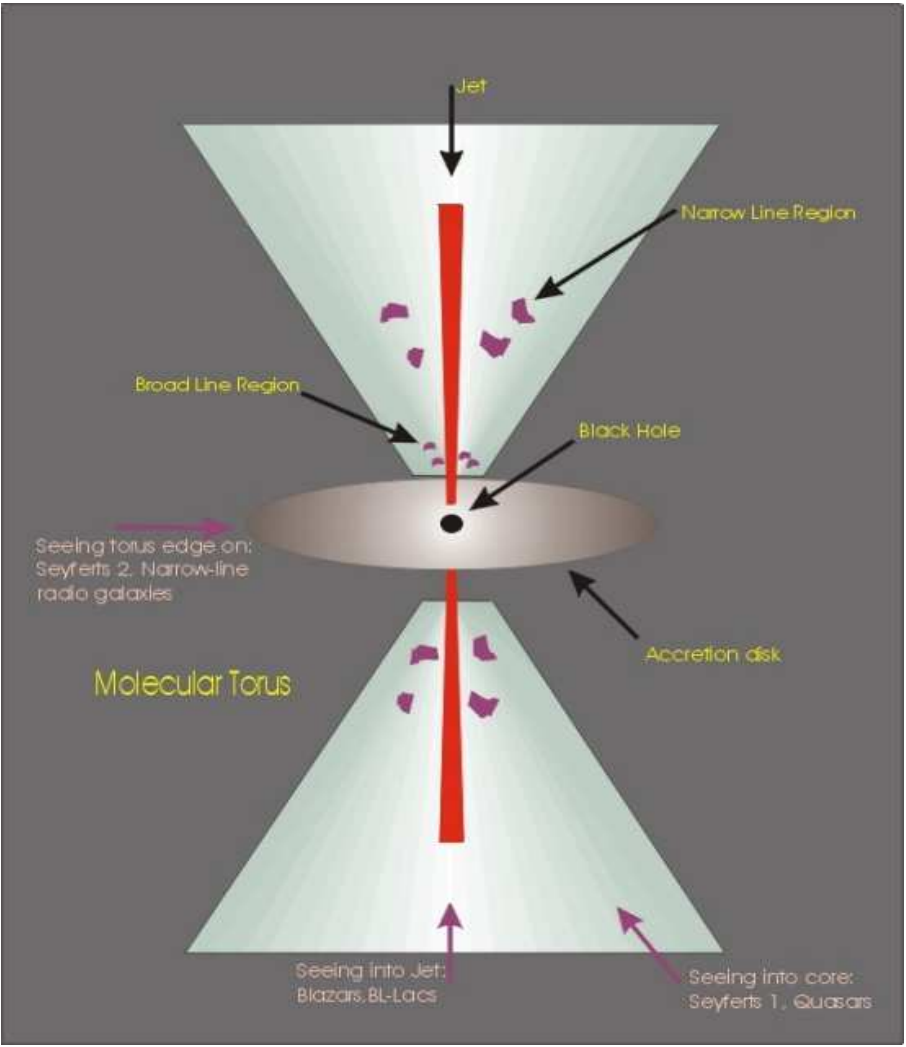}
\caption[Schematic of AGN]{Schematic of AGN (not to scale). Different
AGN types are believed to be due to different viewing angles.}
\label{fig:AGN}
\end{figure}

The current idea of what AGN are is shown in Fig.~\ref{fig:AGN}.  A
supermassive black hole ($M_{\rm \bullet}$ between $10^6\,M_{\odot}$
and $10^9\,M_{\odot}$) accretes surrounding material that settles in a
circumnuclear accretion disc.  Some of the material's potential energy
in the gravitational field of the AGN is turned into radiation by
viscous friction in the accretion disc, but most of both matter and
energy ends up in the black hole.  Some of the material is expelled
into two jets in opposite directions.  Whether the jets are made of an
electron-proton or an electron-positron plasma is still controversial,
but as the observed radiation is certainly synchrotron radiation,
given the high brightness temperatures and degrees of polarization, it
must be ionized material circulating in magnetic fields. The jets are
not smooth, and in many cases new components are observed as they are
ejected from the AGN and travel outwards into the jet direction, and
jet bends of any angle are observed. Though not very well determined,
around 10\,\% of the material's rest mass is turned into energy, the
process thus being incredibly efficient compared to hydrogen burning
in stars, where only 0.7\,\% of the hydrogen mass is turned into
energy in the production of helium. Thermal gas from the jet
surroundings may also be entrained into the jet.  The jets can
propagate very large distances, e.g.  more than 100\,kpc in the case
of Cygnus~A, thus forming the largest physically connected structures
in the universe (after cluster galaxies). At distances of less than a
parsec, high-density ($10^{8}\,{\rm cm^{-3}}$) gas clouds orbit the
AGN and form the broad line region, or BLR.  These clouds have speeds
of several thousand kilometres per second and hence cause the
linewidths observed by Seyfert in 1943. Further out, slower gas clouds
constitute the narrow line region, or NLR, with lower densities
($10^3\,{\rm cm^{-3}}$) and speeds of only a few hundred kilometres per
second. Surrounding the AGN and its constituents in the polar plane is
a toroidal agglomeration of material that can deeply hide the AGN and
its activity.\\

After two decades of hassle with AGN phenotypes it became clear in the
1990s that there are basically two separate kinds of AGN
classification: each object either is radio loud or radio quiet, and
either belongs to the type 1 or 2 AGN.  The former classification is
based on the ratio of radio luminosity to optical luminosity, $R$,
with the dividing line being at around $R=1$
(\citealt{Kellermann1989}). The latter classification is based on whether
the optical emission lines are broad (type 1) or narrow(type 2).

The cause of the bimodality in radio luminosity is still unknown, but
several suggestions have been made, involving either ``intrinsic''
differences in the central engine or ``extrinsic'' differences in the
surrounding medium. Intrinsic differences that have been suggested
include (1) systematically lower black hole masses
(\citealt{Laor2000}), (2) lower black hole spins
(\citealt{Wilson1995}), (3) a ``magnetic switch'' that was identified
by \cite{Meier1997} during numerical modelling of jets, (4) the
production of buoyant plasmons that bubble up through the density
gradient of the NLR instead of a collimated relativistic jet
(\citealt{Pedlar1985}, \citealt{Whittle1986}, \citealt{Taylor1989}),
(5) a large thermal plasma fraction in the jet
(\citealt{Bicknell1998}), or (6) radiative inefficiency
(\citealt{Falcke1995}). Extrinsic differences generally invoke the
rapid deceleration of initially relativistic jets by collisions in a
dense surrounding BLR or interstellar medium
(e.g. \citealt{Norman1984}). Unfortunately, there are still only very
few observational constraints, especially because the radio-weak
objects are difficult to observe, and with so many possible causes,
the question remains open why they have such low absolute
luminosities.

Unlike the difference between radio-loud and radio-quiet objects, the
separation into type 1 and 2 AGN is understood as being due to an
orientation effect. It depends on whether one can look into the
central regions and see the innermost few tenths of a parsec, where
the BLR clouds are, or whether the circumnuclear material shadows the
BLR, in which case only narrow emission lines from the NLR are
observed. Although each individual object has its peculiarities,
virtually all AGN belong to one of the radio loud/quiet and type 1/2
classes.\\

\begin{figure*}[htpb!]
\centering
\includegraphics[width=0.4\linewidth,angle=270]{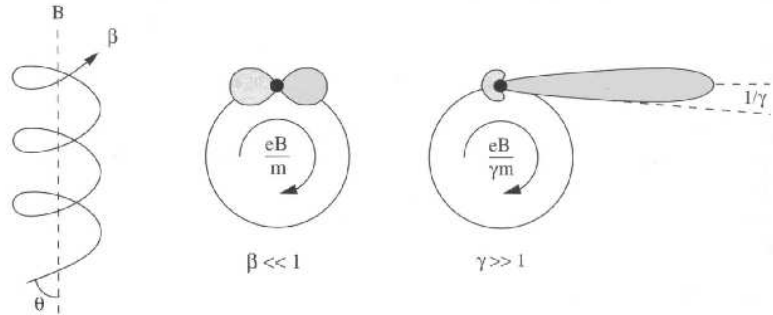}
\caption[Sketch of an electron moving in a magnetic field]{Sketch of an
electron moving in a magnetic field (reproduced from \citealt{Peacock1999}).}
\label{fig:synchrotron}
\end{figure*}

Before going into the details of classification, a brief description
is needed of three relativistic effects which are important in the
understanding of AGN: relativistic beaming, Doppler boosting and
apparent superluminal motion. The former two terms describe a
directional anisotropy of synchrotron emission arising from charges
moving at large fractions, $\beta$, of the speed of light, $c$,
towards the observer. If an object is moving at an angle $\theta$
towards the observer with speed $\beta c$ so that the Lorentz factor,
$\gamma$, with

\begin{equation}
\gamma=\frac{1}{\sqrt{1-\beta^2}}
\end{equation}
is $>>1$, and if the object emits an isotropic flux density
$S_0(\nu)$, then the radiation is confined to a cone with half opening
angle $1/\gamma$ (Fig.~\ref{fig:synchrotron}) and the observer
measures

\begin{equation}
S(\nu)=S(\nu/\delta)\delta^3 = S_0(\nu)\delta^{3-\alpha},
\end{equation}
where $\alpha$ is the source spectral index defined here as
\begin{equation}
\alpha=\frac{{\rm log}(S_{\nu_1}/S_{\nu_2})}{{\rm log}(\nu_1/\nu_2)}
\end{equation}
and $\delta$ is the Doppler factor

\begin{equation}
\delta=\gamma^{-1}(1-\beta\,{\rm cos}\,\theta)^{-1}.
\end{equation}

Thus, the radiation is not only confined to a smaller cone (beaming)
but the observer also measures an increased flux density (Doppler
boosting) because the source is moving nearly as fast as its own
radiation. Both effects increase the flux density radiated in the
forward direction and decrease it in the opposite direction by the
same amount, so that even in relatively modest relativistic sources
($\gamma\approx4$), the jet to counter-jet ratio of flux densities can
reach $10^6$.

In the same geometric source configuration (small angle between the
line of sight and the jet, relativistic speed of jet material),
components can be seen moving at speeds exceeding the speed of
light. This is a purely geometric effect: because the component is
travelling almost as fast as the radiation it emits, the time interval
between the emission of two photons appears shortened to us observers,
and the apparent transverse speed can therefore exceed $c$.

\subsection{Seyfert Galaxies}

Seyfert galaxies belong to the class of radio-quiet AGN. Unlike their
powerful siblings, they rarely show clearly defined, linear radio jet
structures on pc scales. They are mostly spirals and, although
supermassive black holes have been found in some of them (e.g.,
$3.6\times10^7\,{\rm M_{\odot}}$ in NGC\,4258,
\citealt{Miyoshi1995}), their luminosities are only tiny fractions of
that observed in quasars and radio galaxies. The sub-division into
type 1 and type 2 Seyferts indicates the viewing angle: when the angle
between the accretion disc normal and the line of sight is
intermediate or small (type 1), one has a direct view onto the BLR,
and broad emission lines are visible.  If the angle is large,
circumnuclear material blocks the direct view onto the central engine
and the BLR, and only NLR clouds are seen in the optical. Compelling
evidence for this picture comes from galaxies like NGC\,1068 which are
classified as type 2 Seyferts, but show typical type 1 spectra in
polarized light (\citealt{Antonucci1985, Antonucci1994b}).  In these
cases, the type 1 spectrum is hidden by a foreground absorber, but
reflections outside the absorber direct the light into our sight.

\subsection{Radio Galaxies}

Radio galaxies mostly exhibit single-sided, and only rarely two-sided
jets on pc scales, but they almost always show two-sided radio
emission on kpc scales, mostly in the form of bubble-like, irregularly
shaped ``radio lobes''. Fifteen percent to 20\,\% of radio galaxies
are radio-loud AGN (e.g., \citealt{Kellermann1989}). On sub-pc to tens
of pc scales, material is transported outwards in collimated jets
which exhibit bright knots. On scales of kpc to hundreds of kpc, the
jet flow becomes unstable, and either gently fades or abruptly stops
in hot spots. Here, huge radio lobes evolve, and the jet material is
slowly flowing back into the host galaxy. Because the material in the
lobes is no longer relativistic, its emission is isotropic and the
visibility of radio lobes is (to zeroth order) independent of the
inclination angle. The single-sidedness on pc scales is due to
relativistic effects, and only in a few sources where the jets have
angles close to $90^{\circ}$ with the line of sight, a double-sided
structure on pc scales is seen. Another interpretation of the
double-sidedness is that jets are always intrinsically single-sided,
and that double-sidedness occurs when the direction of the jet rapidly
flips from one side of the accretion disc to the other (e.g.,
\citealt{Rudnick1984},
\citealt{Feretti1993}).  Similar to Seyfert galaxies, radio galaxies
are divided into type 1 and type 2 based on optical appearance and as
a result of the same geometric configuration. However, the terms
Narrow Line Radio Galaxies (NLRG) and Broad Line Radio Galaxies (BLRG)
are also established. A further sub-classification was established for
radio galaxies based on their kpc-scale appearance by
\cite{Fanaroff1974}.  Analysing a sample of 57 radio galaxies and
quasars from the 3CR catalogue, they discovered that the relative
positions of regions of high and low surface brightness in the lobes
of extragalactic radio sources are correlated with their radio
luminosity. Fanaroff and Riley divided the sample into two classes
using the ratio $R_{\rm FR}$ of the distance between the regions of
highest surface brightness on opposite sides of the central galaxy or
quasar, to the total extent of the source up to the lowest brightness
contour in the map. Sources with $R_{\rm FR}<0.5$ were placed in class
I and sources with $R_{\rm FR}>0.5$ in class II. It was found that
nearly all sources with luminosity

\begin{equation}
L(178\,{\rm MHz}) \le 2\times 10^{25}\,h_{100}^{-2}\,{\rm W\,Hz^{-1}\,sr^{-1}}
\end{equation}
were of class I while the brighter sources were nearly all of class
II. The boundary between them is not very sharp, and there is some
overlap in the luminosities of sources classified as FR-I or FR-II on
the basis of their structures. The physical cause of the FR-I/II
dichotomy probably lies in the type of flow in the jets. FR-I jets are
thought to be subsonic, possibly due to mass entrainment, which makes
them amenable to distortions in the interaction with the ambient
medium, while the jets in FR-II sources are expected to be highly
supersonic, allowing them to travel large distances.

\subsection{Quasars, BL~Lacs, OVVs}

The objects in this section all belong to the radio-loud class. If the
angle between the jet axis and the line of sight is small, one can see
the BLR and the accretion disc directly. Relativistic effects are now
dominating the phenotypical properties of the AGN. The jet emission is
focused into a narrow cone, and rapid (intra-day) variability might
occur. Quasars reveal single-sided pc-scale jets and apparent
superluminal motion of knots that travel down the jets.  BL~Lac
objects are highly beamed, they show strong variability from radio to
optical wavelengths and they have almost no optical emission lines,
neither broad nor narrow. This happens when, at very small inclination
angles, the thermal emission from the AGN is superimposed with the
optical synchrotron emission from the jet base. At larger inclination
angles, the optical synchrotron emission is mostly beamed away from
the observer, revealing the underlying thermal emission, and emission
lines start to show up. This effect does not occur in the radio regime
due to the absence of radio emission lines. The lack of optical
emission lines makes a distance determination of BL~Lacs
difficult. The Optically Violently Variables, or OVVs, are a subclass
of the BL~Lacs, showing broad emission lines and rapid optical
variability.

\subsection{Compact Symmetric Objects (CSO)}

CSOs are double-sided, sub-kpc scale, symmetric radio sources whose
spectra frequently have a peak in the GHz regime. Based on size and
proper motion measurements, they are commonly regarded as being young
($<10^4\,{\rm yr}$) radio galaxies whose jets have not yet drilled
their way through the host galaxy's interstellar medium.

\section{The Gas Around AGN}

The circumnuclear material in AGN is thought to settle in a probably
rotationally symmetric body around the black hole. This gas is
commonly referred to as the circumnuclear torus, although the toroidal
shape has been established in only a few cases. Width, height and
radius of this ``torus'' are not necessarily well constrained and are
probably very different from object to object. In any case, however,
the torus is expected to shield the central engine from the observer's
view if the inclination angle is right. Observational evidence for
tori comes from radio observations of total intensity and H\,{\small
I} and molecular absorption and molecular emission, from ${\rm H_2O}$
maser observations, from the shape of ionized [O\,{\small III}] and
${\rm H\,\alpha}$+[N\,{\small II}] regions, from X-ray observations and
from the detection of thermal emission. A description of the physical
properties of the circumnuclear gas can be found in, e.g.,
\cite{Krolik1989}.

\subsection{Radio Absorption Measurements}

In those few cases where double-sided pc-scale radio jets are
observed, a rather narrow ``gap'' in the emission across the source is
frequently detected towards lower frequencies (e.g., in NGC\,1052,
\citealt{Vermeulen2003}, and in NGC\,4261,
\citealt{Jones2001}). These gaps are mostly due to free-free
absorption, as identified by its characteristic frequency-dependence
of the optical depth:

\begin{equation}
\tau_{\nu}\propto \nu^{-2.1}.
\end{equation}

In these cases, the UV radiation from the AGN ionizes the inner parts
of a circumnuclear absorber, and the ionized medium then gives rise to
free-free absorption.  This situation is seen in, e.g., NGC\,1052
(\citealt{Kameno2001, Vermeulen2003}), NGC\,4261 (\citealt{Jones2000,
Jones2001}) and Centaurus~A (\citealt{Tingay2001}). A reliable
detection of free-free absorption requires flux density measurements
at three frequencies at least. If, however, such measurements exist at
only two frequencies, the spectral index $\alpha$ can be used to
exclude synchrotron self-absorption (SSA). In SSA, the synchrotron
radiation produced by the relativistic electrons is absorbed in the
source, and if the distribution of the electron energies follows a
power law, the spectral index cannot exceed +2.5. Thus, whenever
$\alpha>2.5$ is observed at cm wavelengths, the absorption process is
most likely free-free absorption (in the cases presented here, the
Razin-Tsytovich effect has no effect, see Chapter~\ref{ch:sample}).\\

{\sc H\,i} absorption has been found on sub-pc scales in, e.g.,
NGC\,1052 (\cite{Vermeulen2003}), Cygnus~A (\citealt{Conway1995}) and
in NGC\,4261 (\citealt{Langevelde2000}). In the latter, the {\sc H\,i}
absorption was found no closer than 2.5\,pc away from the core,
supporting the idea that the material is ionized at smaller distances
to the AGN. This was proposed by \cite{Gallimore1999}, who detected
{\sc H\,i} absorption in a number of Seyfert galaxies almost exclusively
towards off-core radio components.\\

A variety of molecular lines is also detected in AGN: NGC\,1052 shows
OH in absorption and emission (\cite{Vermeulen2003}), in NGC\,4261,
\cite{Jaffe1994} detected CO in absorption and \cite{Fuente2000} found
${\rm CO^+}$ absorption in the core of Cygnus~A. This list is by no
means complete and many other line measurements exist.\\

\subsection{Masers}

The most compelling evidence for the partly molecular nature of at
least parts of the circumnuclear material comes from the detection of
${\rm H_2O}$ ``megamasers'' in the vicinity of AGN. A few prominent
examples are NGC\,3079 (\citealt{Henkel1984}, and, more recently,
\citealt{Kondratko2003}), NGC\,2639 (\citealt{Wilson1995b}) and
NGC\,4258 (\citealt{Miyoshi1995}). In the last case, the velocity
distribution of the masers yielded a model-independent measure for the
enclosed mass with high accuracy. In NGC\,1068, both the alignment and
the velocity gradient of the masers found by \cite{Gallimore1996} are
oriented perpendicular to the radio jet axis, hence suggesting the
presence of a circumnuclear, relatively dense region of material. The
physical properties of ${\rm H_2O}$ in AGN have been described by
\cite{Neufeld1994}.

\subsection{Ionization Cones}

In the last 15 years, optical observations of AGN showed cone-shaped
regions of line emission (\citealt{Pogge1988}), frequently in two
opposite directions (\citealt{Storchi-Bergmann1992, Wilson1993}) and
sometimes aligned with linear radio structures
(\citealt{Falcke1998}). The shape of the line emitting regions was
immediately interpreted as being due to shadowing by circumnuclear
material, i.e., the nuclear UV emission can only escape towards the
poles of the torus, where it ionizes the gas.

\subsection{X-ray Observations}

Evidence for circumnuclear absorbers also comes from X-ray
observations. \cite{Marshall1993} showed that the X-ray spectrum of
NGC\,1068 is best modelled as a continuum source seen through Compton
scattering. In Cygnus~A, \cite{Ueno1994} found evidence for an
absorbed power-law spectrum. Seyfert 2 galaxies are mostly heavily
absorbed in the X-ray regime, with column densities of the order of
$>10^{24}\,{\rm cm^{-2}}$ (\citealt{Krolik1988}).

\subsection{Thermal Emission}

In NGC\,1068, \cite{Gallimore1997} have discovered a region of
flat-spectrum radio continuum emission with brightness temperatures
too low to be due to self-absorbed synchrotron emission. The shape and
orientation of the region is suggestive of a circumnuclear disc or
torus, and they conclude that the emission is either due to unseen
self-absorbed synchrotron emission that is reflected by a torus into
the line of sight or free-free emission from the torus itself.

\section{Feeding Gas Into the AGN}

\subsection{Magnetic Fields}
\label{sec:MagneticFields}

A lot of material is seen in the vicinity of AGN, but how is the
material moved closer to and into the black hole and into the jets?
\cite{Shlosman1989} suggested that a stellar bar in a galaxy sweeps gas
into the central few hundred pc. The gas forms a disc which also
develops a bar potential and funnels the gas further in to scales of a
hundred parsecs. Processes that transport the gas further in to parsec
scales are largely unknown, and no theory exists that could be tested
by observations. To move gas into smaller radius orbits requires a
mechanism to shed angular momentum, and the candidates are viscosity
and magnetic fields. Viscosity, responsible for angular momentum
transport in the innermost regions of AGN, would form the gas into a
disc, which is not generally observed. Only in few objects have such
discs with diameters of $~\sim100\,{\rm pc}$ been found (e.g., in
NGC\,4261, \citealt{Jaffe1996, Ferrarese1996} and in Mrk\,231,
\citealt{Kloeckner2003}), but the role that the discs in these objects
play in gas transport is unclear. 

Magnetic fields, on the other hand, play an essential role in models
of tori, accretion discs and jet formation in the central parsec and
sub-parsec scale regions (e.g., \citealt{Koide2000, Meier2001,
Krolik1988}). The basic idea is that the accretion disc is threaded
with magnetic fields, and that differential rotation of the disc
twists the fields to a spiral structure, which applies a breaking
torque on the inner disc and accelerates material in the outer disc,
thus transporting angular momentum outwards.

The origin of the required magnetic fields is not clear. They are
probably frozen into the accreted material on scales of $>10\,{\rm
pc}$, although parts of the field may be lost in reconnection.  Dynamo
processes in the disc can also produce magnetic fields (e.g.,
\citealt{Rekowski2003}) which participate in the generation of an
outflow, but the details are not well known. Estimates of magnetic
field strengths in AGN and their surroundings mainly come from
equipartition arguments, assuming that the energy in particles equals
that in magnetic fields.  Whether this assumption is justified or not
is not known, and so equipartition arguments are unsatisfactory.

Magnetic fields are routinely observed on the largest scales in
galaxies and are modelled on the smallest, but are largely unknown on
intermediate scales of a few parsecs to several tens of parsecs. As
they are probably involved in the transport of gas into the AGN, their
strength and orientation is of considerable interest.

\subsection{Faraday Rotation and Free-Free Absorption}

One way to measure magnetic field strengths is by means of Faraday
rotation. When an electromagnetic wave travels through an ionized
medium that is interspersed with a magnetic field with a component
parallel to the direction in which the wave is travelling, then the
plane of polarization of the wave is rotated. The amount of Faraday
rotation cannot be measured directly because the intrinsic position
angle of the polarization is not known and the rotation has
ambiguities of $\pi$. Instead, one exploits the frequency dependence
of the Faraday rotation, and the change of the effect with frequency
yields the constant of proportionality, the rotation measure $RM$.\\

The physics behind this effect is the birefringence of the magnetized
plasma. A linearly polarized beam of radiation with electric vector
position angle $\theta$ can be considered as the superposition of two
circularly polarized waves with equal amplitudes but opposite senses
of rotation. The circular polarizations have different indices of
refraction in the plasma which causes one polarization to be retarded
with respect to the other, and the plane of linear polarization,
composed of the two circular polarizations, rotates. 

Faraday rotation is of particular importance in this thesis so we give
a brief derivation of the formula here, following \citealt{Kraus1973},
p. 737f.

In a magnetized plasma, where the magnetic field is parallel to the
direction of propagation of the waves, the phase constants $\beta^+$
and $\beta^-$ of the two waves are given by

\begin{equation}
\beta^\pm=\omega\sqrt{\mu_0 (\epsilon_{11} \pm \epsilon_{12})},
\label{eq:beta}
\end{equation}
where $\omega$ is the angular frequency of the wave, $\mu_0$ is the
vacuum permeability, and $\epsilon_{11}$ and $\epsilon_{11}$ are
elements of the permittivity tensor $\overline{\epsilon}$, with

\begin{equation}
\begin{split}
\epsilon_{11}&=\left(1+\frac{\omega_0^2}{\omega_{\rm g}^2-\omega^2}\right)\epsilon_0\\
\epsilon_{12}&=\frac{-i\omega_0^2\omega_{\rm g}\epsilon_0}{\omega(\omega_{\rm g}^2-\omega^2)}\\
\end{split}
\end{equation}

Here, $\omega_0=e\sqrt{n_{\rm e}/\epsilon_0 m}$ denotes the angular
plasma frequency ($e$ is the particle charge in C, $n_{\rm e}$ the
particle density in ${\rm m^{-3}}$ and $m$ the particle mass in kg),
$\omega_{\rm g}=(e/m)B$ the angular gyrofrequency and $\epsilon_0$ the
vacuum permittivity. Travelling a distance $dl$ through the plasma
changes $\theta$ by

\begin{equation}
d\theta=\frac{\beta^- - \beta^+}{2}dl.
\end{equation}

Inserting the expressions for $\omega_0$ and $\omega_{\rm g}$ into
Eq.~\ref{eq:beta} and integrating over the line of sight then yields

\begin{equation}
\begin{split}
\theta&=\lambda^2\times\frac{e^3}{8\pi^2c^3\epsilon_0 m^2}\int_0^L n_{\rm e}\,B_{\parallel}\,dl\\
      &=\lambda^2\times 2.63 \times 10^{-13}\int_0^L n_{\rm e}\,B_{\parallel}\,dl\\
      &=\lambda^2\times RM,
\end{split}
\label{eq:rm}
\end{equation}
where $\lambda$ is the wavelength in m, $e$ is the elementary charge
in C, $c$ is the speed of light in ${\rm m\,s^{-1}}$, $\epsilon_0$ is
the vacuum permittivity in ${\rm F\,m^{-1}}$, $m$ is the electron mass
in kg, $n_{\rm e}$ is the electron density in ${\rm m^{-3}}$,
$B_{\parallel}$ is the line-of-sight component of the magnetic flux
density in T, and $l$ is the path length in m.

Observations of Faraday rotation yield a measure of the integral of
the product $n_{\rm e}\,B_{\parallel}$. To separate the magnetic field from
the electron density requires an independent measurement of the
electron density and the path length through the ionized gas.\\

Ionized gases also produce free-free absorption which causes an
exponential decrease of intensity with path length through the
absorber. The spectral energy distribution of synchrotron radiation
which is absorbed by free-free absorption, is given by

\begin{equation}
S_{\nu}=S_0\,\nu^{\alpha_0}\times \exp(-\tau_\nu^{\rm ff}),
\end{equation}
where

\begin{equation}
\tau_\nu^{\rm ff}=8.24\times10^{-2}\,T^{-1.35}\,\nu^{-2.1}\,\int{N_+N_-ds}
\label{eq:ffa}
\end{equation}
(e.g., \citealt{Osterbrock1989}, eq. 4.32). Here, $S_{\nu}$ is the
observed flux density in mJy, $S_0$ is the intrinsic flux density
(before the radiation passes the absorber) in mJy, $\nu$ is the
observing frequency in GHz, $\alpha_0$ is the dimensionless intrinsic
spectral index, $T$ is the gas temperature in K, $N_+$ and $N_-$ are
the number densities of positive and negative charges, respectively,
in ${\rm cm^{-3}}$, and $s$ is the path length in pc.

Faraday rotation depends linearly on the electron density, $n_{\rm
e}$, whereas the optical depth in free-free absorption goes as $n_{\rm
e}^2$ (assuming $N_+=N_-=n_{\rm e}$). An analysis of the
frequency-dependence of $\tau_\nu^{\rm ff}$, together with a diameter
measurement of the absorber from VLBI images, yields $n_{\rm e}$.
Putting $n_{\rm e}$ and the diameter into Eq.~\ref{eq:rm} then allows
one to solve for the magnetic field strength, $B_{\parallel}$.

\subsection{Jet Collimation}

As was mentioned in the section \S\ref{sec:MagneticFields}, models
exist for the innermost regions of AGN where the jets are launched and
collimated. Some possible mechanisms for the acceleration of jets are
the ``magnetic slingshot'' model (\citealt{Blandford1982}), through
the extraction of Poynting flux from the black hole, the so-called
Blandford-Znajek mechanism (\citealt{Blandford1977}), or by radiation
or thermal pressure (\citealt{O'Dell1981, Livio1999}).

Unfortunately, although today's radio interferometers routinely yield
angular resolutions of 0.1\,mas, these regions are still not resolved,
except for a few nearby objects with high black hole masses, like M87
(\citealt{Junor1999}). Jet collimation is expected to happen on scales
of 10 to 1000 Schwarzschild radii ($R_{\rm s}$), and in a moderately
distant radio galaxy with redshift $z=0.1$ and a typical black hole
with $10^8\,M_{\odot}$, these scales are still factors of 20 to 2000
smaller than the synthesized beam. Hence, only little observational
evidence exists so far to constrain jet formation models. In a few
nearby AGN, however, the highest resolution VLBI observations at
86\,GHz can in principle resolve scales of $<1000\,R_{\rm s}$. These
observations are challenging because most nearby sources are weak at
high frequencies and the antenna sensitivities decrease, and most
attempts to detect the targets have failed.

\section{The Aim of This Thesis}

Nearby AGN provide unique opportunities to study the circumnuclear
environments of supermassive black holes. VLBI observations have long
been concentrated on the brightest, but unfortunately more distant
objects, yielding rather low linear resolutions of more than one
parsec. In this thesis, I present VLBI observations primarily of
nearby objects, yielding some of the highest linear resolution images
ever made. The goal was to investigate the distribution of gas and
magnetic fields in those objects, and to probe the jet collimation
region of a nearby radio galaxy with highest linear resolution. This
required the development of fast frequency switching for phase
calibration of 86\,GHz VLBI observations of weak sources.

\subsection{Polarimetric Observations of Six Nearby AGN}

Six nearby AGN were selected because they show good evidence for
circumnuclear free-free absorbers that shadow the radiation of the
AGN. If such absorbers are interspersed with magnetic fields and the
radio emission from the AGN is polarized, Faraday rotation is expected
to occur. A joint analysis of the free-free absorption and Faraday
rotation then allows one to determine the magnetic field strength in
the absorber and so yields a measurement of a quantity which is
difficult to determine by other means. I present the results of pilot
VLBI observations that were made to look for polarized emission before
making time-consuming Faraday rotation measurements. 

\subsection{Case Study of NGC\,3079}

In addition to the polarimetric observations of NGC\,3079, I present
an analysis of multi-epoch, multi-frequency observations of this
nearby Seyfert 2 galaxy to investigate the nature and origin of the
radio emission. Furthermore, I present a statistical analysis of VLBI
observations of Seyfert galaxies reported in the literature to compare
NGC\,3079 to other Seyferts and to compare Seyferts to radio-loud
objects to investigate the difference between powerful and weak AGN.

\subsection{Phase Calibration Strategies at 86\,GHz}

We have explored the feasibility of a new phase calibration strategy
for VLBI observations, in which one cycles between a lower reference
frequency, at which the source is strong enough for self calibration,
and the target frequency.  The phase solutions from the reference
frequency are scaled by the frequency ratio and interpolated onto the
target frequency scans to remove the atmospheric phase fluctuations.
The result is a phase-referenced image at the target frequency, and
indefinitely long coherent integrations can be made on sources that
are too weak for self-calibration. The primary use of the technique is
to image nearby, weak AGN at 86\,GHz to obtain highest linear
resolutions. Using data from a pilot project, we have improved the
observing and data calibration strategy to a ready-for-use level, and
we have obtained the first detection of NGC\,4261 which could not be
imaged previously at 86\,GHz because it is too weak.

\chapter{Special VLBI Techniques}

Very Long Baseline Interferometry, or VLBI, has become a standard
observing technique in the radio astronomy community during the last
20 years. Although frequently referred to as ``experiments'', VLBI
observations have left the experimental stage, and especially
observations with the U.S. 10-element Very Long Baseline Array, or
VLBA, are relatively easy to prepare and analyse, and now the first
automated data calibration methods exist (\citealt{Sjouwerman2003}).
A good description of VLBI principles and the data calibration steps
required has recently been given by \cite{Klare2003}, and details can
be found in, e.g., \cite{Thompson1986}. I therefore will not go into
the details.

There are, however, non-standard VLBI observations that require
special observing procedures and/or data calibration steps. Two of
these techniques, phase-referencing and polarimetry, have been used to
gather data for this thesis and therefore are described in detail.
Furthermore, a short section is dedicated to a description of
ionospheric phase noise in phase referencing and how it can be
calibrated.

\section{Phase Referencing}

In VLBI, the true visibility phase is altered by errors from numerous
sources. They can be divided into geometric, instrumental and
atmospheric errors. Geometric errors arise from inaccurate antenna
positions, motion of antennas with tectonic plate motion, tidal
effects, ocean loading and space curvature due to the mass of the sun
and the planets close to the targeted position.  Instrumental errors
include clock drifts, changes in antenna geometry due to gravity
forces, cable length changes, and electronic phase errors due to
temperature variations. Most of these errors are sufficiently well
known or are measured continuously and are accounted for in the
correlator model of the array.  Atmospheric and ionospheric phase
noise, however, is difficult to predict and is therefore the largest
source of error in VLBI observations.  The following description of
the effect of atmospheric and ionospheric phase noise on VLBI
observations closely follows \cite{Beasley1995}.

\subsection{Tropospheric Phase Noise}

At cm wavelengths, the largest source of error in VLBI observations
comes from fluctuations of the tropospheric water vapour content along
the line of sight of the telescopes. Changes in the water vapour
content cause phase changes of the observed visibilities and thus
limit the atmospheric coherence time. This is the time over which data
can be coherently averaged, and is taken to be the average time it
takes for the phase to undergo a change by one radian ($\sim57^\circ$).

The tropospheric excess delay, i.e., the additional time it takes for
the waves to travel through the atmosphere compared to vacuum, can be
divided into two components, the dry troposphere and the wet
troposphere. This means that the mixture of air and water vapour would
have the same effect on the visibility phases as a layer of dry air
and a layer of water vapour, both of thicknesses equivalent to the
tropospheric content of air and water vapour, respectively. The zenith
excess delay due to the dry troposphere, $l_{\rm z}$, can be modelled
quite accurately using measurements of pressure and antenna latitude
and altitude (\citealt{Davis1985}):

\begin{equation}
l_{\rm z}=\frac{0.228\,{\rm m/mbar}\,P_0}{1-0.00266\cos(2\lambda)-0.00028\,{\rm km^{-1}}h}.
\end{equation}

Here, $l_{\rm z}$ is the zenith excess path length in cm, $P_0$ is the
the total pressure at the surface in mb and $h$ is the altitude of the
antenna above the geoid in km. With this equation, the dry
troposphere, contributing around 2.3\,m, can be modelled to an
accuracy of $\approx0.5\,{\rm mm}$. The equation can be extended to
include the tropospheric water vapour, yielding the Saastamoinen model
(\citealt{Ronnang1989})

\begin{equation}
l_{\rm z}=\frac{0.228\,{\rm m/mbar}\left[P_0+P_{\rm w}\left(\frac{1255}{T}+0.05\right)\right]}
               {1-0.00266\cos(2\lambda)-0.00028\,{\rm km^{-1}}h}.
\end{equation}

Here, $P_{\rm w}$ is partial pressure of water vapour and $T$ is the
temperature in K. However, the effects of the wet troposphere are much
more difficult to describe because the water vapour is not well mixed
with the dry air, and turbulence makes the delay due to the wet
troposphere highly variable. In general, tropospheric water vapour
contributes up to 0.3\,m of excess path delay, but to determine the
exact value requires precise measurements of the water vapour in front
of each telescope. Water vapour radiometers therefore have become
increasingly popular, especially at radio telescopes operating at mm
wavelength. The new Effelsberg Water Vapour Radiometer
(\citealt{Roy2003}) is now able to measure the delay to an accuracy of
0.12\,mm ($1\,\sigma$), corresponding to, e.g., $2.2^{\circ}$ at
15.4\,GHz. Further improvements will shortly increase the accuracy to
about 0.04\,mm and hence allow one to use the water vapour radiometer
to calibrate VLBI data at frequencies of up to 86\,GHz.

The path length through the atmosphere scales as $\sin^{-1}(\beta$),
where $\beta$ is the antenna elevation. Thus, the amount of
troposphere along the line of sight has doubled at an elevation of
$30^{\circ}$, and has increased to five times its zenith value at
$12^{\circ}$ elevation. As a consequence, the phase noise dramatically
increases towards low antenna elevations. At frequencies above
$\sim5\,{\rm GHz}$, tropospheric phase noise is the dominant source
of error in VLBI observations.\\

\subsection{Ionospheric Phase Noise}

The ionosphere is a region of free electrons and protons at altitudes
of 60\,km to 10000\,km above the earth's surface. It adds phase and
group delays to the waves and causes Faraday rotation of linearly
polarized waves. The excess zenith path in m is
(\citealt{Thompson1986}, eq.~13.128)

\begin{equation}
l_0\approx-\frac{40.3\,{\rm Hz^2\,m^3}}{\nu^2}{\rm TEC},
\end{equation}
where $\nu$ is the frequency in Hz and TEC is the vertical total
electron content in ${\rm m^{-2}}$. The TEC changes on various
timescales. Long-term variations are caused predominantly by changing
solar radiation (solar cycle, seasonal and diurnal variations), but
short-term variations are mostly caused by atmospheric gravity waves
through the upper atmosphere, oscillations of air caused by buoyancy
and gravity. The biggest effect on radio observations are medium-scale
travelling ionospheric disturbances (MSTIDs) with horizontal speeds of
$100\,{\rm m\,s^{-1}}$ to $300\,{\rm m\,s^{-1}}$, periods of 10\,min
to 60\,min and wavelengths of several hundred km.  The ionosphere
introduces a delay that scales as $\nu^{-2}$, whilst the phase scales
as $\nu^{-1}$, and this contribution is the dominant source of phase
error in VLBI observations at frequencies below $\sim5\,{\rm GHz}$.
Ionospheric Faraday rotation is negligible at cm wavelengths, with a
maximum of 15 turns of the electric vector position angle at 100\,MHz
(\citealt{Evans1968}) and hence at most 9 turns, 3 turns and 1 turn of
phase at 1.7\,GHz, 5.0\,GHz and 15.4\,GHz, respectively.

\subsection{Phase Referencing}

In standard VLBI observations at cm wavelengths, the so-called phase
self-calibration is used to solve for phase errors that cannot be
accounted for by the correlator model and mostly are tropospheric.
Starting with a point source model in the field centre, and refining
the model iteratively, correction phases are derived that make the
visibility phases compliant with the model. This procedure is known as
self-calibration, or hybrid mapping (\citealt{Cornwell1981}). But in
two cases, this procedure does either not work or is not desirable. In
the first case, the target source is weak and cannot be detected
reliably within the atmospheric coherence time. As an illustration, in
a typical 15\,GHz VLBA experiment with ten stations in moderate
weather, the coherence time is 60\,s and the $7\,\sigma$ detection
limit follows to 21\,mJy, and weaker sources cannot be detected (a
detailed calculation of sensitivity limits is given in
Chapter~\ref{ch:ffs}).  In the second case, one is interested in
absolute astrometry of the target source. A priori phase
self-calibration destroys that information because the visibility
phases are initially adjusted to fit a point source at a position
provided by the observer.  Literally, one can ``move around'' the
source in the field of view and one is always able to properly adjust
the visibility phases. Phase-referencing solves both of these
problems.\\

In brief, phase referencing uses interleaved, short observations
(``scans'') of a nearby calibrator to measure the tropospheric phase
noise which is subtracted from the target source visibilities. It
works only if the interval between two calibrator scans is shorter
than the atmospheric coherence time and the calibrator structure is
well known. The first demonstration of this technique was published by
\cite{Alef1988} (a similar approach had already been proven to work by
\citealt{Marcaide1984}, but in their case, the calibrator and target
were in the telescope primary beams, and no source switching was needed).\\

Consider an observing run in which each target source scan of several
minutes is sandwiched between short scans on a nearby calibrator. Let
us further assume that all a priori calibration information has been
applied, i.e., the amplitudes are calibrated and the bandpass shape
and instrumental delay and phase offsets are corrected for. The
measured visibility phases can then be described with

\begin{equation}
\begin{split}
\phi_{\rm cal}(t_1)&=
\phi_{\rm c}(t_1)+
\phi_{\rm ins}^{\rm c}(t_1)+
\phi_{\rm pos}^{\rm c}(t_1)+
\phi_{\rm ant}^{\rm c}(t_1)+
\phi_{\rm atm}^{\rm c}(t_1)+
\phi_{\rm ion}^{\rm c}(t_1)\\
\phi_{\rm tar}(t_2)&=
\phi_{\rm t}(t_2)+
\phi_{\rm ins}^{\rm t}(t_2)+
\phi_{\rm pos}^{\rm t}(t_2)+
\phi_{\rm ant}^{\rm t}(t_2)+
\phi_{\rm atm}^{\rm t}(t_2)+
\phi_{\rm ion}^{\rm t}(t_2)\\
\phi_{\rm cal}(t_3)&=
\phi_{\rm c}(t_3)+
\phi_{\rm ins}^{\rm c}(t_3)+
\phi_{\rm pos}^{\rm c}(t_3)+
\phi_{\rm ant}^{\rm c}(t_3)+
\phi_{\rm atm}^{\rm c}(t_3)+
\phi_{\rm ion}^{\rm c}(t_3).\\
\end{split}
\label{eq:vis1}
\end{equation}

$\phi_{\rm c}$ and $\phi_{\rm t}$ are the true visibility phases on
the calibrator and the target source, $\phi_{\rm ins}$ is the residual
instrumental phase error due to clock drifts and other electronics,
$\phi_{\rm pos}$ and $\phi_{\rm ant}$ are geometric errors arising
from source and antenna position errors, and $\phi_{\rm atm}$ and
$\phi_{\rm ion}$ are tropospheric and ionospheric phase noise
contributions. Using self-calibration, the measured visibility phase
is decomposed into the source structure phase and the difference in
antenna-based phase errors at $t_1$ and $t_3$, and interpolation
yields the calibrator visibilities at $t_2$:

\begin{equation}
\tilde{\phi}_{\rm cal}(t_2)=
\tilde{\phi}_{\rm c}(t_2)+
\tilde{\phi}_{\rm ins}^{\rm c}(t_2)+
\tilde{\phi}_{\rm pos}^{\rm c}(t_2)+
\tilde{\phi}_{\rm ant}^{\rm c}(t_2)+
\tilde{\phi}_{\rm atm}^{\rm c}(t_2)+
\tilde{\phi}_{\rm ion}^{\rm c}(t_2)\\
\label{eq:vis2}
\end{equation}
(a tilde denotes interpolated values). Subtracting the interpolated
calibrator visibility phases from the observed target visibility
phases at time $t_2$ gives

\begin{equation}
\begin{split}
\phi_{\rm tar}-\tilde{\phi}_{\rm cal}&=
(\phi_{\rm t}-\tilde{\phi_{\rm c}})+
(\phi_{\rm ins}^{\rm t}-\tilde{\phi}_{\rm ins}^{\rm c})+
(\phi_{\rm pos}^{\rm t}-\tilde{\phi}_{\rm pos}^{\rm c})\\
&+(\phi_{\rm ant}^{\rm t}-\tilde{\phi}_{\rm ant}^{\rm c})+
(\phi_{\rm atm}^{\rm t}-\tilde{\phi}_{\rm atm}^{\rm c})+
(\phi_{\rm ion}^{\rm t}-\tilde{\phi}_{\rm ion}^{\rm c}).\\
\label{eq:vis3}
\end{split}
\end{equation}

In this equation, most terms are zero. Residual instrumental errors
vary slower than the lag between the calibrator and target source
scan, and therefore $\phi_{\rm ins}^{\rm t}-\tilde{\phi}_{\rm
ins}^{\rm c}=0$. Atmospheric and ionospheric noise will also be the
same in adjacent scans if the separation between calibrator and target
is less than a few degrees and the lines of sight pass through the
same isoplanatic patch, the region across which tropospheric and
ionospheric contributions are constant. This yields $\phi_{\rm
atm}^{\rm t}-\tilde{\phi}_{\rm atm}^{\rm c}=0$ and $\phi_{\rm
ion}^{\rm t}-\tilde{\phi}_{\rm ion}^{\rm c}=0$. Antenna position
errors have the same effect on both the calibrator and the target if
their separation on the sky is small, and hence $\phi_{\rm ant}^{\rm
t}-\tilde{\phi}_{\rm ant}^{\rm c}=0$. The calibrator visibility phases
change slowly, and hence $\tilde{\phi}_{\rm cal}=\phi_{\rm
c}$.  Also, when using either a compact calibrator or a good
calibrator model in phase self-calibration, the phase errors derived
from phase self-calibration are those of a point source, and hence
$\tilde{\phi_{\rm c}}=0$. This yields

\begin{equation}
\phi_{\rm tar}-\tilde{\phi}_{\rm cal}=
\phi_{\rm t}+
(\phi_{\rm pos}^{\rm t}-\tilde{\phi}_{\rm pos}^{\rm c})+
\phi_{\rm int},
\label{eq:vis4}
\end{equation}
in which $\phi_{\rm int}$ denotes interpolation errors. This equation
expresses that after interpolation, the difference between the
calibrator and the target source phases is the target source
structural phases plus the position error. Hence, phase-referencing
not only allows one to calibrate the visibility phases, but also to
precisely measure the target source position with respect to the
calibrator position.\\

In general, the ionospheric phase component can be calibrated in
self-calibration, together with the tropospheric component. But
especially at low elevations and frequencies below 5\,GHz, the
assumption that $\phi_{\rm ion}^{\rm t}-\tilde{\phi}_{\rm ion}^{\rm
  c}=0$ is no longer valid, and a correction has to be applied. One
approach to correct for the ionospheric delays is to use TEC models
derived from GPS data. They are provided by several working groups,
yielding global TEC maps every two hours, and giving the TEC on a grid
with $2.5^{\circ}$ spacings in latitude and $5^{\circ}$ spacings in
longitude. From this grid, the TEC at each antenna can be
interpolated. The error in these maps, however, can be quite high, up
to 20\,\% when the TEC is as high as a few tens of TEC units
(1\,TECU=$10^{16}\,{\rm e^-\,m^{-2}}$), and up to 50\,\% or higher
when the TEC is of the order of a few TEC units. Details on how the
TEC is derived from GPS data are discussed by \cite{Ros2000}, and a
set of tests is described in
\cite{Walker2000}.\\

\section{Polarimetry}

Measuring the linear or circular polarization of radio waves is a
relatively young technique in VLBI. The first polarization-sensitive
VLBI observation of an AGN jet was published by \cite{Cotton1984}
(unfortunately, their 3C\,454.3 image was rotated by $180^{\circ}$
because they had mistaken the phase signs). But VLBI polarimetry has
become increasingly popular, leading to extensive surveys in the last
few years (e.g., \citealt{Zavala2003, Pollack2003}). This section
describes technical details and calibration of polarization-sensitive
VLBI observations.

\subsection{Stokes Parameters}

\begin{figure}[htpb!]
\centering
\includegraphics[width=0.7\linewidth,angle=0]{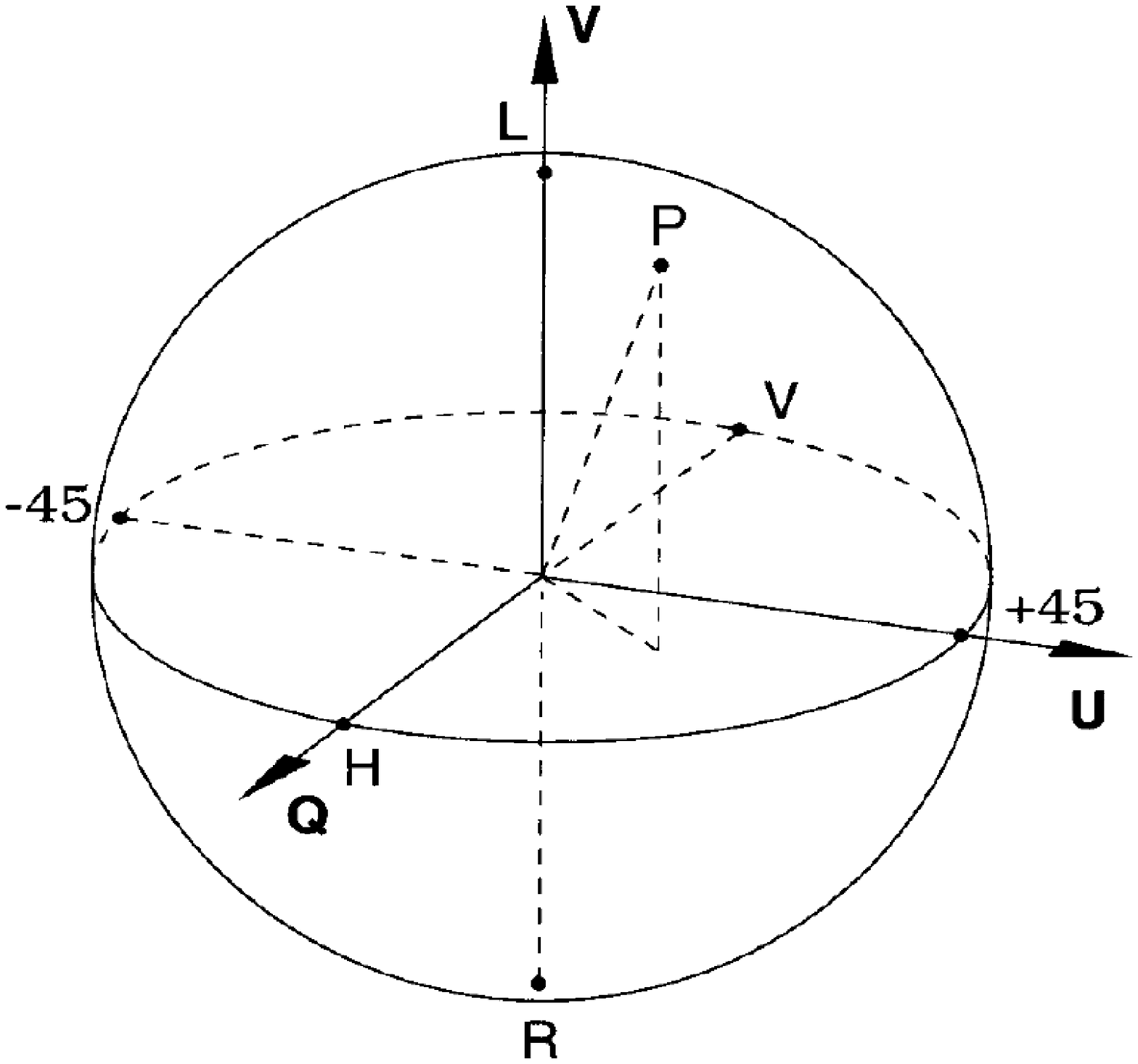}
\caption[The Poincar\'e sphere]{The Poincar\'e sphere, reproduced from
\cite{Scott2001}.}
\label{fig:poincare}
\end{figure}

The polarization of any source can be described by means of the Stokes
parameters $I$, $Q$, $U$ and $V$. $I$ denotes the total intensity of
the source, $Q$ and $U$ the fractions of linear polarization
perpendicular to the propagation of the wave and in directions that
enclose angles of $45^{\circ}$, and $V$ denotes the fraction of
circular polarization. A convenient way to look at this is the
Poincar\'e sphere (Fig.~\ref{fig:poincare}). Consider a Cartesian
coordinate system in which $Q$ is on the $x$ axis, $U$ is on the $y$
axis and $V$ is on the $z$ axis. The radius of the sphere is the
degree of polarization,

\begin{equation}
I_{\rm p}=\sqrt{Q^2+U^2+V^2},
\end{equation}
and the points on the sphere represent the different states of
polarization. The intersection points of the $Q$, $U$ and $V$ axes
with the sphere represent the following states of polarization,
respectively: linear, horizontal ($+Q$), linear, vertical ($-Q$),
linear, $+45^{\circ}$ ($+U$), linear $-45^{\circ}$ ($-U$), circular,
left-handed ($+V$) and circular, right-handed ($-V$). In general, the
polarization is a combination of all three parameters, and the
polarization is elliptical.\\

\subsection{Interferometer Response to a Polarized Signal}

Determination of an extended source's polarization characteristics
requires measuring the Stokes parameters over the source region. An
interferometer with coordinates $(u,v)$ in the plane perpendicular to
the line of sight measures the Fourier transform of the sky brightness
distribution, given in angular coordinates $(\zeta, \eta)$. Similar to
the Fourier transform of the total intensity brightness distribution,
$I$, the interferometer response to the sky distribution of the other
Stokes parameters, $Q$, $U$ and $V$ can be defined:

\begin{equation}
\begin{split}
\mathscr{I}(u,v)&=\int\int I(\zeta,\eta)e^ {-j2\pi(\zeta u+\eta v)}d\zeta d\eta\\
\mathscr{Q}(u,v)&=\int\int Q(\zeta,\eta)e^ {-j2\pi(\zeta u+\eta v)}d\zeta d\eta\\
\mathscr{U}(u,v)&=\int\int U(\zeta,\eta)e^ {-j2\pi(\zeta u+\eta v)}d\zeta d\eta\\
\mathscr{V}(u,v)&=\int\int V(\zeta,\eta)e^ {-j2\pi(\zeta u+\eta v)}d\zeta d\eta.\\
\end{split}
\end{equation}

Starting with the interferometer response to a point source, I now
describe how the quantities $I$, $Q$, $U$ and $V$ are restored from
the visibility measurements. I then describe the calibration of
polarization-sensitive interferometer data. The polarization
calibration procedure used in this thesis was developed by
\cite{Leppanen1995}. I follow their notation to explain the
interferometer response and the correlation to the Stokes parameters;
the full derivation can be found in \cite{Leppanen1995b}.

\subsubsection{Response to a Point Source}

The voltages $A$ present at an antenna output can be described by

\begin{equation}
\begin{split}
A^R &= g^R(E^R e^{-j\alpha}+E^L D^R e^{ j\alpha})\\
A^L &= g^L(E^L e^{ j\alpha}+E^R D^L e^{-j\alpha}).\\
\label{eq:antenna-output}
\end{split}
\end{equation}

Here, the $g$'s are complex quantities proportional to the antenna
gains, $E$'s are the RCP and LCP parts of the electric field,
respectively, $D$'s are the complex leakage of power from one
polarization into the other, and $e^{\pm j\alpha}$ describe the
rotation of the wave with respect to the feed horns due to the
parallactic angle of the source, $\alpha$. These terms are commonly
referred to as $D$-terms. Because virtually all sources exhibit
fractional polarizations that are of the same order as the $D$-terms on
the scales probed with VLBI, the significance of the detection of
polarized emission depends critically on the $D$-terms being properly
calibrated. All quantities, except for $D$, are functions of time.

The cross-correlation function, $\rho$, between two antennas is given
by

\begin{equation}
\rho^{pq}_{mn}(\tau)=\langle A^p_m(t) A^q_n(t-\tau)\rangle.
\label{eq:cross-correlation}
\end{equation}

This equation expresses an average of the multiplication of the
antenna voltages, $A$, from two antennas, $m$ and $n$, with a certain
delay, $\tau$. The indices $p$ and $q$ represent either LCP or RCP,
yielding four possible combinations for each baseline. The functions
$A$ are real, and hence their cross-correlation is equal to their
convolution. One gets the cross-power spectrum, $r(\omega)$, by
Fourier-transforming Eq.~\ref{eq:cross-correlation}. Applying the
convolution theorem then yields

\begin{equation}
r^{pq}_{mn}(\omega)=\mathscr{A}^p_m(\omega)\mathscr{A}^{q*}_n(\omega),
\label{eq:cross-power}
\end{equation}
where the $\mathscr{A}(\omega)$ are the Fourier transforms of the terms in
Eq.~\ref{eq:antenna-output}. Eq.~\ref{eq:cross-power} is valid for any
antenna pair because we still assume a point source in the field
centre, and the interferometer amplitude therefore is independent of
baseline length and orientation.\\

When Eq.~\ref{eq:antenna-output} is inserted into
Eq.~\ref{eq:cross-correlation} before Fourier transformation,
Eq.~\ref{eq:cross-power} contains the terms $E^R E^{R*}$, $E^L E^{L*}$
$E^R E^{L*}$ $E^L E^{R*}$, which we denote by $RR$, $LL$, $RL$ and
$LR$, and their Fourier transforms by $\mathscr{RR}$, $\mathscr{LL}$,
$\mathscr{RL}$ and $\mathscr{LR}$.  The four cross-power spectra then
take on the form

\begin{equation}
\begin{split}
r^{RR}_{mn}=g^R_m g^{R*}_n
                    &[ e^{-j(\alpha_m-\alpha_n)}\times \mathscr{RR} +
                       D^{R }_m e^{ j(\alpha_m+\alpha_n)} \times \mathscr{LR}\\
                    &+ D^{R*}_n e^{-j(\alpha_m+\alpha_n)} \times \mathscr{RL}
                     + D^{R }_m D^{R*}_n e^{ j(\alpha_m-\alpha_n)} \times \mathscr{LL}]\\
r^{LL}_{mn}=g^L_m g^{L*}_n
                    &[ e^{ j(\alpha_m-\alpha_n)}\times \mathscr{LL} +
                       D^L _m e^{-j(\alpha_m+\alpha_n)} \times \mathscr{RL}\\
                    &+ D^{L*}_n e^{ j(\alpha_m+\alpha_n)} \times \mathscr{LR}
                     + D^L _m D^{L*}_n e^{-j(\alpha_m-\alpha_n)} \times \mathscr{RR}]\\
r^{RL}_{mn}=g^R_mg^{L*}_n
                    &[ e^{-j(\alpha_m+\alpha_n)}\times \mathscr{RL} +
                       D^R _m e^{ j(\alpha_m-\alpha_n)} \times \mathscr{LL}\\
                    &+ D^{L*}_n e^{-j(\alpha_m-\alpha_n)} \times \mathscr{RR}
                     + D^R _m D^{L*}_n e^{ j(\alpha_m+\alpha_n)} \times \mathscr{LR}]\\
r^{LR}_{mn}=g^L_mg^{R*}_n
                    &[ e^{ j(\alpha_m+\alpha_n)}\times \mathscr{LR} +
                       D^L   _m e^{-j(\alpha_m-\alpha_n)} \times \mathscr{RR}\\
                    &+ D^{R*}_n e^{ j(\alpha_m-\alpha_n)} \times \mathscr{LL}
                     + D^L   _m D^{R*}_n e^{-j(\alpha_m+\alpha_n)} \times \mathscr{RL}].\\
\end{split}
\label{eq:leakage-term-model}
\end{equation}

This particular representation of cross-power spectra, electric
vectors and leakage factors is called the ``leakage-term model''
(\citealt{Cotton1993}). The equations become a lot simpler considering
that sources observed with VLBI are only weakly polarized (thus $\mathscr{LR}$
and $\mathscr{RL}$ terms are small) and that the $D$-terms are small (thus $DD$
terms are even smaller). Then, in Eq.~\ref{eq:leakage-term-model}, all
terms involving products of $\mathscr{LR}$ or $\mathscr{RL}$ and $D$-terms, as well as those
involving $DD$ terms, can be neglected, yielding

\begin{equation}
\begin{split}
r^{RR}_{mn}&=g^R_m g^{R*}_n e^{-j(\alpha_m-\alpha_n)}\times \mathscr{RR}\\
r^{LL}_{mn}&=g^L_m g^{L*}_n e^{ j(\alpha_m-\alpha_n)}\times \mathscr{LL}\\
r^{RL}_{mn}&=g^R_m g^{L*}_n
                      [ e^{-j(\alpha_m+\alpha_n)}\times \mathscr{RL} +
                       D^R _m e^{ j(\alpha_m-\alpha_n)} \times \mathscr{LL}\\
                     &+ D^{L*}_n e^{-j(\alpha_m-\alpha_n)} \times \mathscr{RR}]\\
r^{LR}_{mn}&=g^L_m g^{R*}_n
                      [ e^{ j(\alpha_m+\alpha_n)}\times \mathscr{LR} +
                       D^L _m e^{-j(\alpha_m-\alpha_n)} \times \mathscr{RR}\\
                     &+ D^{R*}_n e^{ j(\alpha_m-\alpha_n)} \times \mathscr{LL}].\\
\end{split}
\label{eq:leakage-term-model_2}
\end{equation}

\subsubsection{Response to an Extended Source}

For an extended source, the cross-power spectrum is the superposition
of incoherent contributions from all parts of the source, and
integration over the source region is required:

\begin{equation}
r^{pq}(u,v,\omega)=\int\int I^{pq}(\zeta,\eta,\omega)
                                e^ {-j2\pi(\zeta u+\eta v)}d\zeta d\eta.
\end{equation}

This equation shows that the cross-power spectra of extended sources
observed by an interferometer are the Fourier transforms of the
brightness distribution in sky coordinates:

\begin{equation}
r^{pq}(u,v,\omega)=\mathscr{F} \{ I^{pq}(\zeta,\eta,\omega) \}.
\end{equation}

It is straightforward to enhance Eq.~\ref{eq:leakage-term-model_2} for
extended sources, yielding

\begin{equation}
\begin{split}
r^{RR}_{mn}(u,v, \omega)&=g^R_mg^{R*}_n e^{-j(\alpha_m-\alpha_n)}\times \mathscr{\mathscr{RR}}\\
r^{LL}_{mn}(u,v, \omega)&=g^L_mg^{L*}_n e^{ j(\alpha_m-\alpha_n)}\times \mathscr{LL}\\
r^{RL}_{mn}(u,v, \omega)&=g^R_mg^{L*}_n
                          [ e^{-j(\alpha_m+\alpha_n)}\times \mathscr{RL} +
                            D^R _m e^{ j(\alpha_m-\alpha_n)} \times \mathscr{LL}\\
                          &+ D^{L*}_n e^{-j(\alpha_m-\alpha_n)} \times \mathscr{\mathscr{RR}}]\\
r^{LR}_{mn}(u,v, \omega)&=g^L_mg^{R*}_n
                          [ e^{ j(\alpha_m+\alpha_n)}\times \mathscr{LR} +
                            D^L _m e^{-j(\alpha_m-\alpha_n)} \times \mathscr{\mathscr{RR}}\\
                          &+ D^{R*}_n e^{ j(\alpha_m-\alpha_n)} \times \mathscr{LL}].\\
\end{split}
\label{eq:leakage-term-model_3}
\end{equation}

We now need to establish a relation between the cross-power spectra in
Eq.~\ref{eq:leakage-term-model_3} and the Stokes parameters. The
Stokes parameters and the cross-correlation of the electric field
components are connected through

\begin{equation}
\begin{split}
E^RE^{R*} &= RR = I+V\\
E^LE^{L*} &= LL = I-V\\
E^RE^{L*} &= RL = Q+jU\\
E^LE^{R*} &= LR = Q-jU\\
\end{split}
\end{equation}
(\citealt{Thompson1986}), from which the Stokes parameters can be
separated into

\begin{equation}
\begin{split}
I & =  \frac{1}{2} (RR + LL)\\
Q & =  \frac{1}{2} (RL + LR)\\
U & = j\frac{1}{2} (LR - RL)\\
V & =  \frac{1}{2} (RR - LL).\\
\end{split}
\end{equation}

Applying the inverse Fourier transform to
Eq.~\ref{eq:leakage-term-model_3} then yields the desired sky
distribution of the Stokes parameters.

\subsection{Calibration of Instrumental Effects}

Given the usually low degrees of polarization in extragalactic
sources, measuring the four Stokes parameters requires an accurate
calibration of the array. There are six sources of error in VLBI
polarization observations: 1) $D$-term calibration errors; 2) errors
in the relative phase between RCP and LCP (R-L phase offset); 3)
thermal noise; 4) gain calibration errors; 5) deconvolution errors and
6) closure errors. 4), 5) and 6) are usually small, because the
dynamic ranges in polarization images are low, and therefore are
neglected, and 3) cannot be calibrated, but integration time is
planned to make this small. This leaves 1) and 2) as the dominant
sources of error. 2) is easily calibrated (in the case of the VLBA)
using the station monitoring data, and 1) requires significant effort.

\subsubsection{The R-L Phase Offset}
\label{sec:pulsecals}

\begin{figure}[htpb!]
\centering
 \subfigure[Parallel-hand spectrum of 3C345 at 5.0\,GHz without pulse
 calibration applied, showing a 1\,min-average of the visibility phase
 (deg, upper panels) and amplitude (Jy, lower panels) in 14 out of 16
 observed frequency channels. The two highest frequency channels in
 each IF have been flagged due to the bandpass
 limitations. Instrumental delays dominate the phase slopes across the
 pass band and the phase offsets between the IFs.]{
 \includegraphics[width=0.6\linewidth,angle=270]{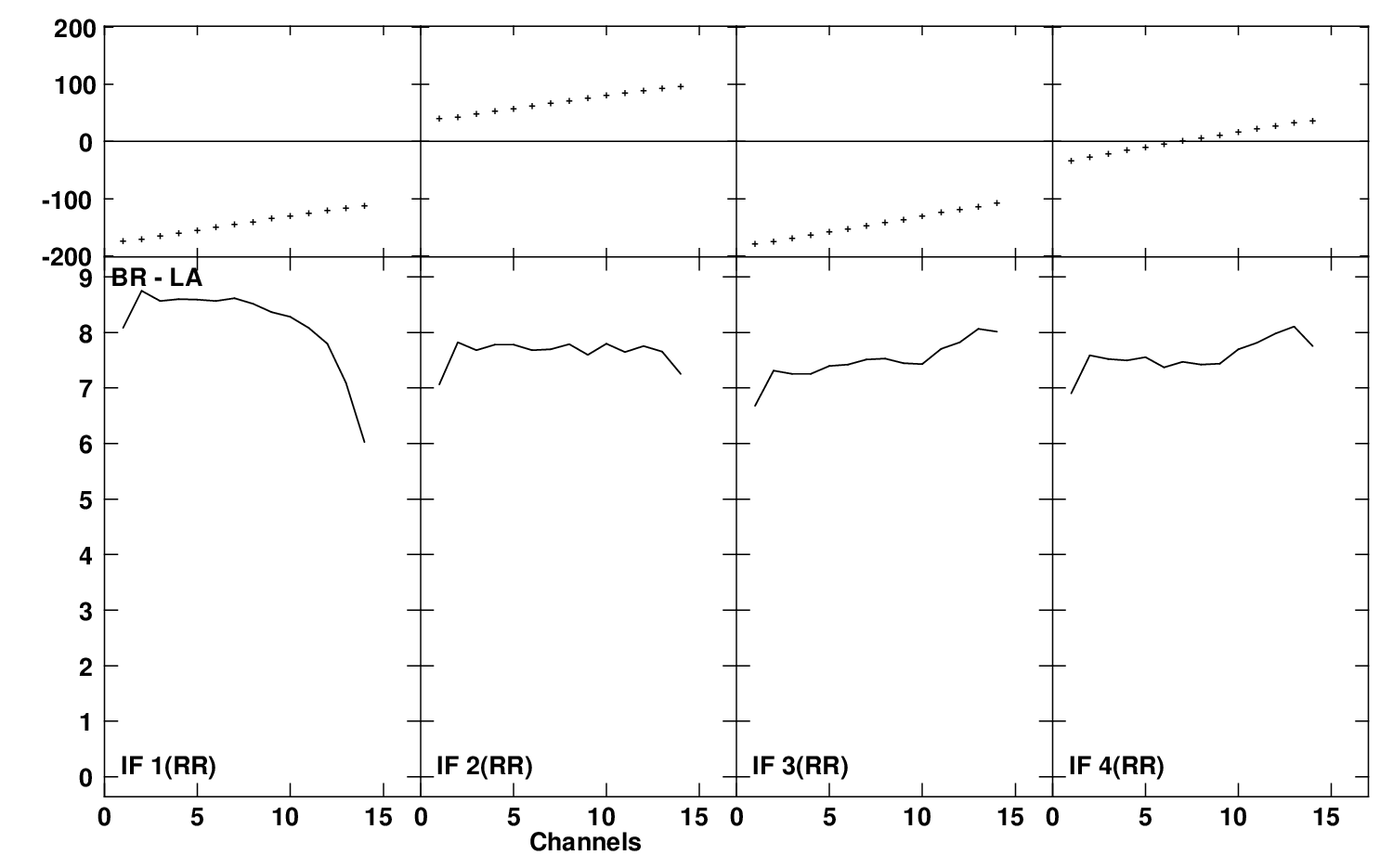}
 \label{fig:pccor1}}\\ \subfigure[The same spectrum as above with
 phase corrections derived from the pulse calibration system. The
 phase slopes have been removed and the phases across the band have
 been aligned, allowing coherent integration over the observing
 bandwidth.]{
 \includegraphics[width=0.6\linewidth,angle=270]{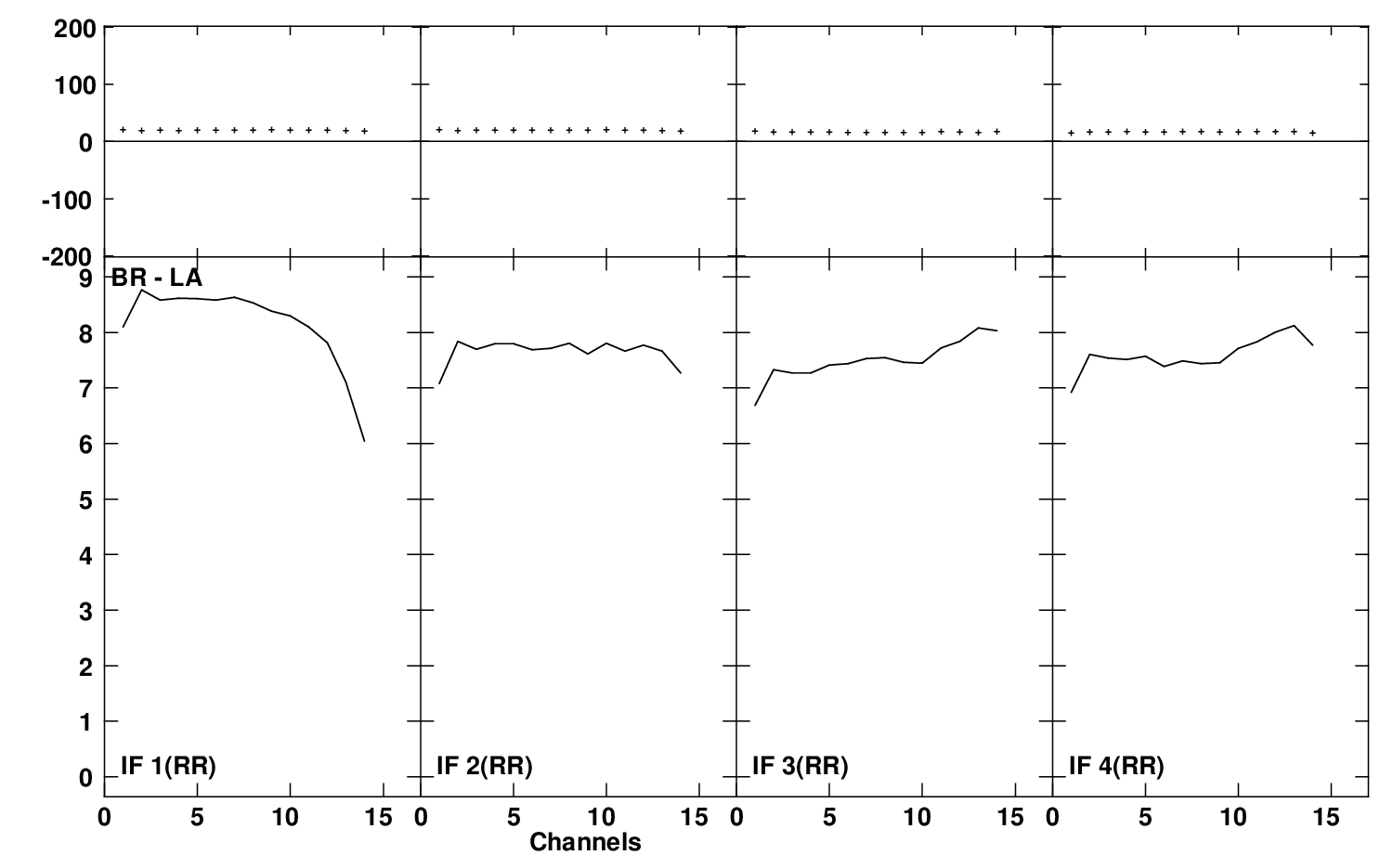}
 \label{fig:pccor2}}
\caption[Illustration of the effect of the pulse calibration
system]{Illustration of the effect of the pulse calibration system}
\label{fig:pccor}
\end{figure}

The relative phase between the two circular polarization receiving
channels affects the apparent position angle of the linear
polarization on the sky, or electric vector position angle (EVPA).
Also, the relative phases of adjacent baseband (or IF) channels need
to be lined up to allow averaging over the observing bandwidth. At the
VLBA, pulses are injected at the start of the signal path with a
period of $10^{-6}\,{\rm s}$. In the frequency domain, the pulses
produce a comb of lines spaced by 1\,MHz, bearing a fixed, known phase
relationship to each other. They pass through the receiver and
downconversion chain along with the radio astronomical signal and the
phases of the pulses are measured in the back-end before the signal is
digitized and written on tape. This allows one to derive the
time-dependent, instrumental phase changes over the band (the
instrumental delay) and the offsets between the IFs.  Basically, the
pulse calibration system moves the delay reference point from the
samplers to the pulse calibration injection point at the feeds,
reducing instrumental phase changes almost to zero. This also holds
for the phase offsets between the two parallel hands of circular
polarization and between the cross hands.  \cite{Leppanen1995b} has
shown that, after application of the pulse calibration, the
residual R-L phase errors are of the order of $1^{\circ}$.\\

In practice, applying the pulse calibration is straightforward. A
pulse calibration, or PC table is generated at the correlator and is
attached to the $(u,v)$ data. The AIPS task PCCOR is then used to
generate a calibration table with phase corrections. PCCOR takes two
phase measurements per IF, at the upper and lower edge of the band,
and computes the delay. To resolve phase ambiguities, one needs to
specify a short scan on a strong source from which the delay is
measured using a Fourier transform with a finer channel spacing than
that used by the pulse calibration system. Once the ambiguity is
resolved and assuming that the instrumental phase changes are small
throughout the observation (which is virtually always true), PCCOR
computes phase corrections from the PC table. The data are then
prepared for averaging in frequency to increase the signal-to-noise
ratio, and if polarization measurements are desired, the EVPA is
calibrated.

\subsubsection{Calibration of $D$-terms}
\label{sec:d-term-cal}

In brief, the calibration of $D$-terms uses the effect that the
linearly polarized emission from the source rotates in a different way
with respect to the feed horns than does the leakage polarization as
the earth rotates and the antennas track the source. Let us multiply
each line in Eq.~\ref{eq:leakage-term-model_3} by a power of $e$ such
that those terms unaffected by the $D$-terms are unrotated and the
power of the $e$-factor is zero (e.g., $e^{j(\alpha_{\rm
    m}+\alpha_n)}$ for the third line). This yields

\begin{equation}
\begin{split}
r^{RR}_{mn}(u,v)&=g^R_mg^{R*}_n \times \mathscr{RR}\\
r^{LL}_{mn}(u,v)&=g^L_mg^{L*}_n \times \mathscr{LL}\\
r^{RL}_{mn}(u,v)&=g^R_mg^{L*}_n
                          [ \mathscr{RL} +
                            D^R _m e^{2j\alpha_m} \times \mathscr{LL}
                          + D^{L*}_n e^{2j\alpha_n} \times \mathscr{RR}]\\
r^{LR}_{mn}(u,v)&=g^L_mg^{R*}_n
                          [ \mathscr{LR} +
                            D^L _m e^{-2j\alpha_m} \times \mathscr{RR}
                          + D^{R*}_n e^{-2j\alpha_n} \times \mathscr{LL}],\\
\end{split}
\label{eq:leakage-term-model_4}
\end{equation}
in which only the leakage terms rotate twice as fast with the
parallactic angle of the source, and the visibility is independent of
parallactic angle. This situation can be understood as a vector of a
certain length and position angle, representing the linearly polarized
emission from the source, to which the vector of the polarization
leakage, rotating with the parallactic angle of the source, is added.
Separation of these vectors yields the $D$-terms.\\

Two direct consequences can be drawn from this picture: the smaller
the $D$-terms, the more difficult they are to determine because the
relative contribution of thermal noise to the polarization leakage
vector increases, and the larger the range of parallactic angles over
which a source has been observed is, the more accurate the $D$-term
calibration is possible.\\

Separation of the true linear polarization and the leakage
polarization is implemented in the AIPS task LPCAL, which works as
follows. One can think of the leakage polarization as the convolution
of a ``leakage beam'' with the total intensity structure. In such an
image, unphysical features would appear as linear polarization.
Unfortunately, unlike in the total intensity images, one cannot use a
positivity constraint for Stokes $Q$ and $U$ because they can have
either sign. However, polarized emission should usually appear only at
locations in the image where total intensity emission is seen, and
LPCAL uses this constraint (the ``support'') to derive the $D$-terms.
The support is provided to the algorithm in terms of a total intensity
model derived from self-calibration. In general, the polarized
intensity structure differs considerably from the total intensity
structure, and so the model must be divided into sub-models, small
enough that the polarization of each can be described by a single
complex number $p_{\rm c}=(Q_{\rm c}+jU_{\rm c})/I_{\rm c}$.\\

The visibilities of each of the sub-models, $i$, can now be written as
$p_{\rm i}\mathscr{I}_{\rm i}$. To derive the model visibilities from
Eq.~\ref{eq:leakage-term-model_4} requires two more steps: the gains
$g$ can be set to unity (because model component visibilities do not
require calibration) and the {\it observed} parallel-hand visibilities
$r^{RR}_{mn}(u,v)$ and $r^{LL}_{mn}(u,v)$ are used
instead of the {\it true} visibilities $\mathscr{RR}$ and
$\mathscr{LL}$, respectively. The last step requires that the
visibilities are sufficiently well calibrated in self-calibration. The
predicted cross-hand visibilities $\tilde{r}^{RL}_{mn}(u,v)$
and $\tilde{r}^{LR}_{mn}(u,v)$ can now be written as the sum
of the sub-model visibilities $p_{\rm i}\mathscr{I}_{\rm i}$ and the
leakage of the observed parallel-hand visibilities, rotated by the
parallactic angle:

\begin{equation}
\begin{split}
\tilde{r}^{RL}_{mn}(u,v)&=\sum_{\rm i} p_{\rm i}\mathscr{I}_{\rm i}
                                  +D^R _m e^{2j\alpha_m} r^{LL}_{mn}(u,v)
                                  +D^{L*}_n e^{2j\alpha_n} r^{RR}_{mn}(u,v)\\
\tilde{r}^{LR}_{mn}(u,v)&=\sum_{\rm i} p^*_{\rm i}\mathscr{I}_{\rm i}
                                  +D^L _m e^{-2j\alpha_m} r^{RR}_{mn}(u,v)
                                  +D^{R*}_n e^{-2j\alpha_n} r^{LL}_{mn}(u,v)\\
\end{split}
\label{eq:leakage-term-model_5}
\end{equation}

A least squares fit can now be used to solve for the $D$ and $p$
terms.\\

The results from least squares fits are not necessarily unique, and
parts of the true polarization of the source may be assigned to the
polarization leakage of the feeds. \cite{Leppanen1995b} shows that the
stability of LPCAL with respect to source structure changes is high,
the rms of the $D$-term solutions being less than 0.002.
\chapter{The Sample}
\label{ch:sample}

The aim of the observations presented here is to study the sub-pc
structure of jets in AGN, with particular focus on measuring the
magnetic field strength in the AGN's close vicinity. The tool we chose
is Faraday rotation, which occurs in magnetized, thermal
plasmas. Evidence for Faraday rotation, however, requires the
observation of the electric vector position angle, or EVPA, of
linearly polarized emission at at least three, but preferably more,
frequencies. Hence, the observations require a lot of telescope time,
and we decided to look for polarized emission in our candidate sources
at a single frequency before making Faraday rotation
measurements. \\

The following few considerations led to the selection criteria for the
observing instrument and the sources.

\begin{itemize}

\item To observe scales of 1\,pc or smaller in an AGN at a distance of
100\,Mpc requires a resolution of 2\,mas or better, or baselines of
$100\,{\rm M\lambda}$ or longer.  The VLBA provides an angular
resolution of less than 2\,mas at 8.4\,GHz and less than 1\,mas at
15.4\,GHz.  The nominal sensitivities after a 2-hour integration are
$0.10\,{\rm mJy\,beam^{-1}}$ at 8.4\,GHz and $0.18\,{\rm
mJy\,beam^{-1}}$ at 15.4\,GHz.  Any birefringent effects causing
depolarization due to wave propagation in ionized media, either in the
source, in our galaxy or in the earth's ionosphere, decrease as
$\nu^{-2}$, i.e., are reduced by a factor of 3.4 at 15.4\,GHz compared
to 8.4\,GHz.  From these arguments, we decided to use the VLBA at
15.4\,GHz as it provides an extra margin of resolution, sufficient
sensitivity within a reasonable integration time and little
depolarization.

\item Faraday rotation occurs in magnetized, thermal plasmas, but as our
project aimed at {\it measuring} the magnetic fields and only little
is known about AGN magnetic field strengths and structures, we
confined our sample compilation to those sources that have clear
evidence for a thermal plasma in front of the AGN core or jet. An
unambiguous signpost of thermal plasma is free-free absorption which
can be traced by radio spectra. It causes an exponential cutoff
towards low frequencies, and the observed spectral index, $\alpha$,
can be arbitrarily high. In contrast, synchrotron self-absorption can
produce a maximum spectral index of $\alpha=+2.5$ (e.g.,
\citealt{Rybicki1979}, chap. 6.8). Another effect that can cause
exponential cutoffs at low frequencies is the Razin-Tsytovich
effect. It becomes important at frequencies below

\begin{equation}
\nu_R=20\frac{n_{\rm e}}{B}
\end{equation}
(\citealt{Pacholczyk1970}, eq.~4.10), where $n_{\rm e}$ is the particle
density in ${\rm cm^{-3}}$ and $B$ is the magnetic field strength in
gauss. Typical values for circumnuclear absorbers are $n_{\rm
e}=10^4\,{\rm cm^{-3}}$ and $B=0.1\,{\rm G}$, so that $\nu_R=2\,{\rm
MHz}$. As the particle density is usually reasonably well defined by
observations, the magnetic fields would need to be three order of
magnitude less than currently expected to cause measurable effects in
the GHz regime. Thus, whenever a spectral index of $\alpha>+2.5$ is
observed in the central regions of AGN, it is almost inevitably caused
by free-free absorption.\\

This allows us to re-state the selection criteria in the following
way.

\item The sources must show clear evidence for a foreground free-free
absorber.

\item The sources need to be bright enough to be observed with the VLBA
at 15.4\,GHz within a few hours. Especially the absorbed parts, where
Faraday rotation is expected to occur, need to be strong enough so
that polarized emission of the order of 1\,\% can be observed.

\item The sources need to be closer than 200\,Mpc to achieve a linear
resolution of less than 1\,pc.

\end{itemize}

The sample listed in Table~\ref{tab:sample} is, we believe, a complete
list of objects selected from the literature that have good evidence
for a parsec-scale free-free-absorber seen against the core or jet. We
excluded 3C\,84 and OQ\,208 because they are unpolarized
(\citealt{Homan1999, Stanghellini1998}) and the Compact Symmetric
Object (CSO) 1946+708 because this class of objects is known to
exhibit very low polarization (\citealt{Pearson1988}), NGC 4258
because the jet is too weak (\citealt{Cecil2000}), and the Seyferts
Mrk\,231, Mrk\,348 and NGC\,5506 because their parsec-scale structures
do not show a clear continuous jet (\citealt{Ulvestad1999,
Middelberg2004}). For the remaining candidates there have been no VLBI
polarization observations yet made.\\

\begin{table*}[tbp]
\scriptsize
\begin{center}
\begin{tabular}{lcccc}
\hline
\hline
Source          & Dist. / Mpc   & Scale / ${\rm pc\,mas^{-1}}$ & $S$(15\,GHz) / mJy     & $n_{\rm e}$ / ${\rm cm^{-3}}$\\
\hline
NGC\,3079       & 15.0$^a$  & 0.07  & 50$^g$        & -                     \\
NGC\,1052       & 19.4$^b$  & 0.09  & 489$^h$       & $3\times10^4$ to $10^5$  \\
NGC\,4261       & 35.8$^c$  & 0.17  & 156$^i$       & $3\times10^4$         \\
Hydra A         & 216$^d$   & 1.05  & 127$^d$       & $8\times10^2$         \\
Centaurus A     & 4.2$^e$   & 0.02  & 1190$^j$      & $>3\times10^4$        \\
Cygnus A        & 224$^f$   & 1.09  & 600$^k$       & $10^4$                \\
\hline
\end{tabular}
  \caption[Summary of source parameters]{Summary of source
  parameters. References: $a$ - \cite{Irwin1991}, $b$ -
  \cite{Tonry2001}, $c$ - \cite{Nolthenius1993}, $d$ -
  \cite{Taylor1996}, $e$ - \cite{Tonry2001}, $f$ - \cite{Owen1997}, $g$
  - Krichbaum et al., priv. comm., $h$ - \cite{Kameno2001}, $i$ -
  extrapolated from \cite{Jones2001}, $j$ - interpolated from
  \cite{Tingay2001}, $k$ - \cite{Kellermann1998}.}
\label{tab:sample}
\end{center}
\end{table*}

\section{NGC\,3079} 

\begin{figure}[htpb!]
\centering
\includegraphics[width=0.7\linewidth, angle=270]{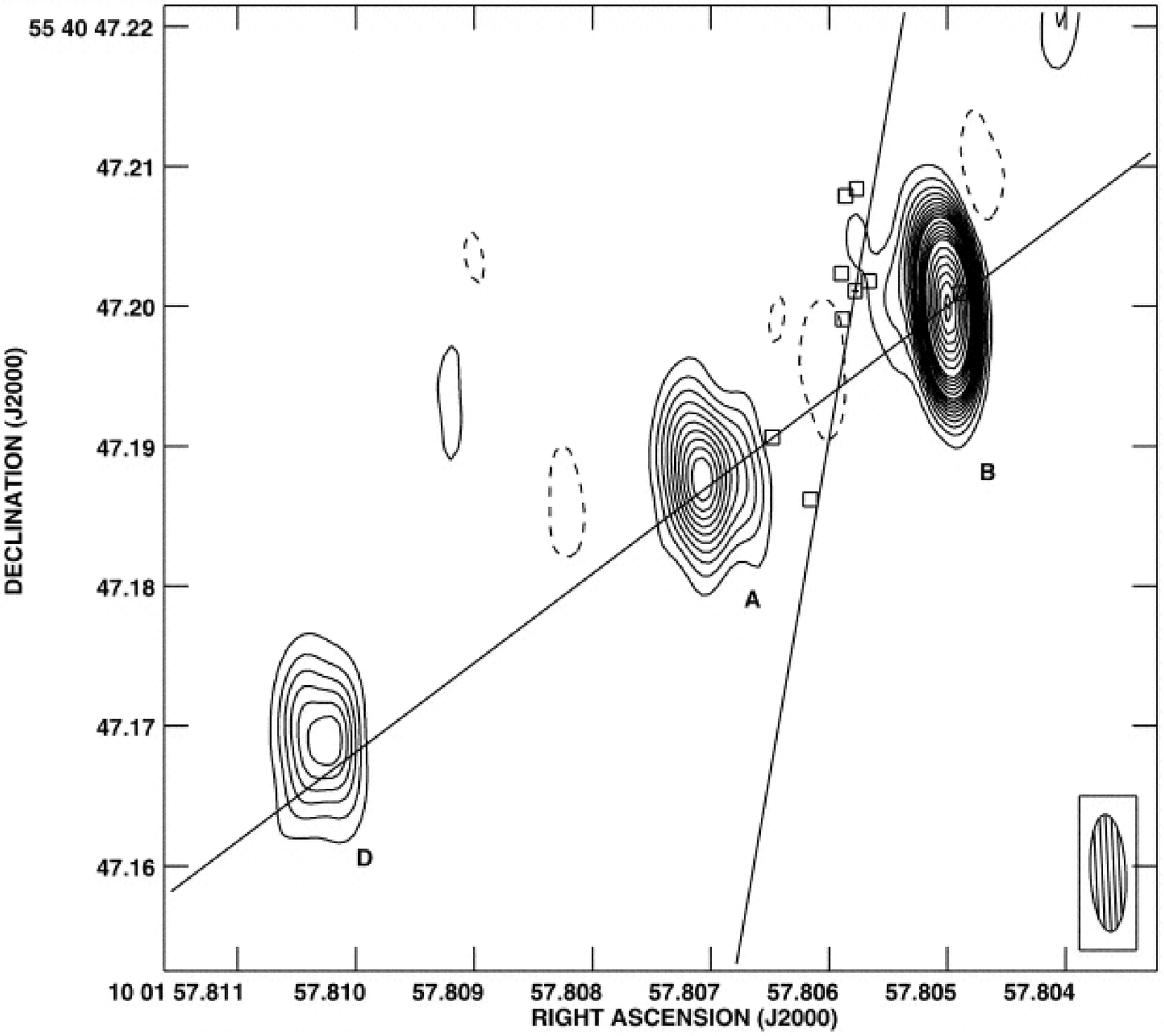}
\caption[VLBI image of NGC\,3079 by \cite{Trotter1998}]{VLBI image of
NGC\,3079 by \cite{Trotter1998}.}
\label{fig:3079_trotter1998}
\end{figure}

NGC\,3079 is a nearby (15.0\,Mpc, \citealt{Vaucouleurs1991}, ${\rm
H_0=75\,km\,s^{-1}\,Mpc^{-1}}$) LINER (\citealt{Heckman1980}) or
Seyfert~2 (\citealt{Sosa-Brito2001}) galaxy in an edge-on, dusty
spiral. It does not formally satisfy our selection criteria as it is
relatively weak ($S_{\nu}\lesssim{\rm 50\,mJy}$ at 15\,GHz) and the
location of the core is not known. We have observed NGC\,3079 as part
of another project and will analyse it considering the same aspects as
in the analysis of the radio galaxies, but a separate section is
dedicated to a detailed interpretation of the observational results of
NGC\,3079.

Radio: VLA B-array observations at 5\,GHz show a bright core and two
remarkable radio lobes 2\,kpc in length ejected from the nucleus
perpendicular to the disc of the galaxy (\citealt{Duric1983}). Pilot
VLBI observations by \cite{Irwin1988} resolved the core into two
strong components, $A$ and $B$, separated by 20.2\,mas. They also
found emission along a line between these components (component
$C$). These three components do not align with either the kpc-scale
lobes or the galaxy disc. The spectra of both $A$ and $B$ were found
by \cite{Trotter1998} to peak between 5\,GHz and 22\,GHz, with
spectral indices of $\alpha=1.8$ and $\alpha=1.6$ between 5\,GHz and
8.4\,GHz, respectively. Their 5\,GHz image is reproduced in
Fig.~\ref{fig:3079_trotter1998} for convenience.  Spectral-line and
continuum observations by \cite{Trotter1998} and
\cite{Sawada-Satoh2000} have resolved the ${\rm H_2O}$ maser emission
in NGC\,3079 into numerous aligned spots at an angle of $30^{\circ}$
to $35^{\circ}$ with respect to the jet axis.
\cite{Trotter1998} suggest that the location of the AGN is on the
$A$-$B$ axis, where it intersects with the line of masers, although
there is no radio emission from that location. They propose that this
point is the centre of a circumnuclear disc of approximately 2\,pc
diameter. \cite{Sawada-Satoh2000} identify component $B$ as the core
and propose that the masers are located in a circumnuclear torus.
Component $D$ was found by \cite{Trotter1998} further down the
$A$-$B$ axis but was not confirmed by
\cite{Sawada-Satoh2000}. $A$ and $B$ were found by
\cite{Sawada-Satoh2000} to separate at $0.16\,c$, while $B$ appeared
to be stationary with respect to the brightest maser clumps. VLA
H\,{\small I} line observations of NGC\,3079 and its neighbouring
galaxy NGC\,3073 by \cite{Irwin1987} revealed a tail of emission in
NGC\,3073 pointing away from NGC\,3079. This tail is remarkably
aligned with the axis of the radio lobes in NGC\,3079 seen by
\cite{Duric1983}, and was explained to be gas stripped away by the
outflow of NGC\,3079. \cite{Irwin1991} detected an H\,{\small I}
absorbing column density, $N_{\rm HI}$, of $10^{21}\,{\rm cm^{-2}}$ to
$10^{22}\,{\rm cm^{-2}}$ (assuming $T_{\rm spin}=100\,{\rm K}$) using
the VLA. \cite{Sawada-Satoh2000} resolved the H\,{\small I} absorption
spatially into two components coincident with what they identified as
continuum emission from $A$ and $B$ at 1.4\,GHz. From observations of
the CO (J=1-0) transition,
\cite{Koda2002} identify, among other features, a rotating disc of gas
600\,pc in diameter. Recent VLBI observations of the red- and
blueshifted ${\rm H_2O}$ maser emission by \cite{Kondratko2003}
indicate a black hole mass of $2\times10^6\,M_{\odot}$.

IR / Optical / UV: NGC\,3079 is host to one of the clearest examples
of a galactic superwind. Using long-split spectra,
\cite{Filippenko1992} found evidence for gas moving at speeds of up to
$2000\,{\rm km\,s^{-1}}$ with respect to the systemic velocity at
distances of a few hundred pc from the galactic nucleus. This is
considerably higher than in any other galaxies of a sample by
\cite{Heckman1990}, and therefore might not exclusively be starburst,
but also AGN driven. This is supported by \cite{Cecil2001}, who find a
gas filament at the base of the wind that aligns with the axis of the
VLBI scale radio jet.

X-ray: \cite{Cecil2002} report on X-ray observations with the Chandra
satellite and on $I$ band and [N\,{\small II}] $\lambda\lambda$\,6548,
6583+${\rm H}\,\alpha$ line emission. They find striking similarities
in the patterns of the X-ray and ${\rm H}\,\alpha$ emission which they
ascribe either to standoff bow shocks in the wind or to the conducting
surface between a hot, shocked wind and cooler ISM.

\section{NGC\,1052} 

NGC\,1052 is a nearby ($D=19.4~{\rm Mpc}$, \citealt{Tonry2001},
independent of redshift) elliptical LINER galaxy
(\citealt{Heckman1980}).

Radio: VLBI observations at 2.3~GHz to 43.2~GHz (\citealt{Kameno2001},
\citealt{Vermeulen2003}) show a two-sided, continuous jet, inclined at
$\approx90^{\circ}$ to the line of sight
(\citealt{Vermeulen2003}). VLBI observations by
\cite{Kellermann1998} at 15.4\,GHz show a well-defined double-sided jet
structure, the western jet of which was later found to be receding
(\citealt{Vermeulen2003}). In a trichromatic VLBI observation,
\cite{Kameno2001} found the spectrum of the brightest component at
15.4\,GHz to be $\alpha_{8.4}^{15.4}>+3$
($S_{\nu}\propto\nu^{\alpha}$). This exceeds the theoretical maximum
of $+2.5$ for optically-thin synchrotron self-absorption in uniform
magnetic fields, and \cite{Kameno2001} conclude that the absorption is
due to free-free absorption. They estimate the path length through the
absorber, $L$, to be $\approx0.7\,{\rm pc}$, the electron density,
$n_{\rm e}$, to be $>3.1\times10^4\,{\rm cm^{-3}}$ and the magnetic
field strength, $B$, to be larger than $7\times10^{-4}\,{\rm
G}$. These values yield a maximum $RM$ of the order of $10^7\,{\rm
rad\,m^{-2}}$.  Observing NGC\,1052 at seven frequencies,
\cite{Vermeulen2003} find numerous components with a low-frequency
cutoff which they ascribe to free-free absorption. They derive $n_{\rm
e}=10^5\,{\rm cm^{-3}}$ if the absorber is uniform and has a thickness
of 0.5\,pc. H\,{\small I} absorption was first reported by
\cite{Shostak1983} and \cite{Gorkom1986}, and more recent VLBA
observations of the H\,{\small I} line by \cite{Vermeulen2003} reveal
numerous velocity components and a highly clumpy structure of
H\,{\small I} absorption in front of the pc and sub-pc scale radio
continuum of both the jet and the counter-jet. They derive H\,{\small
I} column densities of the order of $10^{20}\,{\rm cm^{-2}}$ to
$10^{21}\,{\rm cm^{-2}}$ ($T_{\rm spin}=100\,{\rm K}$) for various
parts of the source, and hence H\,{\small I} densities of $100\,{\rm
cm^{-3}}$ to $1000\,{\rm cm^{-3}}$. A deficit in H\,{\small I}
absorption was found to be coincident with the location of the
strongest free-free absorption, indicating that the gas in the central
regions is ionized by the AGN. \cite{Vermeulen2003} also report on OH
absorption and emission. The ${\rm H_2O}$ maser emission found by
\cite{Braatz1994} using the Effelsberg 100-m telescope was resolved
into multiple spots along the counter-jet by
\cite{Claussen1998}. The velocity dispersion of the maser spots
perfectly agrees with the single-dish data, which were showing an
unusual broad line width of $85\,{\rm km\,s^{-1}}$ compared to typical
line widths of much less than $10\,{\rm km\,s^{-1}}$. OH absorption
measurements using the VLA (\citealt{Omar2002}) yielded velocities
equal to the centroid of the 22\,GHz maser emission, but the
connection between OH, ${\rm H_2O}$ and H\,{\small I} is still not
understood.

IR / Optical / UV: NGC\,1052 was the first LINER in which broad ${\rm
H}\,\alpha$ emission lines were detected in polarized light
(\citealt{Barth1999}), indicating that LINERs, similar to Seyfert~2
galaxies (e.g., NGC\,1068, \citealt{Antonucci1994}), harbour an
obscured type 1 AGN. This is supported by \cite{Barth1998}, who
discovered that LINERs are most likely to be detected in
low-inclination, high-extinction hosts, suggestive of considerable
amounts of gas. There is a disagreement whether the gas in NGC\,1052
is predominantly shock-ionized (\citealt{Sugai2000}) or photoionized
(\citealt{Gabel2000}).

X-ray: Using the Chandra X-ray observatory, \cite{Kadler2004}
find an X-ray jet and an H\,{\small I} column density towards the core
of $10^{22}\,{\rm cm^{-2}}$ to $10^{23}\,{\rm cm^{-2}}$, in agreement
with the radio observations. \cite{Guainazzi2000} use the bulge B
magnitude versus black hole mass relation of \cite{Magorrian1998} to
derive a black hole mass of $10^{8\pm1}\,M_{\odot}$.\\

\section{NGC\,4261 (3C\,270)} 
\label{sec:ngc4261}

NGC\,4261 is an elliptical LINER (\citealt{Goudfrooij1994}) galaxy
located at a distance of $35.8\,{\rm Mpc}$ (\citealt{Nolthenius1993},
${\rm H_0=75\,km\,s^{-1}\,Mpc^{-1}}$).

Radio: VLBI radio continuum images at frequencies from 1.6\,GHz to
43\,GHz by \cite{Jones1997}, \cite{Jones2000} and \cite{Jones2001}
show a double-sided jet structure (the counter-jet pointing towards
the east) with a clear gap 0.2\,pc wide across the counter-jet at low
frequencies only. The spectral index in the gap exceeds $+3$ between
4.8\,GHz and 8.4\,GHz, and so it is clearly due to a free-free
absorber. \citet{Jones2001} infer $n_{\rm e}=3\times10^4\,{\rm
cm^{-3}}$ for $L=0.3\,{\rm pc}$ and $T_{\rm e}=10^4\,{\rm K}$. VLBI
observations by \cite{Langevelde2000} revealed H\,{\small I}
absorption towards the counter-jet, 18\,mas (3.1\,pc) east of the
core, with a column density of $N_{\rm HI}=2.5\times10^{21}\,{\rm
cm^{-2}}$ ($T_{\rm spin}=100\,{\rm K}$), yielding a density of $n_{\rm
HI}=600\,{\rm cm^{-3}}$. No H\,{\small I} absorption was found towards
the core, yielding an upper limit on the H\,{\small I} column density
of $<2.2\times10^{20}\,{\rm cm^{-2}}$. This is much lower than that
towards the counter-jet as expected, because the gas close to the core
is ionized by UV radiation from the AGN, which ionizes a Str\"omgren
sphere out to a radius of $r_{\rm S}=31.8\,{\rm pc}$ (assuming an
H\,{\small I} density of $600\,{\rm cm^{-3}}$ and a temperature of
$10^4\,{\rm K}$). \cite{Langevelde2000} ascribe the inconsistency
between the Str\"omgren sphere radius and the H\,{\small I} absorption
close to the core to a presumably higher $T_{\rm spin}$, causing
larger densities and hence smaller ionized spheres. No water maser
emission was found in observations by \cite{Braatz1996} and
\cite{Henkel1998}.

IR / Optical / UV: NGC\,4261 has gained some fame through the
beautiful HST images made by \cite{Jaffe1996} showing a well-defined
disc around a central object with a diameter of 125\,pc and a
thickness $<40\,{\rm pc}$. They also find symmetrically broadened
forbidden lines which are likely to be caused by rotation in the
vicinity of a $4\times10^7\,{\rm M_{\odot}}$ compact object in the
central 0.2\,pc. \cite{Ferrarese1996} report on continuum and spectral
HST observations from which they conclude that this disc has a
pseudospiral structure and provides the means by which angular
momentum is transported outwards and material inwards. They also
report that the disc is offset by $(16\pm8)\,{\rm pc}$ with respect to
the isophotal centre and by $(4\pm2)\,{\rm pc}$ with respect to the
nucleus, which they interpret as being remnant of a past merging event
after which an equilibrium position has not yet been
reached. \cite{Ferrarese1996} derive a black hole mass of
$(4.9\pm1.0)\times10^8\,{\rm M_{\odot}}$, 12 times more than
\cite{Jaffe1996}.

X-ray: \cite{Chiaberge2003} observed NGC\,4261 with the Chandra
X-ray observatory and detected both the jet and the counter-jet, but
could not establish whether the emission was synchrotron emission from
the jet or scattered radiation from a ``misaligned'' BL\,Lac-type
AGN. They derive an H\,{\small I} column density of
$(6.33\pm0.78)\times10^{22}\,{\rm cm^{-2}}$. X-ray observations by
\cite{Sambruna2003} using the XMM-Newton satellite showed significant
flux variations on timescales of 1\,h. This result challenges ADAF
accretion models because they predict the X-ray emission to come from
larger volumes.

\section{Hydra~A (3C\,218)} 

Hydra~A is the second luminous radio source in the local ($z\,<\,0.1$)
Universe, surpassed only by Cygnus~A. It is hosted by an optically
inconspicuous cD2 galaxy (\citealt{Matthews1964}) at a distance of
216\,Mpc (\citealt{Taylor1996}, ${\rm H_0=75\,km\,s^{-1}\,Mpc^{-1}}$).

Radio: \citet{Taylor1996} observed this source with the VLBA between
1.3\,GHz and 15\,GHz and found symmetric parsec-scale jets and a
spectral index of +0.8 between 1.3\,GHz and 5 GHz towards the core and
inner jets.  The core was resolved and, for the measured brightness
temperature of $10^8\,{\rm K}$ and equipartition conditions,
synchrotron self-absorption would produce a turnover at around
100\,MHz.  Free-free absorption fits the spectrum more naturally, and
assuming that $L = 15$ pc, equal to the width of the absorbed region,
yields $n_{\rm e}=800\,{\rm cm^{-3}}$. \citet{Taylor1996} also
found neutral hydrogen absorption with $N_{\rm
HI}=1.4\times10^{22}\,{\rm cm^{-2}}$ towards the core and along the
jets out to 30\,pc, which he interpreted as a circumnuclear disc with
thickness $\sim30\,{\rm pc}$. This result agrees well with X-ray
observations by
\citet{Sambruna2000}, who discovered nuclear emission that is best
fitted with a heavily absorbed power law with an intrinsic H\,{\small
I} column density of $N_{\rm HI}=3\times10^{22}\,{\rm cm^{-2}}$ and
photon index $\Gamma\approx1.7$. Hydra~A did not show any ${\rm H_2O}$
maser emission in a survey by \cite{Henkel1998}.

IR / Optical / UV: Hydra~A appears to have a double optical nucleus
(\citealt{Dewhirst1959}). Observations by \cite{Ekers1983} revealed
two independently rotating regimes in Hydra~A: a fast rotating, inner
disc and an extended, infalling component.

X-ray: Hydra~A has an associated type II cooling-flow nebula
(\citealt{Heckman1989}), characterized by high ${\rm H}\alpha$ and
X-ray luminosities, but relatively weak N\,{\small II} and
S\,{\small II} and strong O\,{\small I} $\lambda 6300$
emission lines, usually found in LINERs.

\section{Centaurus~A (NGC\,5128)} 

The giant elliptical galaxy NGC\,5128, at a distance of 4.2\,Mpc
(\citealt{Tonry2001}), hosts Centaurus~A, by far the nearest AGN and a
strong radio source. Its proximity allowed detailed observations at
almost every observable wavelength, showing a wealth of detail (see
\citealt{Israel1998} for a comprehensive review).

Radio: In the radio, Centaurus~A covers an area of
$10^{\circ}\times5^{\circ}$ in the sky (\citealt{Cooper1965}). VLA
H\,{\small I} observations by \cite{Gorkom1990} showed the neutral
hydrogen to be well associated with the dust lane seen in optical
wavelengths, and filling the gap between the kpc-scale X-ray emission
and the inner parts of the galaxy (\citealt{Karovska2002}). Centimetre
radio observations show a complex jet-lobe structure from the largest
(500\,kpc) to the smallest (0.1\,pc) scales. \citet{Tingay2001} found
the compact VLBI core to be strongly absorbed below 8.4\,GHz with a
spectral index of $+3.8$ between 2.2\,GHz and 5\,GHz, clearly ruling
out synchrotron self-absorption, even though $T_{\rm
b}>9\times10^9\,{\rm K}$ at 8.4 GHz at the core. Absorption is seen
against the core and not against the jet 0.1\,pc away, so they let
$L<0.016\,{\rm pc}$ (i.e.  less than their limit on the core diameter)
and $T_{\rm e}=10^4\,{\rm K}$ and derive $n_{\rm e}>3\times10^4\,{\rm
cm^{-3}}$ in the free-free absorber. Observations in the millimetre
regime by
\cite{Hawarden1993} revealed a circumnuclear disc perpendicular to the
centimetre jet, but at an angle to the optical dust lane. They find
flat spectrum at wavelengths as short as $800\,\mu{\rm m}$ and
conclude that Centaurus~A is a ``misaligned'' blazar, in agreement
with
\cite{Bailey1986}, who made this assessment based on IR polarimetry.

IR / Optical / UV: The appearance of Centaurus~A is remarkably
different from other elliptical galaxies due to its prominent dust
lane. Moreover, the dust lane is oriented along the minor axis of the
optical galaxy and is well inside it, as can be seen in the deep
images of \cite{Haynes1983}.

X-ray: Observations with the Chandra X-ray satellite by
\cite{Karovska2002} show a kpc-scale structure, suggestive of a ring
seen in projection. It possibly arises from infalling material where
it hits cooler dust in the galaxy centre. They speculate that
arc-shaped, axisymmetric emission on scales $>>\,{\rm kpc}$ is due to
transient nuclear activity $10^7\,{\rm yr}$ ago. They also find a
jet-like feature which is coincident with the arcsecond-scale radio
jet. Assuming optically thin thermal emission, they derive $N_{\rm
HI}=3.9\times10^{21}\,{\rm cm^{-2}}$, in contrast to
\cite{Turner1997}, who found $N_{\rm HI}=10^{23}\,{\rm cm^{-2}}$ using
the ROSAT satellite.

\section{Cygnus~A (3C\,405)} 

Cygnus~A at a distance of 224\,Mpc (\citealt{Owen1997}, ${\rm
H_0=75\,km\,s^{-1}\,Mpc^{-1}}$) is the prototypical FR\,II radio
galaxy, hosted by an elliptical cD galaxy with complex optical
structure, located in a poor cluster.

Radio: Observations in the radio bands display finest details from kpc
to sub-pc scales (\citealt{Perley1984, Krichbaum1998}). Proper motions
of jet components seen with VLBI were first reported by
\cite{Carilli1994}, who did not significantly detect a
counter-jet. Detailed multi-frequency and multi-epoch VLBI imaging by
\cite{Krichbaum1998} show a pronounced double-sided jet structure
with rich internal structure and indicate that the (eastern)
counter-jet is obscured by a circumnuclear absorber. The spectral
index of the pc-scale counter-jet is systematically higher than the
spectral index of the jet, and the frequency dependence of the
jet-to-counter-jet ratio, $R$, shows the characteristics expected to
arise from a free-free absorber rather than from synchrotron
self-absorption. Supported by H\,{\small I} column densities inferred
independently from X-ray observations by
\citet{Ueno1994} and from VLA H\,{\small I} absorption measurements by
\citet{Conway1995}, \citet{Krichbaum1998} derive $n_{\rm e}=10^4\,{\rm
cm^{-3}}$ (using $T_{\rm e}=10^4\,{\rm K}$ and $L<15\,{\rm pc}$) for
the absorber.

IR / Optical / UV: The optical emission from Cygnus~A exhibits narrow
lines in total intensity and very broad ($26\,000\,{\rm km\,s^{-1}}$,
\citealt{Ogle1997}) lines in polarized light, supporting the presence of a
hidden broad-line quasar core. HST near-infrared observations by
\cite{Tadhunter1999} revealed an edge-brightened biconcical structure
around a bright point source. They could not establish whether this
structure is due to an outflow or a radiation-driven wind. HST
near-infrared polarization measurements by \cite{Tadhunter2000}
suggest an intrinsic anisotropy in the radiation fields in these
cones, and a high degree of nuclear polarization (25\,\%) indicates
that most of the radiation is re-processed in scattering and that
previous works have substantially underestimated the nuclear
extinction.

X-ray: From X-ray spectroscopy with the Ginga satellite,
\cite{Ueno1994} conclude that the nucleus in Cygnus~A is absorbed by
atomic hydrogen with a column density of $N_{\rm
HI}=(3.75\pm0.73)\times10^{23}\,{\rm cm^{-2}}$. This is two orders of
magnitude more than the column density of $N_{\rm
HI}=(2.54\pm0.44)\times10^{21}\,{\rm cm^{-2}}$ ($T_{\rm
spin}=100\,{\rm K}$) inferred from VLA H\,{\small I} line observations
by \cite{Conway1995}. Although \cite{Conway1995} cite the
\cite{Ueno1994} measurements, they do not comment on the obvious
discrepancy.

\chapter{Observations}
\label{ch:obs}

The observations of the sample have been carried out with the Very
Long Baseline Array, or VLBA, partly supplemented by the Effelsberg
100\,m telescope and a single VLA antenna. The VLBA comprises 10
identical antennas of 25\,m diameter that have been constructed
exclusively for VLBI, in contrast to arrays composed of already
existing antennas like the European VLBI Network. Due to the
homogeneous antenna characteristics, simplified scheduling and user
support, the VLBA has made interferometric observations much easier
than with any other VLBI array. Some of the key advantages of the VLBA
antennas are their frequency agility, good polarization properties and
high slewing rates. The frequencies are changed within 10\,s by moving
the subreflector, allowing almost uninterrupted observing at many
frequencies. The polarization leakage of the feed horns, i.e., the
fraction of LCP power received by the RCP receiver channel and vice
versa, are frequently less than two or three percent. Fast slewing
($90^{\circ}\,{\rm min^{-1}}$ in azimuth and $30^{\circ}\,{\rm
min^{-1}}$ in elevation) enables rapid switching between sources,
improving phase-referenced observations by decreasing coherence losses
due to tropospheric phase noise. These three advantages have
contributed much to the quality of the data presented here. The use of
the Effelsberg 100\,m antenna during some of the observations provided
the highest sensitivity on intercontinental baselines. This is
especially valuable at low frequencies because one can match the
synthesized beam to that at higher frequencies by using approximately
scaled arrays, without losing too much data to tapering. A single VLA
antenna was also used during parts of the observations for better
$(u,v)$ coverage, particularly for short spacings.\\

\section{Observations}

The observations were made in five observing campaigns, the details of
which are summarized in Table~\ref{tab:observations}. Polarization
calibrators were observed four or more times at each frequency,
covering a range of parallactic angles larger than $90^{\circ}$.

\begin{itemize}

\item NGC\,3079 was observed with the VLBA, a single VLA antenna (Y1)
and Effelsberg (5.0\,GHz and 15.4\,GHz only) at 1.7\,GHz, 5.0\,GHz
and 15.4\,GHz during three epochs: on November 20, 1999, March 6, 2000
and November 30, 2000. A total bandwidth of 32\,MHz was recorded using
2-bit sampling.  This campaign was intended as a pilot monitoring
program over a wide range of frequencies. The results --~new components
at 5.0\,GHz and only few detections at 1.7\,GHz~-- led to a different
experiment design for the fourth epoch.

\item NGC\,3079 was re-observed with the VLBA and Effelsberg on
September 22, 2002 at 1.7\,GHz, 2.3\,GHz and 5.0\,GHz. Effelsberg
observed only at 1.7\,GHz because the receiver is in the prime focus
whereas all other receivers are in the secondary focus, and to match
the 1.7\,GHz synthesized beam to the beams of the higher
frequencies. A total bandwidth of 64\,MHz was recorded using 2-bit
sampling. Denser frequency spacing at lower frequencies allowed us to
measure the components' spectra. Phase-referencing was used because
the source was previously found to be weak below 5.0\,GHz and provided
absolute positions for image registration and later proper motion
measurements. Polarization calibrator scans were inserted to calibrate
the $D$-terms. The angular separation to the phase-referencing source
J0957+5522 is $41^{\prime}$.

\item NGC\,1052 was observed with the VLBA on December 28, 2001 at
13.4\,GHz and 15.4\,GHz. A total bandwidth of 32\,MHz was recorded
using 2-bit sampling. The experiment aimed at detecting polarized
emission, but we were assigned enough observing time to observe at two
frequencies with sufficient sensitivity. We decided to distribute the
time equally among 13.4\,GHz and 15.4\,GHz so that we could look for a
change in EVPA with frequency had polarized emission been detected.

\item NGC\,4261, Hydra~A and Centaurus~A were observed with the VLBA on
July 5, 2002, at 13.4\,GHz and 15.4\,GHz using several combinations of
recording bandwidths and frequencies, but all used 2-bit
sampling. NGC\,4261 was observed with 64\,MHz bandwidth due to the
weakness of the absorbed gap, and Centaurus~A was strong enough to be
observed at 13.4\,GHz and 15.4\,GHz to look for a possible EVPA change
with frequency. Hydra~A was observed at 15.4\,GHz only using 32\,MHz
bandwidth.

\item Cygnus~A was observed with the VLBA, Effelsberg and a single VLA
antenna on July 7, 2002, at 15.4\,GHz, and a total bandwidth of
64\,MHz was recorded using 2-bit sampling.

\end{itemize}

\begin{sidewaystable*}[hp!]
\scriptsize
\begin{center}
\begin{tabular}{l.cc|clcccccc}
\hline
\hline
Source & \multicolumn{1}{c}{\rm Position} & {\rm Ref.} & Purpose &                               & \multicolumn{6}{c}{Integration times / minutes}\\
       & \multicolumn{1}{c}{\rm (J2000)}  &            &         & $\nu$/GHz~$\rightarrow$       & 1.7 & 1.7 & 2.3 & 5.0 & 5.0 & 15.4 & \\
       &                                  &            &         & $\Delta\nu$/MHz~$\rightarrow$ & 32  & 64  & 64  & 32  & 64  &  32  & \\
(1)    & (2)                              & (3)        & (4)     &                               & (5) & (6) & (7) & (8) & (9) & (10) & \\
\hline
\multicolumn{3}{l}{\it{Observations on November 20, 1999}}&&&&&\\
%
NGC\,3079     & 10~01~57.7906     & 1 & target       && $108^{\rm Y1}$ &&& $144^{\rm EB,Y1}$ && $144^{\rm EB,Y1}$\\
              & 55~40~47.1788     &   &              &\\
\hline
\multicolumn{3}{l}{\it{Observations on March 6, 2000}}&&&&&\\
NGC\,3079     &                   & 1 & target       && $108^{\rm Y1}$ &&& $144^{\rm EB,Y1}$ && $136^{\rm EB,Y1}$\\
\hline
\multicolumn{3}{l}{\it{Observations on November 30, 2000}}&&&&&\\
NGC\,3079     &                   & 1 & target       && $108^{\rm Y1}$ &&& $144^{\rm EB,Y1}$ && $144^{\rm EB,Y1}$\\
\hline
\multicolumn{3}{l}{\it{Observations on September 22, 2002}}&&&&&\\
NGC\,3079     &                   & 1 & target       &&& $114^{\rm EB}$ & 107 && 112\\
J0957+5522    & 09~57~38.1849709  & 2 & Phase        \\
              & 55~22~57.769241   &   & calibrator   \\
J1146+3958    & 11~46~58.29791    & 3 & polarization &&& $18^{\rm EB}$  & 18  && 18\\
              & 39~58~34.30461    &   & calibrator   \\
J1159+2914    & 11~59~31.83391    & 3 & polarization &&& $18^{\rm EB}$  & 18  && 18\\
              & 29~14~43.82694    &   & calibrator   \\
\hline
\end{tabular}
\caption[Sources and observational parameters]{Sources and
observational parameters. (1) - source names, (2) - phase centre
(J2000), (3) - position reference, (4) - purpose of source, (5)-(13) -
frequencies, bandwidths and integration times used in the
observations. References: 1 - Taken from Sawada-Satoh et al.'s key
file for their 1996 observations, 2 - VLBA calibrator survey, 3 -
\cite{Gambis1999}, 4 - 43 GHz map centre in \cite{Jones2000}, 5 -
G. B. Taylor (priv. comm.), 6 -
\cite{Ma1998}, 7 - T. P. Krichbaum, priv. comm. Superscript ``EB'' and
``Y1'' indicate the participation of Effelsberg and a single VLA
antenna, respectively.}
\label{tab:observations}
\end{center}
\end{sidewaystable*}

\begin{sidewaystable}[hp!]
\scriptsize
\begin{center}
\begin{tabular}{l.cc|clccc}
\hline
\hline
Source & \multicolumn{1}{c}{\rm Position} & {\rm Ref.} & Purpose &                               & \multicolumn{4}{c}{Integration times / minutes}\\
       & \multicolumn{1}{c}{\rm (J2000)}  &            &         & $\nu$/GHz~$\rightarrow$       & 13.4 & 13.4 & 15.4 & 15.4\\
       &                                  &            &         & $\Delta\nu$/MHz~$\rightarrow$ & 32   & 64   &  32  &  64 \\
(1)    & (2)                              & (3)        & (4)     &                               & (11) & (12) & (10) & (13)\\
\hline
\multicolumn{3}{l}{\it{Observations on December 28, 2001}}&&&&&\\           
NGC\,1052     & 02~41~04.798520   & 3 & target       && 228 &   & 234\\ 
              &-08~15~20.75184    &   &              \\ 
DA\,193       & 05~55~30.805609   & 3 & polarization && 14 &   & 18\\  
              & 39~48~49.16500    &   & calibrator   \\  
0420-014      & 04~23~15.800727   & 3 & polarization && 18 &   & 17\\ 
              &-01~20~33.06531    &   & calibrator   \\  
\hline                                                                      
\multicolumn{3}{l}{\it{Observations on July 5, 2002}}&&&&&\\                
NGC\,4261     & 12~19~23.221      & 4 & target       &&&&& 251\\  
              & 05~49~29.795      &   &              \\  
Hydra\,A      & 09~18~05.6740     & 5 & target       &&&& 148\\
              &-12~05~43.92       &   &              \\  
Centaurus\,A  & 13~25~27.615217   & 6 & target       && 50 &   & 48\\ 
              &-43~01~08.80528    &   &              \\  
0927+3902     & 09~27~03.0139160  & 3 & polarization && 35 & 11 & 33 & 10\\  
              & 39~02~20.851950   &   & calibrator   \\  
1146+3958     & 11~46~58.2979088  & 3 & polarization &&&& 12 & 31\\  
              & 39~58~34.304611   &   & calibrator   \\  
\hline                                                                      
\multicolumn{3}{l}{\it{Observations on July 7, 2002}}&&&&&\\                
Cygnus\,A     & 19~59~28.35668    & 2 & target       &&&&& $231^{\rm EB,Y1}$\\
              & 40~44~02.09658    &   &              \\  
BL\,Lac       & 22~02~43.29137    & 2 & polarization &&&&& $22^{\rm EB,Y1}$\\
              & 42~16~39.97994    &   & calibrator   \\  
2013+370      & 20~15~28.7126     & 7 & polarization &&&&& $27^{\rm EB,Y1}$\\
              & 37~10~59.694      &   & calibrator   \\  
\hline
\end{tabular}
\contcaption{(continued)}
\end{center}
\end{sidewaystable}

\section{Calibration}

The initial data calibration was carried out in the Astronomical Image
Processing System, or AIPS, a software package developed and
maintained by the U.S. National Radio Astronomy Observatory.

The visibility amplitudes were calibrated using $T_{\rm sys}$
measurements and the known elevation-dependent antenna gains. $T_{\rm
sys}$ was measured at each antenna every two minutes and typically
varied on timescales of hours.

The absolute value of the complex gain of a station (i.e., how much it
contributes to the visibility amplitude) is affected by fluctuations
in the thresholds of the digitizer samplers, especially when the
signal is digitized with two bits (four levels) per sample instead of
one bit (two levels). These errors can be corrected using the
autocorrelation data of an antenna, and the amplitudes of these
corrections were typically less than 5\,\%.

At low source elevations, the opacity of the atmosphere becomes
important and the amplitudes are decreased. This effect needed to be
considered in the case of Centaurus~A, because its declination of
$-43^{\circ}$ caused it to be below $17^{\circ}$ elevation at the
central five stations of the VLBA. The opacity was estimated and
corrected as follows.  System noise at a station can be expressed as a
temperature equivalent (the system temperature), $T_{\rm sys}$, which
is the sum of many contributions from the telescope, e.g., the
receiver temperature, $T_{\rm rec}$, the antenna temperature, $T_{\rm
ant}$, the atmospheric temperature, $T_{\rm atm}$, and the microwave
background temperature, $T_{\rm CMB}$. In this simplified treatment,
we refer to all of these contributions as $T_0$, except for the
atmospheric noise contribution, $T_{\rm atm}$. The amount of noise
from the atmopshere is the product of the physical temperature of the
air in front of the telescope, $T_{\rm air}$, its opacity, $\tau$, and
the air mass in front of the telescope, which is proportional to
$1/\sin(\theta)$ and is unity for an antenna elevation of
$\theta=90^\circ$:

\begin{equation}
T_{\rm sys}=T_0+T_{\rm air}\tau_{\rm zen}A=T_0+\frac{T_{\rm air}\tau_{\rm zen}}{\sin(\theta)}.
\end{equation}

Measuring $T_{\rm sys}$ at various antenna elevations and plotting it
against $1/\sin(\theta)$ yields a linear function with intercept
$T_{\rm sys}$ and slope $T_{\rm air}\tau_{\rm zen}$. To solve for
$\tau_{\rm zen}$ hence requires an estimate of $T_{\rm air}$, but the
atmospheric temperature profile is a very complicated matter and would
require a long treatment. Assuming that $T_{\rm atm}$ is 273.15\,K and
constant with altitude and time, we have derived and applied a
first-order approximation of $\tau$. 

In the next step, the visibility phases were corrected for parallactic
angle changes of the sources with respect to the antenna feed
horns. This step is of particular importance for polarization
observations as it affects the calibration of the antenna feed horn
polarization leakage, and also for phase referencing.

To account for ionospheric delays, we derived corrections using the
AIPS task TECOR from the TEC maps provided by the Center for
Orbit Determination in Europe,
CODE\footnote{\url{http://www.aiub.unibe.ch/ionosphere.html}}. We applied
corrections only to the September 2002 observations of NGC\,3079 at
all frequencies. We did not apply corrections to any other
observations because the effects are negligible at 15.4\,GHz and the
1999/2000 1.7\,GHz observations were not phase-referenced and the
source was too weak for self-calibration at 1.7\,GHz.

To allow coherent averaging over the recorded bandwidth requires one
to remove the phase and delay offsets between the different baseband
(or IF) channels. At the VLBA, this alignment of IFs can be done using
the pulse calibration system (see \S\ref{sec:pulsecals}). The system
provided excellent results for all data sets, except for Effelsberg
and the VLA antenna, which do not provide such pulses, and
self-calibration on a short scan of a strong source was used to
measure the phase offsets.

The final step in source-independent data calibration is the so-called
fringe-fitting to solve for and remove residual delays and
fringe-rates. The residuals remain due to errors in the correlator
model and changing weather conditions, especially changes in the wet
troposphere. Fringe-fitting is the first step in self-calibration. The
visibility data are divided into short segments. Their length may not
exceed the time scale on which the phase rate does not change,
typically 30\,s to 60\,s at 15.4\,GHz, but this needs to be estimated
from plots of visibility phase with time for each data set
separately. These segments are Fourier-transformed from lag-time into
delay-rate space, where a maximum is searched. The derived residual
phase, rate and delay errors are subtracted from the data and can then
be used to form an image by a Fourier transform from the $(u,v)$ plane
to the image plane.

Fringe-fitting requires the assumption of a point source in the field
centre, a condition which is satisfied by virtually all AGN to zeroth
order. The sum of the visibility phases in each antenna triangle,
known as closure phase (\citealt{Jennison1958}), is calculated. In the
case of a point source, this sum should be zero. Under this
assumption, deviations from the closure phase must be due to errors,
and antenna-based errors can be derived. In a closure triangle, two
baselines determine the phase of the third, and so there are two
independent measurables for three antenna-based phase errors. To solve
this underdetermined set of equations requires setting one of the
antenna phases to zero and solving for the phase at the other two
stations. This assumption destroys the absolute position information
of the visibility phases, and phase-referencing is required if
astrometric information is to be preserved.

The September 2002 observations of NGC\,3079 were calibrated using
phase-referencing. Each scan of 4\,min (5.0\,GHz) or 5.5\,min
(2.3\,GHz and 1.7\,GHz) length on NGC\,3079 was sandwiched between
short (1\,min at all frequencies) scans on a strong, nearby
calibrator, the fringe-fit delays, rates and phases of which were then
interpolated onto the NGC\,3079 scans. We had to use this observing
technique because the 1999 and 2000 observations had shown that
NGC\,3079 is too weak for fringe-fitting below 5.0\,GHz. In addition,
phase-referencing preserves absolute coordinates, allowing one to look
for absolute proper motion when multiple epochs are observed.

The data were then exported to Difmap (\citealt{shepherd1997}), a
software package for imaging, calibration and modelling of radio
interferometer data. Difmap implements hybrid mapping
(\citealt{Cornwell1981}, equivalent to self-calibration) and easy and
fast data editing. After initial editing of obvious outliers, the data
were averaged into 10\,s bins to speed up computing and to calculate
errors based on the scatter in each bin. The data were then
Fourier-transformed into a first image, the so-called ``dirty
map''. The dirty map is the sky brightness distribution convolved with
the point spread function of the array, which is the Fourier transform
of the sampling function (i.e., the tracks in the $(u,v)$ plane).

The Difmap implementation of the ``Clean'' algorithm by
\cite{Hogbom1974} was used to obtain a first, coarse model, in which
the brightness distribution is represented with delta components. This
model is Fourier transformed back to the $(u,v)$ plane to calculate
the model visibilities, the phases of which are subtracted from the
measured visibility phases. The residual antenna-based phase errors
can then be solved for by using the closure phase condition
(self-calibration) and a new image is formed from the measured
visibilities with the refined calibration. Finally, the set of Clean
model components is convolved with the ``clean beam'', a
two-dimensional Gaussian which has been fitted to the Fourier
transform of the sampling function. Adding the residuals from the
Clean algorithm then yields the ``clean image''.

This loop --~applying Clean and subsequent refinement of visibility
phases to fit the model~-- is known as phase self-calibration or
hybrid mapping. Because the process is underdetermined, many different
solutions exist that are compatible with the data and so care must be
taken not to build non-existing components into the model. This is
especially difficult with data that have low SNR or sparse $(u,v)$
coverage. Difmap also implements amplitude self-calibration
(\citealt{Readhead1980}), but this was only used with long ($>30\,{\rm
min}$) solution intervals to correct slowly changing station-based
gain errors.

The calibration was considered to be complete when either the hybrid
mapping had converged and no further improvements were possible or
when the rms noise in the image was within a few tens of percent of
the theoretically calculated image noise. The former limit was mostly
reached in the calibration processes of the strong ($S_{\rm
max}>100\,{\rm mJy\,beam^{-1}}$) sources, the noise level in images of
which are dominated by residual sidelobes, whereas the latter limit
was reached in the case of NGC\,3079, which has $S_{\rm max}<60\,{\rm
mJy\,beam^{-1}}$ at all observed frequencies, causing the residual
sidelobes to be at or below thermal noise.

\begin{figure*}[htpb!]
\centering
\includegraphics[width=\linewidth]{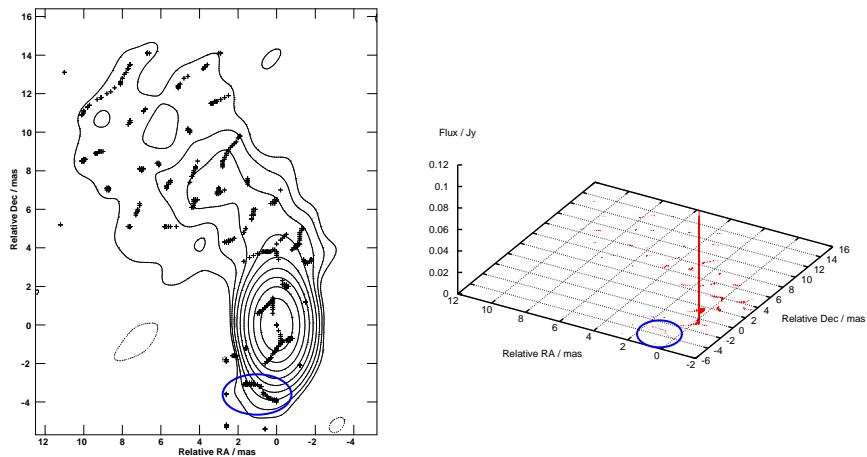}
\caption[Plotting Clean components to reveal their relative
strengths.]{Plotting Clean components to reveal their relative
strengths. {\it Left panel:} 5.0\,GHz contour map of the polarization
calibrator 1159+2914 observed in September 2002. Crosses indicate the
locations of Clean components. {\it Right Panel:} The Clean components
from the same image plotted to reveal their relative strengths. This
representation reveals that almost the entire flux is concentrated in
the central peak, and, e.g., that the group of components in the blue
ellipses can be neglected in the $D$-term calibration. As a consequence,
the calibration was done using a single component model.}
\label{fig:CCs}
\end{figure*}

Polarization measurements require the calibration of the $D$-terms. A
few scans on two polarization calibrators were carried out during all
observing runs except for the 1999/2000 observations to determine
them. After imaging in Difmap, the calibrated data were re-imported to
AIPS. An image was formed using the same Clean loop gain, number
of iterations and Clean windows as in Difmap. The implementation
of the \cite{Leppanen1995b} calibration algorithm in AIPS (see
\S\ref{sec:d-term-cal}) requires the source model to be
divided into less than ten components, i.e., one has to define windows
in each of which the Clean model components are thought to
represent one physical source component with uniform polarization. A
plot of the Clean components in a three-dimensional plot can help
doing this, revealing which of them are strong and which represent
only noise (Fig.~\ref{fig:CCs}). The new merged components are used by
the AIPS task LPCAL to derive the $D$-term solutions which are stored in
a calibration table. This table was copied to the target source data
files, where they were used to form Stokes I, Q and U images, from
which the polarized intensity and polarization angle maps were
calculated.

\begin{figure}[htpb!]
\centering
\subfigure[NL $D$-terms at 1.7\,GHz on September 22, 2002.]
{\includegraphics[width=0.7\linewidth]{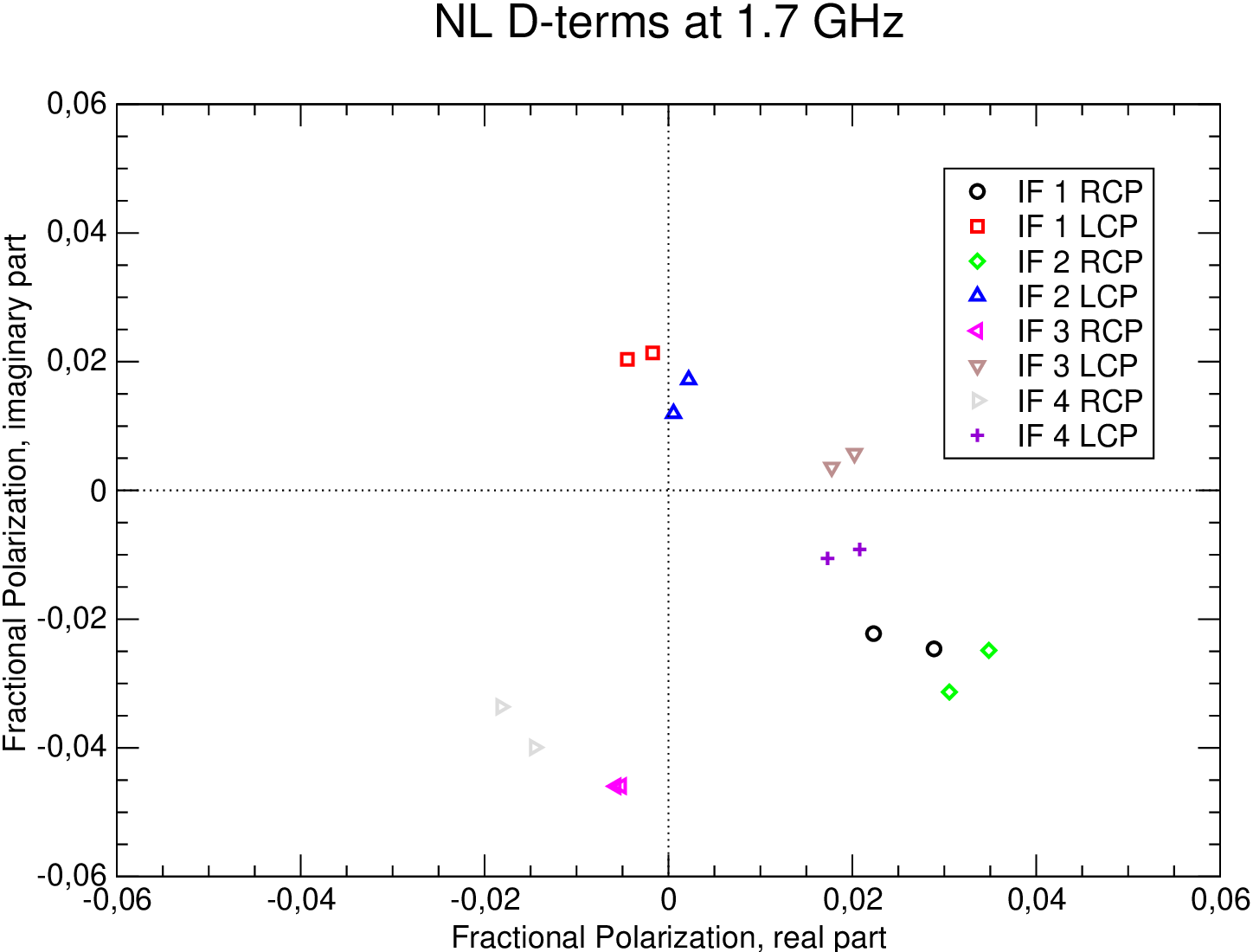}
\label{fig:NL_Dterms_1.7GHz}}\\
\subfigure[NL $D$-terms at 15.4\,GHz on July 5, 2002.]
{\includegraphics[width=0.7\linewidth]{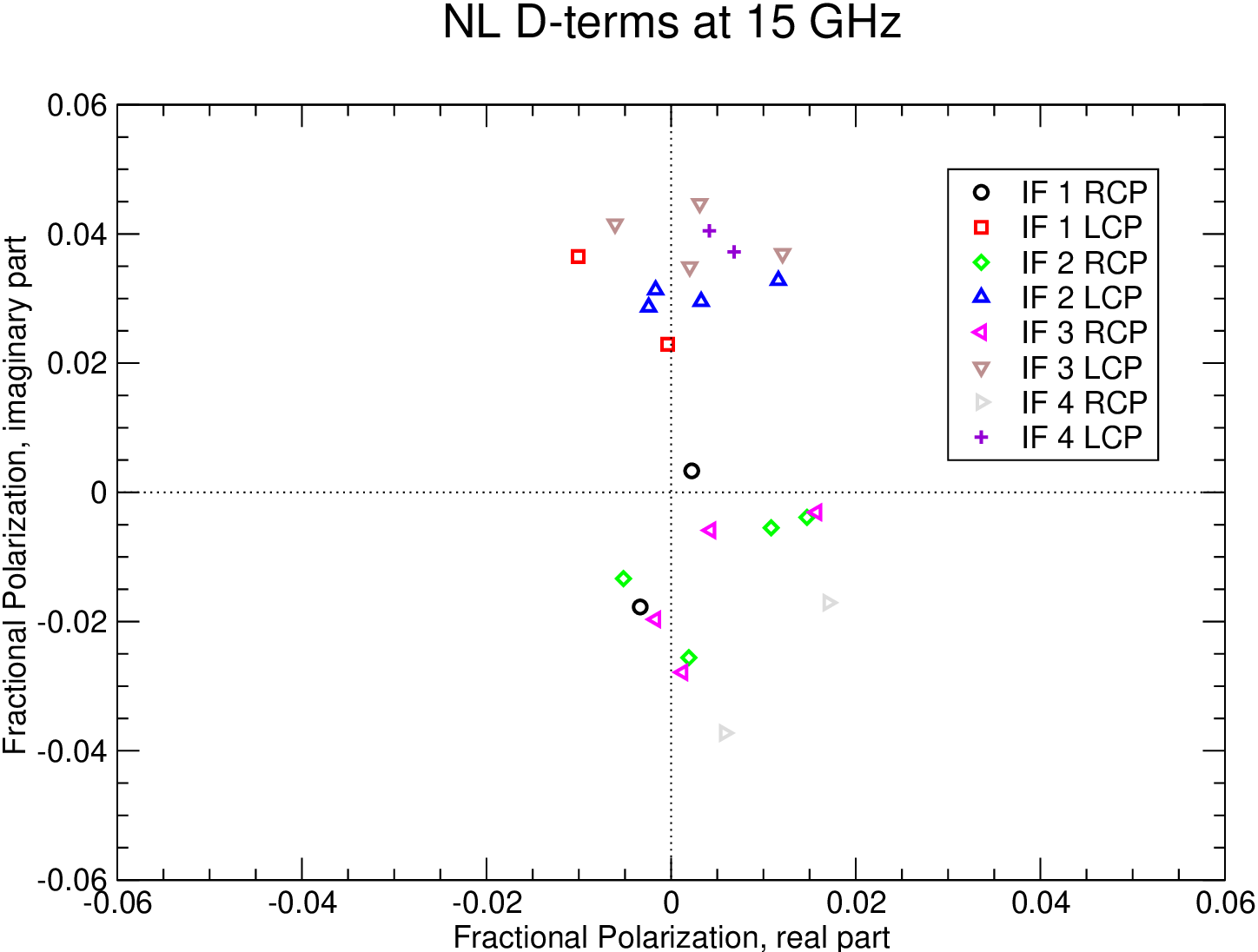}
\label{fig:NL_Dterms_15GHz}}\\
\caption[North Liberty $D$-terms at 1.7\,GHz and 15.4\,GHz]{North Liberty
$D$-terms at 1.7\,GHz and 15.4\,GHz derived from the September 22, 2002
and July 5, 2002 observations, respectively. Each colour represents the
measurements of a $D$-term in a specific IF, derived from observations
of two calibrators. The scatter is representative of the polarization
calibration of the array on these days (i.e., there are antennas with
both smaller and larger scatter). The $D$-terms change systematically
between the IFs at 1.7\,GHz, whereas there is no such trend at
15.4\,GHz. This reflects the different fractional bandwidths, 0.47\,\%
at 1.7\,GHz and 0.05\,\% at 15.4\,GHz.}
\label{fig:NL_Dterms}
\end{figure}

The quality of the calibration can be estimated by comparing the
$D$-term solutions from the two polarization calibrators; two examples
are shown in Fig.~\ref{fig:NL_Dterms}. The $D$-terms derived from the
September 2002 observations of NGC\,3079 (with the exception of the
2.3\,GHz $D$-terms, see below) and from the July 2002 observations of
Cygnus~A were very good and consistent among the two calibrators. The
$D$-terms derived from the July 2002 observations of NGC\,4261, Hydra~A
and Centaurus~A were a little less well constrained, but still of good
quality. To obtain the most sensitive polarization images, however,
images were made from this epoch without the stations at Mauna Kea,
Hancock and Saint Croix, whose $D$-terms showed large scatter. These
images did not differ significantly from those that were made from all
data, and the polarization calibration was considered to be final. We
have estimated the detection limit of polarized emission as follows:
two polarization images of a calibrator were made from the same set of
calibrated $(u,v)$ data and using the same imaging parameters
(weighting, number of iterations of Clean, etc), but in one case using
the $D$-terms derived from the calibrator itself and in the other case
using the $D$-terms derived from the other calibrator. The polarized
intensity images were divided by the associated total intensity images
to obtain two images of fractional polarization of the
calibrator. Finally, the fractional polarization images were
differenced, and the average difference was measured over the source
region. The average differences in fractional polarization were found
to be 0.3\,\% for the observations of NGC\,3079 at 1.7\,GHz and
5.0\,GHz; 0.5\,\% for NGC\,1052; 1\,\% for NGC\,4261, Hydra~A and
Centaurus~A; and 0.3\,\% for Cygnus~A.

A special problem occurred during the $D$-term calibration of the
NGC\,3079 2.3\,GHz observations. The $D$-term amplitudes were unusually
large, around 5\,\% in both RCP and LCP, and clustered around
$-160^{\circ}$ (RCP) and $-100^{\circ}$ (LCP) degrees, whilst the
1.7\,GHz and 5.0\,GHz were of the order of 1\,\% to
2\,\%. Surprisingly, the $D$-terms were similar at all antennas,
suggesting a connection to the antenna structure. A feature of the
2.3\,GHz system at the VLBA is the presence of a dichroic beamsplitter
in the optical path for simultaneous observations at 2.3\,GHz and
8.4\,GHz, the so-called geodetic S/X-band mode. The beamsplitter is a
permanently installed piece of circuit board with an imprinted copper
pattern that efficiently reflects emission at 8.4\,GHz away from the
feed onto a mirror over the 8.4\,GHz feed horn.

Applying the $D$-terms to the calibrator data yielded fractional
polarizations of 0.8\,\% and 1.7\,\%, respectively. Applying the
$D$-terms to the NGC\,3079 data, however, yielded a fractional
polarization of about 5\,\%, whilst it was unpolarized at 1.7\,GHz and
5.0\,GHz. A fractional polarization of the same order as the $D$-terms
is highly unlikely. The beamsplitter in the 2.3\,GHz optical path
might explain the unusual similarities of the $D$-terms and their
particularly high amplitudes. But it does not explain why the
systematic effect has not been calibrated on NGC\,3079. We have
explored numerous variations to the calibration strategy, e.g.,
changing the partitioning of the calibrator sources into sub-models,
repeating the calibration without exporting the data to Difmap and
different data weightings, but without success. We therefore do not
present a 2.3\,GHz polarization image of NGC\,3079.\\

\begin{figure}[ht!]
\centering
 \includegraphics[width=11cm]{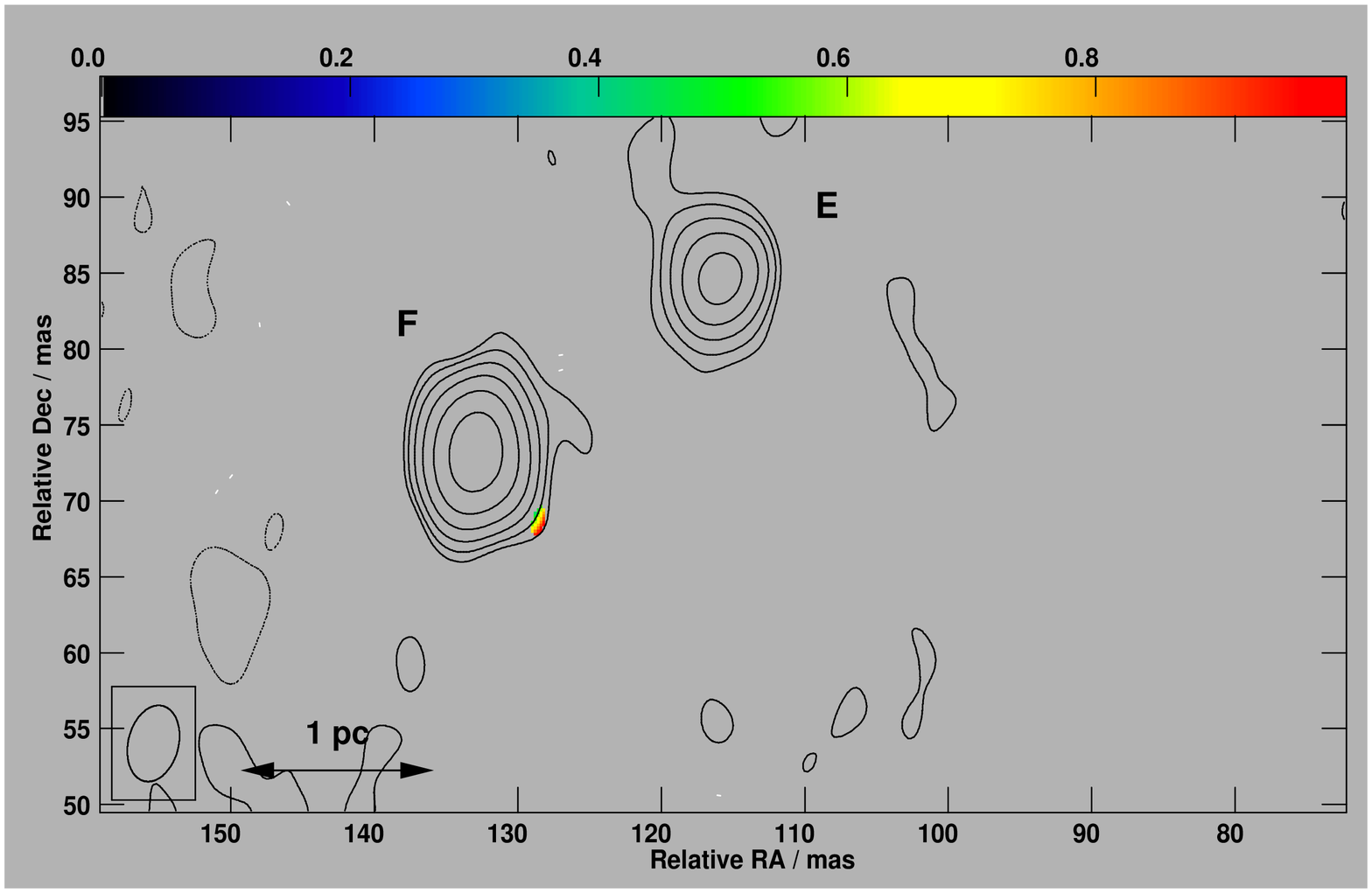}
 \caption[NGC\,3079 1.7\,GHz image]{NGC\,3079 1.7\,GHz VLBA+EB
 image from September 2002. Peak $9.70\,{\rm mJy\,beam^{-1}}$, contours
 $0.19\,\rm{mJy\,beam^{-1}}\times2^N$, nat. weighting, beam
 $5.12\,\mas\times3.49\,\mas$ in P.A. $-22^{\circ}$, pol. ticks
 $1\,{\rm mJy\,beam^{-1}\,mas^{-1}}$ where $>5\,\sigma=0.22\,{\rm
 mJy\,beam^{-1}}$. Axes relative to RA 10~01~57.7906 Dec 55~40~47.1788
 (J2000), colours show fractional polarization.}
 \label{fig:NGC3079L_it2_polmap}
\end{figure}

\begin{figure}[ht!]
\centering
 \includegraphics[width=11cm]{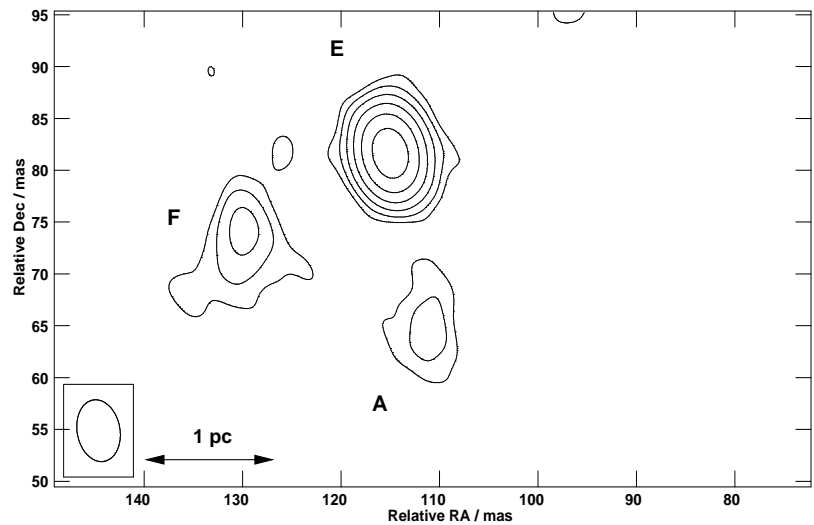}
 \caption[NGC\,3079 2.3\,GHz  image]{NGC\,3079 2.3\,GHz VLBA
 image from September 2002. Peak is $19.8\,{\rm mJy\,beam^{-1}}$, contours are at
 $0.39\,\rm{mJy\,beam^{-1}}\times2^N$, natural weighting, beam is
 $6.02\,\mas\times4.34\,\mas$ in P.A. $12^{\circ}$. Axes relative to
 RA 10~01~57.7906 Dec 55~40~47.1788 (J2000).}
 \label{fig:NGC3079S_it1}
\end{figure}

\begin{figure}[ht!]
\centering
 \includegraphics[width=11cm]{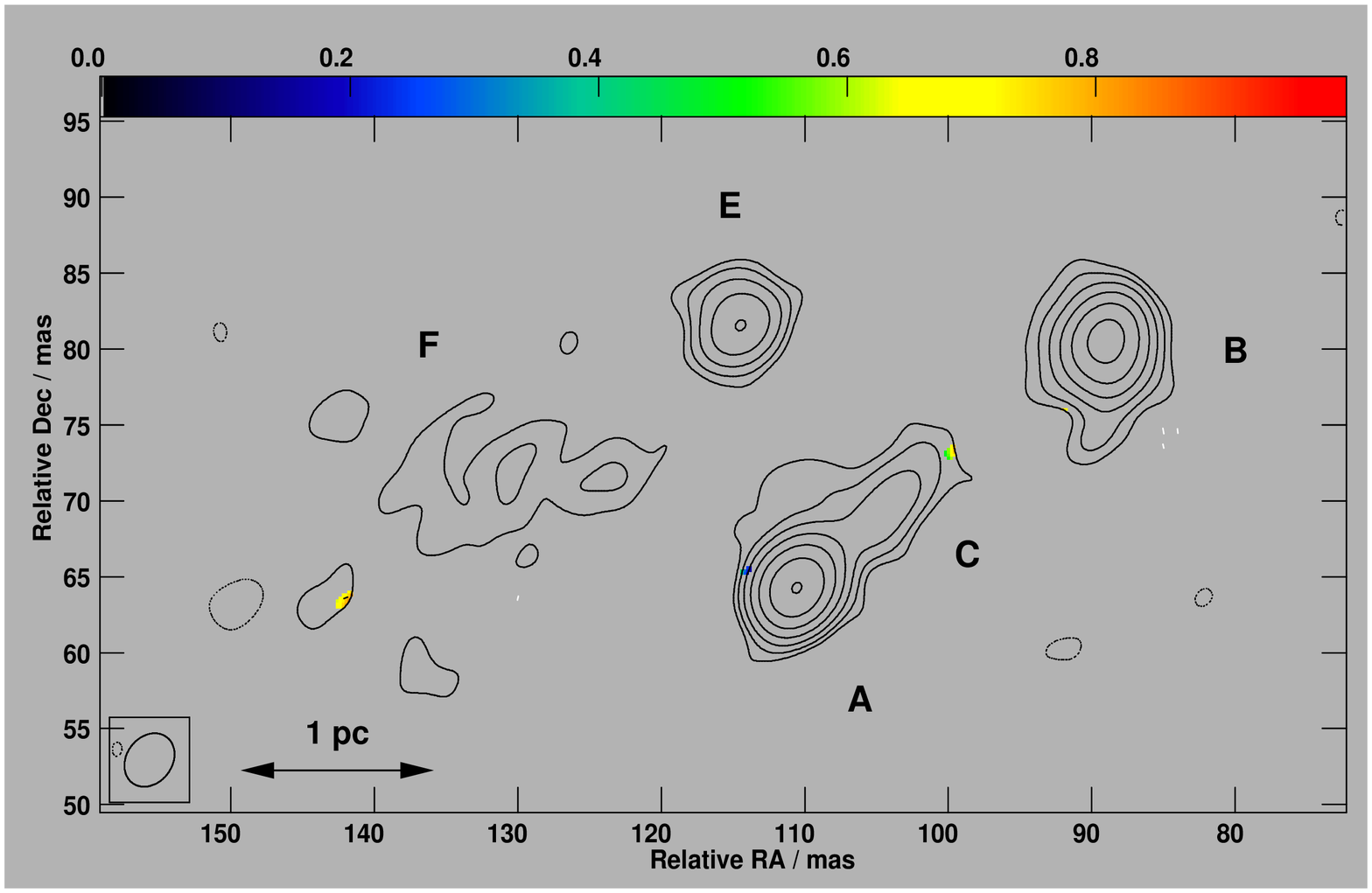}
 \caption[NGC\,3079 5.0\,GHz  image]{NGC\,3079 5.0\,GHz VLBA image from September 2002.
 Peak $21.64\,{\rm mJy\,beam^{-1}}$, contours
 $\rm{0.33\,mJy\,beam^{-1}}\times2^N$, nat. weighting,
 beam $3.86\,\mas\times3.11\,\mas$ in P.A. $-40^{\circ}$, 
 pol. ticks  $1\,{\rm mJy\,beam^{-1}\,mas^{-1}}$ where
 $>5\,\sigma=0.27\,{\rm
 mJy\,beam^{-1}}$. Axes relative to RA 10~01~57.7906 Dec
 55~40~47.1788 (J2000), colours show fractional polarization.}
 \label{fig:NGC3079C_it3_polmap}
\end{figure}

\begin{figure}[ht!]
\centering
\includegraphics[width=11cm]{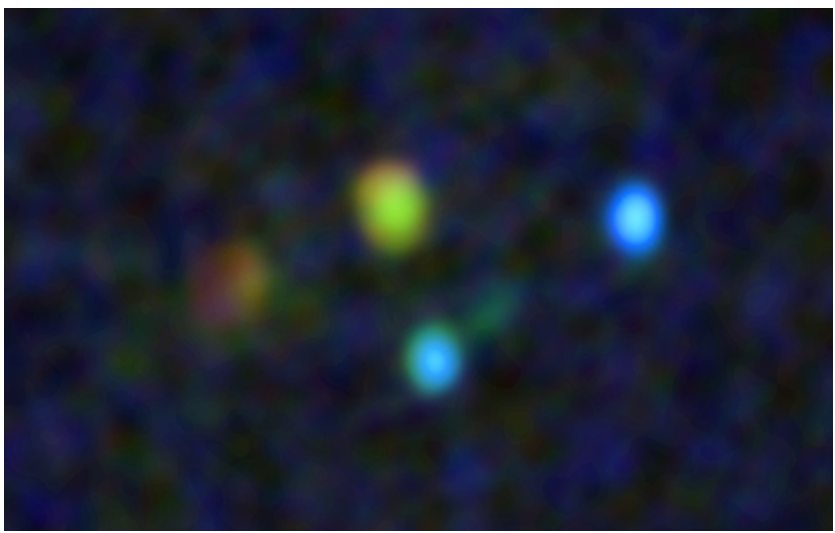}
 \caption[Four-colour image of NGC\,3079]{Four-colour image of
 NGC\,3079, composed of the observations from 2002 and the 15.4\,GHz
 observations from November 2000. 1.7\,GHz  is represented
 as red, 2.3\,GHz as yellow, 5.0\,GHz as green and 15.4\,GHz as
 blue.}
 \label{fig:NGC3079_4colors}
\end{figure}

\begin{figure}[ht!]
 \centering
 \includegraphics[width=12cm]{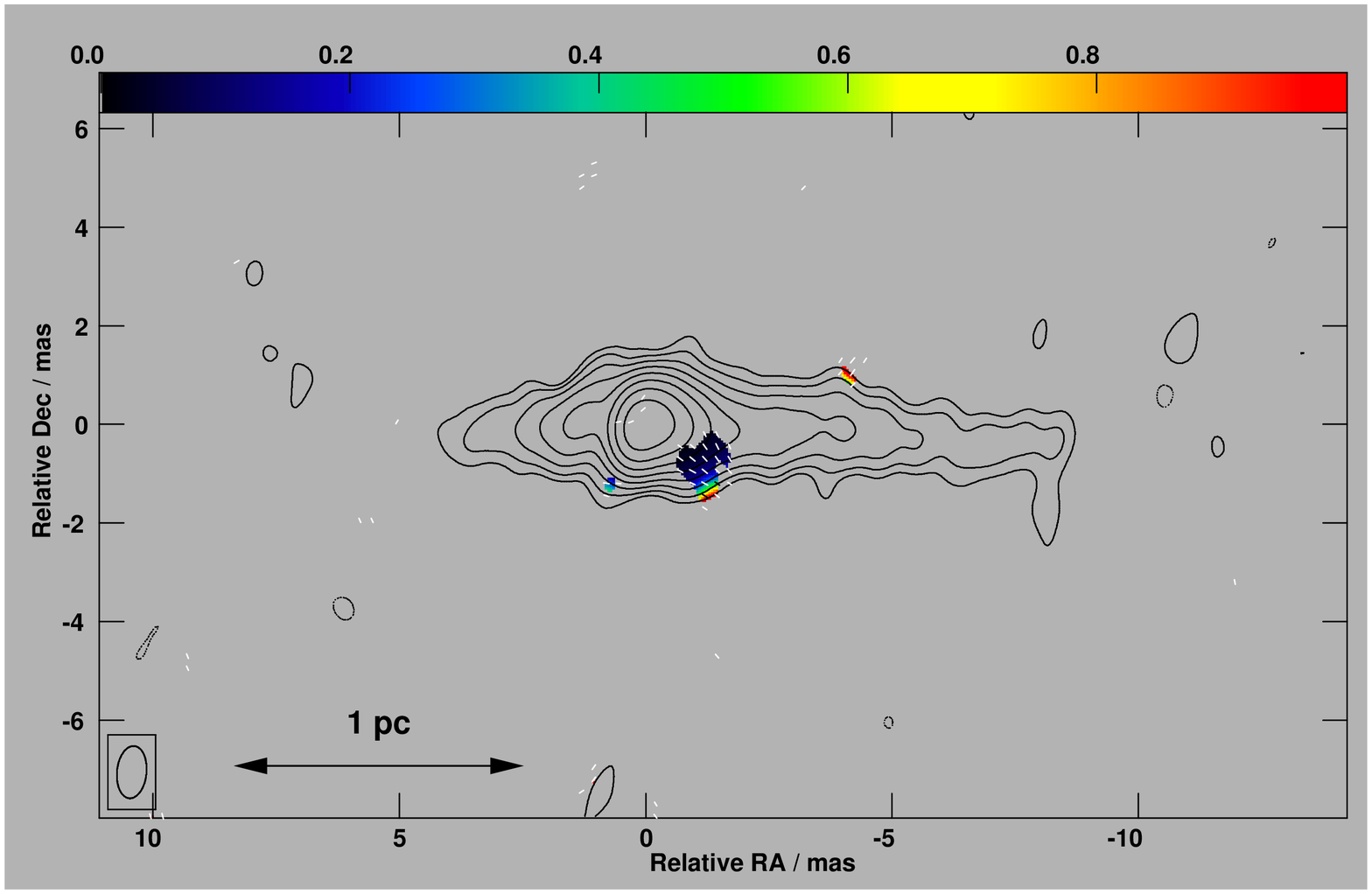}
 \caption[NGC\,4261 15.4\,GHz image.]{NGC\,4261 15.4\,GHz VLBA image. 
 Peak is $129\,{\rm mJy\,beam^{-1}}$, contours are
 at $\rm{0.54\,mJy\,beam^{-1}}\times2^N$, uniform weighting,
  beam is $1.07\,\mas\times0.60\,\mas$ in P.A. $-6^{\circ}$,
 polarization ticks are $5\,{\rm mJy\,beam^{-1}\,mas^{-1}}$ 
 where $>5\,\sigma=0.45\,{\rm mJy\,beam^{-1}}$, colours show fractional polarization.}
 \label{fig:4261CL10it2_1146}
\end{figure}

\begin{figure}[ht!]
 \centering
 \includegraphics[width=10cm]{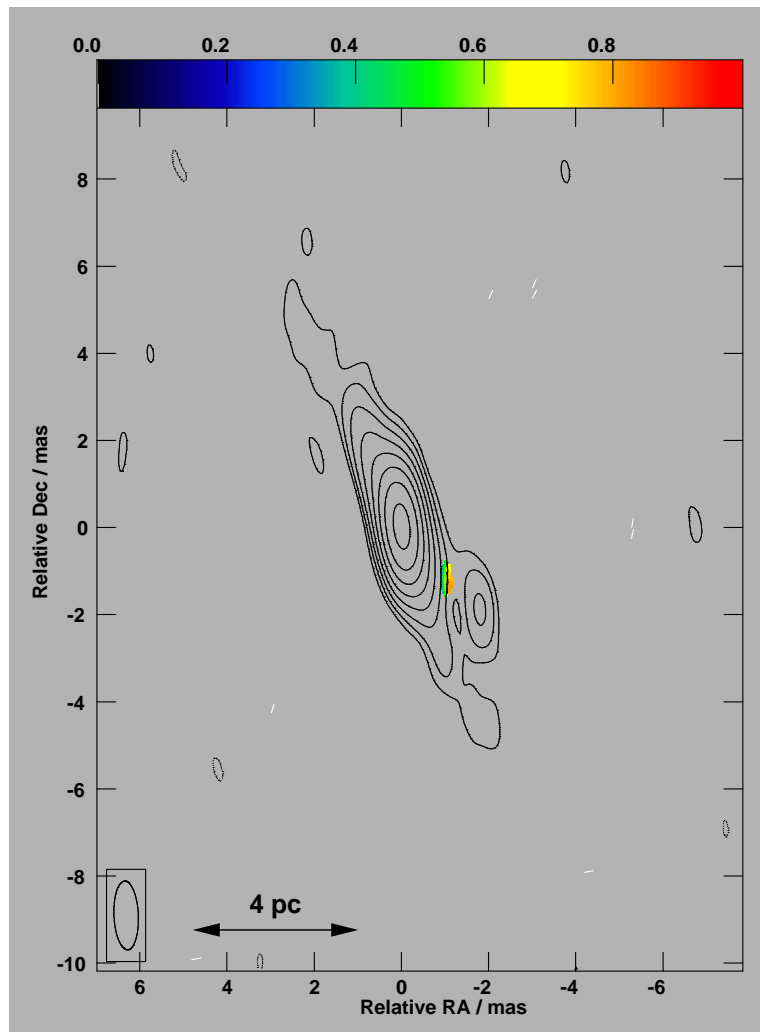}
 \caption[Hydra~A 15.4\,GHz  image]{Hydra~A 15.4\,GHz VLBA image. Peak is
 $144\,{\rm mJy\,beam^{-1}}$, contours are at
 $\rm{0.87\,mJy\,beam^{-1}}\times2^N$,  uniform weighting,
  beam is $1.58\,\mas\times0.56\,\mas$ in P.A. $3^{\circ}$. The length of the
 polarization ticks is $5\,{\rm mJy\,beam^{-1}\,mas^{-1}}$ and start
 at the polarization $5\,\sigma$ level of $1.00\,{\rm mJy\,beam^{-1}}$.
 Colours show fractional polarization.}
 \label{fig:hyd_a_it3_1146}
\end{figure}

The final images are shown in Figs.~\ref{fig:NGC3079AC} to
\ref{fig:Cyg_A}. Figs.~\ref{fig:NGC3079L_it2_polmap}-\ref{fig:Cyg_A} have
superimposed polarization ticks, except for the NGC\,3079 2.3\,GHz
image in Fig.~\ref{fig:NGC3079S_it1}, and display the fractional
polarization as colours. The lowest contours have been set to three
times the image rms noise, and the polarization ticks have been cut
off below five times the polarization image rms, except for the
Cygnus~A image, where the polarization has been cut off below seven
times the polarization image rms to emphasise the detection.

\section{The Images}

\begin{sidewaystable*}[htpb]
\scriptsize
\begin{center}
\begin{tabular}{lccccccccccc}
\hline
\hline
                                  &              &                                & \multicolumn{3}{c}{Beam parameters}\\
\cline{4-6}
\multicolumn{1}{c}{Source}        & Epoch        &  \multicolumn{1}{c}{Frequency} & \multicolumn{1}{c}{$b_{\rm maj}$} & \multicolumn{1}{c}{$b_{\rm min}$} & \multicolumn{1}{c}{P.A.}    & \multicolumn{1}{c}{Peak}                   & \multicolumn{1}{c}{rms}   & Dyn.         & \multicolumn{1}{c}{P-Peak}                 & \multicolumn{1}{c}{P-rms} &   Dyn.\\
\cline{7-8}\cline{10-11}
                                  &              &  \multicolumn{1}{c}{GHz}       & \multicolumn{1}{c}{mas}           & \multicolumn{1}{c}{mas}           & \multicolumn{1}{c}{degrees} &                   \multicolumn{2}{c}{${\rm mJy\,beam^{-1}}$}           & range        &                    \multicolumn{2}{c}{${\rm mJy\,beam^{-1}}$}          &  range\\
\multicolumn{1}{c}{(1)}&
\multicolumn{1}{c}{(2)}&
\multicolumn{1}{c}{(3)}&
\multicolumn{1}{c}{(4)}&
\multicolumn{1}{c}{(5)}&
\multicolumn{1}{c}{(6)}&
\multicolumn{1}{c}{(7)}&
\multicolumn{1}{c}{(8)}&
\multicolumn{1}{c}{(9)}&
\multicolumn{1}{c}{(10)}&
\multicolumn{1}{c}{(11)}&
\multicolumn{1}{c}{(12)}\\
\hline
NGC\,3079     & Nov 20, 1999 &   5.0  & 1.92 & 1.07 & -30  &  20     &  0.08  & 270:1 &          &       &      \\ 
              &              &  15.4  & 3.10 & 1.66 &  36  &  45     &  0.34  & 140:1 &          &       &      \\ 
NGC\,3079     & Mar  6, 2000 &   5.0  & 1.20 & 0.85 & -29  &  17.1   &  0.06  & 270:1 &          &       &      \\ 
              &              &  15.4  & 2.01 & 1.33 & -56  &  51     &  0.35  & 150:1 &          &       &      \\ 
NGC\,3079     & Nov 30, 2000 &   5.0  & 1.40 & 0.87 & -23  &  17.8   &  0.06  & 290:1 &          &       &      \\ 
              &              &  15.4  & 2.25 & 1.69 & -24  &  51     &  0.27  & 210:1 &          &       &      \\     
NGC\,3079     & Sep 22, 2002 &  1.7   & 5.12 & 3.49 & -22  &  9.1    &  0.06  & 140:1 & 0.29$^a$ & 0.04  & 7.3  \\ 
              &              &  2.3   & 6.02 & 4.34 &  12  &  19.8   &  0.13  & 150:1 \\          
              &              &  5.0   & 3.86 & 3.11 & -40  &  22     &  0.11  & 200:1 & 0.43$^a$ & 0.05  & 8.6  \\ 
NGC\,1052     & Dec 28, 2001 &  13.4  & 1.49 & 0.58 &  -4  &  344    &  0.36  & 960:1 & 1.36$^a$ & 0.15  & 9.1  \\ 
              &              &  15.4  & 1.33 & 0.50 &  -5  &  245    &  0.32  & 770:1 & 1.39$^a$ & 0.16  & 8.7  \\ 
NGC\,4261     & Jul 5, 2002  &        & 1.07 & 0.60 &  -6  &  129    &  0.18  & 720:1 & 0.83     & 0.09  & 9.2  \\ 
Hydra\,A      & Jul 5, 2002  &        & 1.58 & 0.56 &   3  &  144    &  0.29  & 500:1 & 1.81$^a$ & 0.20  & 9.1  \\  
Centaurus\,A  & Jul 5, 2002  &  13.4  & 2.09 & 0.47 &   0  &  1080   & 10.87  & 100:1 & 7.98     & 0.71  & 11.2 \\ 
              &              &  15.4  & 2.15 & 0.42 &   1  &  907    &  7.46  & 120:1 & 9.91     & 1.05  & 9.4  \\ 
Cygnus\,A     & Jul 7, 2002  &  15.4  & 0.59 & 0.34 & -21  &  316    &  0.23  & 1400:1& 1.39     & 0.11  & 12.6 \\
\hline
\end{tabular}
\caption[Image parameters]{Image parameters. $a$: the peak in polarized
intensity is outside the image shown.}
\label{tab:imageparms}
\end{center}
\end{sidewaystable*}

\subsection{NGC\,3079}

\begin{figure}[ht!]
\centering
\includegraphics[width=11cm]{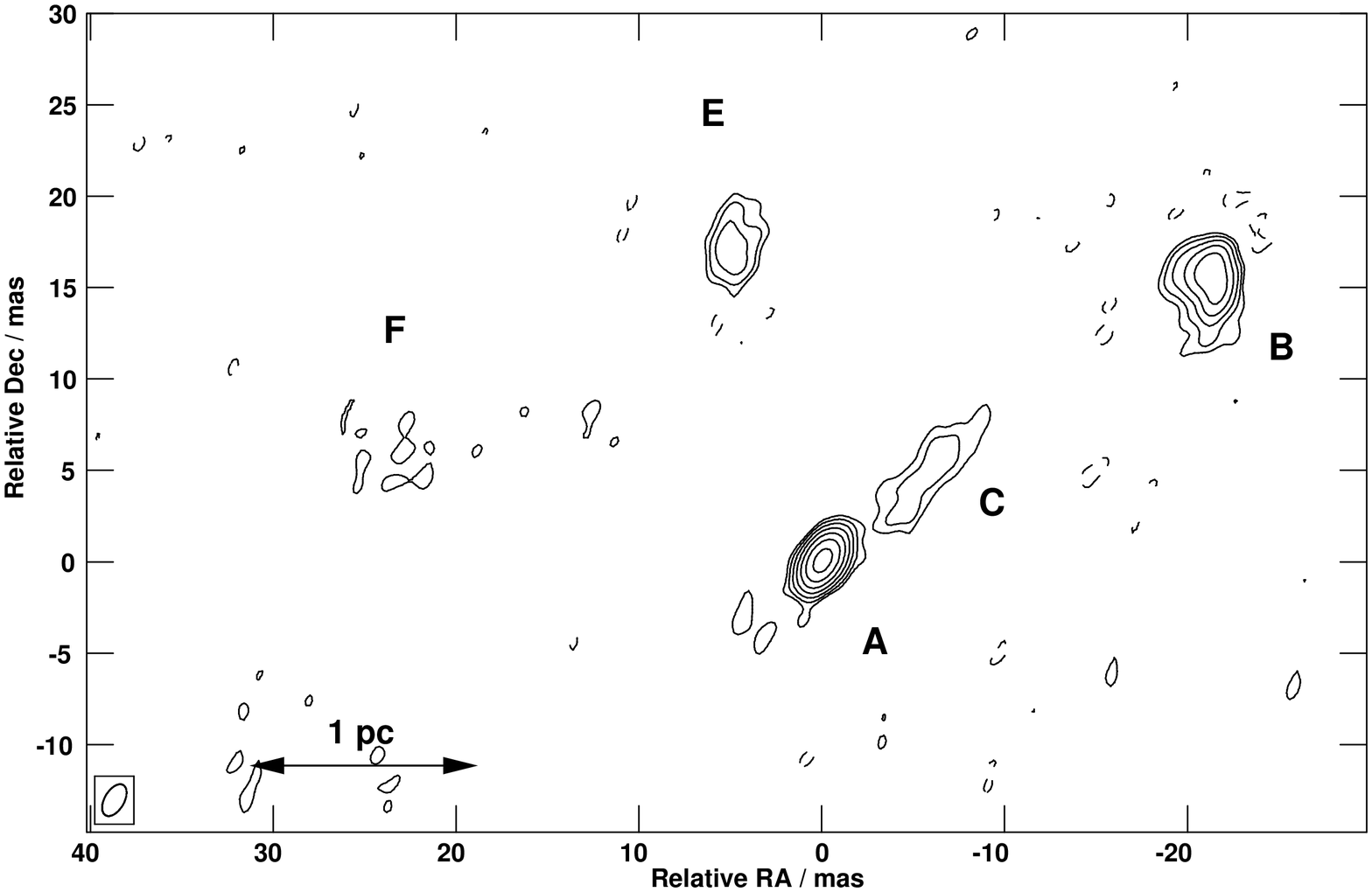}
\caption[NGC\,3079 5.0\,GHz image from November 20, 1999]{NGC\,3079 5.0\,GHz
VLBA+EB+Y1 image from November 20, 1999. Peak is $20\,{\rm mJy\,beam^{-1}}$, contours are at
$\rm{0.23\,mJy\,beam^{-1}}\times2^N$, natural weighting, beam is
$1.92\,\mas\times1.07\,\mas$ in P.A. $-30^{\circ}$.}
\label{fig:NGC3079AC}
\end{figure}

\begin{figure}[ht!]
\centering
\includegraphics[width=11cm]{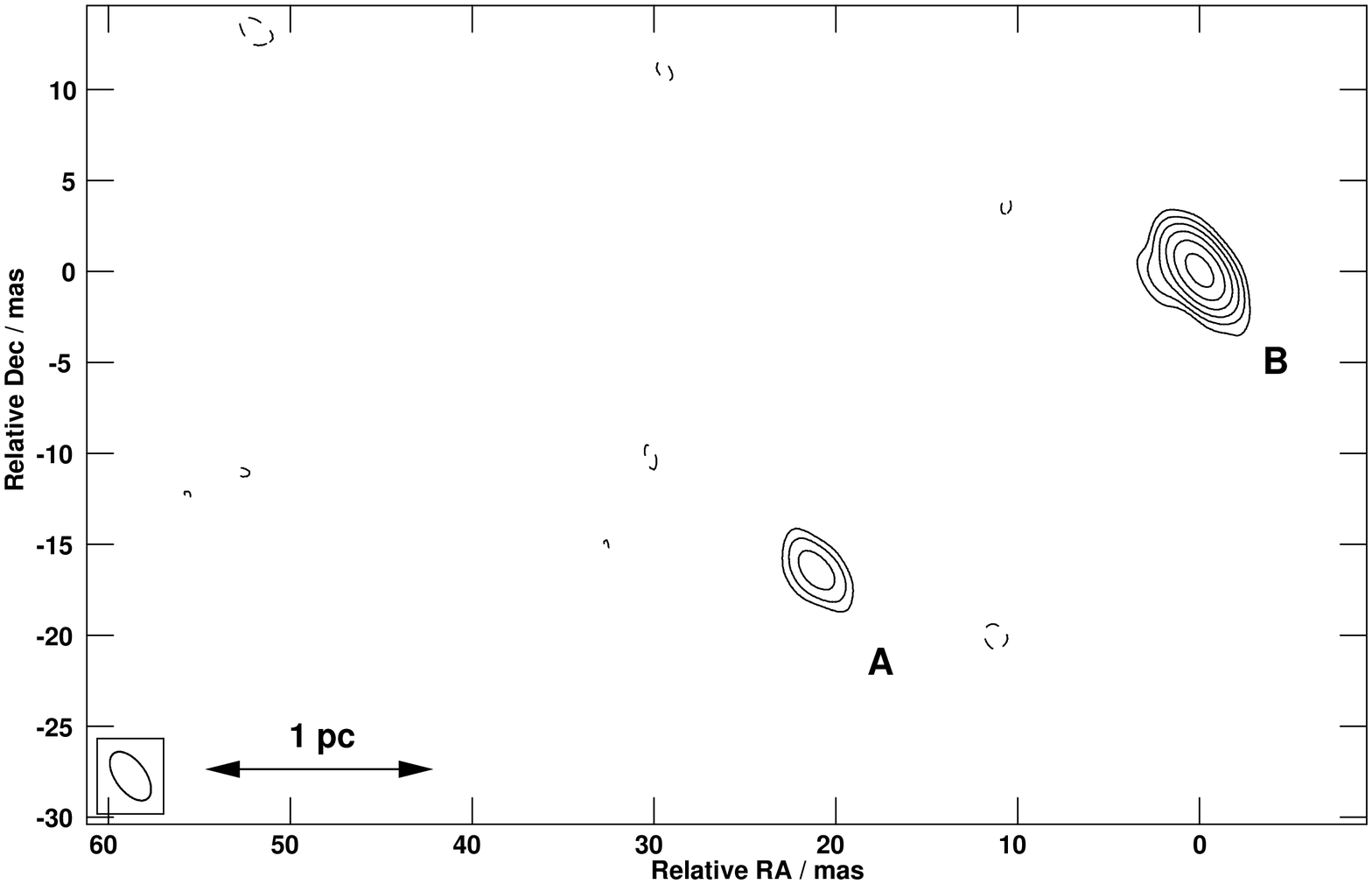}
\caption[NGC\,3079 15.0\,GHz image from November 20, 1999]{NGC\,3079 15.0\,GHz VLBA+EB+Y1
image from November 20, 1999. Peak is $45\,{\rm mJy\,beam^{-1}}$, contours are at
$\rm{1.02\,mJy\,beam^{-1}}\times2^N$, natural weighting, beam is
$3.10\,\mas\times1.66\,\mas$ in P.A. $36^{\circ}$.}
\label{fig:NGC3079AU}
\end{figure}

\begin{figure}[ht!]
\centering
 \includegraphics[width=11cm]{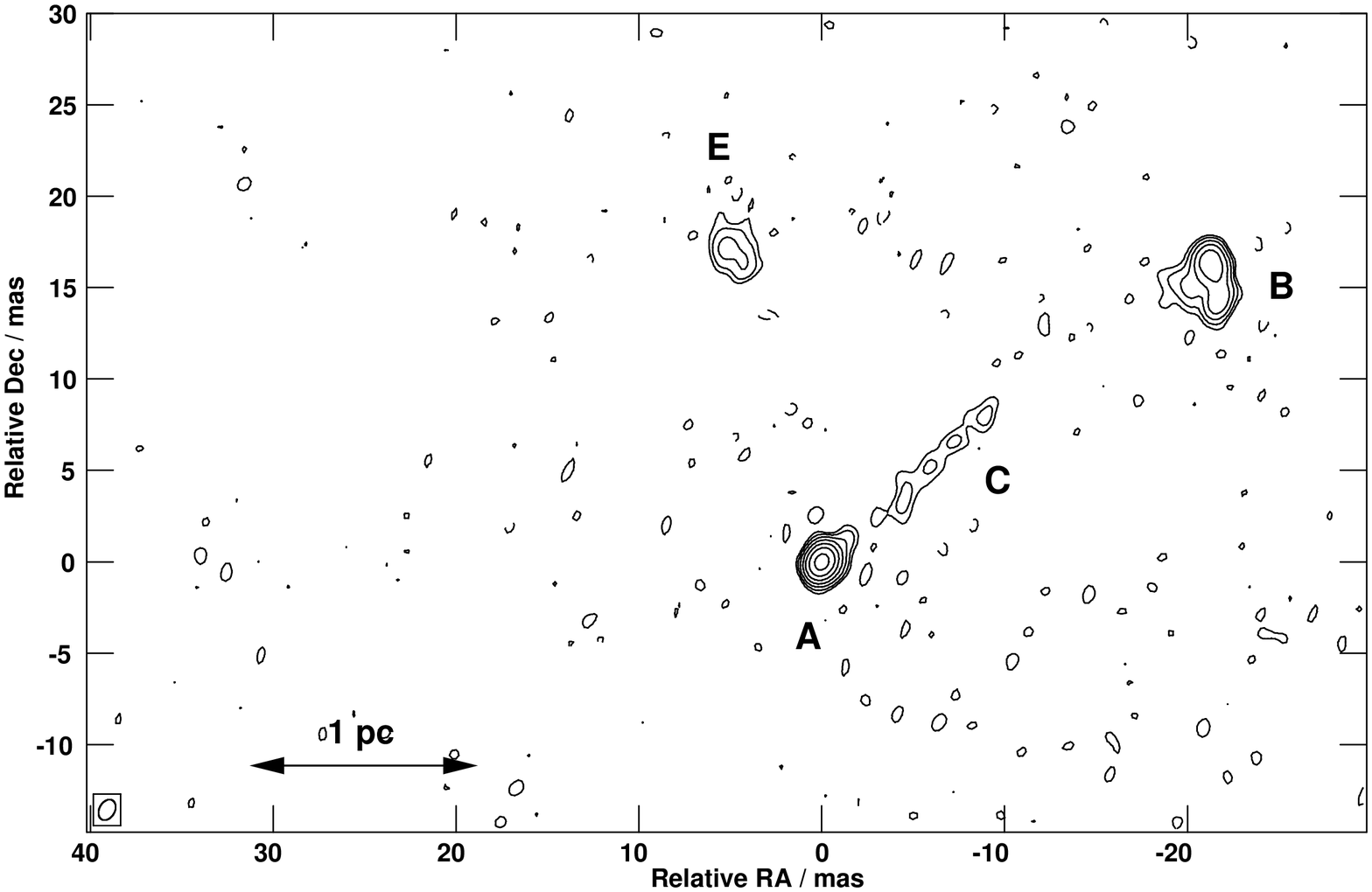}
 \caption[NGC\,3079 5.0\,GHz  image from March 6, 2000]{NGC\,3079 5.0\,GHz
 VLBA+EB+Y1 image from March 6, 2000. Peak is $17.1\,{\rm mJy\,beam^{-1}}$, contours are
 at $\rm{0.19\,mJy\,beam^{-1}}\times2^N$, natural weighting, beam is
 $1.20\,\mas\times0.85\,\mas$ in P.A. $-29^{\circ}$.}
 \label{fig:NGC3079BC}
\end{figure}

\begin{figure}[ht!]
\centering
 \includegraphics[width=11cm]{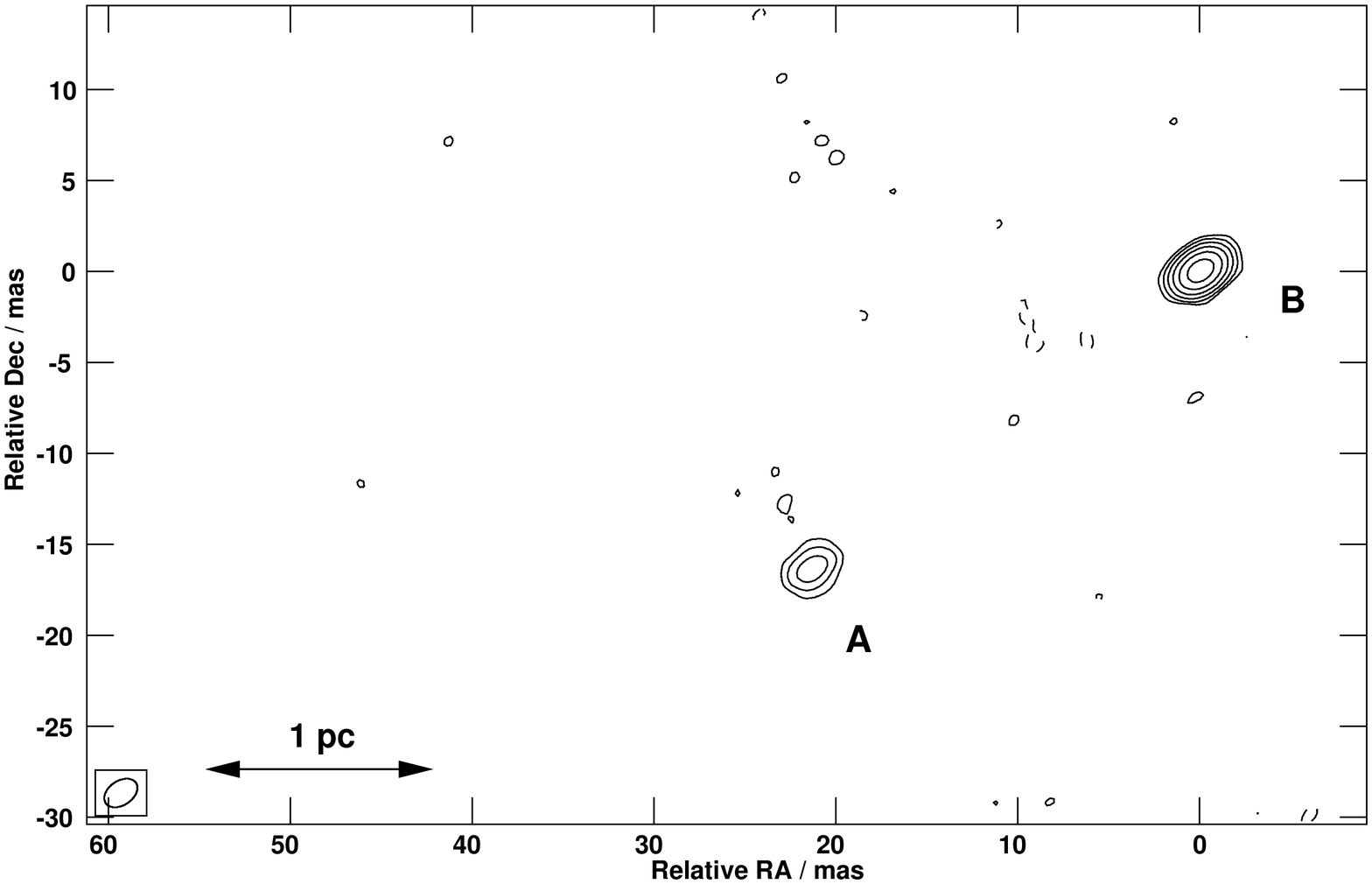}
 \caption[NGC\,3079 15.0\,GHz image from March 6, 2000]{NGC\,3079 15.0\,GHz VLBA+EB+Y1
 image from March 6, 2000. Peak is $51\,{\rm mJy\,beam^{-1}}$, contours
 are at $\rm{1.05\,mJy\,beam^{-1}}\times2^N$, natural
 weighting, beam is $2.01\,\mas\times1.33\,\mas$ in
 P.A. $-56^{\circ}$.}
 \label{fig:NGC3079BU}
\end{figure}

\begin{figure}[ht!]
\centering
 \includegraphics[width=11cm]{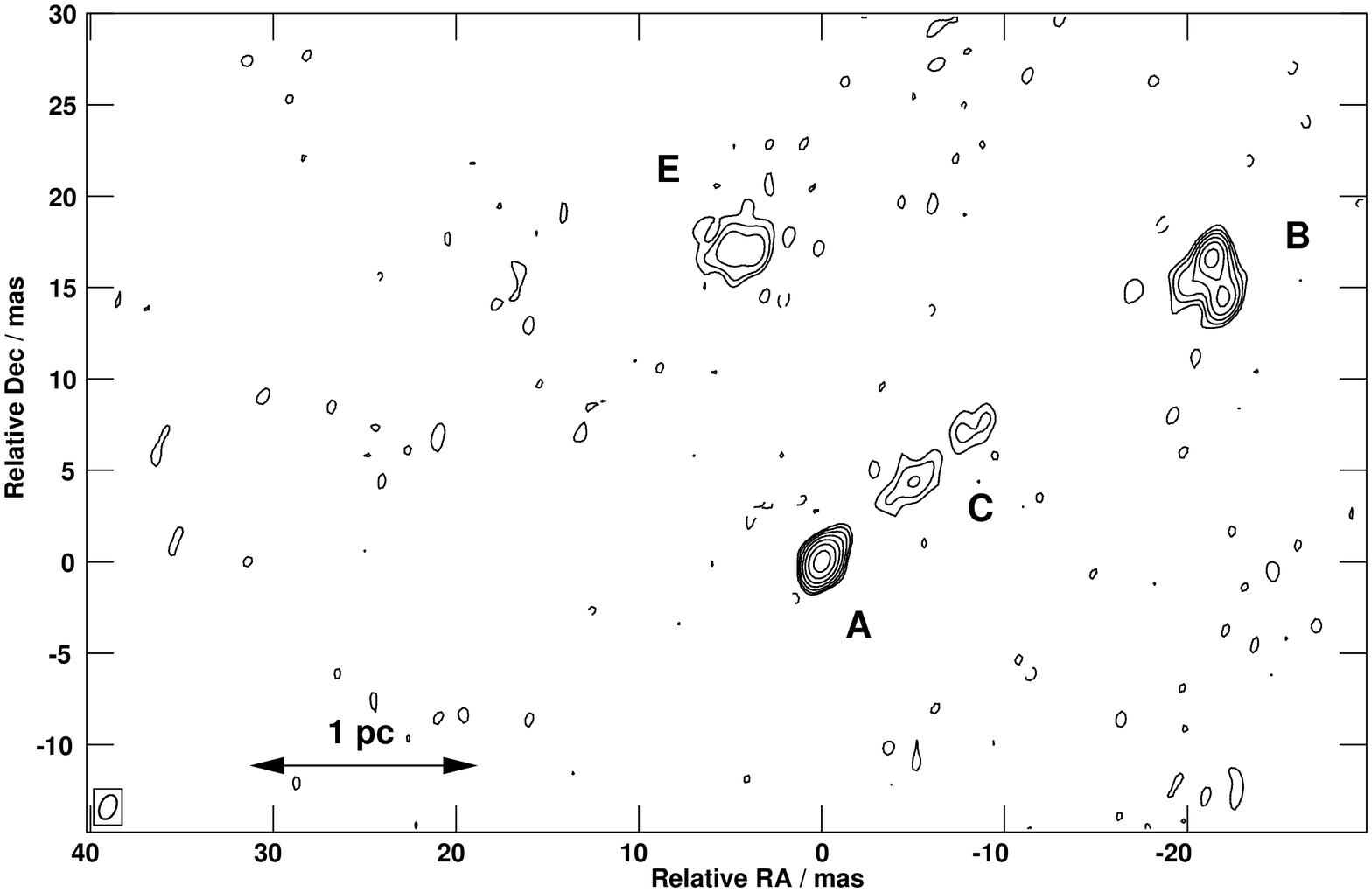}
 \caption[NGC\,3079 5.0\,GHz image from November 30, 2000]{NGC\,3079 5.0\,GHz
 VLBA+EB+Y1 image from November 30, 2000. Peak is $17.8\,{\rm mJy\,beam^{-1}}$, contours are
 at $\rm{0.18\,mJy\,beam^{-1}}\times2^N$, natural weighting, beam is
 $1.40\,\mas\times0.87\,\mas$ in P.A. $-23^{\circ}$.}
 \label{fig:NGC3079CC}
\end{figure}

\begin{figure}[ht!]
\centering
 \includegraphics[width=11cm]{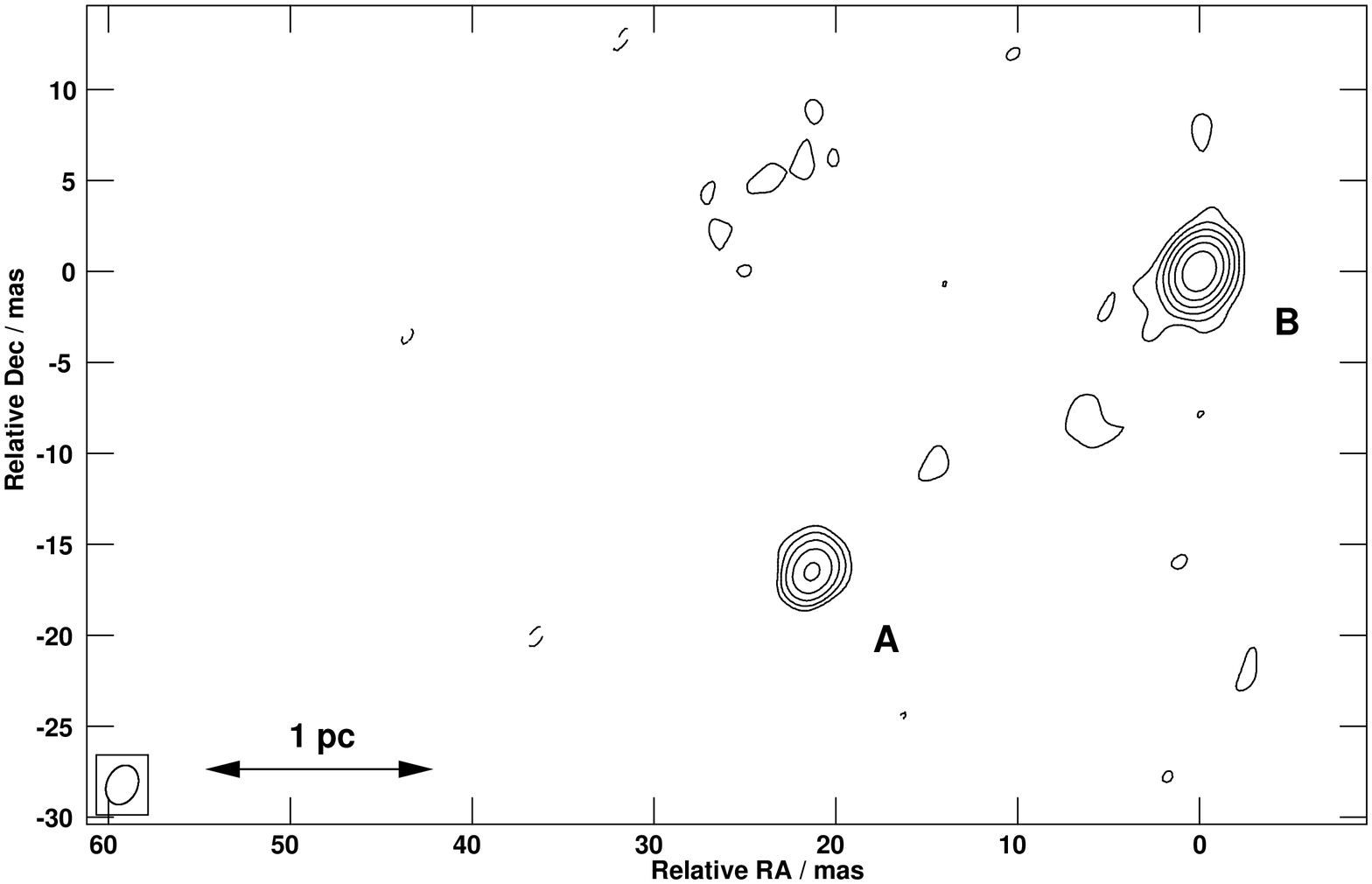}
 \caption[NGC\,3079 15.0\,GHz image from November 30, 2000]{NGC\,3079 15.0\,GHz
 VLBA+EB+Y1 image from November 30, 2000. Peak is $51\,{\rm mJy\,beam^{-1}}$, contours are at
 $\rm{0.82\,mJy\,beam^{-1}}\times2^N$, natural weighting, beam is
 $2.25\,\mas\times1.69\,\mas$ in P.A. $-24^{\circ}$.}
 \label{fig:NGC3079CU}
\end{figure}

The first three epochs of NGC\,3079 shown in
Figs.~\ref{fig:NGC3079AC}-\ref{fig:NGC3079CU} were observed using
identical setups and have comparable quality. They show three
components at 5.0\,GHz and two components at 15.4\,GHz. These epochs
have about twice the resolution of the fourth epoch, because they
included Effelsberg at 5.0\,GHz and 15.4\,GHz.

\begin{itemize}

\item{5.0\,GHz}~~Our images display three compact components and
an elongated feature between the southern and the western
feature. Based on this morphology and adopting the notation from
\cite{Irwin1988}, we identify the components as $A$ (south-east), $B$
(west) and $C$ (in between $A$ and $B$). The compact component north
of $A$ was not detected by earlier 5\,GHz observations, and we name it
$E$. Component $A$ is the strongest component at 5.0\,GHz, causing
fringe-fitting to shift it to the image centre. It is compact and
shows only a little extension to the north-west, in direction of $B$
and the jet-like feature $C$ on the $A$-$B$ axis. Component $B$ is
resolved, which is clearly visible in the second and third epoch,
where it is resolved into three sub-components. Component $E$ is also
resolved, and its major axis has rotated from north-south in the first
epoch (Fig.~\ref{fig:NGC3079AC}) to east-west in the third epoch
(Fig.~\ref{fig:NGC3079CC}). The jet-like component $C$ is continuous
in the first and second epoch and resolved into sub-components in the
third epoch.

\item{15.4\,GHz}~~Our images show two compact components. Based on the
separation and relative P.A. of the two matching that of $A$ and $B$
at 5.0\,GHz, we identify the component towards the east with $A$ and
the component towards the west with $B$. Our identification is further
supported by 22\,GHz observations by \cite{Trotter1998} who detected
$A$ and $B$ with similar separation and relative P.A., and by 8.4\,GHz
observations by \cite{Kondratko2003}, who noticed that $E$ is fading
towards higher frequencies and probably becoming invisible at 15\,GHz
and that $A$ and $B$ were still strong at this frequency.  Thus, our
identification appears to be correct, although we detect neither $E$
nor $C$ at 15.4\,GHz. $A$ and $B$ are compact in all epochs, and
component $B$ is the strongest at 15.4\,GHz, so it is in the image
centre. Fringe-fitting flags data that have too low SNR, and the
remaining data of the first two epochs (Figs.~\ref{fig:NGC3079AU} and
\ref{fig:NGC3079BU}) were distributed over a pronounced elliptical
$(u,v)$ coverage, causing an almost $90^{\circ}$ change in position
angle of the synthesized beam. This also caused the change in position
angle of the major axes of $A$ and $B$. No detections were made on
baselines to Effelsberg, Mauna Kea, Hancock and Y1 during the first
epoch, to Effelsberg, Mauna Kea and St. Croix during the second epoch
and to Brewster, Effelsberg, Hancock, Mauna Kea and St. Croix in the
third epoch. These are predominantly long ($>100\,{\rm M\lambda}$)
baselines, so that the resolution at 15.4\,GHz during these epochs is
lower than at 5.0\,GHz.

The flux densities of $A$ and $B$ changed on a significant level,
especially between epoch two and three (Figs.~\ref{fig:NGC3079BU} and
\ref{fig:NGC3079CU}), where $A$ brightened remarkably. The ratio of
the peak flux densities, $S_{\nu, A}/S_{\nu, B}$, is $0.19\pm0.01$ in
epoch one, $0.11\pm0.01$ in epoch two and $0.28\pm0.02$ in epoch
three. The changes have a significance of $9\,\sigma$ and $8\,\sigma$,
respectively, and cannot be due to amplitude calibration errors
because they scale the flux density of both components by the same
factor, leaving the ratio constant.

\end{itemize}

In the fourth epoch of NGC\,3079, Effelsberg did not observe at
2.3\,GHz and 5.0\,GHz, and so the beam size was significantly larger
than in the first three epochs.

\begin{itemize}

\item{1.7\,GHz (Fig.~\ref{fig:NGC3079L_it2_polmap})}~~Two components
are visible, a more compact one to the north-west and an extended one
to the south-east. The south-eastern component, we named $F$, is the
strongest at 1.7\,GHz but is barely visible at 5.0\,GHz. Due to their
similar separation and relative position angle, $E$ and $F$ are easily
misidentified with $A$ and $B$ (as unfortunately happened to
\citealt{Sawada-Satoh2000}), but phase-referencing allowed us to
identify the north-western component with component $E$ seen at
5.0\,GHz during the same epoch. All components are unpolarized, and
neither $A$, nor $B$ nor $C$ were detected at 1.7\,GHz.

\item{2.3\,GHz (Fig.~\ref{fig:NGC3079S_it1})}~~Three components
are visible. Component $E$ is the strongest, while $F$ has become much
fainter compared to the detection at 1.7\,GHz. Component $A$, although
weak, is reliably detected, whereas component $B$ is not seen.

\item{5.0\,GHz (Fig.~\ref{fig:NGC3079C_it3_polmap})}~~The well-known
triple structure appears to be almost unchanged since the previous
epoch in 2000. Component $A$ is compact, component $B$ is slightly
extended in the north-south direction, and $E$ is extended
east-west. Due to the lower resolution, $C$ appears to be connected to
$A$ and is not resolved into sub-components. $F$ is very weak at
5.0\,GHz. All components are unpolarized.

\end{itemize}

Component $D$ seen by \cite{Trotter1998} was not detected in any of
our observations. 

The observations from 2002 and the 15.4\,GHz observation from November
2000 were used to compose a four-colour image of NGC\,3079, in which
red corresponds to 1.7\,GHz, yellow to 2.3\,GHz as, green to 5.0\,GHz
and blue to 15.4\,GHz (Fig.~\ref{fig:NGC3079_4colors}). We prefer this
somewhat arbitrary transfer function of frequencies to colours because
a linear transfer function would map the frequencies of 1.7\,GHz
to 5.0\,GHz into those 35\,\% of the visible wavelengths (380\,nm to
780\,nm) that appear as red. This is due to the large gap between
5.0\,GHz and 15.4\,GHz. The colour representation that we chose
emphasises the spectral differences of the components.

\subsubsection{Additional Results from NGC\,3079}

The observations presented in this thesis allow a more detailed study
of NGC\,3079 than for the other five objects. To parameterize the
observations, integrated flux densities and component positions have
been measured (Table~\ref{tab:ngc3079_parms1}. The images in this
chapter display the ``best'' images with full resolution obtained from
each particular dataset. To minimise resolution effects and peak
shifts due to changing imaging parameters, integrated flux densities
and positions have been measured from uniformly weighted images
tapered to a common resolution of $4\,\mas\times3\,\mas$ (not shown).

The integrated flux densities have then been measured by summing over
the pixels of the source region. This method is also sensitive to very
weak and possibly extended emission which is below the lowest contour
in the images shown. If a well-defined component was visible,
positions have been measured by fitting a two-dimensional parabola to
the brightest and the eight surrounding pixels from the images tapered
to $4\,\mas\times3\,\mas$ resolution. The fits also yielded peak flux
densities. This procedure was found superior to fitting Gaussians
because the sources were not always well described by a Gaussian.

In the cases of component $F$ at 5.0\,GHz and component $B$ at
2.3\,GHz, no component is visible in the contour plots (i.e., the peak
flux density was below $3\,\sigma$), but significant flux density was
measured over the regions where they are visible at other
frequencies. In the case of $F$ at 5.0\,GHz, the emission is weak and
probably also resolved due to the lack of short $(u,v)$ spacings. We
therefore give the flux densities in Table~\ref{tab:ngc3079_parms1},
but do not give coordinates, because our procedure to measure peak
flux densities did not yield meaningful results in these cases.  We
estimate the flux density errors to be of the order of 5\,\%, in
agreement with the VLBA's observational status
summary\footnote{\url{http://www.aoc.nrao.edu/vlba/obstatus/obssum/obssum.html}}
and with what is commonly adopted in the literature for VLBA
observations. Exploring five images from a reasonable imaging
parameter space revealed a position uncertainty of 0.15\,mas
($1\,\sigma$).\\

The data presented in Table~\ref{tab:ngc3079_parms1} can be used to
derive spectral indices and component positions. Only the fourth epoch
was phase-referenced and provides absolute coordinates. For the proper
motion analysis, we therefore give the relative separations among the
three components $A$, $B$ and $E$, because they are well-defined and
visible during all epochs. Spectral indices and relative positions are
given in Tables~\ref{tab:ngc3079_spectra} and
\ref{tab:ngc3079_separations}. Assuming a $1\,\sigma$ error of 5\,\% for
the integrated flux densities yields spectral index errors of 0.09 for
the 5.0\,GHz/15.4\,GHz and 1.7\,GHz/5.0\,GHz pairs, 0.33 for the
1.7\,GHz/2.3\,GHz pair and 0.13 for the 2.3\,GHz/5.0\,GHz pair. Each
position has an error of 0.15\,mas, so that separations have errors of
$\sqrt{2}\times0.15\,{\rm mas}=0.21\,{\rm mas}$.


\begin{table}[ht!]
\scriptsize
\begin{center}
\begin{tabular}{lr|ccc|ccccc}
\hline
\hline
             &        &\multicolumn{3}{c|}{full resolution}& \multicolumn{5}{c}{tapered}        \\ 
Epoch        & Freq.  & $S$  & $S_{\rm max}$ & $x$  & $S$    & $S_{\rm max}$ & $x$   & RA       & Dec     \\
\hline                                         	                                                        
&&\multicolumn{8}{c}{Component $A$}\\    
Nov 20, 1999 &   5.0  & 27   & 21            & 0.78 & 28     & 25            & 0.89  & -        & -	  \\   
             &  15.4  & 9.59 & 6.30          & 0.66 & 9.03   & 6.21          & 0.69  & -        & -	  \\
Mar  6, 2000 &   5.0  & 26   & 17            & 0.65 & 26     & 24            & 0.92  & -        & -	  \\
             &  15.4  & 14.8 & 6.81          & 0.46 & 6.75   & 7.02          & 1.04  & -        & -	  \\
Nov 30, 2000 &   5.0  & 23   & 17.8          & 0.77 & 24     & 23            & 0.96  & -        & -	  \\
             &  15.4  & 19.0 & 15.1          & 0.79 & 18.4   & 17.5          & 0.95  & -        & -	  \\  
Sep 22, 2002 &  1.7   & -    & -             &      & -      & -             & -     & -        & -       \\
             &  2.3   & 1.88 & 1.24          & 0.66 &  2.45  & 1.13          & 0.46  & 57.80373 & 47.2436 \\
             &  5.0   & 25   & 22            & 0.88 & 26     & 22            & 0.85  & 57.80368 & 47.2430 \\
\hline                                         	                                                        
&&\multicolumn{8}{c}{Component $B$}\\                                          
Nov 20, 1999 &   5.0  & 14.9 & 6.92          & 0.46 & 16.9   & 13.9          & 0.82  & -	& -	  \\
             &  15.4  & 51   & 45            & 0.88 & 48     & 46            & 0.96  & -	& -	  \\  
Mar  6, 2000 &   5.0  & 19.4 & 5.86          & 0.30 & 17.9   & 14.6          & 0.82  & -	& -	  \\  
             &  15.4  & 59   & 51            & 0.86 & 59     & 54            & 0.92  & -	& -	  \\
Nov 30, 2000 &   5.0  & 19.2 & 8.05          & 0.42 & 18.7   & 14.6          & 0.78  & -	& -	  \\
             &  15.4  & 68   & 51            & 0.75 & 66     & 62            & 0.94  & -	& -	  \\
Sep 22, 2002 &  1.7   & -    & -             &      & -      & -             & -     & -        & -       \\
             &  2.3   & -    & -             &      &  0.64  & -             & -     & -        & -       \\
             &  5.0   & 22   & 14.4          & 0.66 & 22     & 14.8          & 0.67  & 57.80112 & 47.2594 \\ 
\hline                                                                                                           
&&\multicolumn{8}{c}{Component $E$}\\                                                 
Nov 20, 1999 &   5.0  & 4.14 & 1.61          & 0.39 & 4.12   & 3.45          & 0.84  & -        & -	  \\
             &  15.4  & -    & -             &      & -      & -             & -     & -        & -	  \\
Mar  6, 2000 &   5.0  & 3.28 & 1.04          & 0.32 & 2.99   & 3.22          & 1.08  & -        & -	  \\
             &  15.4  & -    & -             &      & -      & -             & -     & -        & -	  \\
Nov 30, 2000 &   5.0  & 7.18 & 1.42          & 0.20 & 5.14   & 3.96          & 0.77  & -        & -	  \\
             &  15.4  & -    & -             &      & -      & -             & -     & -        & -       \\
Sep 22, 2002 &  1.7   & 5.43 & 4.23          & 0.78 &  3.75  & 2.94          & 0.78  & 57.80429 & 47.2633 \\
             &  2.3   & 21   & 19.8          & 0.94 & 18.0   & 19.8          & 1.10  & 57.80420 & 47.2605 \\
             &  5.0   & 7.78 & 5.41          & 0.70 &  7.77  & 5.57          & 0.72  & 57.80413 & 47.2603 \\  
\hline                                         	                                                        
&&\multicolumn{8}{c}{Component $F$}\\                                                 
Nov 20, 1999 &   5.0  & -    & -             &      & 4.12   & -             & -     & -        & -       \\
             &  15.4  & -    & -             &      & -      & -             & -     & -        & -       \\
Mar  6, 2000 &   5.0  & -    & -             &      & 3.80   & -             & -     & -        & -       \\
             &  15.4  & -    & -             &      & -      & -             & -     & -        & -       \\
Nov 30, 2000 &   5.0  & -    & -             &      & 2.88   & -             & -     & -        & -       \\
             &  15.4  & -    & -             &      & -      & -             & -     & -        & -       \\
Sep 22, 2002 &  1.7   & 15.3 & 9.71          & 0.64 & 17.1   & 9.30          & 0.54  & 57.80632 & 47.2521 \\  
             &  2.3   & 7.18 & 2.11          & 0.29 &  9.31  & 2.18          & 0.23 & 57.80596 & 47.2530 \\
             &  5.0   & 6.89 & -             &      &  8.32  & -             & -     & -        & -       \\
\hline
\end{tabular}
\caption[NGC\,3079 component data]{NGC\,3079 component
data. ``Tapered'' flux densities have been measured from images
tapered to the same resolution and convolved with a beam of
$4\,\mas\times3\,\mas$. ``Full resolution'' flux densities have been
measured from images using all available data. In September 2002, the
tapered images differed only marginally from the full resolution
images.  Integrated flux densities, $S$, are given in mJy, peak flux
densities, $S_{\rm max}$, are given in mJy\,beam$^{-1}$, the ratio
$x=S_{\rm max}/S$ is dimensionless, coordinates are seconds of right
ascension relative to 10:01:00 and arcseconds of declination relative
to 55:40:00. Only the the last epoch used phase-referencing and hence
provides coordinates measured with respect to the phase calibrator
J0957+5522 at RA 09:57:38.1849709 and Dec +55:22:57.769241. When a
flux density, but no position is given, the emission is very extended
and possibly also faint, sometimes not being visible in the
images. The relative separations of the components measured during all
four epochs are listed in Table~\ref{tab:ngc3079_separations}.}
\label{tab:ngc3079_parms1}
\end{center}
\end{table}

\renewcommand{\arraystretch}{1.25}
\begin{table*}[thpb]
\scriptsize
\begin{center}
\begin{tabular}{lc|cccc}
\hline
\hline
Epoch      & SI                    & $A$     & $B$     & $E$      & $F$ \\
\hline                                                             
1999-11-20 & $\alpha^{5.0}_{15.4}$ & -0.99   &   0.92  & $<-1.24$ & -           \\  
2000-03-06 & $\alpha^{5.0}_{15.4}$ & -1.19   &   1.06  & $<-0.93$ & -           \\
2000-11-30 & $\alpha^{5.0}_{15.4}$ & -0.25   &   1.13  & $<-1.64$ & -           \\
2002-09-22 & $\alpha^{1.7}_{2.3}$  & $>8.52$ & $>4.14$ &  5.11    & $-1.64$     \\
           & $\alpha^{2.3}_{5.0}$  &  3.02   & $ 4.52$ & $-1.07$  & $-0.28$     \\      
           & $\alpha^{1.7}_{5.0}$  & $>4.56$ & $>4.41$ &  0.67    & $-0.67$     \\ 
\hline
\end{tabular}
\caption[Spectral indices of NGC\,3079 components.]{Spectral indices of
NGC\,3079 components. Limits have been calculated using three times
the image rms.  Assuming a $1\,\sigma$ error of 5\,\% for the
integrated flux densities yields spectral index errors of 0.09 for the
5.0\,GHz/15.4\,GHz and 1.7\,GHz/5.0\,GHz pairs, 0.33 for the
1.7\,GHz/2.3\,GHz pair and 0.13 for the 2.3\,GHz/5.0\,GHz
pair. Although our observations were designed to avoid resolution
effects, we cannot rule out that the spectral index of $F$ is affected
by its extend and by some of the 2.3\,GHz flux density missing due to
the lack of short baselines. }
\label{tab:ngc3079_spectra}
\end{center}
\end{table*}
\renewcommand{\arraystretch}{1.0}

\begin{table}[thpb]
\scriptsize
\begin{center}
\begin{tabular}{lc|ccc|ccc|ccc}
\hline
\hline
Epoch                          & Freq.  & \multicolumn{3}{c}{$A-B$} & \multicolumn{3}{c}{$B-E$} & \multicolumn{3}{c}{$A-E$}\\
                               &        & 
$\Delta{\rm RA}$  &
$\Delta{\rm Dec}$ &
$\Theta$          &
$\Delta{\rm RA}$  &
$\Delta{\rm Dec}$ &
$\Theta$          &
$\Delta{\rm RA}$  &
$\Delta{\rm Dec}$ &
$\Theta$          \\
\hline
\multirow{2}{14mm}{1999-11-20} &   5.0  & 21.1 & 15.5 & 26.2  & 26.2 & 1.5  & 26.3 & 5.1  & 17.0 & 17.7 \\
                               &  15.4  & 21.1 & 16.5 & 26.8  & -    & -    & -    & -    & -    & -    \\
\hline                                                                                                  
\multirow{2}{14mm}{2000-03-06} &   5.0  & 21.3 & 15.6 & 26.4  & 26.1 & 1.4  & 26.2 & 4.8  & 17.0 & 17.7 \\
                               &  15.4  & 21.2 & 16.3 & 26.8  & -    & -    & -    & -    & -    & -    \\
\hline                                                                                                  
\multirow{2}{14mm}{2000-11-30} &   5.0  & 21.3 & 15.8 & 26.5  & 25.9 & 1.2  & 25.9 & 4.6  & 17.0 & 17.6 \\
                               &  15.4  & 21.3 & 16.2 & 26.8  & -    & -    & -    & -    & -    & -    \\
\hline                                                                                                  
\multirow{3}{14mm}{2002-09-22} &  1.7   & -    & -    & -     & -    & -    & -    & -    & -    & -    \\
                               &  2.3   & -    & -    & -     & -    & -    & -    & -    & -    & -    \\
                               &  5.0   & 21.7 & 16.4 & 27.2  & 25.5 & 0.9  & 25.5 & 3.8  & 17.3 & 17.7 \\
\hline
\end{tabular}
\caption[Relative positions of NGC\,3079 components.]{Relative
positions of NGC\,3079 components. All separations are in mas and have
errors of 0.21\,mas} 
\label{tab:ngc3079_separations}
\end{center}
\end{table}

\begin{table}[thpb]
\scriptsize
\begin{center}
\begin{tabular}{lc|cccccc}
\hline
\hline        
Epochs      & 
Freq. & 
\multicolumn{2}{c}{$A-B$} &
\multicolumn{2}{c}{$B-E$} &
\multicolumn{2}{c}{$A-E$}\\
            &       & $\Delta\Theta$ & $v$ & $\Delta\Theta$ & $v$ & $\Delta\Theta$ & $v$ \\
            &       & mas & $c$  & mas & $c$  & mas & $c$  \\
\hline
1999-11-20/ &  5.0  & 0.2 & 0.16 & 0.1 & 0.08 & 0.0 & 0.0  \\
2000-03-06  & 15.4  & 0.0 & 0.0  & -   & -    & -   & -    \\
\hline                             
2000-03-06/ &  5.0  & 0.1 & 0.03 & 0.2 & 0.07 & 0.1 & 0.03 \\
2000-11-30  & 15.4  & 0.0 & 0.0  & -   & -    & -   & -    \\
\hline                             
2000-11-30/ &  5.0  & 0.7 & 0.09 & 0.4 & 0.05 & 0.1 & 0.01 \\
2002-09-22  &       &     &      &     &      &     &      \\       
\hline
\end{tabular}
\caption[Proper motions derived from separations in
Table~\ref{tab:ngc3079_separations}]{Proper motions derived from
separations in Table~\ref{tab:ngc3079_separations}. $\Delta\Theta$ has
errors of 0.21\,mas, and hence $v$ has errors of $0.17\,c$, $0.07\,c$
and $0.03\,c$ in the first, second and third row, respectively.}
\label{tab:propmotions}
\end{center}
\end{table}

\subsection{NGC\,1052}

\begin{figure}
\centering
\includegraphics[width=10cm]{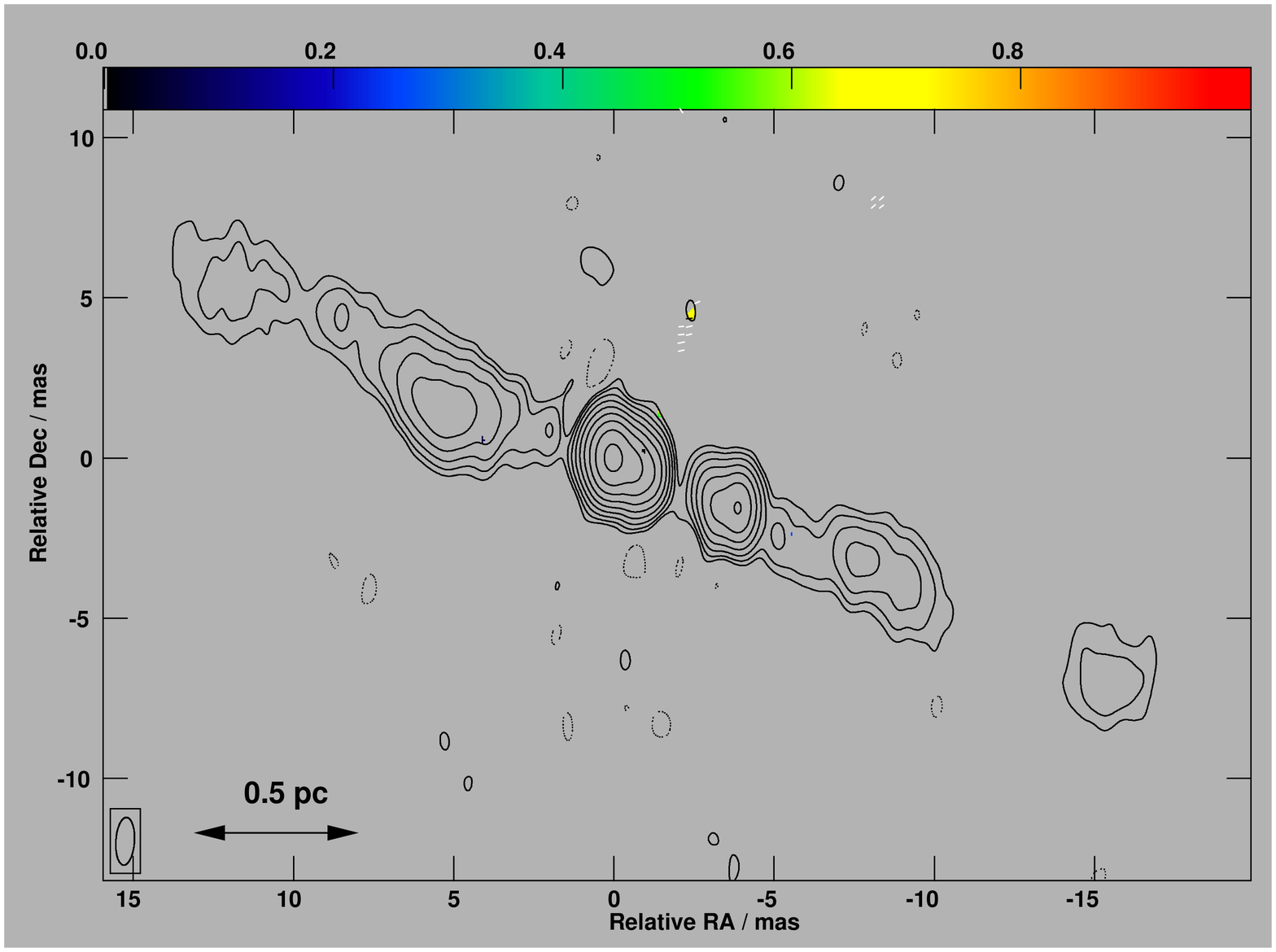}
 \caption[NGC\,1052 13.4\,GHz image]{NGC\,1052 13.4\,GHz VLBA
 image. Peak is $344\,{\rm mJy\,beam^{-1}}$, contours are at
 $\rm{1.08\,mJy\,beam^{-1}}\times2^N$, uniform weighting, beam is
 $1.49\,\mas\times0.58\,\mas$ in P.A. $-4^{\circ}$, pol. ticks are
 $5\,{\rm mJy\,beam^{-1}\,mas^{-1}}$ where $>5\,\sigma=0.74\,{\rm
 mJy\,beam^{-1}}$, colours show fractional polarization.}
 \label{fig:1052L_0420}
\end{figure}

\begin{figure}
\centering
\includegraphics[width=10cm]{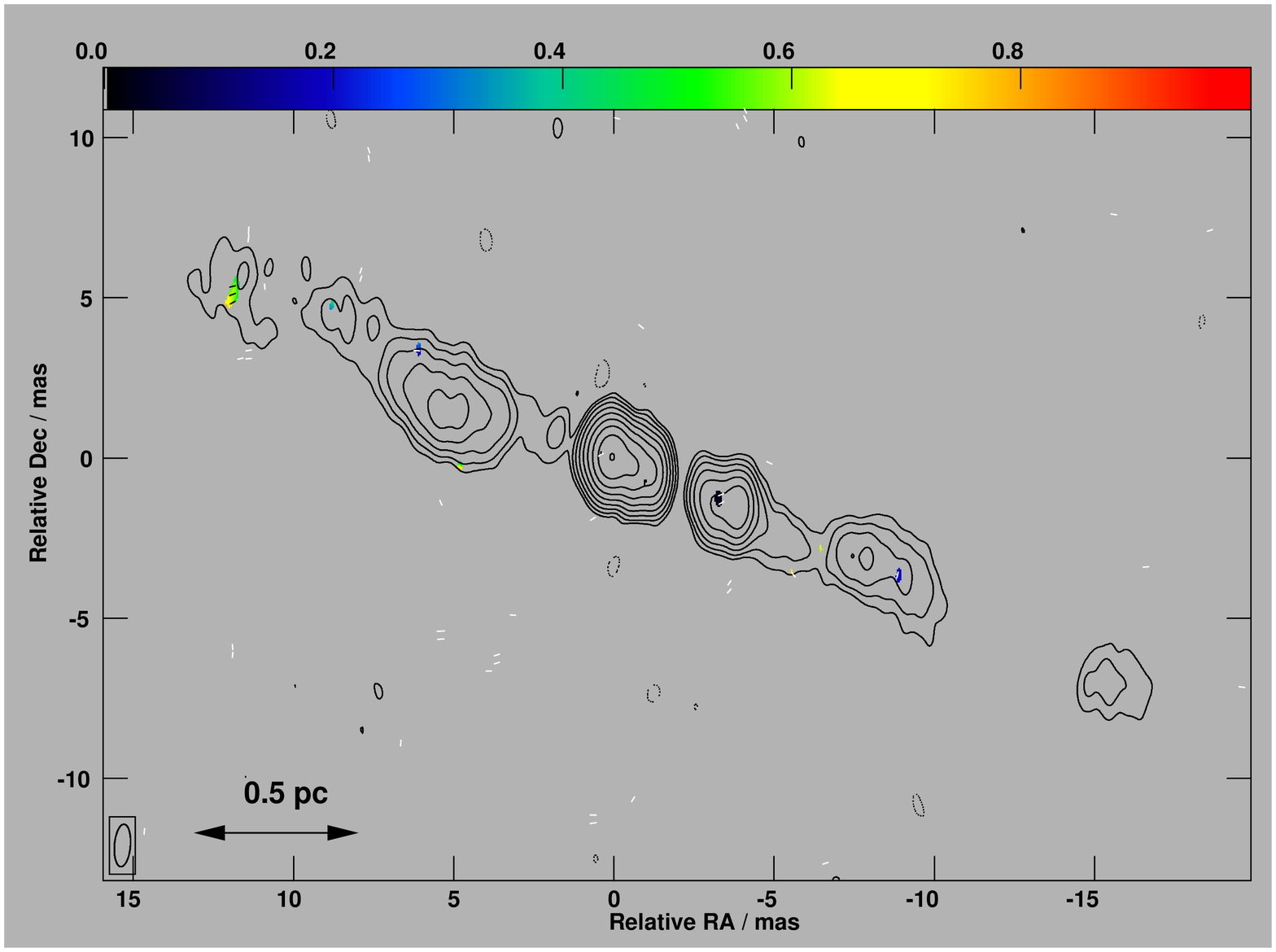}
\caption[NGC\,1052 15.4\,GHz  image]{NGC\,1052 15.4\,GHz VLBA image.
 Peak is $245\,{\rm mJy\,beam^{-1}}$, contours are at
 $\rm{0.94~mJy\,beam^{-1}}\times2^N$, uniform weighting, beam is
 $1.33\,\mas\times0.50\,\mas$ in P.A. $-5^{\circ}$, pol. ticks are
 $5~{\rm mJy\,beam^{-1}\,mas^{-1}}$ where $>5\,\sigma=0.80\,{\rm
 mJy\,beam^{-1}}$, colours show fractional polarization.}
\label{fig:1052U_0420}
\end{figure}

The 13.4\,GHz and 15.4\,GHz images of NGC\,1052 are shown in
Figs.~\ref{fig:1052L_0420} and \ref{fig:1052U_0420}. The excellent
agreement between these two images confirms that the data calibration
and the imaging process were correct. Furthermore, our 15.4\,GHz image
nicely fits into the time series of images shown by
\cite{Vermeulen2003}. Compared to their last epoch observed in 2001.21
(ours being observed at 2001.99), some significant evolution is
visible in the two inner and most compact components, whereas the
outer extended components appear to be unchanged.

\begin{itemize}

\item{13.4\,GHz (Fig.~\ref{fig:1052L_0420})}~~The image shows the
well-known pronounced double-sided jet structure, dominated by a
central component with an extension to the west. The eastern jet is
stronger than the western jet and widens at 5\,mas distance from the
strongest feature in the image centre. It then gently narrows to
2\,mas width before it fades in an extended, low surface-brightness
feature. The western jet displays a bright, compact component with a
clear extension in the direction of the jet axis and is connected by a
very weak bridge of emission to the main component. There is a sharp
transition to a weak, continuous feature that bridges 3\,mas to the
next, extended jet feature. Further down the western jet is a weak,
isolated component. NGC\,1052 is entirely unpolarized at 13.4\,GHz.

\item{15.4\,GHz (Fig.~\ref{fig:1052U_0420})}~~This image differs only
little from the 13.4\,GHz image. Due to the slightly higher
resolution, the main component is more clearly resolved into two
components, and, probably due to a negative spectral index, the
connection to the eastern jet has become weaker. The weak object at
the end of the eastern jet has dimmed to the one-contour level and is
disconnected from the jet. The connection between the two brightest
components was not detected, now leaving a gap between them. Compared
to the 13.4\,GHz image, the changes in the western jet are
marginal. NGC\,1052 is entirely unpolarized at 15.4\,GHz.

\end{itemize}

\subsection{NGC\,4261}

The 15.4\,GHz image in Fig.~\ref{fig:4261CL10it2_1146} displays a
double-sided, continuous structure. The central feature gently fades
into the western jet that extends out to 9\,mas from the
core. Compared to this transition, there is a rather sharp drop-off
between the central peak and the eastern jet due to the circumnuclear
disc. The polarization ticks 2\,mas south-west of the core have a
maximum of $0.68\,{\rm mJy\,beam^{-1}}$ at a position where the total
intensity is $14.4\,{\rm mJy\,beam^{-1}}$ (4.7\,\% fractional
polarization). The polarization was found to change considerably when
the imaging parameters (weighting, tapering) were changed. They are
probably due to residual deconvolution errors. We therefore conclude
that NGC\,4261 is entirely unpolarized.

NGC\,4261 has previously not been observed at 15.4\,GHz, and we
therefore compare our image to the 8.4\,GHz and 22.2\,GHz images by
\cite{Jones1997} and \cite{Jones2000}. NGC\,4261 is undergoing
considerable spectral evolution between these two frequencies, and our
image nicely fits inbetween. At 8.4\,GHz, the absorbed gap in front of
the counter-jet is clearly visible. At 15.4\,GHz, the absorber is
manifest by somewhat denser isophotes $<1\,{\rm mas}$ east of the
core, and at 22.2\,GHz, there is almost no indication of the
absorber. The lengths of the jet and counter-jet are shorter at
15.4\,GHz than at 8.4\,GHz, and shorter at 22.2\,GHz than at
15.4\,GHz. This is due to the negative spectral index, the longer
integration time used at 8.4\,GHz and the higher sensitivity at this
frequency.

\subsection{Hydra~A}

The 15.4\,GHz image in Fig.~\ref{fig:hyd_a_it3_1146} shows a
relatively small, double-sided core-jet structure, the transitions
from the core into the jets being equally smooth on both sides. Aside
from the jets, a weak component was detected 3\,mas south-west of the
core which is connected to the jets via a weak bridge of
emission. Hydra~A is entirely unpolarized.

We see no significant structural difference between our image and the
15.4\,GHz image shown by \cite{Taylor1996}. The peak flux density in
our image, $(144\pm7)\,{\rm mJy\,beam^{-1}}$, is higher than the
integrated flux density reported by \cite{Taylor1996},
$(127\pm6)\,{\rm mJy}$, but almost in agreement within the errors.

\subsection{Centaurus~A}

\begin{figure}[ht!]
\centering
\includegraphics[width=10cm]{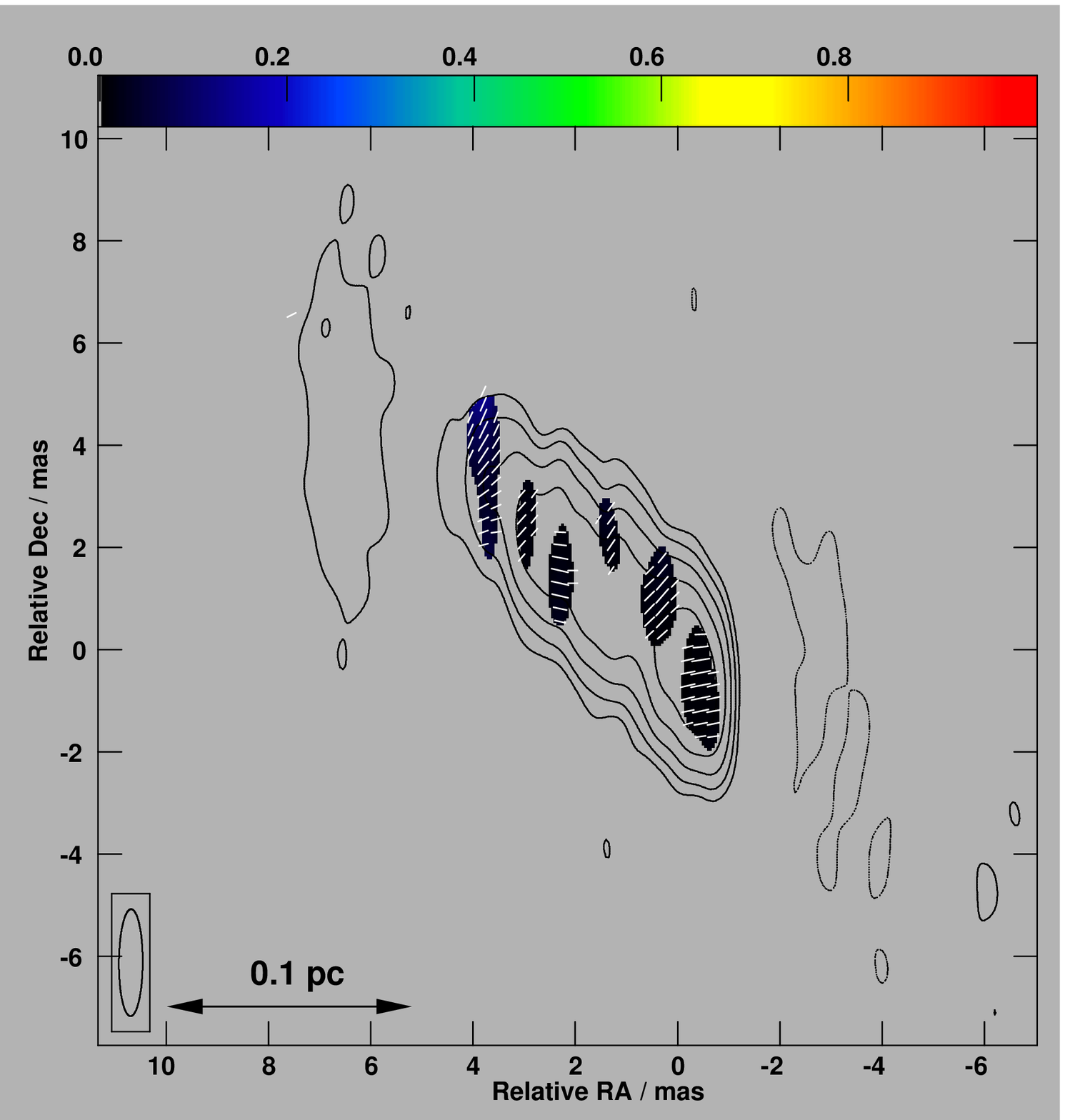}
 \caption[Centaurus~A 13.4\,GHz  image]{Centaurus~A 13.4\,GHz VLBA
 image. Peak is $1.08\,{\rm Jy\,beam^{-1}}$, contours are at
 $\rm{32.61\,mJy\,beam^{-1}}\times2^N$, uniform weighting, beam is
 $2.09\,\mas\times0.47\,\mas$ in P.A. $0^{\circ}$, polarization ticks
 are $20\,{\rm mJy\,beam^{-1}\,mas^{-1}}$ where $>5\,\sigma=3.55\,{\rm
 mJy\,beam^{-1}}$, colours show fractional polarization.}
 \label{fig:cen_13_0927}
\end{figure}

\begin{figure}[ht!]
\centering
\includegraphics[width=10cm]{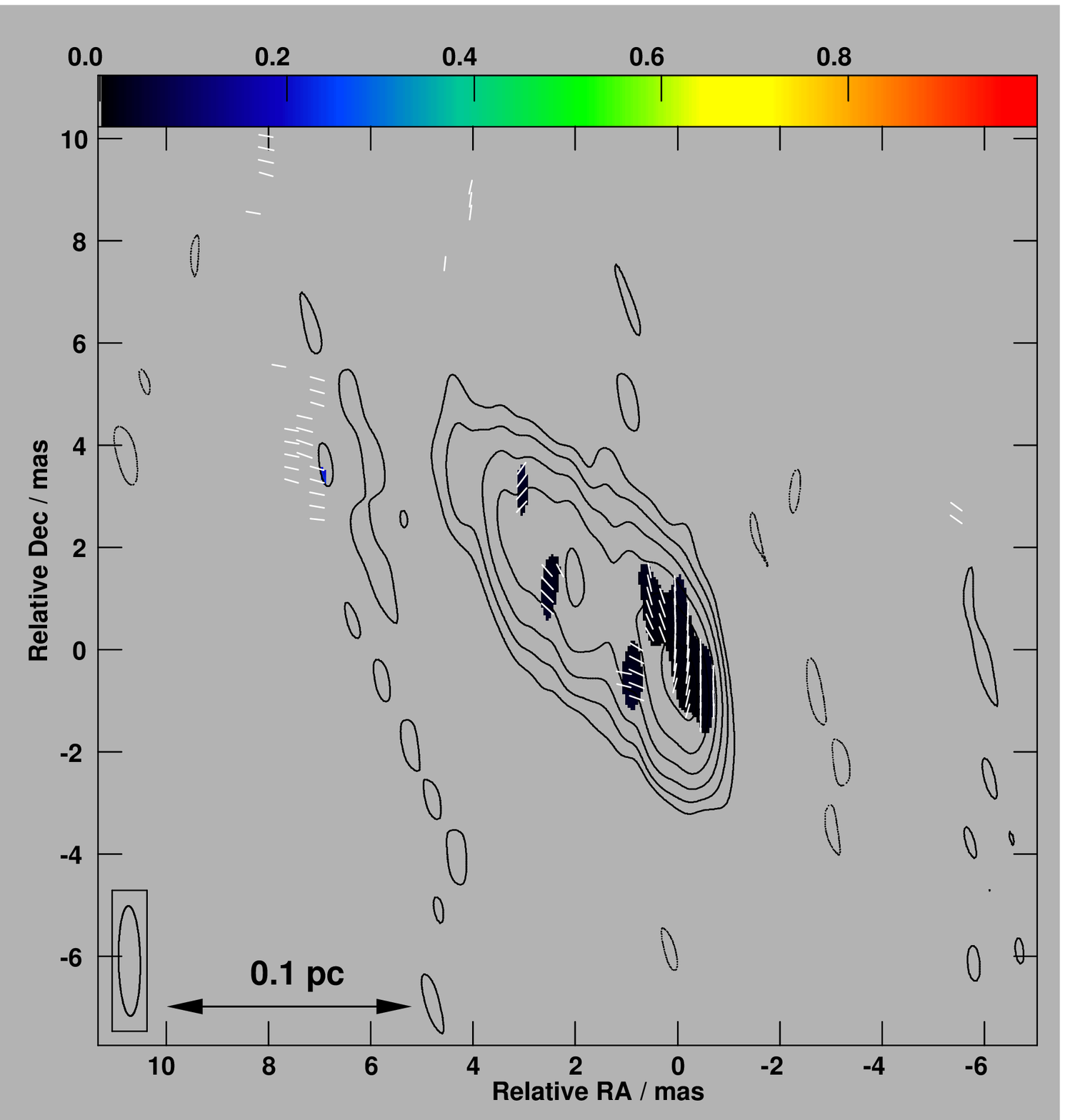}
\caption[Centaurus~A 15.4\,GHz image]{Centaurus~A 15.4\,GHz VLBA
image. Peak is $907\,{\rm mJy\,beam^{-1}}$, contours are at
$\rm{22.38\,mJy\,beam^{-1}}\times2^N$, uniform weighting, beam is
$2.15\,\mas\times0.42\,\mas$ in P.A. $1^{\circ}$, polarization ticks
are $20\,{\rm mJy\,beam^{-1}\,mas^{-1}}$ where $>5\,\sigma=5.25\,{\rm
mJy\,beam^{-1}}$, colours show fractional polarization.}
\label{fig:cen_15_1146}
\end{figure}

The 13.4\,GHz and 15.4\,GHz images are shown in
Figs.~\ref{fig:cen_13_0927} and \ref{fig:cen_15_1146}. The iterative
self-calibration and cleaning was found to give multiple solutions due
to the poor $(u,v$) coverage at the source declination of
$-43^{\circ}$. The similarity of the final images, however, indicates
that the calibration of both frequencies has converged on the most
likely source structure.

\begin{itemize}

\item{13.4\,GHz}~~The image in Fig.~\ref{fig:cen_13_0927} shows a core
and a single-sided jet structure extending over almost 10\,mas to the
north-east. The flux density drops off to 25\,\% of the peak flux
density within the first 1.5\,mas, reaches a plateau and fades at
8\,mas core distance. Large noise patches surrounding the source and a
high image rms indicate significant sidelobes. Centaurus~A
has a formally significant degree of polarization of 4.9\,\%, but
there are also formally significant regions of polarization outside
the source area which obviously are due to deconvolution errors. These
lie outside the region shown in Fig.~\ref{fig:cen_13_0927}, but
similar effects can be seen in the 15.4\,GHz image in
Fig.~\ref{fig:cen_15_1146}. We therefore consider Centaurus~A
to be unpolarized at 13.4\,GHz.

\item{15.4\,GHz}~~The image in Fig.~\ref{fig:cen_15_1146} is
similar to the 13.4\,GHz image, the only difference being a barely
visible bend in the jet base close to the core. A detection of this
bend has been claimed by \cite{Fujisawa2000} from VLBI observations at
$4.8\,{\rm GHz}$, but their $(u,v)$ coverage was extremely
sparse. \cite{Tingay2001} have monitored the source during eight
epochs at frequencies of up to 22\,GHz and did not detect such a bend
in any of their epochs, and we interpret the bend claimed
by \cite{Fujisawa2000} as an imaging artefact. We consider the
polarization seen in this image as being due to deconvolution errors,
like in the 13.4\,GHz image.

\end{itemize}

Our images have considerably higher resolution
($2\,\mas\times0.5\,\mas$) than the 8.4\,GHz images
($13\,\mas\times3\,\mas$) and the 22.2\,GHz images
($5\,\mas\times1\,\mas$) presented by
\cite{Tingay2001}. Hence, most of the jet structure and also the weak, extended
counter-jet feature seen by these authors is resolved out in our
images.

\subsection{Cygnus~A}

\begin{figure}
\centering
 \includegraphics[width=12cm]{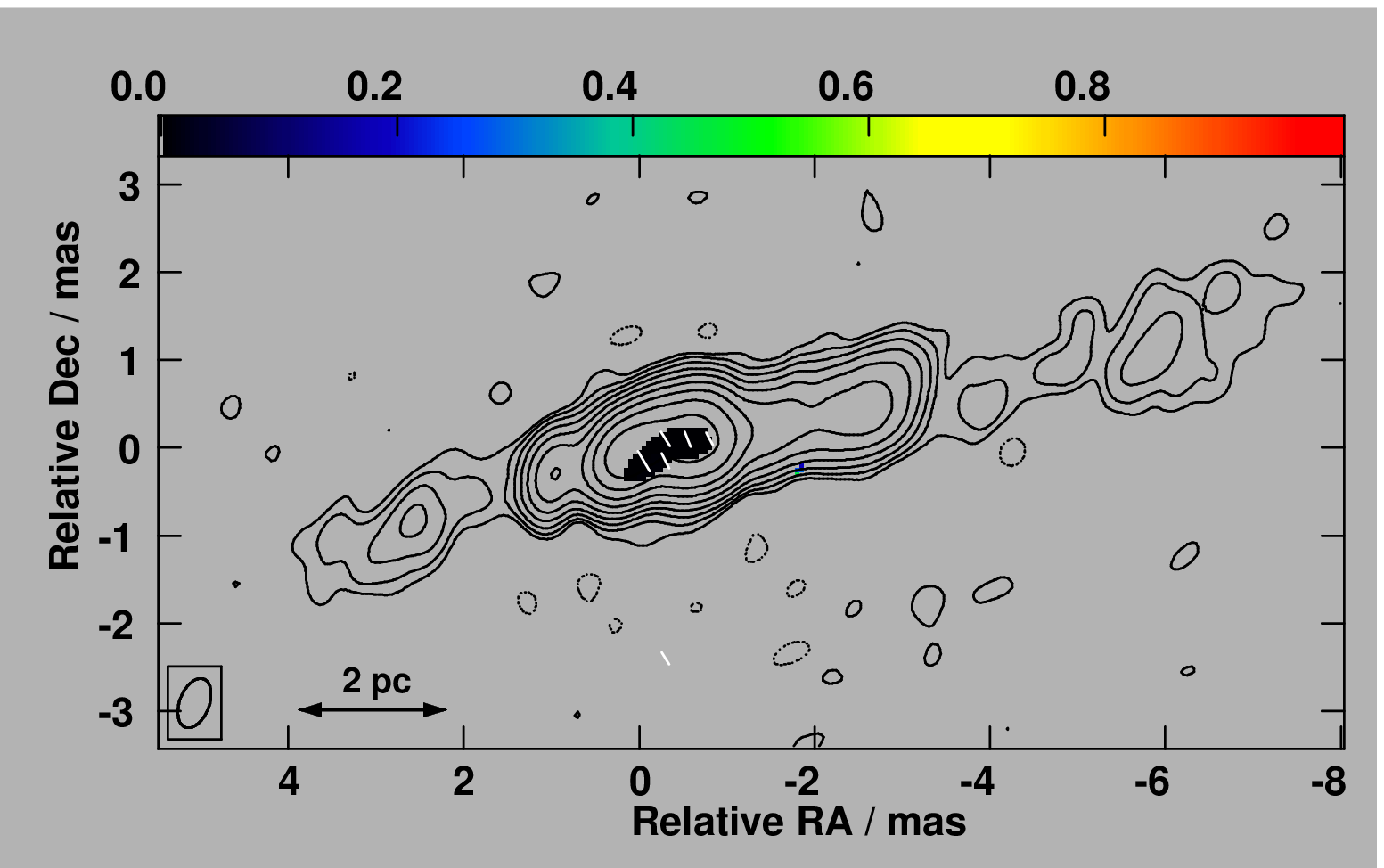}
 \caption[Cygnus~A 15.4\,GHz image]{Cygnus~A 15.4\,GHz VLBA+EB+Y1
 image. Peak is $316\,{\rm mJy\,beam^{-1}}$, contours are at
 $\rm{0.69\,mJy\,beam^{-1}}\times2^N$, uniform weighting, beam is
 $0.59\,\mas\times0.34\,\mas$ in P.A. $-21^{\circ}$, polarization
 ticks are $5\,{\rm mJy\,beam^{-1}\,mas^{-1}}$ where
 $>7\,\sigma=0.77\,{\rm mJy\,beam^{-1}}$, colours show fractional
 polarization.}
 \label{fig:Cyg_A}
\end{figure}

The 15.4\,GHz image in Fig.~\ref{fig:Cyg_A} shows a continuous
double-sided jet structure spanning more than 12\,mas. The core
appears to be elongated and slightly bean-shaped. The eastern jet
comprises a compact component of 1\,mas extend and a weak, extended
component 2\,mas to 4\,mas east of the core. The western jet has a
brightness temperature similar to that of the compact component in the
eastern jet, but extends further out. At a position 3.5\,mas west of
the core, the flux density sharply drops to less than 1\,\% of the
peak flux density, and a diffuse, irregular structure of almost 5\,mas
length marks the continuation of the jet. Cygnus~A is the only source
presented here that shows significant, believable polarization. The
central region shows $1.4\,{\rm mJy\,beam^{-1}}$ of polarized emission
on top of a Stokes I peak of $316\,{\rm mJy\,beam^{-1}}$, i.e., the
region is 0.44\,\% polarized. The $D$-term calibration in this data
set was very good, and the image has the highest dynamic range of all
observations presented here (1400:1). Furthermore, the data
calibration was repeated independently by a colleague who found very
similar results (U. Bach, priv. comm.). We therefore conclude that
Cygnus~A is polarized.

Our image displays basically the same structure as has been found by
\cite{Bach2003}, although a comparison is difficult due to the higher
resolution in our image that was caused by the participation of
Effelsberg. The faint and extended jet and counter-jet structures in
our image are therefore resolved and weaker than in the image by
\cite{Bach2003}.

\clearpage

\chapter{Discussion}

The discussion of the observational results presented in the previous
chapters is divided into two sections: one treating the polarimetry
results and one treating the multi-epoch observations of NGC\,3079,
considering spectral indices, proper motions and general properties of
nuclear radio components in Seyfert galaxies.

\section{Polarimetry Results}

\subsection{Introduction}

The jet in M87 was the first extragalactic jet in which polarized
emission was found (\citealt{Baade1956}), and was immediately
identified as synchrotron emission. With synchrotron emission being
established as the process by which AGN jets radiate in the radio
regime, polarization measurements allowed investigations of magnetic
field orientations and strengths, shock properties, and optical
depths.\\

These investigations are based on the degree, orientation, location in
the source and temporal evolution of polarized emission. A {\it lack}
of polarized emission, however, is a fundamental observational
difference between the sources observed for this thesis and most other
AGN, and it indicates a process which is not commonly found. It would
be an unnecessary complication to postulate a different intrinsic
emission mechanism in polarized and unpolarized sources, given their
similarities in so many other respects. We therefore believe that the
emission process in the source is also synchrotron emission and
therefore is intrinsically significantly polarized, and that the
observed lack of polarized emission therefore is due to a
depolarization mechanism or a blending effect. The mechanism(s)
reprocess(es) the radiation either inside or outside the emitting
regions, and we refer to them as ``intrinsic'' and ``extrinsic''
depolarization mechanisms, respectively. Because blending effects
would also occur inside the sources, they are treated as an intrinsic
depolarization mechanism.

\subsection{Significance of Lack of Polarization}

We first demonstrate the significance of our findings (significant
polarized emission in only one out of six sources) by comparing our
results to the samples by \cite{Kellermann1998} and
\cite{Zensus2002}. They have observed 171 compact sources at 15\,GHz
to investigate their jet structure and kinematics, and to derive
statistical properties. The continuation of the project, known as the
MOJAVE prgramme, has also polarimetric information on a statistically
complete subsample (\citealt{Lister2003}). This sample comprises all
sources with declination $\delta>-20^\circ$, galactic latitude
$|b|>2.5^\circ$ and total 15.4\,GHz VLBA flux density of more than
1.5\,Jy. The sources are core-dominated, beamed quasars because one of
its goals was to test Monte-Carlo simulations of beaming in a
flux-limited sample of jets. Preliminary results are available on the
MOJAVE prgramme
homepage\footnote{\url{http://www.physics.purdue.edu/~mlister/MOJAVE/}}. As
of October 2003, 129 sources with polarization measurements were
available (Lister, priv. comm.), with a detection limit of 0.3\,\% of
fractional polarization. Given this limit, the sample has 10
unpolarized sources. In our sample, not considering Centaurus~A due to
the low image fidelity and NGC\,3079 because it is clearly not a radio
galaxy, three out of four sources are unpolarized. In a
difference-of-two-proportions test with Yates correction for
continuity, we find a probability of 0.03\,\% that the two samples
have been drawn from the same parent distribution. If we require a
detection limit of more than 0.6\,\% for the MOJAVE sample sources,
equal to the averaged detection limit of our three 15\,GHz observing
runs in December 2001 and July 2002, 19 of the 129 sources in the
MOJAVE sample are unpolarized. This subsample has a probability of
1.2\,\% of being drawn from the same parent distribution as our
sources, and so the polarization properties of the two samples are
significantly different.

Furthermore, the fraction of double-sided jets (sources with structure
on more than one side of the brightest component) in our sample is
four out of six. If we exclude NGC\,3079 because it is not a radio
galaxy, then the fraction is four out of five, and if Centaurus~A is
treated as a double-sided source, because a counter-jet is not
prominent but some evidence has been found (\citealt{Tingay2001}),
then the fraction is five out of five. In the MOJAVE sample, the
number of double-sided sources is six out of 129. A
difference-of-two-proportions test yields a probability of less than
0.01\,\% that the two samples have been drawn from the same parent
distribution (the probability remains less than 0.01\,\% when
Centaurus~A is treated as a single-sided source).

Our sample selection criteria (free-free absorbers in front of a jet,
sources bright enough for VLBI polarimetry at 15.4\,GHz and closer
than 200\,Mpc), therefore selected sources with significantly
different polarization characteristics compared to those satisfying
the MOJAVE selection criteria (basically, sources brighter than 1.5\,Jy
at 15.4\,GHz). However, the statements about the fraction of polarized
sources and the single- and double-sidedness are not strong because
our sample is small. Furthermore, the MOJAVE programme is flux-density
limited at 15.4\,GHz and so is biased towards highly beamed quasars.

\subsection{Intrinsic Causes}

\subsubsection{Tangled Internal Magnetic Fields}
\label{sec:tangled_internal_fields}

If the emitting regions are optically thin, the radiation that one
receives is the superposition of radiation from along the line of
sight through the source. In this case, for a homogeneous magnetic
field and a population of relativistic electrons with a power-law
distribution of the form $N(E)\,dE=N_0\,E_0^{-p}\,dE$, the theoretical
maximum of polarization is

\begin{equation}
m(p)= 100\times\frac{3\,p + 3}{3\,p + 7}
\end{equation}
(\citealt{LeRoux1961}), where $m$ is the percentage of
polarization. For $p=2$, $m(p)=70\,\%$, and these high degrees of
polarization are in fact found in radio lobes (e.g., in Cygnus~A,
\citealt{Carilli1989}) and also in BL~Lac objects (e.g.,
\citealt{Gabuzda1994}). In these cases, the magnetic fields must be
very homogeneous on angular scales similar to or larger than the
resolution of the instrument. In general, magnetic fields are not
homogeneous, but are composed of a uniform component, $B_0$, and a
random component, $B_{\rm r}$, which varies on a typical length scale,
thus making ``cells'', in which the magnetic field is
homogeneous. Provided that $B_{\rm r}$ varies on scales much less than
the source diameter, the intrinsic degree of polarization in each
cell, $m_{\rm i}(p)$, is averaged over the source to the observed
degree of polarization, $m_{\rm o}(p)$, given by

\begin{equation}
\begin{split}
m_{\rm o}(p) &= m_{\rm i}(p)\times\frac{B^2_0}{B^2_0 + B^2_{\rm r}}\\
             &\approx\frac{\rm Energy~in~uniform~field}{\rm Energy~in~total~field}\\
\end{split}
\end{equation}
(\citealt{Burn1966}). For $p=2$, the fractional polarization in each
cell is 70\,\%. To depolarize this below our detection threshold of
0.3\,\%, the magnetic field energy would need to be dominated by the
random component by a factor of more than 200. The jets would have to
be turbulent, and very little ordering of the magnetic fields by the
overall outward motion of the jet flow would be permitted.

The fractional polarization decreases as $i^{-1/2}$, where $i$ is the
number of independent magnetic cells across the emitting region
(\citealt{Jones1977}). To depolarize an intrinsic fractional
polarization of 70\,\% to 0.3\,\% requires 54000 cells, and to
depolarize an intrinsic fractional polarization of 10\,\% still
requires 1100 cells. Thus, the jets would need to be interspersed
throughout with randomly oriented magnetic fields, a situation which
does not agree with two observational facts. First, AGN jets are
shaped by magnetic fields from sub-pc to kpc scales, requiring at
least some ordering of the fields in all parts of the jets. Second,
jets evolve along their axes, which has been observed in many radio
jets on parsec scales. Shocks in jets compress the magnetic fields and
so leave behind regions of ordered magnetic fields
(\citealt{Laing1980}). Evolving jets have been observed in NGC\,1052
(\citealt{Vermeulen2003}) and Cygnus~A (\citealt{Krichbaum1998}), and
it is therefore unlikely that these jets are dominated by random
magnetic fields. Furthermore, turbulent magnetic fields require a
source of energy to sustain the disorder because ordered fields have
lower energy.\\

In the case of an optically thick source, the maximum intrinsic degree
of polarization is 10\,\% to 12\,\% (\citealt{Jones1977}), and one
receives emission only from the surface. The same considerations as
for the optically thin case apply, and to depolarize the radiation
below the detection threshold requires more than 1100 cells across the
observing beam, so that in this scenario, the magnetic fields inside
the jets must be ordered to confine the jet, but must be turbulent in
the radiating surface.  The idea of such an outer boundary layer
(also called shear layer) which is physically different from the jet
interior has been raised for the first time by \cite{Owen1989} to
explain their VLA observations of the kpc-scale jet of M\,87. In the
case of M\,87, however, the boundary layer is optically thin, as
indicated by up to 50\,\% of polarization (\citealt{Perlman1999}), and
the magnetic field in the boundary layer is highly ordered, indicated
by similar P.A. of the polarized emission over regions $>100\,{\rm
pc}$ in size. On smaller spatial scales, evidence for boundary layers
has been found in some BL~Lac objects and FR\,I radio galaxies by,
e.g., \cite{Gabuzda2002} and \cite{Attridge1999}. FR\,I radio
galaxies and BL~Lacs are thought to be similar objects, the difference
being due to different inclination angles.
\cite{Chiaberge2000} find that FR\,I radio galaxies are more luminous
by factors of 10 to $10^4$ than predicted from spectral energy
distributions of BL~Lacs seen edge-on. They explain the discrepancy
with a stratified jet structure, in which a highly relativistic spine
is surrounded by slower (but still relativistic) layers. When viewed
from smaller angles, emission from the beamed spine dominates the jet
emission, and the jet morphology is that of a BL~Lac. When viewed from
larger angles, the beaming cone points away from earth, but emission
from the unbeamed boundary layer remains, preventing the full 10 to
$10^4$-fold decrease of luminosity, and the morphology is that of an
FR\,I. Thus, there is theoretical and observational evidence for a
region around radio jets that has different physical parameters than
the jet inside.\\

However, there are considerable objections to the depolarization being
due to jet boundary layers in the sources presented here. First, even
though the physical properties of jet boundary layers are very much
different from the embedded spine, that does not necessarily mean that
the magnetic fields are turbulent. Observations of the distribution of
polarized emission have yielded evidence {\it for} the existence of
boundary layers, not against it (like the abovementioned observations
by \citealt{Perlman1999} and \citealt{Attridge1999}), and so the
boundary layers must be ordered. Second, numeric models of jets by
\cite{Aloy2000} show that the boundary layer magnetic field is well
aligned with the jet direction, and third, \cite{Laing1980} showed
that compression of a volume with randomly oriented magnetic fields
onto a plane almost always causes the radiation to be polarized by
more than 10\,\%. Thus, only if the initial boundary-layer random
field is uncompressed and the jet does not evolve and stretch the
boundary layer magnetic fields, is there the possibility that the
radiation from the boundary layer is unpolarized. Given the works
mentioned above, we deem this situation very unlikely.

\subsubsection{Internal Faraday Rotation}

In the case of internal Faraday rotation, polarized emission from
various depths along the line of sight through the source is
Faraday-rotated by the source itself, the degree of rotation depending
on the depth of the emitting region. Internal Faraday rotation is only
significant at the transition from frequencies where the source is
optically thick to where it becomes optically thin, because
self-absorption goes as $\sim\nu^{-3}$, whereas Faraday rotation goes
as $\nu^{-2}$, shallower than the self-absorption. Towards low
frequencies, the $\tau=1$ surface (the depth in the source from which
most emission is received) approaches the surface of the source, and
only few material is left that participates in Faraday
rotation. Towards high frequencies, the source becomes transparent,
and all material contributes to the Faraday rotation, but the rotation
measure quickly drops as $\nu^{-2}$ and the effect is
small. Furthermore, internal Faraday rotation requires a significant
fraction of ``cold'' electrons in the jet with low Lorentz factors
($\gamma_{\rm min}\approx 1$ to $10$), because the relativistic mass
growth of the electrons at higher Lorentz factors reduces their
response to the electric fields. As electrons in jets of radio
galaxies are expected to be highly relativistic ($\gamma_{\rm
min}\approx 100$), we do not consider this explanation
further. However, there is the possibility that the situation in
NGC\,3079 is different. Modelling the NLR line emission to be powered
by energy and momentum transfer from radio jets, \cite{Bicknell1998}
find that electrons in Seyfert radio components may be only mildly
relativistic.

Given the careful tuning of conditions that is required to explain the
lack of polarized emission with intrinsic causes, and given how
unlikely they are to be fulfilled, we prefer to reject those
explanations, and the depolarization must have a different origin.\\

Both intrinsic and extrinsic depolarization mechanisms impose an upper
limit on the typical length scale on which the magnetic field changes
direction. Assuming that this scale is similar for all sources, the
best limit is given by NGC\,1052 at a distance of 19.4\,Mpc, since it
is the closest of the double-sided jet sources in which a solid
non-detection was made. In NGC\,1052, our highest resolution image at
15.4\,GHz resolved 0.5\,mas, corresponding to 0.047\,pc in the
source. Within a volume of 0.047\,pc on a side, $i=54000$ independent
cells are required to depolarize an intrinsic degree of 70\,\%
polarization to the observed upper limit of 0.3\,\%. Thus, the typical
length scale of a cell is $<1/\sqrt[3]{54000}$ of the observing beam,
or 0.0012\,pc. In the case of the source being optically thick, the
typical cell size must still be $<1/10$ of the beam, or 0.0047\,pc.

\subsection{External Causes}

\subsubsection{Bandwidth Depolarization} 

At each wavelength, the electric vector position angle of the waves
after passing through the medium is different, and integrating over
the observing bandwidth yields an apparent net polarization that is
lower than the initial state. The depolarization across the observing
bandwidth can be calculated as follows. Consider an ensemble of waves
with wavelengths $\lambda$ in the observed bandwidth $\Delta\lambda$,
so that $\lambda_0<\lambda<\lambda_0+\Delta\lambda$. Let us assume
that all waves have the same electric vector position angle. If the
waves pass through a magnetized plasma, the electric vectors
$\vec{E}(\lambda)$ are rotated by Faraday rotation.

Assuming that a homogeneously magnetized slab of plasma has a rotation
measure $RM$, the change of EVPA with wavelength is given by

\begin{equation}
\begin{split}
\theta(\lambda_0) &= RM\times\lambda_0^2+\theta_0\\
\theta(\lambda_1) &= RM\times\lambda_1^2+\theta_0\\
\Rightarrow\Delta\theta(\lambda_1, \lambda_0) &= RM(\lambda_1^2-\lambda_0^2),\\
\end{split}
\end{equation}
where $\theta_0$ is the EVPA of the waves before passing through the
medium and $\theta(\lambda)$ is the wavelength-dependent EVPA after
passing through the medium. With respect to the reference wavelength
$\lambda_0$, the electric vectors $\vec{E}(\lambda)$ are rotated by
$\Delta\theta(\lambda_1, \lambda_0)$:

\begin{equation}
\begin{split}
\vec{E}(\lambda) &= \vec{E}(\lambda_0)  e^{i\Delta\theta(\lambda, \lambda_0)}\\
&= \vec{E}(\lambda_0) e^{iRM(\lambda^2-\lambda_0^2)}.\\
\end{split}
\end{equation}

Integrating all vectors in the observing band yields

\begin{equation}
\begin{split}
<\vec{E}>(\lambda_1, \lambda_0)&= \frac{1}{\lambda_1-\lambda_0}\int_{\lambda_1}^{\lambda_0}\vec{E}(\lambda)\,d\lambda\\
&= \frac{1}{\lambda_1-\lambda_0}\int_{\lambda_1}^{\lambda_0}\vec{E}(\lambda_0) e^{iRM(\lambda^2-\lambda_0^2)}\,d\lambda\\
&= \frac{\vec{E}(\lambda_0) e^{-iRM\lambda_0^2}}{\lambda_1-\lambda_0}\int_{\lambda_1}^{\lambda_0} e^{iRM(\lambda^2)}\,d\lambda\\
\end{split}
\end{equation}

\begin{figure}[htpb!]
\includegraphics[width=0.7\linewidth, angle=270]{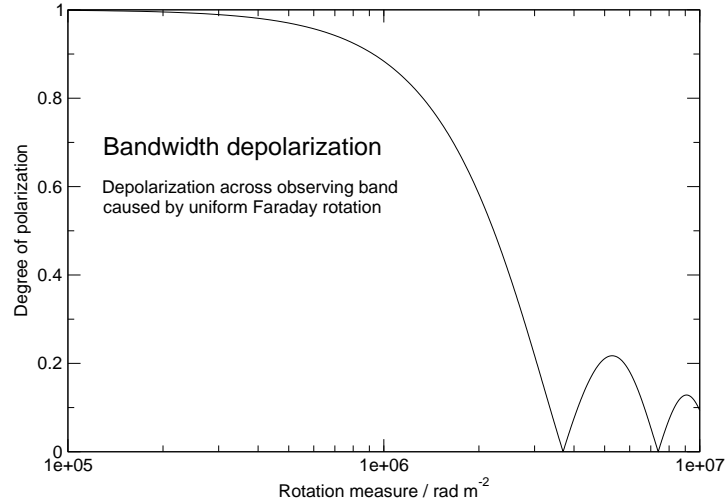}
\caption[Bandwidth depolarization in 15\,GHz band]{Bandwidth
depolarization in the 15\,GHz band.}
\label{fig:bandwidth-depol}
\end{figure}

This integral can be solved numerically
(Fig.~\ref{fig:bandwidth-depol}), and one finds that for the radiation
to be depolarized to $50\,\%$ of initial value requires
$RM=2.2\times\,10^6$ in a 32\,MHz wide band at 15.0\,GHz. The largest
values of RM that have been measured in quasar cores are a few
thousands (e.g., \citealt{Zavala2003}). Assuming a rotation measure of
$RM=10^4\,{\rm rad\,m^{-2}}$ causes $0.00012\,\%$
depolarization. Bandwidth depolarization is unlikely to be important
because unprecedently high and uniform (to rotate the electric vector
position angle in one direction only) rotation measures would be
needed to depolarize the sources, requiring very uniform conditions in
the Faraday screen.

\subsubsection{Beam Depolarization}

If the rotation measure in a foreground Faraday screen changes on
angular scales much smaller than the observing beam, regions with
similar degrees of polarization but opposite signs average out and the
observed degree of polarization is decreased. Thus, spatially highly
variable $RM$s can in principle depolarize the source. Estimates of
$RM$ and the length scale on which the magnetic field needs to be
tangled or the electron density changes follow similar approaches as
in the previous sections. The electric vectors must be rotated by
$180^\circ$ on average to remove any residual polarized emission,
requiring the rms of the variations in the Faraday screen,
$\sigma_{RM}$, to be

\begin{equation}
\begin{split}
\theta(\lambda)&=\lambda^2\times RM\\
 \Rightarrow RM &=\theta(\lambda)/\lambda^2\\
 \Rightarrow \sigma_{RM} &= \frac{\pi\,{\rm rad}}{(c/15.4\,{\rm GHz})^2} = 8290\,{\rm rad\,m^{-2}}
\end{split}
\end{equation}
or larger at 15.4\,GHz. If the $RM$ changes are caused only by density
fluctuations in the medium ($B_\parallel$ and $L$ constant), then the
rms of the fluctuations is of the order of $10^4\,{\rm cm^{-3}}$.  For
NGC\,3079, the constraints on the Faraday rotation are weaker because
the highest frequency at which no polarized emission has been observed
is 5.0\,GHz, and so the rms of $RM$ must be $872\,{\rm rad\,m^{-2}}$
or more. The estimate of the rotation measure cell size was calculated
in \S\ref{sec:tangled_internal_fields}, yielding upper limits of
0.0012\,pc to 0.0047\,pc for NGC\,1052, the closest double-sided
source.

\subsubsection{Free-Free Absorption}

Conditions that cause such high rotation measures can also produce
strong free-free absorption, which is seen only towards comparatively
small parts of the jets, due to which these sources were selected,
while most of the jet parts appear to be unabsorbed (with the
exception of NGC\,1052, where free-free absorption has been detected
towards larger parts of the jets (\citealt{Kameno2001,
Vermeulen2003}). The rotation measure depends linearly on the electron
density, whereas free-free absorption depends on the electron density
squared:

\begin{equation}
RM \propto \int{n_{\rm e}\,dl}
\end{equation}

\begin{equation}
\tau_\nu^{\rm ff}\propto \int{n_{\rm e}^2\,dl}.
\end{equation}

Thus $\tau^{\rm ff}_\nu$ decreases more rapidly than $RM$ with
decreasing electron density. Furthermore, $\tau_\nu^{\rm ff}\propto
T^{-1.35}$, so a turbulent magnetized plasma can depolarize the radio
emission without causing too much free-free absorption if the absorber
is either extended (so the density is low) or hot, or both.\\

We can test the feasibility of such a gas by estimating the free-free
absorption and Faraday rotation caused by conditions typically
inferred for circumnuclear regions in AGN. The free-free optical depth
(Eq.~\ref{eq:ffa}) depends on the temperature, $T$, the observing
frequency, $\nu$, and the density of positive and negative charges,
$N_+$ and $N_-$, respectively. We assume that any circumnuclear gas is
dominated by hydrogen, and that therefore, $N_+=N_-=n_{\rm e}$, where
$n_{\rm e}$ is the electron density. We further assume $T=10^4\,{\rm
K}$, $n_{\rm e}=3\times10^4\,{\rm cm^{-3}}$ and $L=0.5\,{\rm pc}$. The
temperature of gas exposed to ionizing UV radiation is
thermostatically regulated to a temperature of $10^4\,{\rm K}$. This
is because the most efficient cooling is by low excitation levels
(1.9\,eV to 3.3\,eV) in metals (\citealt{Spitzer1978}). The
probability of electrons in a Maxwellian distribution having the
average energy of 2.6\,eV increases by a factor of 32, when the
temperature of the gas increases from 5000\,K to 15000\,K, and the gas
therefore cools much more efficiently when a temperature of more than
10000\,K is reached. The electron density and the path length are
typical values for the well-studied absorbed gaps in NGC\,1052
($n_{\rm e}>3.1\times10^4\,{\rm cm^{-3}}$ and $L=0.7\,{\rm pc}$,
\citealt{Kameno2001}, $n_{\rm e}=10^5\,{\rm cm^{-3}}$ and $L=0.5\,{\rm
pc}$, \citealt{Vermeulen2003}) and NGC\,4261 ($n_{\rm
e}>3\times10^4\,{\rm cm^{-3}}$ and $L=0.3\,{\rm pc}$,
\citealt{Jones2001}), and Eq.~\ref{eq:ffa} yields $\tau_\nu^{\rm
ff}=0.95$ at 15.4\,GHz. The same conditions would give rise to a
rotation measure of $RM=1.2\times10^6\,{\rm rad\,m^{-2}}$, if
$B_\parallel=0.1\,{\rm mG}$ (comparable to what \cite{Lobanov1998}
derived for Cygnus~A).

Whilst the abovementioned conditions are a good fit for the
well-defined free-free absorbers in front of the jets, the conditions
along the unabsorbed but still depolarized parts of the jets must
differ to produce an optical depth close to zero and maintain the
rotation measure still reasonably high. To satisfy $RM=10^4\,{\rm
rad\,m^{-2}}$, the electron density needs to be a factor 120 lower
than in the example above, and the optical depth would then be
$\tau_\nu^{\rm ff}=0.00$.\\

We can now summarize the absorber properties as follows. To yield
$RM=10^4\,{\rm rad\,m^{-2}}$ within 5\,pc of the AGN (so that the jets
fit into the absorber) with a tangled magnetic field of an average
strength of $B_\parallel=0.1\,{\rm mG}$ requires an electron density
of only $25\,{\rm cm^{-3}}$. An important consequence is that the
depolarization cannot be due variable Faraday rotation caused by
density fluctuations. The fluctuations would need to be of the order
of $10^4\,{\rm cm^{-3}}$ as shown above, and assuming a Gaussian
distribution of these fluctuations, the average density then is
$5\times10^3\,{\rm cm^{-3}}$, yielding too high optical depths.\\

\subsection{Origin of the Faraday Screen}

Let us assume that a uniform absorber is present within several
parsecs of the AGN. The absorber must be ionized on each line of sight
to parts of the jets which were found to be unpolarized, i.e., the
ionized regions in the absorber must be larger than the jet
lengths. The radius of a sphere which can be completely ionized by a
source of ionizing photons with $\lambda<91.1\,{\rm nm}$ is called the
Str\"omgren sphere radius. It is given by, e.g., \cite{Osterbrock1989}
(in units convenient for AGN astrophysics):

\begin{equation}
R_{\rm H\,II}=10.6\,T_{\rm e}^{1/4}\, N^{1/3}\,n_{\rm e}^{-2/3}.
\end{equation}

$T_{\rm e}$ is the electron temperature in K, $N$ the number of UV
photons with $\lambda<91.1\,{\rm nm}$ emitted per second, and $n_{\rm
e}$ is the electron density in ${\rm cm^{-3}}$. We estimate the number
of ionizing photons as follows. The spectral energy distribution of
most AGN can be approximated by a power-law over the entire spectrum.
The flux density at a certain frequency is then given by

\begin{equation}
S_{\nu}=S_0 \left(\frac{\nu}{\nu_0}\right)^\alpha
\end{equation}

The number of photons required to produce this flux density is

\begin{equation}
N_{\nu}=\frac{S_{\nu}}{\rm h\,\nu}
       =\frac{S_0}{{\rm h\,\nu}}\left(\frac{\nu}{\nu_0}\right)^\alpha
       =\frac{S_0}{\nu_0^{\alpha}\,{\rm h}}\nu^{\alpha-1}
\end{equation}

Integrating over a given frequency range yields

\begin{equation}
N=\int_{\nu_1}^{\nu_2}{N_{\nu}d\nu}
 =\int_{\nu_1}^{\nu_2}\frac{S_0}{\nu_0^{\alpha}\,{\rm h}}\nu^{\alpha-1}
 =\frac{S_0}{\alpha\,\nu_0^{\alpha}\,\rm h}\left(\nu_2^{\alpha}-\nu_1^{\alpha}\right).
\end{equation}

For example, NGC\,4261 has $S_0=S(8.2\times10^{14}\,{\rm
Hz})=3.1\times10^{-28}{\rm W\,m^{-2}\,Hz^{-1}}$
(\citealt{Vaucouleurs1991}), and $S(5.3\times10^{17}\,{\rm
Hz})=2.5\times10^{-33}{\rm W\,m^{-2}\,Hz^{-1}}$
(\citealt{Fabbiano1992}). The optical-to-$\gamma$ spectral index
therefore is $\alpha=-1.81$. Integrating from the ionization limit of
${\rm 13.6\,eV\corr3.3\times10^{15}\,Hz}$ to ${\rm 10^{20}\,Hz}$ yields
$21\,000\,{\rm m^{-2}\,s^{-1}}$. At the distance of NGC\,4261
(35.8\,Mpc, \citealt{Nolthenius1993}), this corresponds to a total
production rate of ionizing photons of ${\rm
3.2\times10^{53}\,s^{-1}}$. $R_{\rm H\,II}$ then follows to be
260\,pc, when $n_{\rm e}=25\,{\rm cm^{-3}}$ and $T_{\rm e}=10^4\,{\rm
K}$ are assumed. Thus, the Str\"omgren sphere radius exceeds the
required minimum of $\sim10\,{\rm pc}$ imposed by the longest
unpolarized jets in Cygnus~A and Hydra~A by two orders of magnitude. A
Faraday screen which is not seen as a free-free absorber therefore can
be ionized by the AGN. Note that UV radiation is easily absorbed in
galaxies and that the intrinsic UV fluxes which contribute the bulk of
the ionizing photons might be considerably higher. Similar estimates
for the other sources are listed in Table~\ref{tab:stromgren-spheres}.\\

We conclude that the non-detection of polarized emission is due to a
foreground screen in which non-uniform Faraday rotation causes beam
depolarization. The low electron density in the screen imposed by
constraints on the optical depth requires the changes in Faraday
rotation to be caused by turbulent magnetic fields and not by density
fluctuations. Such a low-density screens can be completely ionized by
the AGN radiation.

\begin{table}[htpb]
\scriptsize 
\begin{center}
\begin{tabular}{lccccccc}
\hline
\hline
Source          & $\nu_1$            & $S(\nu_1)$                 & Ref. & $\nu_2$            & $S(\nu_2)$                 & Ref. & $r_{\rm S}$\\
                & Hz                 & ${\rm W\,m^{-2}\,Hz^{-1}}$ &      & Hz                 & ${\rm W\,m^{-2}\,Hz^{-1}}$ &      & pc\\
\hline            
NGC\,3079       & $8.2\times10^{14}$ & $4.3\times10^{-28}$        & 1    & $5.3\times10^{17}$ & $1.1\times10^{-33}$        & 2    & 144\\
NGC\,1052       & $8.2\times10^{14}$ & $3.4\times10^{-28}$        & 1    & $5.3\times10^{17}$ & $1.0\times10^{-33}$        & 2    & 155\\
NGC\,4261       & $8.2\times10^{14}$ & $3.1\times10^{-28}$        & 1    & $5.3\times10^{17}$ & $2.5\times10^{-33}$        & 2    & 264\\
Hyd~A           & $9.6\times10^{ 8}$ & $6.5\times10^{-25}$        & 3    & $3.3\times10^{17}$ & $1.4\times10^{-31}$        & 4    & 373\\
Cen~A           & $8.4\times10^{ 8}$ & $3.9\times10^{-24}$        & 5    & $6.0\times10^{19}$ & $1.3\times10^{-32}$        & 6    &  83\\
Cyg~A           & $1.4\times10^{ 9}$ & $1.6\times10^{-23}$        & 7    & $1.8\times10^{14}$ & $4.0\times10^{-29}$        & 8    & 436\\
\hline
\end{tabular}
\caption[Str\"omgren sphere radii]{Str\"omgren sphere radii for the
observed sources. References: 1 - \cite{Vaucouleurs1991}, 2 -
\cite{Fabbiano1992}, 3 - \cite{Kuehr1981}, 4 - \cite{Brinkmann1994}, 5
- \cite{Jones1992}, 6 - \cite{Steinle1998}, 7 - \cite{Kellermann1969},
8 - \cite{Lebofsky1981}}
\label{tab:stromgren-spheres}
\end{center}
\end{table}

\subsection{Possible Absorbers}

In this section, we investigate the absorption and depolarization to
be expected from various regions of ionized gas surrounding AGN.

\paragraph{Broad line region clouds} 

Forbidden lines have almost never been observed in BLR clouds, and so
they must have densities of at least $10^8\,{\rm cm^{-3}}$, because
the forbidden transitions are collisionally de-excited before they can
radiate (e.g., \citealt{Osterbrock1989}).  Adopting $n_{\rm
e}=10^8\,{\rm cm^{-3}}$, $B_\parallel=0.1\,{\rm mG}$ and $L=0.5\,{\rm
pc}$, the rotation measure is $4\times10^8\,{\rm rad\,m^{-2}}$,
sufficient to depolarize a background radio source, but the optical
depth of the absorber would be $\tau^{\rm ff}_{\nu}=1.1\times10^7$ and
one would not see the background source. Also, BLR clouds are found in
only the innermost tens of light-days of the AGN and have filling
factors of the order of a few percent, whereas the depolarization in
the jets extends at least out to 1\,pc and is continuous. The
depolarization therefore cannot be due to BLR clouds.

\paragraph{Narrow line region clouds} 

NLR clouds have much lower densities than BLR clouds, typically
between $10^4\,{\rm cm^{-3}}$ and $10^5\,{\rm cm^{-3}}$. They can
extend to scales of kpc and have volume filling factors of $10^{-4}$
to $10^{-2}$. The free-free optical depths at 15.4\,GHz following from
these parameters cover the range of 0.01 to 100, and at 5.0\,GHz cover
0.11 to 1100. In NGC\,1052 free-free absorption has been detected
towards the radio jets (\citealt{Kameno2001, Vermeulen2003}). Adopting
$\tau=1$ for the highest observed turnover frequency of 43\,GHz,
\cite{Vermeulen2003} derive $n_{\rm e}=10^5\,{\rm cm^{-3}}$ if the gas is
uniformly distributed over 0.5\,pc. The optical depth decreases with
turnover frequency towards larger distances from the core. If a
magnetic field of $10\,{\rm \mu G}$ is adopted, comparable to
interstellar magnetic fields, we find a rotation measure of
$4\times10^6\,{\rm rad\,m^{-2}}$. We conclude that NLR gas cannot be
excluded as the cause of depolarization in NGC\,1052. The situation is
similar in NGC\,3079. Our observations yield good evidence for a
foreground free-free absorber (see \S\ref{sec:spectra}) with optical
depths of order unity at 5.0\,GHz and hence may be due to NLR gas. The
expected rotation measure is of the order of a few thousand and so
exceeds the $870\,{\rm rad\,m^{-2}}$ required to depolarize NGC\,3079
at 5.0\,GHz.

No free-free absorption has been detected along the jets in the other
four sources, and so their NLR gas is unlikely to be the cause of
depolarization.

\paragraph{H\,{\small II} Regions and Interstellar Gas} 

The line of sight to the radio jets intersects H\,{\small II} regions
and interstellar gas in the sources. The H\,{\small II} clouds are
ionized by UV radiation from stars, and so may contribute to the
depolarization of background radio emission. Although the interstellar
medium, with densities of 0.1 to 10 atoms per cubic centimetre, is
mostly neutral, it has a spatially variably ionized component. Both
H\,{\small II} regions and interstellar gas therefore contribute to
depolarization. However, one would expect the rotation measures to be
similar to those on lines of sight through our Galaxy. Galactic
rotation measures have been well determined using pulsar observations,
(e.g., \citealt{Weisberg2004}, \citealt{Mitra2003} and
\citealt{Han1999}), and these authors find rotation measures of mostly
$<100\,{\rm rad\,m^{-2}}$, even for galactic latitudes below
$10^\circ$ and pulsar distances of several kiloparsecs. Only few
pulsars exceed $100\,{\rm rad\,m^{-2}}$, but no pulsar shows
$\ge1000\,{\rm rad\,m^{-2}}$. Only the highest rotation measures
therefore are of the order of magnitude required to depolarize
NGC\,3079. The argument is strengthened considering that NGC\,3079
shows more star forming activity than our galaxy and so the
interstellar gas has a higher ionized fraction. We therefore cannot
rule out that the radio emission from NGC\,3079 is depolarized by
H\,{\small II} regions and interstellar gas. The interstellar medium
is unlikely to be important in the other five objects which have
elliptical host galaxies. First, ellipticals generally show fewer star
forming activity than spirals and second, the interstellar gas in
these objects has a lower density.

\paragraph{Spherical accretion onto AGN}

The origin of the material which settles around AGN in accretion discs
is largely unknown. We present here a simple model with a spherical
infall of gas towards the AGN (Bondi accretion). The gas is ionized by
the UV radiation from the accretion disc and therefore causes Faraday
rotation.

The mass distribution in the sphere can be approximated with a
rotationally symmetric density profile, and the expected rotation
measure can then be obtained by integrating over the density
profile. 

Let us assume that gas with a density $\rho$ (in ${\rm
kg\,m^{-3}}$) is freely falling radially inwards in the direction of
the AGN, so that its density is a function of the radius $R$. The
free-fall velocity is given by $v_{\rm ff} =
\alpha(2\,G\,M_{\bullet}/R)^{1/2}$, in which $M_{\bullet}$ is the
black hole mass in kg, $R$ is the radius in m, the dimensionless
viscosity parameter $\kappa$ is commonly set to 0.05 and $G$ is the
universal gravitation constant. The density profile is then given by

\begin{equation}
\rho(R) = \frac{\dot{M}}{4\pi\,R^2\,v_{\rm ff}}
        = \frac{\dot{M}}{4\pi\,\kappa(2\,G\,M_{\bullet})^{1/2}}\,R^{-3/2},
\end{equation}
where $\dot{M}$ is the accretion rate in ${\rm kg\,s^{-1}}$. We
estimate the magnetic field in the freely falling gas by means of
equipartition:

\begin{equation}
\begin{split}
\frac{B^2}{2\mu_0} &= \frac{1}{2}\,\rho\,v_{\rm ff}^2
                    = \frac{1}{2}\,\frac{\dot{M}\,R^{-3/2}}{4\pi\,\kappa(2\,G\,M_{\bullet})^{1/2}}\,
                      \kappa^2(2\,G\,M_{\bullet}/R)\\
\Leftrightarrow B  &= \left(\frac{\kappa\mu_0\,(2G)^{1/2}}{4\pi}\right)^{1/2}\,\dot{M}^{1/2}\,
                      M_\bullet^{1/4}\,R^{-5/4},\\
\end{split}
\label{eq:equipartition}
\end{equation}
where $\mu_0$ denotes the permeability of free space.

For the rotation measure, we need the particle density, which can be
derived by dividing the density $\rho(R)$ by $2\mu m_{\rm H}$, where
$m_{\rm H}$ is the proton mass and the factor $2\mu$ with $\mu=0.622$
accounts for the fraction of helium.

\begin{equation}
n_{\rm e} = \frac{\rho(R)}{2\mu m_{\rm H}}
          = \frac{\dot{M}}{8\mu m_{\rm H}\pi\kappa(2GM_{\bullet})^{1/2}}R^{-3/2}
\label{eq:electron_density}
\end{equation}

Inserting both $B$ and $n_{\rm e}$ into Eq.~\ref{eq:rm} now yields

\begin{equation}
\begin{split}
RM &= 2.63\times10^{-13}\int \frac{\dot{M}R^{-3/2}}{8\mu m_{\rm H}\pi\kappa(2GM_{\bullet})^{1/2}}\\
   &  \times \left(\frac{\kappa\mu_0\,(2G)^{1/2}}{4\pi}\right)^{1/2}\,\dot{M}^{1/2}\,M^{1/4}\,R^{-5/4}\,dR\\ 
   &= \frac{2.63\times10^{-13}\mu_0^{1/2}}
           {\kappa^{1/2}(2G)^{1/4}16\mu m_{\rm H}\pi^{3/2}}\,\frac{\dot{M}^{3/2}}{M^{1/4}}
      \times \left(R_{\rm min}^{-7/4}-R_{\rm max}^{-7/4}\right)\\
   &= 4.19\times10^{12}\frac{\dot{M}^{3/2}}{M^{1/4}}\left(R_{\rm min}^{-7/4}-R_{\rm max}^{-7/4}\right).\\
\end{split}
\end{equation}

We exemplify the calculation with data from NGC\,4261. The black hole
mass was estimated by \cite{Ferrarese1996} to be
$4.9\times10^8\,M_\odot=9.8\times10^{38}\,{\rm kg}$, and assuming that
the accretion rate is 1\,\% of the Eddington accretion rate (see
below) one gets $0.01\,M_\odot\,{\rm yr^{-1}} = 6.9\times10^{20}\,{\rm
kg\,s^{-1}}$. Assuming that the infall starts at $10\,{\rm
pc}=3.09\times10^{16}\,{\rm m}$ (the results do not change much when
$R_{\rm max}>10\,{\rm pc}$) and stops at $0.2\,{\rm
pc}=6.17\times10^{15}\,{\rm m}$ (where the disc is seen in NGC\,4261),
one gets $RM=3.1\times10^6\,{\rm rad\,m^{-2}}$. According to
Eq.~\ref{eq:electron_density}, the electron densities at the inner and
outer edges are $n_{\rm e}(0.2\,{\rm pc})=3.0\times10^9\,{\rm m^{-3}}$
and $n_{\rm e}(10\,{\rm pc})=8.5\times10^{7}\,{\rm m^{-3}}$,
corresponding to $3.0\times10^3\,{\rm cm^{-3}}$ and $85\,{\rm
cm^{-3}}$, respectively.

We also need to verify that the optical depth is not too
high. Equation~\ref{eq:ffa}, converted from units of ${\rm cm^{-3}}$
for $n_{\rm e}$ and pc for $L$ to ${\rm m^{-3}}$ and m, respectively,
takes on the form

\begin{equation}
\tau_\nu^{\rm ff} = 2.67\times10^{-30}\,T^{-1.35}\,\nu^{-2.1}\,\int n_{\rm e}^2\,dL
\end{equation}

Inserting Eq.~\ref{eq:electron_density} into this equation,
substituting the path length, $L$, by the radius, $R$, and integrating
over $R$ from $R_{\rm min}$ to $R_{\rm max}$ yields

\begin{equation}
\begin{split}
\tau_\nu^{\rm ff} &= 2.67\times10^{-30}\,T^{-1.35}\,\nu^{-2.1}\\
                  & \times \int_{R_{\rm min}}^{R_{\rm max}}{\left(\frac{\dot{M}\,R^{-3/2}}{8\mu m_{\rm H}\pi\kappa(2GM_{\bullet})^{1/2}}\right)^2 dR}\\
                  &=\frac{2.67\times10^{-30}\,T^{-1.35}\,\nu^{-2.1} \dot{M}^2}{(8\mu m_{\rm H}\pi\kappa)^2\,2GM_{\bullet}}
                    \times \int_{R_{\rm min}}^{R_{\rm max}}{R^{-3} dR}\\
                  &=-\frac{1}{2}\times\frac{2.67\times10^{-30}\,T^{-1.35}\,\nu^{-2.1} \dot{M}^2}{(8\mu m_{\rm H}\pi\kappa)^2\,2GM_{\bullet}}
                    \times \left(R_{\rm max}^{-2}-R_{\rm min}^{-2} \right)\\
                  &= 5.85\times10^{33}\,T^{-1.35}\,\nu^{-2.1}\frac{\dot{M}^2}{M_{\bullet}}
                    \times \left(R_{\rm min}^{-2}-R_{\rm max}^{-2} \right).\\
\end{split}
\end{equation}

Converting to convenient units, the equation becomes

\begin{equation}
\tau_\nu^{\rm ff}  = 1.23\times10^{16}\,T^{-1.35}\,\nu^{-2.1}\frac{\dot{M}^2}{M_{\bullet}}
                    \times \left(R_{\rm min}^{-2}-R_{\rm max}^{-2} \right).\\
\end{equation}

In this equation, $T$ is in K, $\nu$ is in GHz, $\dot{M}$ and
$M_{\bullet}$ are in $M_\odot\,{\rm yr^{-1}}$ and $M_\odot$,
respectively, and $R$ is in pc. For NGC\,4261, the free-free optical
depth then is 0.000\,94, so that radiation at 15.4\,GHz can pass
through almost unaffected by free-free absorption. Thus, a spherical
accretion halo around the AGN can depolarize the synchrotron radiation
without being seen as an absorber. As this is a promising approach, we
calculate $RM$ and $\tau$ for each galaxy in the sample. Accretion
rates are difficult to determine, but they can be estimated using the
black hole masses as follows. The luminosity at which the force
generated by radiation pressure balances the gravitational force is
called the Eddington luminosity. It can be derived by equating the
force due to radiation pressure and the gravitational force

\begin{equation}
\begin{split}
\frac{L_{\rm Edd}\sigma_{\rm T}}{4\pi r^2c}&=\frac{GM_\bullet m}{r^2}\\
L_{\rm Edd}&=\frac{4\pi cGM_\bullet m}{\sigma_{\rm T}},
\end{split}
\end{equation}
where $L_{\rm Edd}$ is the luminosity in ${\rm J\,s^{-1}}$, $c$ is the
speed of light in ${\rm m\,s^{-1}}$, $G$ is the gravitational
constant, $M_\bullet$ is the black hole mass in kg, $m$ is the mass of a
particle in kg (here, the proton mass) and $\sigma_{\rm T}$ is the
Thompson scattering cross-section in ${\rm m^2}$. 

The Eddington accretion rate can then be estimated as the mass
equivalent of the Eddington luminosity to place an upper limit on the
accretion rate.

\begin{equation}
\dot{M}=\frac{L_{\rm Edd}}{\eta c^2}.
\end{equation}

The fraction $\eta$ describes the actual accretion rate in terms of
the Eddington accretion rate, but unfortunately is only poorly
constrained by models. \cite{Brunner1997} have modelled IUE and ROSAT
data from observations of 31 radio-quiet quasars as thermal emission
from a thin accretion disc. The median black hole mass in their sample
is $6.1\times10^{8}\,M_\odot$ ($1.2\times10^{39}\,{\rm kg}$), and they
derive accretion rates of the order of 10\,\% to 30\,\% of the
Eddington accretion rate, with a median of 12\,\%. \cite{O'Dowd2002}
studied the relation between AGN power and host galaxy using 40 BL~Lac
objects and 22 radio-loud quasars. They find black hole masses between
$10^8\,M_{\odot}$ and $10^{10}\,M_{\odot}$ and derive accretion rates
of ($2\times10^{-4}$ to 1)$\times L_{\rm Edd}$. We use accretion rates
of 1\,\% of the Eddington accretion rate because this is the average
accretion rate found by \cite{O'Dowd2002} for the BL~Lac
objects. BL~Lacs are considered ``misaligned'' FR\,I radio galaxies,
the class to which NGC\,4261, Hydra~A and probably also Centaurus~A
belong. NGC\,3079 and NGC\,1052 are classified as Seyferts or LINERs,
and it is not clear what the accretion rates in these types of objects
are. NGC\,1052, however, displays a double-sided radio jet which marks
considerable nuclear activity.  Unfortunately, the black hole mass is
only poorly constrained but the suggested range overlaps with the
black hole masses suggested fo FR\,I galaxies, and for now, we assume
the same accretion rate. Cygnus~A is a FR\,II galaxy and so is
probably a powerful quasar viewed edge-on. \cite{O'Dowd2002} find
higher accretion rates for quasars than for BL~Lacs, and so the
accretion rate and hence rotation measure we find in Cygnus~A are
probably only lower limits. The model details for each source are
listed in Table~\ref{tab:depol-parms}. The rotation measures are
higher than the required $8290\,{\rm rad\,m^{-2}}$ in each of the
sources, and is higher than the required $872\,{\rm rad\,m^{-2}}$ (due
to the lower highest frequency of 5.0\,GHz at which no polarized
emission was found) in NGC\,3079. The optical depths in all sources
are small enough to be neglected. We conclude that the spherical
accretion model, if correct to an order of magnitude, can
significantly reduce the observed degree of polarization.\\

Two arguments need to be considered. First, the rotation measures we
derive critically depend on the value of $R_{\rm min}$. We have
adopted the scales on which the well-defined free-free absorbed gaps
were observed in NGC\,1052 and NGC\,4261, indicating where a different
source geometry sets in. The depolarization screens in NGC\,4261,
Hydra~A and Cygnus~A provide enough Faraday rotation as long as
$R_{\rm min}<3\,{\rm pc}$, but the rotation measures in NGC\,3079,
NGC\,1052 and Centaurus~A then drop to $100\,{\rm rad\,m^{-2}}$ and
the spherical accretion model fails. We are, however, convinced that
adopting $R_{\rm min}<1\,{\rm pc}$ is justified, and the model then
predicts enough Faraday rotation.

Second, the accretion rates in NGC\,3079 and NGC\,1052 might be
significantly lower than the 1\,\% of the Eddington accretion
rate. The bolometric luminosities of these two AGN are at least two
orders of magnitude lower than in FR\,I radio galaxies, suggesting
that the accretion rate also is lower. With $\dot{M}=0.0001\times
L_{\rm Edd}$, the model predicts rotation measures of less than
$50\,{\rm rad\,m^{-2}}$ for both objects. We conclude that our model
cannot explain the depolarization in NGC\,3079 and NGC\,1052.

\begin{table}[htpb]
\scriptsize 
\begin{center}
\begin{tabular}{llllll}
\hline
\hline
Source           & $M_{\bullet}$   & $L_{\rm Edd}$ & $\dot{M}$                 & RM                                   & $\tau_{\rm ff}$				  \\
                 & $10^6\,M_\odot$ & $10^{37}$\,W  & $M_\odot\,{\rm yr^{-1}}$  & ${\rm rad\,m^{-2}}$                  &                				  \\
\hline                                                                         					                       
NGC\,3079        & 2$^a$           & 2.5           & 4.44                      & $3.2\times 10^3$                     & $3.9\times 10^{-6}$                         \\
NGC\,1052        & 10 to 1000$^b$  & 13 to 1300    & 22 to 2200                & $2.4\times 10^4$ to $7.6\times 10^6$ & $1.9\times 10^{-5\pm1}$  \\
NGC\,4261        & $490\pm100$$^c$ & $620\pm130$   & 1090                      & $(3.1\pm0.4)\times 10^6$             & $9.5\times 10^{-4}$                         \\
Hyd~A            & 490$^d$         & 620           & 1090                      & $3.1\times 10^6$                     & $9.5\times 10^{-4}$                         \\
Cen~A            & 10$^e$          & 13            & 22                        & $2.4\times 10^4$                     & $1.9\times 10^{-5}$                         \\
Cyg~A            & 170$^f$         & 220           & 377                       & $8.3\times 10^4$                     & $3.3\times 10^{-4}$                         \\
\hline
\end{tabular}
\caption[Rotation measures for the sample sources]{Rotation measures
for the sample sources. The accretion rate has been assumed to be
1\,\% of the Eddington accretion rate, but might be significantly
lower in NGC\,3079 and NGC\,1052. References: $a$ -
\cite{Kondratko2003}, $b$ - \cite{Guainazzi2000}, $c$ -
\cite{Ferrarese1996}, $d$ - \cite{Woo2002}, $e$ - \cite{Israel1998},
$f$ - \cite{Conway1999}}
\label{tab:depol-parms}
\end{center}
\end{table}


\paragraph{King model}

In an alternative model, we assume that the gas distribution follows
the model by (\citealt{King1972}). The model was invented to describe
the density in stellar and galaxy clusters. If we assume that stars
are present even in the very vicinity of the AGN, the gas kinematics
are dominated by the stars due to their comparatively large
mass. However, the black hole begins to dominate the gas kinematics at
radii $1<r<10\,{\rm pc}$ (depending on its mass), and as we are
observing radio emission on scales of the order of 1\,pc, we actually
should not apply the \cite{King1972} model. We nevertheless present
the calculation to estimate the order of magnitude.\\

The gas density is given by the King model as

\begin{equation}
n_{\rm e} = n_0 \left[ 1 + \left(\frac{r}{r_{\rm c}}\right)^2 \right] ^ {-3/2}.
\end{equation}

Here, $n_0$ and $r_{\rm c}$ are normalization constants. We require
$n_{\rm e}=10^9\,{\rm m^{-3}}=10^3\,{\rm cm^{-3}}$ at
$r=6.17\times10^{15}\,{\rm m}=0.2\,{\rm pc}$, and so
$n_0=2.83\times10^9\,{\rm m^{-3}}$ and $r_{\rm
c}=6.17\times10^{15}\,{\rm m}$.

Inverting Eq.~\ref{eq:electron_density} yields the density $\rho$

\begin{equation}
\rho=2\mu m_{\rm H}n_{\rm e},
\end{equation}
which we use to derive the equipartition magnetic field:

\begin{equation}
\begin{split}
\frac{B^2}{2\mu_0} &=  2\mu m_{\rm H} n_0             \left[ 1 + \left( \frac{r}{r_{\rm c}} \right)^2 \right] ^ {-3/2}\\
\Leftrightarrow B  &= (4\mu_0\mu m_{\rm H} n_0)^{1/2} \left[ 1 + \left( \frac{r}{r_{\rm c}} \right)^2 \right] ^ {-3/4}.\\
\end{split}
\end{equation}

Inserting this and the electron density into Eq.~\ref{eq:rm} yields

\begin{equation}
\begin{split}
RM   &= 2.64~10^{-13}\times\int_{R_{\rm min}}^{R_{\rm max}} n_0 \left[ 1 + \left( \frac{r}{r_{\rm c}} \right)^2 \right] ^ {-3/2}\\
     & \times (4\mu_0\mu m_{\rm H} n_0)^{1/2} \left[ 1 + \left( \frac{r}{r_{\rm c}} \right)^2 \right] ^ {-3/4}dr\\
     &= 2.64~10^{-13} (4\mu_0\mu m_{\rm H} n_0^3)^{1/2}
       \times \int_{R_{\rm min}}^{R_{\rm max}} \left[ 1 + \left( \frac{r}{r_{\rm c}} \right)^2 \right] ^ {-9/4}dr\\
     &= a\times \int_{R_{\rm min}}^{R_{\rm max}} \left[ 1 + \left( \frac{r}{r_{\rm c}} \right)^2 \right] ^ {-9/4}dr\\
     &= a\biggl(R_{\rm max}\,{_2}F_1 \left[ \frac{1}{2},~\frac{9}{4};~\frac{3}{2};~-\frac{R^2_{\rm max}}{r^2_{\rm c}} \right]
         -R_{\rm min} \,{_2}F_1 \left[ \frac{1}{2},~\frac{9}{4};~\frac{3}{2};~-\frac{R^2_{\rm min}}{r^2_{\rm c}} \right]\biggr),\\
\label{eq:isothermal_king}
\end{split}
\end{equation}
where $a=2.64~10^{-13} (4\mu_0\mu m_{\rm H}
n_0^3)^{1/2}=2.872\times10^{-15}$ and ${_2}F_1[k,~l;~m;~n]$ is the
Hypergeometric2F1 function

\begin{equation}
{_2}F_1[k,~l;~m;~n]=\sum^{\infty}_{i=0}(k)_i(l)_i/(m)_in^i/i!,
\end{equation}
in which brackets denote

\begin{equation}
\begin{split}
(x)_0&=1\\
(x)_i&=x(x+1)(x+2)...(x+i).\\
\end{split}
\end{equation}

The integral can be evaluated numerically and yields a rotation
measure of $1.9\,{\rm rad\,m^{-2}}$, much too low to depolarize the
radiation. Even choosing $n_0$ to be one order of magnitude higher
than in the previous case yields a rotation measure of only
$59\,{\rm rad\,m^{-2}}$, because Eq.~\ref{eq:isothermal_king} goes
with $n_0^{3/2}$. We conclude that the \cite{King1972} model
underestimates the Faraday rotation by several orders of magnitude.

\paragraph{Isothermal gas distribution}

Let us assume that the circumnuclear gas is isothermal and has a
rotationally symmetric power-law density profile with index $-\iota$. Its
pressure can then be written as

\begin{equation}
p(r)=\rho_0\,r^{\iota}\,k\,T
\end{equation}

The gas pressure has to balance the gravitational potential of the
black hole:

\begin{equation}
\begin{split}
\nabla_r\Phi    &= \frac{1}{\rho_0\,r^{\iota}}\frac{\partial p}{\partial r}\\
-\frac{GM}{r^2} &= \frac{1}{\rho_0\,r^{\iota}}\rho_0\,k\,T\,(\iota)\,r^{\iota-1}\\
-\frac{GM}{r^2} &= \frac{k\,T}{r}
\end{split}
\end{equation}

The two sides of the equation can be equal only at a single point
because of the different dependencies on the radius. An isothermal gas
is therefore not able to exist in the gravitational field of a black
hole.

\section{Additional Results from NGC\,3079}
\label{sec:res_3079}

\begin{figure}[htpb!]
\centering
\includegraphics[width=\linewidth]{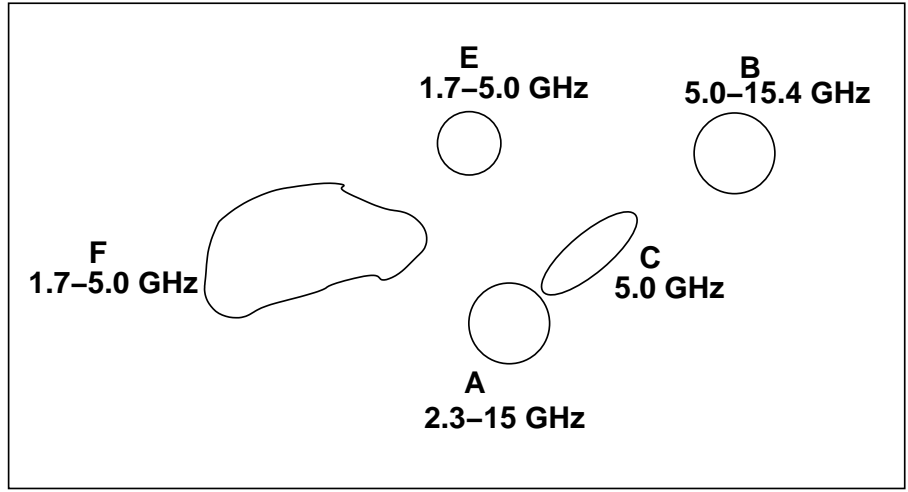}
\caption[Sketch of all components seen in NGC\,3079]{Sketch of all
components seen in NGC\,3079,  showing the frequencies at which
each component was detected.}
\label{fig:NGC3079-sketch}
\end{figure}

This section discusses more aspects of the multi-epoch,
multi-frequency observations of NGC\,3079. As an orientation, we have
sketched NGC\,3079 in Fig.~\ref{fig:NGC3079-sketch} together with the
frequencies at which the various components have been detected.

\subsection{Spectra}
\label{sec:spectra}

\begin{figure}[htpb!]
\centering
\includegraphics[width=0.8\linewidth, clip]{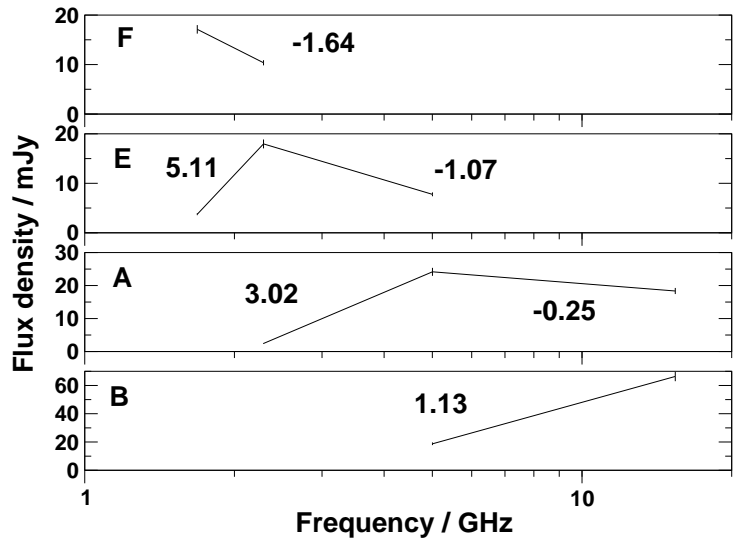}
\caption[Spectra of components in NGC\,3079]{Spectra of components in
NGC\,3079 with the spectral indices labelled. The turnover frequency
of each component increases with decreasing right ascension,
suggestive of a wedge-shaped absorber in front of the AGN. All
spectral indices were calculated from the September 2002 data, except
for $\alpha^{5.0}_{15.4}$ for component $A$ and $B$, which were
calculated from the November 2000 data to avoid variability effects.}
\label{fig:spectra}
\end{figure}

The flux densities of components $A$, $B$, $E$ and $F$ measured in
November 2000 and September 2002 are shown in Fig.~\ref{fig:spectra},
displaying three unusual features. First, there are {\it unusually
inverted} spectra, e.g., $\alpha^{2.3}_{5.0}(A)=3.02$ and
$\alpha^{1.7}_{2.3}(E)=5.11$. Second, there is the {\it fairly steep}
spectrum of $F$, $\alpha^{2.3}_{5.0}(F)=-1.64$. Third, there is a
striking correlation between the peak frequencies and the component
positions in the source, namely that that the turnover frequencies are
increasing with decreasing right ascension. This is illustrated in
Fig.~\ref{fig:spectra}. We now address these three points in turn.

\begin{table}[ht!]
\scriptsize
\begin{center}
\begin{tabular}{lrcc}
\hline
\hline
Epoch        & Freq.  & $S_{\rm max}$          & $T_{\rm B}$\\
             & GHz    & ${\rm mJy\,beam^{-1}}$ & $10^6$\,K \\
\hline                                                  
\multicolumn{4}{c}{Component $A$}\\
Nov 20, 1999 &   5.0  & 25         & 140 \\        
             &  15.4  & 6.21       & 3.6 \\
Mar  6, 2000 &   5.0  & 24         & 130 \\
             &  15.4  & 7.02       & 4.0 \\
Nov 30, 2000 &   5.0  & 23         & 130 \\
             &  15.4  & 17.5       &  10 \\ 
Sep 22, 2002 &  1.7   & -          &  -  \\
             &  2.3   & 1.13       &  29 \\
             &  5.0   & 22         & 120 \\
\hline                                    
\multicolumn{4}{c}{Component $B$}\\       
Nov 20, 1999 &   5.0  & 13.9       &  76 \\        
             &  15.4  & 46         &  26 \\
Mar  6, 2000 &   5.0  & 14.6       &  80 \\
             &  15.4  & 54         &  31 \\
Nov 30, 2000 &   5.0  & 14.6       &  80 \\
             &  15.4  & 62         &  36 \\
Sep 22, 2002 &  1.7   & -          &  -  \\
             &  2.3   & -          &  -  \\
             &  5.0   & 14.8       &  81 \\
\hline                                   
\multicolumn{4}{c}{Component $E$}\\      
Nov 20, 1999 &   5.0  & 3.45       &  19 \\        
             &  15.4  & -          &  -  \\
Mar  6, 2000 &   5.0  & 3.22       &  18 \\
             &  15.4  & -          &  -  \\
Nov 30, 2000 &   5.0  & 3.96       &  22 \\
             &  15.4  & -          &  -  \\
Sep 22, 2002 &  1.7   & 2.94       & 140 \\
             &  2.3   & 19.8       & 510 \\
             &  5.0   & 5.57       &  30 \\
\hline                                    
\multicolumn{4}{c}{Component $F$}\\      
Nov 20, 1999 &   5.0  & -          & -   \\        
             &  15.4  & -          & -   \\
Mar  6, 2000 &   5.0  & -          & -   \\
             &  15.4  & -          & -   \\
Nov 30, 2000 &   5.0  & -          & -   \\
             &  15.4  & -          & -   \\
Sep 22, 2002 &  1.7   & 9.30       & 440 \\ 
             &  2.3   & 2.18       &  56 \\
             &  5.0   & -          & -   \\
\hline
\end{tabular}
\caption[Brightness temperatures of NGC\,3079 components]{Brightness
temperatures of NGC\,3079 components}
\label{tab:ngc3079_T_Bs}
\end{center}
\end{table}

\paragraph{Unusually inverted spectra} 

First we demonstrate that the emission from NGC\,3079 is synchrotron
emission. The brightness temperatures $T_{\rm B}$ (the physical
temperature a blackbody would need to have to produce the observed
flux density) can be calculated by inverting the Rayleigh-Jeans
approximation of Planck's law, yielding

\begin{equation}
T_{\rm B}=\frac{c^2B_\nu}{2\nu^2k}
\end{equation}

Here, $T_{\rm B}$ is the brightness temperature in K, $c$ is the speed
of light in ${\rm m\,s^{-1}}$, $B_{\nu}$ is the flux density per unit
solid angle, $\Omega$, in W\,m$^{-2}$\,Hz$^{-1}$\,sr$^{-1}$, $\nu$ is
the observing frequency in Hz and $k$ is Boltzmann's constant in J/K.

$\Omega$ can be computed as the area of a circle intersected by a cone
with opening angle $\beta$ (the FWHM diameter of the observing beam),
yielding $\Omega=4\pi\sin^2(\beta/4)$, or, using $\sin(x)\approx x$
for $x<<1$ and introducing units of mas, $\Omega \approx 1.846\times
10^{-17}\times \beta^2$. Converting $B_{\nu}$ and $\nu$ to Jy/beam and
GHz, respectively, yields

\begin{equation}
T_{\rm B}=1.76\times10^{12}\,{\rm m^2\,K\,s^{-2}\,J^{-1}}\frac{B_{\nu}}{\nu^2\beta^2}.
\end{equation}

The high brightness temperatures that we calculated are shown in
Table~\ref{tab:ngc3079_T_Bs} and provide good evidence that the
emission process of the radio components is synchrotron emission. The
maximum spectral index of synchrotron radiation, however, is $+2.5$
(\citealt{Rybicki1979}, eq. 6.54), if the energies of the electrons in
the jet are power-law distributed. Higher spectral indices can only be
due to external media, and we therefore are confident that the components
with $\alpha>+2.5$ are obscured by a foreground absorber. We identify
the absorption mechanism as free-free absorption with

\begin{equation}
S_{\nu}=S_{\nu,0}\,\nu^{\alpha_0}\,e^{-\tau_\nu^{\rm ff}}
       =S_{\nu,0}\,\nu^{\alpha_0}\,e^{-\tau^{\rm ff}\,\nu^{-2.1}}
\label{eq:ffa_opt_depth}
\end{equation}
where $S_{\nu}$ and $S_{\nu,0}$ are the observed and intrinsic flux
densities of a component, respectively, $\alpha_0$ is the intrinsic
spectral index, $\tau_\nu^{\rm ff}$ is the optical depth as given in
Eq.~\ref{eq:ffa} and $\tau^{\rm ff}$ is the optical depth without the
frequency term. The frequency dependence of $\tau_{\nu}^{\rm ff}$
causes the flux density to decrease exponentially towards lower
frequencies, so that the spectral index can be arbitrarily high.
Eq.~\ref{eq:ffa_opt_depth} has three unknowns: $S_{\nu,0}$, $\alpha_0$
and $\tau^{\rm ff}$. We have detected both $A$ and $E$ at three
frequencies and can therefore solve for the three unknowns (we do not
have three {\it simultaneous} measurements for $A$, and will therefore
take the 5.0\,GHz and 15.4\,GHz data from November 2000 and the 2.3\,GHz
measurement from September 2002). The results are given in
Table~\ref{tab:fits}. The emission measure, $k$, is given by

\begin{equation}
k=n_{\rm e}^2L,
\end{equation}
where $L$ is the line-of-sight path length through the medium with
electron density $n_{\rm e}$, assuming that $T=10^4\,{\rm K}$ and
$N_+=N_-=n_{\rm e}$. For $A$ and $E$, we find emission measures of
$6.4\times10^7\,{\rm pc\,cm^{-6}}$ and $5.2\times10^7\,{\rm
pc\,cm^{-6}}$. For an estimate of the line-of-sight column densities,
$n_{\rm e}L$, we assume that the absorber has the same depth as the
FWHM of the synthesized beam. At the distance of NGC\,3079, 1\,mas
corresponds to 0.073\,pc, and a circular beam with the same area as
the elliptical beam used for the flux density measurements,
$4\,\mas\times3\,\mas$, has a FWHM of 3.46\,mas, or 0.25\,pc. The
column densities then follow to be $1.23\times10^{22}\,{\rm cm^{-2}}$
for $A$ and $1.11\times10^{22}\,{\rm cm^{-2}}$ for $E$. This is
similar to the absorbers that have been found in, e.g., NGC\,1052
($3.7\times10^{22}\,{\rm cm^{-2}}$, \citealt{Kameno2001}), NGC\,4261
($5\times10^{22}\,{\rm cm^{-2}}$, \citealt{Jones2001}) and Hydra~A
($3.7\times10^{22}\,{\rm cm^{-2}}$, \citealt{Taylor1996}) based on the
same technique, and so indicates that the absorber in front of $A$ and
$E$ has a similar width and similar physical conditions.


\begin{table}[htpb]
\scriptsize 
\begin{center}
\begin{tabular}{lccccc}
\hline
\hline
Component & $S_{\nu,0}$ & $\alpha_0$ & $\tau^{\rm ff}_{\rm 1\,GHz}$  & k               & $n_{\rm e}L$ \\
          & mJy         &            &                               & pc\,cm$^{-6}$   & cm$^{-2}$    \\
\hline
$A$       & 180         & -0.8       & 21                            & $6.4\times10^7$ & $1.23\times10^{22}$ \\
$E$       & 10000       & -4.1       & 17                            & $5.2\times10^7$ & $1.11\times10^{22}$ \\
\hline
\end{tabular}
\caption[Intrinsic properties of components $A$ and $E$]{Intrinsic
properties of components $A$ and $E$. $\alpha_0$ and $\tau^{\rm ff}$
are dimensionless quantities.}
\label{tab:fits}
\end{center}
\end{table}

Component $A$ is reasonably well modelled with an intrinsically
optically thin ($\alpha_0=-0.8$) synchrotron source behind a
moderately thick ($\tau^{\rm ff}_{\rm 1\,GHz}=21$, $\tau^{\rm ff}_{\rm
5\,GHz}=0.72$) free-free absorber. The fit results from component $E$
($\tau^{\rm ff}_{\rm 1\,GHz}=17$, $\tau^{\rm ff}_{\rm 5\,GHz}=0.58$),
however, raise an uncomfortable problem: Our model predicts the
intrinsic spectral index to be $-4.1$, much less than commonly
observed in synchrotron sources. This result is robust with respect to
changes of the flux densities at 1.7\,GHz and 2.3\,GHz by as much as
factors of two, and so amplitude calibration errors and measurement
errors from the image would have only little effect. Such extremely
steep spectra are not impossible to explain. One possibility is that
the electrons in the source age and produce an exponential cutoff at
high frequencies, but this process requires typically $10^8\,{\rm yr}$
in an undisturbed environment without further energy injection and so
is unlikely to happen within parsecs of an AGN. Another possibility is
that the electrons are simply not power-law distributed, but have a
relativistic thermal distribution. Such models have been considered
by, e.g., \cite{Jones1979} and \cite{Beckert1997}, and they usually
reveal an exponential cutoff at high frequencies.  Both explanations
are unsatisfactory because they require the electron distribution
cutoff to lie very close to the frequency at which $\tau=1$, but we
see no better solution.

The limits on $\alpha^{1.7}_{2.3}$ and $\alpha^{1.7}_{5.0}$ of
component $B$ might also be due to free-free absorption, but as $B$
was detected only at 5.0\,GHz and 15\,GHz, we cannot solve
Eq.~\ref{eq:ffa_opt_depth}. We can, however, draw a conclusion from
its extent. $B$ and $E$ are resolved in the 5.0\,GHz images from
November 1999, March 2000 and November 2000. If the inverted spectrum
of $B$ was due to synchrotron self-absorption, one would expect it to
be compact, so its extend indicates that it is unlikely to be
synchrotron self-absorbed and the inverted spectrum is more likely to
be due to free-free absorption.

\paragraph{Steep spectra} Component $F$ has a spectral index
of $-1.64$. Its turnover frequency lies below the frequency range
covered by our observations, so we cannot make a strong statement
about physical parameters. But as all other components have spectral
indices $>4.00$ in parts of their spectra, we can assume that this is
also the case for $F$, and consequently, that an east-west density or
path length gradient in the free-free absorber has moved the peak
frequency to frequencies $<1.7\,{\rm GHz}$. 

\paragraph{Correlation of peak frequencies and right ascension} The
spectra shown in Fig.~\ref{fig:spectra} have been arranged in order of
decreasing right ascensions of the components, and one sees a
systematic shift in the turnover frequency towards higher frequencies
from top to bottom. This arrangement suggests a connection between the
peak frequencies and the component position in the source. If this
effect is not accidental (which has a 12.5\,\% probability), it might
be due to the sub-structure of a free-free absorber in front of the
radio source. Free-free absorption has been suggested to be involved
in $A$, $E$, and probably also $B$, so perhaps a single absorber
covers all components, whose optical depth increases towards the
west. As an example, it could be that the absorber has the shape of a
wedge, the thin part of which is in front of $F$ and the thicker part
of which is in front of $B$.

\subsection{Proper Motion}

\begin{figure}[htpb!]
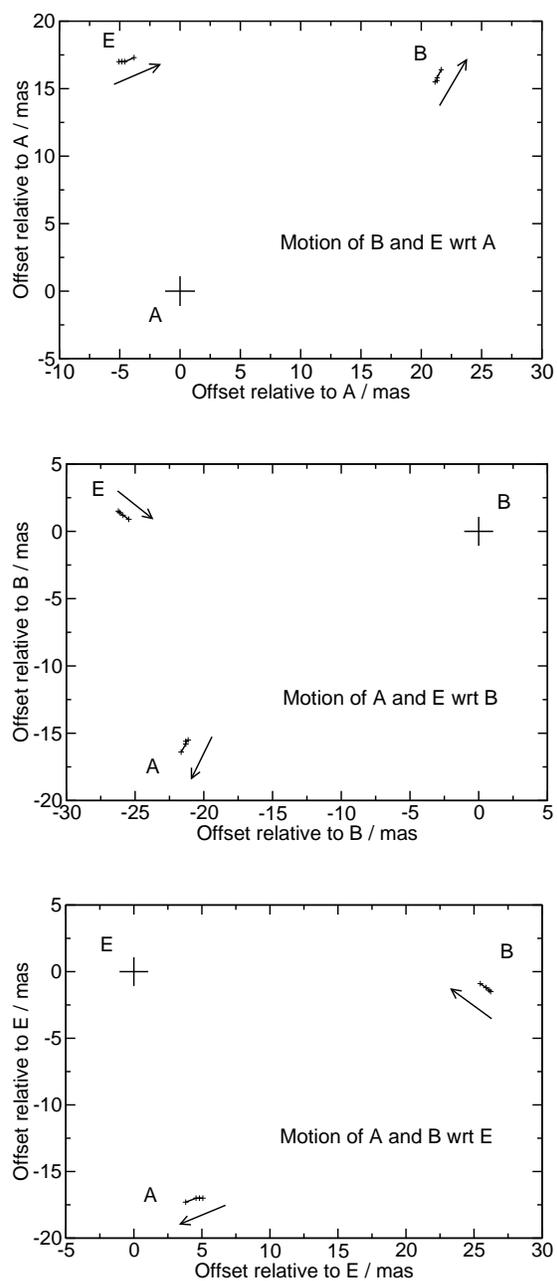

\centering
\subfigure{
\includegraphics[width=0.6\linewidth, clip]{chap5/plots/xy-rel_to_A.eps}}\\
\subfigure{
\includegraphics[width=0.6\linewidth, clip]{chap5/plots/xy-rel_to_B.eps}}\\
\subfigure{
\includegraphics[width=0.6\linewidth, clip]{chap5/plots/xy-rel_to_E.eps}}\\
\caption[Relative positions among components $A$, $B$ and $E$]{Relative
positions among components $A$, $B$ and $E$ in NGC\,3079 at 5.0\,GHz
for all epochs with arrows showing the direction of relative motion.}
\label{fig:3079_positions}
\end{figure}

So far, the location of the AGN in NGC\,3079 has not been accurately
determined. Based on VLBI maser observations, \cite{Trotter1998} have
suggested that the core lies on the $A-B$ axis, where that axis
intersects the line of masers, and this identification has been
supported by \cite{Sawada-Satoh2000} and, recently, by
\cite{Kondratko2003}.\\

The relative positions of components $A$, $B$ and $E$ measured at
5.0\,GHz are listed in Table~\ref{tab:ngc3079_separations} and are
plotted in Fig.~\ref{fig:3079_positions}. In each plot, one component
serves as a stationary reference point, and the arrows indicate in
which direction the other two components are moving. The errors are
$\sqrt{2}\times0.15\,{\rm mas}=0.21\,{\rm mas}$, because the error of
the reference point was added to the other errors.

\paragraph{A stationary} If $A$ is chosen as a stationary reference
point, then $B$ is seen to move radially away from $A$, and $E$ is
moving almost perpendicular to the $A-E$ axis. The motion of $E$ would
not be expected in the standard AGN model, where radio components move
radially away from the AGN. If $E$ was ejected by the AGN located at
or near $A$, then its path must have a large bend that first makes it
travel away from $A$ and then directs it in the same direction in
which $B$ is travelling. We do not consider this scenario further.

\paragraph{B stationary} If $B$ is the reference, then $A$ is seen to
move radially away from $B$, but $E$ is moving towards the $A-B$
axis. This situation, like the previous, does not agree very well with
the standard model, in which one would not expect components to move
in a direction away from the AGN. Nevertheless, this scenario is of
particular importance. \cite{Sawada-Satoh2000}, comparing their maser
and continuum observations to those of \cite{Trotter1998}, found $B$
to be stationary with respect to the brightest maser features. In this
case, plotting the separations relative to $B$ yields the same speeds
and directions as plotting them relative to the masers and hence to
the true (radio-invisible) AGN centre of mass location, and $E$ would
indeed be moving in the direction of the nucleus. Furthermore,
\cite{Kondratko2003} recently found good evidence for a rotating disc
centred on the $A-B$ axis, confirming those results.

\paragraph{E stationary} If $E$ is the reference, then $A$ and $B$ 
move in almost perpendicular directions. This scenario has one
attractive advantage, namely that the back-extrapolation of the
trajectories of $A$ and $B$ intersect in the past. This allows them to
be ejected from the same location, south-west of $B$. $E$ could be an
object that is kinematically disconnected from the nucleus and is not
part of the AGN ejecta. An obvious guess is to identify it as a
supernova, supported by the fact that $E$ has not been observed by
\cite{Trotter1998} in 1992, but was detected by
\cite{Sawada-Satoh2000} in 1996, although they misidentified it with
$B$. Thus, $E$ must have appeared between 1992 and 1996. At 5.0\,GHz,
it has an average flux density of 4.89\,mJy, corresponding to
$1.3\times10^{20}\,{\rm W\,Hz^{-1}}$ at a distance of 15\,Mpc. This is
comparable to the brightest supernova in, e.g., M82, $41.9+58$, having
$4.6\times10^{19}\,{\rm W\,Hz^{-1}}$ at 5\,GHz
(\citealt{Kronberg1985}), although most of the supernovae reported in
that publication have ten times lower luminosity. However, $E$ lacks
the typical decrease in flux density that is seen as the supernova
shells expand. Although \cite{Sawada-Satoh2000} find a flux density of
$(5.7\pm1.7)\,{\rm mJy}$ at 1.4\,GHz, compared to $(3.06\pm0.15)\,{\rm
mJy}$ in our 1.7\,GHz observations, the integrated flux densities were
$(4.15\pm0.21)\,{\rm mJy}$, $(3.08\pm0.15)\,{\rm mJy}$,
$(5.20\pm0.26)\,{\rm mJy}$ and $(7.11\pm0.36)\,{\rm mJy}$ in our
5.0\,GHz observations. This sequence of measurements does not suggest
a gradual decline of flux density with time like has been observed in
M82's $41.9+58$ during 1971-1981, fading with a rate of between
$9\,\%\,{\rm yr^{-1}}$ and $24\,\%\,{\rm yr^{-1}}$
(\citealt{Kronberg1985}).

\paragraph{Core invisible} It is also possible that none of the radio
emitters is coincident with the dynamical centre. This situation is
common in radio galaxies and quasars where one usually sees a $\tau=1$
surface in the jet, and not a component at the location of the black
hole, but as a Seyfert~2 galaxy, NGC\,3079 is certainly an entirely
different object. If the dynamical centre is invisible in the radio, the
components may be shocks in an AGN outflow.

\paragraph{Helical paths} Another interesting interpretation is that
all components are moving on paths that follow a helix with a fairly
wide opening angle. If the axis of the helix is inclined at a small
angle to the line of sight, inwards motion of components would be
easily explained.\\

None of the five scenarios sketched above yields compelling evidence
for the location of the core. However, we think that the model in
which $B$ is stationary (it might be equivalent to the model in which
the core is invisible) and the helical paths model need to be
considered in the future. Maser observations are usually regarded as
one of the ``best possible'' pieces of evidence for the location of
black holes in AGN, but the kinematic centres of circumnuclear discs
do not necessarily need to coincide with these cores. This has been
found in, e.g., NGC\,4261 (\citealt{Ferrarese1996}, see
\S\ref{sec:ngc4261}), although in that case, the multiple centres are
interpreted as being due to past merging events, and spiral galaxies
like NGC\,3079 are generally not considered as descendants of galaxy
mergers.  We plan to monitor NGC\,3079 during the next two years using
phase-referencing with good frequency coverage between 1.7\,GHz and
22\,GHz, allowing us to follow absolute proper motions to resolve the
ambiguous findings and solve for the core position.

\subsection{Remarks on Individual Components}

Component $A$ is the most compact component in NGC\,3079. The ratio of
peak flux density to integrated flux density, $x$, has an average of
$0.85$ when measured from tapered images and of $0.69$ when measured
from the full resolution images from November 1999, March 2000 and
November 2000 (Table~\ref{tab:ngc3079_parms1}). Hence, it is only
slightly resolved.\\

Component $B$ is clearly resolved at 5.0\,GHz in the first three epochs
($x=0.39$) and compact at 15\,GHz ($x=0.83$). This is probably not a
resolution effect because the 5.0\,GHz and 15\,GHz images have similar
resolution (the 15\,GHz resolution even being slightly smaller), but a
combination of steep spectral index and lower sensitivity can in
principle account for the effect. The extension of $B$ to the south in
the 5.0\,GHz images has four to five contours, the lowest contour being
drawn at three times the image rms. The extension therefore has a
surface brightness of 8 to 16 times $3\,\sigma$, which is of the order
of $1.6\,{\rm mJy\,beam^{-1}}$ to $3.2\,{\rm
mJy\,beam^{-1}}$. Assuming a spectral index of $-0.7$ as is common for
optically thin synchrotron radiation then predicts a surface
brightness at 15.4\,GHz of $0.7\,{\rm mJy\,beam^{-1}}$ to $1.5\,{\rm
mJy\,beam^{-1}}$. This is visible as an extension of the lowest
contour in the 15\,GHz images which are drawn at around $1\,{\rm
mJy\,beam^{-1}}$, especially in the November 2000 epoch, where this
contour is at $0.82\,{\rm mJy\,beam^{-1}}$. Furthermore,
\cite{Kondratko2003} finds the same extension of $B$ at 22\,GHz to
have a surface brightness of $({\rm 0.42~to~0.84)\,mJy\,beam^{-1}}$,
consistent with a similar extrapolation of the surface brightness
found by us at 5.0\,GHz. Thus, the northern and compact part of $B$
has an inverted spectrum between 5.0\,GHz and 15.4\,GHz, whereas the
extension to the south fades towards higher frequencies. If these two
different spectral shapes on such small linear scales are interpreted
as being due to free-free absorption, the situation does not agree
well with the idea of a single, wedge-shaped absorber in front of the
radio components as suggested in \S\ref{sec:spectra}. But there are
too few observational constraints to conclusively tell what causes the
spectral gradient in $B$.

Component $E$ is also resolved, with an average peak-to-integrated
ratio of 0.30 in the full-resolution images. Furthermore, the position
angle of its largest extent appears to change between the first three
epochs. We have fitted elliptical Gaussians to the $(u,v)$ data of all
5.0\,GHz epochs, using data with a $(u,v)$ radius of $<100\,{\rm
M\lambda}$ during the first three epochs and using data with a $(u,v)$
radius of $<50\,{\rm M\lambda}$ of the fourth epoch. The limits were
chosen to reduce the ellipticity of the $(u,v)$ coverages while
keeping as many data as possible. The results of the fits are listed
in Table~\ref{tab:PA_E}. The P.A. changes between the first three
epochs are $28.7^{\circ}$ and $55.8^{\circ}$, whereas the P.A. change
between the November 2000 and September 2002 epochs is difficult to
determine due to the $n\pi$ ambiguity of the P.A. However, the rates
at which the P.A. changed is $97.2^{\circ}\,{\rm yr^{-1}}$ between
November 1999 and March 2000 and $76.3^{\circ}\,{\rm yr^{-1}}$ between
March 2000 and November 2000. Extrapolating over the 1.811\,yr period
between the third and fourth epoch yields P.A.s of $231.8^{\circ}$
($\corr 51.8^{\circ}$ when the P.A.s are reflected onto the
range $0^\circ$ to $180^\circ$) for a rate of $97.2^{\circ}\,{\rm
yr^{-1}}$ and $269.8^{\circ}$ ($\corr 89.8^{\circ}$) for a rate of
$76.3^{\circ}\,{\rm yr^{-1}}$, or an average P.A. of
$70.8^{\circ}$. This is very close to the observed P.A. of
$65.0^{\circ}$.

However, we deem this result not very reliable. First, the P.A.s
changed considerably when the maximum $(u,v)$ radius in imaging was
changed and second, the P.A.s predicted for the September 2002
observation, $51.8^{\circ}$ and $89.8^{\circ}$, span $38.0^{\circ}$,
and hence there is a probability of
$100\times(38.0^{\circ}/180^{\circ})=21\,\%$ of the measured
P.A. falling into that range if it was random. The P.A. changes of $E$
are not the result of a changing $(u,v)$ coverage, as the restoring
beam after tapering the images is independent of the P.A. of $E$.

Combining this result with the observed spectra and proper motion, we
find the least exotic explanation for the radio emission of component
$E$ is an evolving shock in a large jet-like outflow.

\begin{table*}[htpb]
\scriptsize
\begin{center}
\begin{tabular}{lccc|ccc}
\hline
\hline
                & \multicolumn{3}{c|}{Beam size of image}           & \multicolumn{3}{c}{Size of $E$}\\
Epoch           & $b_{\rm maj}$ & $b_{\rm min}$ & PA($b_{\rm maj}$) & $b_{\rm maj}$  & $b_{\rm min}$ & PA($b_{\rm maj}$)\\
                & mas           & mas           & deg               & mas            & mas           & deg              \\
\hline
Nov 1999        & 2.69          & 2.20          & -18.3             & 2.50           & 1.42          & 9.2   \\
Mar 2000        & 1.85          & 1.68          & -29.6             & 2.09           & 1.17          & 37.9  \\
Nov 2000        & 2.35          & 1.86          & -24.8             & 2.32           & 1.57          & 93.7  \\
Sep 2002        & 4.16          & 3.81          & -50.6             & 2.28           & 2.07          & 65.0  \\
\hline
\end{tabular}
\caption[Position angle changes of component $E$]{Position angle changes of component $E$.}
\label{tab:PA_E}
\end{center}
\end{table*}

\section{Statistics}

We have carried out a literature review to compare the findings in
NGC\,3079 to other Seyfert galaxies and AGN to address the following
questions.

\begin{itemize}

\item The spectral indices of the radio components in NGC\,3079 were
found to be unusually inverted. Is this peculiar to NGC\,3079 or do
other Seyferts have inverted spectra as well? The result might
indicate whether free-free absorption is common in Seyferts or not.

\item NGC\,3079, like some other Seyferts, shows a kpc-scale outflow,
and if pc-scale components tend to be ejected in the same direction,
this could help in the identification of components in NGC\,3079. Is
there a general correlation between the host galaxy rotation axes in
Seyferts and the pc-scale emission? This might indicate that Seyfert
radio ejecta are undergoing significant interactions with interstellar
gas, or precession in the accretion discs, whereas powerful objects
frequently are aligned from the innermost regions to the outer radio
lobes.

\item How do the VLBI-measured spectral indices in Seyferts generally
compare to those measured with smaller arrays?

\end{itemize}

\subsection{Compilation of the Sample}
\label{sec:stats}

The range of radio properties displayed by Seyfert galaxies has been
explored extensively based on WSRT, VLA or ATCA observations at
1.4\,GHz, 5\,GHz or 8.4\,GHz at arcsec resolution by
\cite{1978AaA....64..433D}, \cite{1984ApJ...285..439U,
1989ApJ...343..659U}, \cite{1987MNRAS.228..521U},
\cite{1990ApJS...72..551G}, \cite{1996ApJ...473..130R},
\cite{1999ApJ...516...97N}, \cite{1999AAS..137..457M},
\cite{2001ApJ...558..561U}, and \cite{VirLal2001}.  They studied the
range of radio luminosities, radio source sizes, morphologies,
alignment between jet and host galaxy, and spectral indices, and
looked for dependence of properties on Seyfert type. With the
exception of \cite{2001ApJ...558..561U}, who found 50\,\% of
low-luminosity Seyferts to have a flat spectrum core, they all found
that most Seyfert galaxies have steep, optically-thin synchrotron
radio spectra, with 5\,\% to 35\,\% (depending on the sample) showing
a flat-spectrum core, much less than the rate of occurrence of
flat-spectrum cores in powerful radio galaxies.  No difference was
found between the spectral index distributions of the two Seyfert
types.

However, the $0.3^{\prime\prime}$ to several arcsec beams used in
those studies do not allow details of nuclear structure to be
separated or the distribution of jet bending angles to be
constructed. Further, they would probably blend together multiple
radio components within the nucleus and could hide the presence of
more absorbed cores.

Many further VLBI observations of Seyfert galaxies are available in
the literature but no statistical summary from those observations has
been presented.  Thus, we made a literature search for VLBI
observations of Seyferts to compare to our observations of
NGC\,3079. The resulting list in Table~\ref{tab:seyferts} contains, we
believe, all VLBI imaging observations of AGN classified as Seyfert 1,
2 or 3 by \cite{Veron2001} (irrespective of any sub-classification),
published as of August 2003 with a linear resolution of $<10~{\rm
pc}$, although it is not impossible that we have inadvertently omitted
some works.  This sample consists mostly of the nearest and best-known
Seyferts.  They were often selected for observing because they show
bright radio nuclei at arcsecond resolution, and so the sample is not
complete in any sense.  Large, uniform samples of Seyfert galaxies are
being observed with the VLBA by \cite{Murray1999} and
\cite{2002AAS...200.4507S}, but the results are not yet
available. Proper motions in Seyferts are being presented by Roy et
al. (in prep.) and we confine ourselves here to spectral indices and
jet bending.

Spectral indices have been taken from the literature if they were
measured simultaneously between L-band and C-band, i.e., between
1.4\,GHz to 1.7\,GHz and 4.9\,GHz to 5.0\,GHz.  To compare pc and
kpc-scale structure, we have searched for interferometric radio
observations with a linear resolution of $>100\,{\rm pc}$, made with,
e.g., the VLA or MERLIN. The angles have been measured from the images
by us, from the identified core or the brightest component to the next
bright component. The complete list is shown in
Table~\ref{tab:seyferts}, those objects used for determining spectral
indices are listed in Table~\ref{tab:seyferts_SI}, and the angles are
shown in Table~\ref{tab:seyferts_angles}. We have plotted the data in
two histograms shown in Fig.~\ref{fig:seyfert-stats}.

\subsection{Spectral Properties of Seyferts}

We selected from Table~\ref{tab:seyferts} a subsample consisting of
the Seyfert galaxies for which dual-frequency VLBI observations have
been made to study the component spectral properties at high
resolution. This subsample is listed in Table~\ref{tab:seyferts_SI}
and includes the spectra of $A$, $B$ and $E$ in NGC\,3079 .

\subsubsection{Spectral Index Distributions}

The distribution of $\alpha^{1.4}_5$ from Table~\ref{tab:seyferts_SI}
is shown in Fig.~\ref{fig:seyfert-stats} (top).  The spectral indices
lie between $-1.8$ and $+4.6$ with a median of $-0.40$ (using the
Kaplan-Meier estimator constructed by ASURV rev 1.2,
\citealt{Lavalley1992} to account for lower limits and
discarding upper limits).  The distribution shows a tail towards
positive spectral index due to four components in NGC~1068, NGC~2639
and NGC~3079 that have strong low-frequency absorption.  

Separating the VLBI measurements by Seyfert type, we found the median
$\alpha^{1.4}_5$ for the Sy 1s was $-0.75$ and for the Sy 2s was
$-0.23$, which were not significantly different under the survival
analysis two-sample tests ($P=6\,\%$ to $12\,\%$ that the
spectral-index distributions came from the same parent distribution)
or under a Kolmogorov-Smirnov test ($P=6\,\%$ to come from the same
parent distribution), treating only lower limits as detections.

Despite the statistical insignificance of a difference between Seyfert
types, we notice that the most inverted spectra all occur in objects
that have evidence for an absorber. NGC\,3079 has good evidence for
free-free absorption, NGC\,1068 and Mrk\,348 are type 2 Seyferts with
type 1 spectra in polarized light and NGC\,2639 shows strong X-ray
absorption (\citealt{1998ApJ...505..587W}). The inner edge of such an
absorber is expected to be ionized by UV emission from the AGN which
would produce free-free absorption and hence produce inverted spectra
preferentially in the type 2s.  All four objects with inverted radio
spectra also host ${\rm H_2O}$ masers, which is also consistent with
an orientation effect if, for example, the radio core is viewed
through a molecular disc or torus, which produces ${\rm H_2O}$ maser
emission.

With spectral indices of $\alpha^{1.4}_5(A)>4.56$,
$\alpha^{1.4}_5(B)>4.41$ and $\alpha^{1.4}_5(E)=0.67$, NGC\,3079 has
contributed the two most inverted spectra seen in any Seyfert on
parsec scales. However, many more radio components in Seyferts have
flat or inverted spectra, and NGC\,3079 seems to be an extreme object,
possibly because it is an edge-on spiral galaxy and so there is more
absorbing gas in the line of sight than in face-on Seyferts.

\subsubsection{Spectral Indices Measured with Smaller Arrays}

The tail of extreme positive spectral indices that we see in our
sample is not seen at arcsec resolution in the works mentioned in
\S\ref{sec:stats}, probably due to component blending.  We made a
statistical comparison of our VLBI spectral indices to arcsec
resolution spectral indices using the Kolmogorov-Smirnov test and
chi-squared homogeneity test by treating the lower and upper limits as
detections, and by applying survival analysis two-sample tests.  The
results are summarized in Table~\ref{tab:spectra-stats2}.  Our
spectral index distribution was significantly different (i.e. level of
significance was less than the critical value of 5\,\%) from all the
comparison samples according to all tests performed, with the
exception of the Palomar Seyferts by
\cite{2001ApJS..133...77H}, which was found to be significantly
different in only the chi-squared test. This excess of flat-spectrum
sources in the Palomar Seyfert sample was already noticed by
\cite{2001ApJS..133...77H}.

\subsubsection{Fraction of Flat-Spectrum Components}

Of the 16 objects in the subsample, a majority of 13 (81\,\%) has at
least one flat or inverted spectrum VLBI component with
$\alpha^{1.4}_5 > -0.3$ (denoted by an asterisk in
Table~\ref{tab:seyferts_SI}).  This fraction is 60\,\% larger than
that measured at lower resolution by \cite{2001ApJS..133...77H} for
the Palomar Seyferts and four or more times larger than the fraction
measured at arcsec resolution for the other samples
(Table~\ref{tab:spectra-stats}).  The difference is highly significant
for most samples, having a probability of $<0.1\,\%$ of occurring at
random.  The larger fraction of flat- and inverted-spectrum components
found with VLBI is not unexpected since increasing resolution reduces
the component blending.\\

\subsubsection{Summary}

Flat- or absorbed- spectrum cores are present in 13 of 16
Seyfert galaxies.  The most strongly absorbed components tend to be in
objects that also show other evidence for absorption, such as water
maser emission or large X-ray absorption columns.  In those cases,
according to the Seyfert unification schemes, the flat-spectrum
components would be viewed through an edge-on obscuring disc or torus,
the ionized inner edge of which could be causing free-free absorption.
If so, the situation in those objects is different from the
Blandford-K\"onigl jet model.  However the case for free-free
absorption is unambiguous in only NGC\,3079 for which spectral indices
are $>+2.5$.  The other 12 Seyferts have $-0.3<\alpha<+2.5$ and so
spectral turnovers could be produced by either synchrotron
self-absorption or free-free absorption.

\subsection{Jet Misalignment}
\label{sec:misalignment}

Kiloparsec-scale linear radio structures in Seyfert galaxies are
thought to be energized by jets or outflows originating at the AGN
core, and so one might expect pc and kpc-scale radio structures to be
aligned.  Misalignments could be caused by changes in the ejection
axis or by pressure gradients in the ISM.  To look for such effects,
we compiled kpc-scale radio observations of the VLBI Seyfert sample,
where available. The resulting sample of 21 objects in
Table~\ref{tab:seyferts_angles} is, we believe, a complete list of
Seyferts with both pc-scale and kpc-scale observations published,
although seven of these objects either have no clear extended
structure on one of the scales or the structure was deemed to be
affected by processes on scales $>1\,{\rm kpc}$, e.g., buoyancy forces
that are unrelated to changes in the jet ejection axis. In one
object, NGC\,1167, the double-sided kpc-scale emission is bent, so we
quote both.  This sample has a large overlap with the sample in
Table~\ref{tab:seyferts_angles} and, likewise, is not complete.
Biases such as selection for bright radio nuclei are probably present,
but whether they affect the distribution of jet bending angle is
unclear.  Doppler boosting would favour jets that point towards us,
but the jet speeds so far measured in Seyferts are low and boosting
effects should be minor.

\begin{SCfigure}[][htpb]
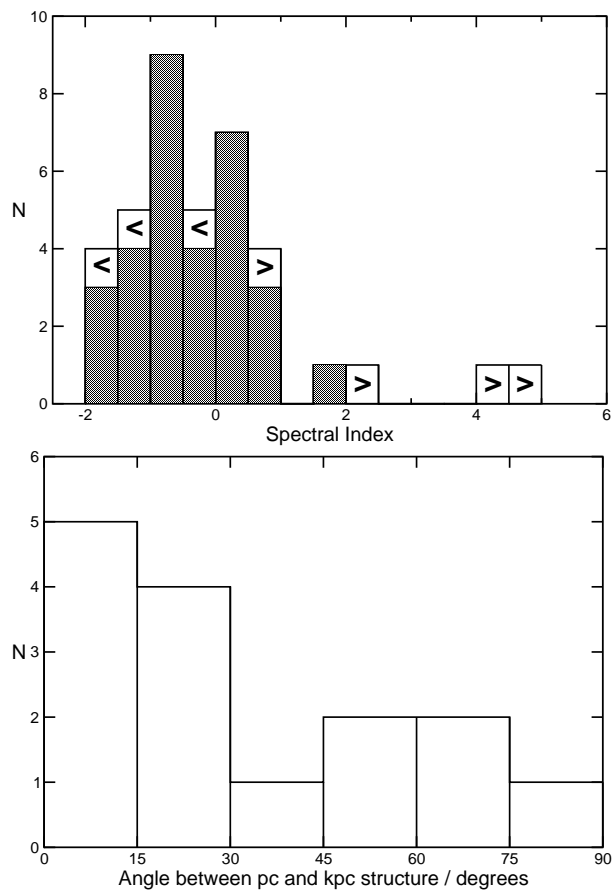

\begin{minipage}[b]{8cm}
 \includegraphics[width=8cm, clip]{chap5/plots/SI-stats.eps}
 \includegraphics[width=8cm, clip]{chap5/plots/PA-stats.eps}
\end{minipage}
\caption[Histograms of Seyfert galaxy literature data]{Histograms of
Seyfert galaxy literature data. {\it Top:} Histogram of spectral
indices with a bin width of 0.5. {\it Bottom}: Histogram of angles
between pc and kpc-scale structure with a bin width of $15^\circ$.}
\label{fig:seyfert-stats}
\end{SCfigure}

The distribution of misalignment angles from
Table~\ref{tab:seyferts_angles} is shown in
Fig.~\ref{fig:seyfert-stats} (bottom).  Misalignment in Seyfert jets
between pc and kpc-scales is common, with 5 out of 15 objects bending
by $45^{\circ}$ or more.  The null hypothesis that the jet
bending-angle distribution was drawn from a parent population of
aligned jets was rejected by a chi-squared test at the 2\,\% level of
significance.  The null hypothesis that the distribution was drawn
from a uniformly distributed parent population could not be rejected
by a chi-squared test, which returned a 25\,\% level of
significance. Thus, the misalignment between the pc-scale and
kpc-scale radio structure in NGC\,3079 is common in Seyfert galaxies.

The frequent misalignments seen in Seyfert galaxy jets could be
evidence either for changes in the jet ejection angle and hence of the
accretion disc, perhaps by a mechanism such as radiation-driven
warping (\citealt{1997MNRAS.292..136P}), or for bending due to
pressure gradients in the ISM (collision or buoyancy) through which
the jets propagate.

We compared the extended pc-scale radio structures to the host galaxy
rotation axes, assuming that the host galaxy is circular and the
rotation axis vector is projected onto the galaxy's minor axis, which
can be measured from optical images. For all spiral galaxies with
pc-scale radio structure, we used the major axes from
\cite{Vaucouleurs1991}, subtracted $90^{\circ}$ and reflected the
angles onto the range $0^{\circ}$ to $90^{\circ}$. Four objects were
not listed in \cite{Vaucouleurs1991}, so we used data from
\cite{Schmitt2000} in two cases and we measured the major axes from
Hubble Space Telescope archival data in the other two cases. We
performed a Spearman rank correlation test and found a correlation
coefficient of 0.07, corresponding to a 82\,\% significance and so
there is no correlation between the directions of the pc-scale
structures and the host galaxy rotation axes. However, the sample was
small (13 objects) and therefore the test is not strong. Thus, the
lack of alignment in Seyferts between pc-scale and kpc-scale structure
does not allow one to identify a pc-scale ejection axis in
NGC\,3079. Precession of the ejection axis is common and may be
reflected in the confusing structure in the core of NGC\,3079.

We compared the bending angles in Seyferts to those in radio-loud
objects using a comparison sample of core-dominated radio-loud objects
compiled from Table~6 in \cite{Pearson1988}, and Table~1 in
\cite{1993ApJ...411...89C}, which have both VLBI and VLA or
MERLIN observations at 5\,GHz.  Those objects show a surprising bimodal
distribution of jet bending angle, in which the pc-scale jets tend to
align with or to be perpendicular to the kpc-scale structure.  We
compared our jet bending angle distribution to the distribution in
radio-loud objects using a Kolmogorov-Smirnov test and found no
significant difference, the probability of being drawn from the same
parent distribution being 14\,\%.  However, the Seyfert sample is
small and the distribution for the Seyferts was shown in the previous
paragraph to be consistent also with a uniform distribution.\\

\subsubsection{Summary}

The position angle differences between pc-scale and kpc-scale radio
emission were found to be uniformly distributed between $0^{\circ}$
and $90^{\circ}$.  Such bends could be due to changes in the jet
ejection axis or due to pressure gradients in the ISM. No correlation
was found between the axis of pc-scale radio structure and the
rotation axis of the host galaxy.\\

We conclude that NGC\,3079 shares most properties with other
Seyferts. Its spectra are extreme, but three other objects also showed
inverted spectra and the transition between these three and NGC\,3079
does not appear to mark a fundamental difference, and misalignment
between pc-scale and kpc-scale radio emission also is a common
finding.

\begin{landscape}
\scriptsize
\begin{longtable}{lllrll}
\caption[Seyferts observed with VLBI]{Seyferts observed with
VLBI. Column 1 lists the names we used, column 2 the names used by
\cite{Veron2001}, column 3 the Seyfert types as described by
\cite{Veron2001} , column 4 the linear resolution in pc, column 5 the
instrument and frequency used, and column 6 the reference to the
publication.}\\
\label{tab:seyferts}\\

\hline
\hline
\multicolumn{1}{c}{Name} 
& \multicolumn{1}{c}{V\'eron name} 
& \multicolumn{1}{c}{Class}
& \multicolumn{1}{c}{Lin. Res.}
& \multicolumn{1}{c}{Instrument}
& \multicolumn{1}{c}{Ref.}
\\
\hline						  
\endfirsthead

\multicolumn{6}{c}{{ \tablename\ \thetable{} -- continued from previous page}} \\
\hline
\hline
\multicolumn{1}{c}{Name} 
& \multicolumn{1}{c}{V\'eron name} 
& \multicolumn{1}{c}{Class}
& \multicolumn{1}{c}{Lin. Res.}
& \multicolumn{1}{c}{Instrument}
& \multicolumn{1}{c}{Ref.}
\\
\hline 
\endhead

\hline \multicolumn{6}{c}{{Continued on next page}} \\
\endfoot

\hline
\endlastfoot

3C 287.1    &                 & S1   & 5.43      & VLBA 5 GHz                    & \cite{2000ApJS..131...95F}\\  
3C 390.3    &                 & S1.5 & 1.09      & global VLBI 5 GHz             & \cite{1996AaA...308..376A}\\  
III Zw 2    &                 & S1.2 & 0.260     & VLBA 15/43 GHz                & \cite{2000AaA...357L..45B}\\  
Ark 564     &                 & -    & 0.598     & global VLBI 5 GHz             & \cite{VirLal2001}         \\  
IC 5063     & PKS 2048-57     & S1h  & 3.30      & LBA 2.3 GHz + LBA HI          & \cite{2000AJ....119.2085O}\\  
MCG 8-11-11 &                 & S1.5 & 0.635     & global VLBI 5 GHz             & \cite{VirLal2001}         \\  
Mrk 1       &                 & S2   & 6.18      & EVN  1.7 GHz                  & \cite{1999ApJ...518..117K}\\  
Mrk 1       &                 & S2   & 0.550     & global VLBI 5 GHz             & \cite{VirLal2001}         \\  
Mrk 3       &                 & S1h  & 5.24      & EVN  1.7 GHz                  & \cite{1999ApJ...518..117K}\\  
Mrk 78      &                 & S2   & 1.75      & global VLBI 5 GHz             & \cite{VirLal2001}         \\  
Mrk 231     &                 & S1.0 & 0.343     & VLBA 15 GHz                   & \cite{Ulvestad1999}       \\  
Mrk 231     &                 & S1.0 & 4.09      & VLBA 1.4/2.3/4.8/8.4/15/22 GHz& \cite{1999ApJ...516..127U}\\  
Mrk 273     &                 & S2   & 7.32      & VLBA 1.4 GHz                  & \cite{2000ApJ...532L..95C}\\  
Mrk 348     &                 & S1h  & 0.131     & VLBA 15 GHz                   & \cite{Ulvestad1999}       \\  
Mrk 348     &                 & S1h  & 6.41      & EVN 1.4 GHz                   & \cite{1983AaA...128..318N}\\  
Mrk 348     &                 & S1h  & 1.49      & VLBA 1.7/5.0/8/15/22 GHz      & \cite{2003ApJ...590..149P}\\  
Mrk 477     &                 & S1h  & 2.39      & global VLBI 5 GHz             & \cite{VirLal2001}         \\  
Mrk 530     & NGC 7603        & S1.5 & 1.45      & global VLBI 5 GHz             & \cite{VirLal2001}         \\  
Mrk 766     &                 & S1.5 & 0.413     & global VLBI 5 GHz             & \cite{VirLal2001}         \\  
Mrk 926     &                 & S1.5 & 0.917     & VLBA 8.4 GHz                  & \cite{2000ApJ...529..816M}\\  
Mrk 1210    &                 & S1h  & 0.392     & EVN 1.6/5 GHz VLBA 1.6/5 GHz  & \cite{Middelberg2004}     \\  
Mrk 1218    &                 & S1.8 & 0.582     & global VLBI 5 GHz             & \cite{VirLal2001}         \\  
NGC 1052    &                 & S3h  & 0.015     & VLBA 5/8.4/22/43 GHz          & \cite{Kadler2002}         \\  
NGC 1052    &                 & S3h  & 0.048     & VLBI 1.4-43.2                 & \cite{Vermeulen2003}      \\  
NGC 1052    &                 & S3h  & 0.038     & VLBA 2.3/8.4/15.4 GHZ         & \cite{Kameno2001}         \\  
NGC 1068    &                 & S1h  & 0.221     & VLBA 1.7,5,15 GHz             & \cite{1998ApJ...504..147R}\\  
NGC 1068    &                 & S1h  & 2.94      & EVN 1.4 GHz                   & \cite{1987AJ.....93...22U}\\  
NGC 1167    &                 & S3   & 0.895     & global VLBI 5 GHz             & \cite{2001ApJ...552..508G}\\  
NGC 1167    &                 & S3   & 0.639     & global VLBI 1.6 GHz           & \cite{1990ApJ...358..159G}\\  
NGC 1275    &                 & S1.5 & 0.416     & VLBA 2.3/5/8.4/15.4/22/43 GHz & \cite{2000ApJ...530..233W}\\  
NGC 2110    &                 & S1i  & 0.151     & VLBA 8.4 GHz                  & \cite{2000ApJ...529..816M}\\  
NGC 2110    &                 & S1i  & 0.966     & EVN 1.6/5 GHz                 & \cite{Middelberg2004}     \\  
NGC 2273    &                 &  -   & 0.342     & global VLBI 5 GHz             & \cite{VirLal2001}         \\  
NGC 2639    &                 & S3   & 0.233     & global VLBI 5 GHz             & \cite{VirLal2001}         \\  
NGC 2639    &                 & S3   & 0.162     & VLBA 1.6/5/15 GHz             & \cite{1998ApJ...505..587W}\\  
NGC 3079    &                 & S2   & 0.044     & VLBA 5/8/22 GHz               & \cite{Trotter1998}        \\  
NGC 3079    &                 & S2   & 0.054     & global VLBI, 5.0 GHz          & \cite{Irwin1988}          \\  
NGC 3079    &                 & S2   & 0.042     & global VLBI 1.4/8.4/15/22     & \cite{Sawada-Satoh2000}   \\  
NGC 3147    &                 & S2   & 0.219     & VLBA 1.6/2.3/5/8.4 GHz        & \cite{2001ApJ...562L.133U}\\  
NGC 3227    &                 & S1.5 & 3.74      & MERLIN 1.6/5 GHz              & \cite{1995MNRAS.275...67M}\\  
NGC 4151    &                 & S1.5 & 0.116     & VLBA 1.6,5 GHz                & \cite{1998ApJ...496..196U}\\  
NGC 4151    &                 & S1.5 & 1.29      & EVN 1.7 GHz                   & \cite{1986MNRAS.218..775H}\\  
NGC 4151    &                 & S1.5 & 4.82      & MERLIN 5 GHz                  & \cite{1993MNRAS.263..471P}\\  
NGC 4168    &                 & S1.9 & 0.236     & VLBA 5 GHz                    & \cite{2002AaA...392...53N}\\  
NGC 4203    &                 & S3b  & 0.084     & VLBA 1.6/2.3/5/8.4 GHz        & \cite{2001ApJ...562L.133U}\\  
NGC 4258    &                 & S2   & 0.029     & VLBA 22 GHz                   & \cite{1997ApJ...475L..17H}\\  
NGC 4258    &                 & S2   & 0.232     & VLBA 1.4/1.6 GHz              & \cite{Cecil2000}          \\  
NGC 4395    &                 & S1.8 & 0.221     & VLBA 1.4 GHz                  & \cite{2001ApJ...553L..23W}\\  
NGC 4565    &                 & S1.9 & 0.207     & VLBA 5 GHz                    & \cite{2000ApJ...542..197F}\\  
NGC 4579    &                 & S3b  & 0.118     & VLBA 1.6/2.3/5/8.4 GHz        & \cite{2001ApJ...562L.133U}\\  
NGC 4579    &                 & S3b  & 0.246     & VLBA 5 GHz                    & \cite{2000ApJ...542..197F}\\  
NGC 5252    &                 & S2   & 0.671     & VLBA 8.4 GHz                  & \cite{2000ApJ...529..816M}\\  
NGC 5506    &                 & S1i  & 0.815     & EVN 1.6/5 GHz                 & \cite{Middelberg2004}     \\  
NGC 5548    &                 & S1.5 & 0.366     & VLBA 8.4 GHz                  & \cite{2000ApJ...531..716W}\\  
NGC 5793    &                 & S2   & 0.135     & VLBA 1.7,8.4,15,22 GHz        & \cite{2001ApJ...560..119H}\\  
NGC 5793    &                 & S2   & 0.316     & VLBA+Y27 1.6/5 GHz            & \cite{2000AaA...360...49H}\\  
NGC 5929    &                 & S3   & 0.290     &  global VLBI 5 GHz            & \cite{VirLal2001}         \\  
NGC 5929    &                 & S3   & 6.44      & MERLIN 0.4/1.6/5              & \cite{1996MNRAS.279.1111S}\\  
NGC 7212    &                 & S1h  & 1.37      & global VLBI 5 GHz             & \cite{VirLal2001}         \\  
NGC 7469    &                 & S1.5 & 0.844     & global VLBI 5 GHz             & \cite{VirLal2001}         \\  
NGC 7674    &                 & S1h  & 2.80      & VLBA+Y27+Arecibo 1.4 GHz      & \cite{Momjian2003}        \\  
NGC 7674    &                 & S1h  & 2.58      & EVN 1.6/5 GHz                 & \cite{Middelberg2004}     \\  
NGC 7674    &                 & S1h  & 2.18      & global VLBI 5 GHz             & \cite{VirLal2001}         \\  
NGC 7682    &                 & S2   & 0.996     & global VLBI 5 GHz             & \cite{VirLal2001}         \\  
T0109-383   & NGC  424        & S1h  & 0.226     & VLBA 8.4 GHz                  & \cite{2000ApJ...529..816M}\\  
\end{longtable}
\end{landscape}

\begin{table}
\scriptsize
\begin{center}
\begin{tabular}{lclrl}
\hline
\hline
\multicolumn{1}{c}{Source} 
& \multicolumn{1}{c}{Flat/Inv.}
& \multicolumn{1}{c}{Comp.}
& \multicolumn{1}{c}{$\alpha^{1.4}_5$}
& \multicolumn{1}{c}{Refs.}
\\
\hline
Mrk  231   & * &   N    &   $-$0.86  &   \cite{1999ApJ...516..127U}    \\ 
           &   &   C    &   $+$0.88  &   \cite{1999ApJ...516..127U}    \\ 
           &   &   S    &   $-$1.54  &   \cite{1999ApJ...516..127U}    \\ 
Mrk  348   & * &        &   $+$0.93  &   \cite{1998AJ....115..885B}    \\ 
Mrk  463E  &   &   L    &   $-$1.06  &   Norris et al., unpublished    \\ 
           &   &   R    &   $-$0.62  &   Norris et al., unpublished    \\ 
           &   &   1    &   $-$0.67  &   Norris et al., unpublished    \\ 
           &   &   2    &   $-$0.77  &   Norris et al., unpublished    \\ 
           &   &   3    &   $-$1.00  &   Norris et al., unpublished    \\ 
Mrk 1210   &   &   NW   &   $-$1.26  &   \cite{Middelberg2004}         \\ 
           &   &   SE   &   $-$0.78  &   \cite{Middelberg2004}         \\ 
NGC 1068   & * &   NE   &   $-$1.5   &   \cite{1998ApJ...504..147R}    \\ 
           &   &   C    &   $>2.2$   &   \cite{1998ApJ...504..147R}    \\ 
           &   &   S1   &   $>0.6$   &   \cite{1998ApJ...504..147R}    \\ 
           &   &   S2   &   $<-1.9$  &   \cite{1998ApJ...504..147R}    \\ 
NGC 2110   & * &   C    &   $+0.12$  &   \cite{Middelberg2004}         \\ 
NGC 2639   & * &        &   $+1.78$  &   \cite{1998ApJ...505..587W}    \\ 
NGC 3079   & * &   E    &   $+0.67$  &   This thesis                   \\ 
           &   &   A    &   $>+4.56$ &   This thesis                   \\ 
           &   &   B    &   $>+4.41$ &   This thesis                   \\ 
NGC 3147   & * &        &   $+0.20$  &   \cite{2001ApJ...562L.133U}    \\ 
NGC 4151   & * &   E    &   $-0.5$   &   \cite{1998ApJ...496..196U}    \\ 
           &   &   D    &   $<-0.2$  &   \cite{1998ApJ...496..196U}    \\ 
           &   &   F    &   $<-1.3$  &   \cite{1998ApJ...496..196U}    \\ 
NGC 4203   & * &        &   $+0.38$  &   \cite{2001ApJ...562L.133U}    \\ 
NGC 4579   & * &        &   $+0.20$  &   \cite{2001ApJ...562L.133U}    \\ 
NGC 5506   & * &   B0   &   $+0.06$  &   \cite{Middelberg2004}         \\ 
           &   &   B1   &   $-0.73$  &   \cite{Middelberg2004}         \\ 
           &   &   B2   &   $-0.30$  &   \cite{Middelberg2004}         \\ 
NGC 5793   & * &   C1C  &   $-0.70$  &   \cite{2000AaA...360...49H}    \\ 
           &   &   C1NE &   $-1.00$  &   \cite{2000AaA...360...49H}    \\ 
           &   &   C2C  &   $-0.46$  &   \cite{2000AaA...360...49H}    \\ 
           &   &   C2W  &   $-0.72$  &   \cite{2000AaA...360...49H}    \\ 
           &   &   C2E  &   $+0.13$  &   \cite{2000AaA...360...49H}    \\ 
NGC 7469   & * &   2    &   $-0.34$  &   Norris et al., unpublished    \\ 
           &   &   3    &   $+0.09$  &   Norris et al., unpublished    \\ 
           &   &   5    &   $-0.12$  &   Norris et al., unpublished    \\ 
NGC 7674   &   &        &   $-1.78$  &   \cite{Middelberg2004}         \\ 
\hline
\end{tabular}
\caption[Spectral indices]{Seyferts from Table~\ref{tab:seyferts} that have
dual-frequency (1.4\,GHz and 5\,GHz) VLBI observations with matching beam
sizes, from which the spectral index is derived. Asterisks denote
objects with at least one component with a flat or inverted spectrum.}
\label{tab:seyferts_SI}
\end{center}
\end{table}

\begin{table}[ht!]
\scriptsize
\begin{center}
\begin{tabular}{llrrrrr}
\hline
\hline
&&\multicolumn{5}{c}{Level of Significance}\\
Sample & &
\multicolumn{1}{c}{$\chi^2$} &
\multicolumn{1}{c}{KS} &
\multicolumn{1}{c}{Logrank} &
\multicolumn{1}{c}{Gehan} &
\multicolumn{1}{c}{Peto-Peto}\\
\hline
\cite{1987MNRAS.228..521U} & X-ray Seyferts      & 0.3\,\%  & 0.7\,\% &  0.3\,\% & 8.3\,\% & 8.3\,\%\\
\cite{1996ApJ...473..130R} & CfA Seyferts        & 0.1\,\%  & 0.1\,\% &  0.0\,\% & 0.5\,\% & 0.5\,\%\\
\cite{1996ApJ...473..130R} & 12\,$\mu$m Seyferts & 0.1\,\%  & 0.6\,\% &  0.0\,\% & 2.0\,\% & 2.0\,\%\\
\cite{1999AAS..137..457M}  & Distance Limited    & 0.1\,\%  & 0.3\,\% &  0.0\,\% & 1.2\,\% & 1.2\,\%\\
\cite{2001ApJS..133...77H} & Palomar Seyferts    & 1.0\,\%  &  27\,\% &   44\,\% &  75\,\% &  77\,\%\\
\cite{1984ApJ...285..439U} & Distance Limited    & 0.1\,\%  & 0.3\,\% &  0.0\,\% & 0.7\,\% & 0.7\,\%\\
\hline
\end{tabular}
\caption[Comparison of the spectral index distribution to other Seyfert
samples]{Comparison of the spectral index distribution of the VLBI
Seyfert sample to various other Seyfert samples.  The level of
significance gives the probability that the spectral index
distribution of the comparison sample and the VLBI seyfert sample in
this thesis were drawn from the same parent population.}
\label{tab:spectra-stats2}
\end{center}
\end{table}

\begin{table}[ht!]
\scriptsize
\begin{center}
\begin{tabular}{llrrrr}
\hline
\hline
Sample & &
\multicolumn{1}{c}{$N_{\alpha>-0.3}$} &
\multicolumn{1}{c}{$N_{\rm total}$} &
\multicolumn{1}{c}{\%} &
\multicolumn{1}{c}{$P$}\\
\hline
\cite{Middelberg2004}      & own obs + lit search  &  17 &  38 &  45\\
\cite{1987MNRAS.228..521U} & X-ray Seyferts        &   1 &  20 &   5 &  $  0.3$\,\%\\
\cite{1996ApJ...473..130R} & CfA Seyferts          &   3 &  46 & 6.5 &  $< 0.1$\,\%\\
\cite{1996ApJ...473..130R} & 12 $\mu$m Seyferts    &   4 &  40 &  10 &  $< 0.1$\,\%\\
\cite{1999AAS..137..457M}  & Distance limited      &  13 &  60 &  22 &  $  1.8$\,\%\\
\cite{2001ApJS..133...77H} & Palomar Seyferts      &  21 &  45 &  47 &  $  50$\,\%\\
\cite{1984ApJ...285..439U} & Distance limited      &   3 &  40 & 7.5 &  $< 0.1$\,\%\\
\hline
\end{tabular}
\caption[Fraction of flat- and inverted-spectrum components in various
Seyfert samples]{Fraction of flat- and inverted-spectrum components in various
Seyfert samples.  The column '$N_{\alpha>-0.3}$' gives the number of
components with spectral indices $\alpha>-0.3$ (i.e., components with
flat or inverted spectra).  The column '$N_{\rm total}$' gives the
total number of components in each sample.  The column \% gives
$(N_{\alpha>-0.3}/N_{\rm total})\times100$.  The column $P$ gives the
level of significance from a comparison of the fraction of flat- and
inverted-spectrum components in the VLBI sample of this thesis (limits
treated as detections), to that of each comparison sample at arcsec
resolution, using the difference-of-two-proportions test with Yates
correction for continuity (e.g. \citealt{Glanz1992}).}
\label{tab:spectra-stats}
\end{center}
\end{table}

\begin{landscape}
\scriptsize
\begin{longtable}{lrrrrllr}
\caption[Position angles of pc-scale and kpc-scale emission]{Subset of
sources from Table~\ref{tab:seyferts} for which both pc and kpc-scale
structures have been observed. The misalignment angle, $\Delta\Theta$,
is calculated as $\Theta({\rm pc})-\Theta({\rm kpc})$, and has then
been reflected onto the range of $0^{\circ}$ to $90^{\circ}$. For the
seven objects at the end of the table, either the P.A. information of
the kpc scale structure was not completely reliable or was
unavailable.  The position angles of the galaxy minor axes in the last
column were drawn from \cite{Vaucouleurs1991}, except for those marked
$a$ taken from \cite{Schmitt2000} and those marked $b$ measured by us
from Hubble Space Telescope archival images (see
\S\ref{sec:misalignment}).}\\
\label{tab:seyferts_angles}\\
\hline
\hline
\multicolumn{1}{c}{Name} 
& \multicolumn{1}{c}{Res.}
& \multicolumn{1}{c}{$\Theta$ (pc)}
& \multicolumn{1}{c}{$\Theta$ (kpc)}
& \multicolumn{1}{c}{$\Delta\Theta$}
& \multicolumn{1}{c}{Array}
& \multicolumn{1}{c}{Refs.}
& \multicolumn{1}{c}{$\Theta$ (opt.)}\\
& \multicolumn{1}{c}{(pc)}
& \multicolumn{1}{c}{(deg)}
& \multicolumn{1}{c}{(deg)}
& \multicolumn{1}{c}{(deg)}
& 
& 
& \multicolumn{1}{c}{deg}\\
\hline						  
\endfirsthead
\multicolumn{8}{c}{{ \tablename\ \thetable{} -- continued from previous page}} \\
\hline
\hline
\multicolumn{1}{c}{Name} 
& \multicolumn{1}{c}{Res.}
& \multicolumn{1}{c}{$\Theta$ (pc)}
& \multicolumn{1}{c}{$\Theta$ (kpc)}
& \multicolumn{1}{c}{$\Delta\Theta$}
& \multicolumn{1}{c}{Array}
& \multicolumn{1}{c}{Refs.}
& \multicolumn{1}{c}{$\Theta$ (opt.)}\\
& \multicolumn{1}{c}{(pc)}
& \multicolumn{1}{c}{(deg)}
& \multicolumn{1}{c}{(deg)}
& \multicolumn{1}{c}{(deg)}
& 
& 
& \multicolumn{1}{c}{deg}\\
\hline						  
\endhead
\hline \multicolumn{8}{c}{{Continued on next page}} \\
\endfoot
\hline
\endlastfoot
\hline
3C 390.3    & 1.09      & 323  &     &   0  & global VLBI 5\,GHz             & \cite{1996AaA...308..376A} &          \\
            & 1210      &      & 323 &      & VLA 5\,GHz                     & \cite{1996AaA...308..376A} &          \\
Mrk 78      & 1.75      & 40   &     &  55  & global VLBI 5\,GHz             & \cite{VirLal2001}          &          \\
            & 504       &      & 275 &      & VLA 5\,GHz                     & \cite{VirLal2001}          &          \\
Mrk 231     & 0.343     & 93   &     & 77   & VLBA 15\,GHz                   & \cite{Ulvestad1999}        & 100      \\
            & 3270      &      & 170 &      & VLA 1.5/8.4/15\,GHz            & \cite{1999ApJ...516..127U} &          \\
Mrk 1218    & 0.582     & 330  &     &  13  & global VLBI 5\,GHz             & \cite{VirLal2001}          & 130$^b$  \\
            & 294       &      & 317 &      & VLA 5\,GHz                     & \cite{VirLal2001}          &          \\
NGC 1052    & 0.015     &  65  &     &  30  & VLBA 5/8.4/22/43\,GHz          & \cite{Kadler2002}          &          \\
            & 162       &      & 275 &      & MERLIN 1.4\,GHz                & \cite{Kadler2003}          &          \\
NGC 1068    & 0.221     &  11  &     &  29  & VLBA 1.7,5,15\,GHz             & \cite{1998ApJ...504..147R} & 160      \\
            & 217       &      & 40  &      & VLA 5\,GHz                     & \cite{2001ApJS..133...77H} &          \\
NGC 1167    & 0.895     & 140  & 313 &   7  & global VLBI 5\,GHz             & \cite{2001ApJ...552..508G} &          \\
            & 131       &      & 66  &  74  & VLA 5\,GHz                     & \cite{1995AaA...295..629S} &          \\
NGC 1275    & 0.416     & 353  &     &  23  & VLBA 2.3 to 43\,GHz            & \cite{2000ApJ...530..233W} &          \\
            & 382       &      & 150 &      & VLA 1.4\,GHz                   & \cite{1990MNRAS.246..477P} &          \\
NGC 3079    & 0.044     & 125  &     &  65  & VLBA 5/8/22\,GHz               & \cite{Trotter1998}         &  75      \\
            & 310       &      &  60 &      & VLA 1.4/4.9                    & \cite{Duric1983}           &          \\
NGC 4151    & 0.116     & 257  &     &   3  & VLBA 1.6,5\,GHz                & \cite{1998ApJ...496..196U} & 140      \\
            & 104       &      &  80 &      & VLA 5\,GHz                     & \cite{2001ApJS..133...77H} &          \\
NGC 4258    & 0.029     &   3  &     &   8  & VLBA 22\,GHz                   & \cite{1997ApJ...475L..17H} &  60      \\
            & 125       &      & 355 &      & VLA 1.5\,GHz                   & \cite{Cecil2000}           &          \\
NGC 7212    & 1.37      & 257  &     &  52  & global VLBI 5\,GHz             & \cite{VirLal2001}          & 133$^a$  \\
            & 609       &      &  25 &      & VLA 5\,GHz                     & \cite{VirLal2001}          &          \\
NGC 7469    & 0.844     &  28  &     & 35   & global VLBI 5\,GHz             & \cite{VirLal2001}          &  35      \\
            & 193       &      & 353 &      & VLA 5\,GHz                     & \cite{VirLal2001}          &          \\
NGC 7674    & 2.80      & 274  &     &  22  & VLBA+Y27+AO 1.4\,GHz           & \cite{Momjian2003}         &  64$^a$  \\
            & 178       &      & 296 &      & MERLIN 1.7                     & \cite{Momjian2003}         &          \\
\hline
III Zw 2    & 0.260     & 303  &     &  63  & VLBA 15/43\,GHz                & \cite{2000AaA...357L..45B} &          \\
            & 3460      &      & 240 &      & VLA 1.5\,GHz                   & Brunthaler, priv. comm.    &          \\
Ark 564     & 0.598     & 282  &     &      & global VLBI 5\,GHz             & \cite{VirLal2001}          &  28      \\
IC 5063     & 3.30      &  66  &     &  49  & LBA 2.3\,GHz + LBA HI          & \cite{2000AJ....119.2085O} &          \\
            & 242       &      & 115 &      & ATCA 8.3/1.4GHz                & \cite{1998AJ....115..915M} &          \\
NGC 2110    & 0.151     & 350  &     &  16  & VLBA 8.4\,GHz                  & \cite{2000ApJ...529..816M} &          \\
            & 240       &      & 186 &      & VLA 1.4/5\,GHz                 & \cite{1983ApJ...264L...7U} &          \\
NGC 3147    & 0.219     & 317  &     &      & VLBA 1.6/2.3/5/8.4\,GHz        & \cite{2001ApJ...562L.133U} &  65      \\
NGC 5506    & 0.815     &  73  &     &  76  & EVN 1.6/5\,GHz                 & \cite{Middelberg2004}      &   1      \\
            & 359       &      & 177 &      & VLA 4.9\,GHz                   & \cite{1996ApJ...467..551C} &          \\
NGC 7682    & 0.996     & 125  &     &      & global VLBI 5\,GHz             & \cite{VirLal2001}          &  68$^b$  \\
\end{longtable}
\end{landscape}

\chapter{Fast Frequency Switching}
\label{ch:ffs}

The invention of Very Long Baseline Interferometry was mainly driven
by the need for higher angular resolution radio
observations. Single-dish observations would never have approached the
resolutions achieved by optical telescopes owing to the long
wavelengths. However, once VLBI was available, it was straightforward
to progressively increase the resolution by increasing the observing
frequencies. In the recent few years, observations at 86\,GHz have
evolved from the experimental stage to a commonly used observing
technique which achieves the highest angular resolutions possible
today, and first detections at even higher frequencies have already
been made (\citealt{Krichbaum2002}). The scientific benefits are
undoubted: even in the closest objects, only millimeter-VLBI can
resolve the linear scales on which extragalactic radio jets are
launched, a process yet not understood. 86\,GHz VLBI observations also
contribute to investigations of internal jet structure, composition and
collimation (e.g., \citealt{Doeleman2001}, \citealt{Gomez1999}).\\

Unfortunately, higher frequency observations involve a number of
serious problems: the sources are usually weaker because the
emissivity of optically thin synchrotron sources drops as
$\nu^{-0.7}$, the efficiencies of typical centimetre radio antennas
drop to about 15\,\% because the surface rms is $\approx \lambda/10$,
the receiver performances become disproportionately worse because the
amplifiers contribute more noise and the atmospheric coherence that
limits the integration time decreases as $1/\nu$. This last problem,
however, is not a hardware property and would be solved if the
atmospheric phase fluctuations could be calibrated. In fact, the
so-called phase-referencing first demonstrated by \cite{Alef1988} is
now an established calibration technique in centimetre VLBI to lower
the detection thresholds to the sub-mJy level and to accurately
determine the positions of sources in VLBI observations. In
phase-referencing, a strong calibrator source is frequently observed
(every few minutes, depending on observing frequency) to calibrate the
visibility phases of the target source integrations, i.e., the
telescopes cycle between the target source and the phase calibrator
source. In millimetre VLBI the technique is not commonly used,
although a successful proof of concept exists
(\citealt{Porcas2002}). Problems arise from the need of a suitable,
strong phase calibrator in the vicinity of target source and from a
combination of short atmospheric coherence and telescope slewing
times.\\

In this chapter, I describe a new phase-referencing technique for
VLBI observations at high frequencies, fast frequency switching.

\section{Introduction}

\begin{figure*}[ht!]
\centering
 \includegraphics[width=0.4\textheight, angle=270]{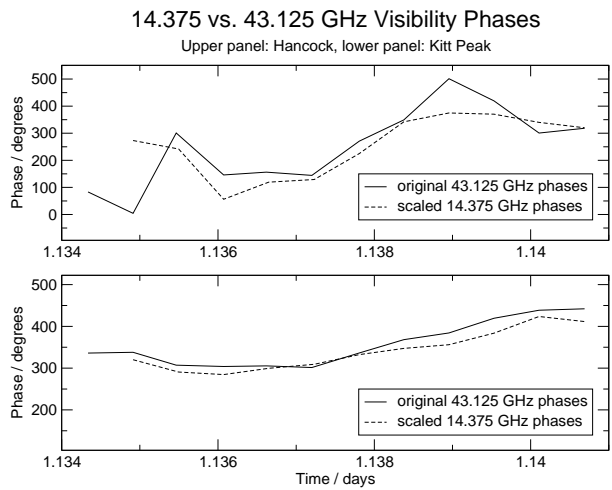}
 \caption[Demonstration of the scalability of phase
 solutions.]{Demonstration of the scalability of phase
 solutions. 43\,GHz fringe-fitted phase solutions (solid lines) on
 3C\,273 compared to 15\,GHz fringe-fitted phase solutions multiplied
 by the frequency ratio (dashed lines) from Hancock (upper panel) and
 Kitt Peak (lower panel). Ten minutes of data are shown, and the
 phases follow each other very well.}  \label{fig:phase}
\end{figure*}

The main source of phase noise in VLBI observations at frequencies
higher than 5\,GHz is turbulence in the troposphere causing refractive
inhomogeneities. These path-length variations are non-dispersive, and
one can self-calibrate the visibility phases at one frequency and use
the solutions to calibrate visibilities at another frequency after
multiplying the phases by the frequency ratio, $r$, provided that the
lag between the two measurements does not exceed half the atmospheric
coherence time (Fig.~\ref{fig:phase}). This is possible with the VLBA
because frequency changes need only a few seconds. After multiplying
the phase solutions by the frequency ratio and applying them to the
target-frequency phases, there remains a constant phase offset,
$\Delta\Phi$, between the signal paths at the two frequencies, which
must be calibrated. It can be monitored with frequent observations of
achromatic, strong calibrators, and must be subtracted from the high
frequency visibility phase.\\

Following the notation from Eqs.~\ref{eq:vis1}ff, the observed
visibility phases using this calibration scheme can be described as

\begin{equation}
\begin{split}
\phi_{\rm \nu_r}(t_1)&=
\phi_{\rm r}(t_1)+
\phi_{\rm ins}^{\rm \nu_r}(t_1)+
\phi_{\rm pos}^{\rm \nu_r}(t_1)+
\phi_{\rm ant}^{\rm \nu_r}(t_1)+
\phi_{\rm atm}^{\rm \nu_r}(t_1)+
\phi_{\rm ion}^{\rm \nu_r}(t_1)\\
\phi_{\rm \nu_t}(t_2)&=
\phi_{\rm t}(t_2)+
\phi_{\rm ins}^{\rm \nu_t}(t_2)+
\phi_{\rm pos}^{\rm \nu_t}(t_2)+
\phi_{\rm ant}^{\rm \nu_t}(t_2)+
\phi_{\rm atm}^{\rm \nu_t}(t_2)+
\phi_{\rm ion}^{\rm \nu_t}(t_2)\\
\phi_{\rm \nu_r}(t_3)&=
\phi_{\rm r}(t_3)+
\phi_{\rm ins}^{\rm \nu_r}(t_3)+
\phi_{\rm pos}^{\rm \nu_r}(t_3)+
\phi_{\rm ant}^{\rm \nu_r}(t_3)+
\phi_{\rm atm}^{\rm \nu_r}(t_3)+
\phi_{\rm ion}^{\rm \nu_r}(t_3).\\
\end{split}
\label{eq:ffs-phase1}
\end{equation}

Here, the indices $\nu_{\rm r}$ and $\nu_{\rm t}$ denote terms of the
reference and target frequency, respectively, and $\phi_{\rm r}$ and
$\phi_{\rm t}$ are the true visibilities at the reference and target
frequency, respectively. Interpolation of the reference frequency
terms to the times of the target frequency scans yields

\begin{equation}
\tilde{\phi}_{\rm \nu_t}(t_2)=
\tilde{\phi}_{\rm r}(t_2)+
\tilde{\phi}_{\rm ins}^{\rm \nu_r}(t_2)+
\tilde{\phi}_{\rm pos}^{\rm \nu_r}(t_2)+
\tilde{\phi}_{\rm ant}^{\rm \nu_r}(t_2)+
\tilde{\phi}_{\rm atm}^{\rm \nu_r}(t_2)+
\tilde{\phi}_{\rm ion}^{\rm \nu_r}(t_2).
\label{eq:ffs-phase2}
\end{equation}

Self-calibration at the reference frequency is then used to obtain a
source model, and self-calibration of the visibilities after dividing
by this model yields the required phase corrections. These correction
terms are interpolated to the times where the target frequency was
observed, and scaled by the frequency ratio, $r$. In practice, the
phase solutions derived in this process comprise contributions from
all five sources of error, and so do the scaled phase corrections. The
difference between the target frequency visibilities and the
interpolated reference frequency visibilities then is

\begin{equation}
\begin{split}
\phi_{\rm \nu_t}-\tilde{\phi}_{\rm \nu_r}= \phi_{\rm t}-\tilde{\phi}_{\rm r}
&+(\phi_{\rm ins}^{\rm \nu_t}-r\tilde{\phi}_{\rm ins}^{\rm \nu_r})
 +(\phi_{\rm pos}^{\rm \nu_t}-r\tilde{\phi}_{\rm pos}^{\rm \nu_r})
 +(\phi_{\rm ant}^{\rm \nu_t}-r\tilde{\phi}_{\rm ant}^{\rm \nu_r})\\
&+(\phi_{\rm atm}^{\rm \nu_t}-r\tilde{\phi}_{\rm atm}^{\rm \nu_r})
 +(\phi_{\rm ion}^{\rm \nu_t}-r\tilde{\phi}_{\rm ion}^{\rm \nu_r}).
\end{split}
\label{eq:ffs-phase3}
\end{equation}

In this equation, $\tilde{\phi}_{\rm r}$ is zero because a model of
the source has been subtracted. The instrumental phase offset is
constant and can be determined from calibrator observations, and so
the term $(\phi_{\rm ins}^{\rm \nu_t}-r\tilde{\phi}_{\rm ins}^{\rm
\nu_r})$ is known and can be removed. The antenna position errors also
cancel out because this type of error causes a shift of the source on
the sky, and to compensate for it requires a phase correction which
scales with frequency. E.g., assume an antenna position error of 1\,cm
along the line of sight on a 1000\,km baseline. The error shifts the
fringe pattern on the sky by $(0.01\,{\rm m}/1000\,{\rm
km})=10^{-8}\,{\rm rad}=2.06\,\mas$. At a frequency of 15\,GHz
($\lambda=2.00\,{\rm cm}$), the fringe spacing on the sky is
$\lambda/D=4.13\,\mas$, and hence a phase correction of $179^{\circ}$
is required to account for the error. At a frequency of 45\,GHz
($\lambda=0.67\,{\rm cm}$), three times higher than the other
frequency, the fringe spacing is 1.38\,mas, and a phase correction of
$537^{\circ}$ is required to compensate for the error, three times
more than at the lower frequency, and so $(\phi_{\rm ant}^{\rm
\nu_t}-r\tilde{\phi}_{\rm ant}^{\rm \nu_r})=0$. Similar to the antenna
position error, the source position error scales with frequency when
measured in units of turns of phase, but a frequency-dependent core
shift, as predicted by conical jet models (e.g.,
\citealt{Lobanov1998}), modulates $(\phi_{\rm pos}^{\rm
\nu_t}-r\tilde{\phi}_{\rm pos}^{\rm \nu_r})$ on each baseline with a
sinusoid. The period of the sinusoid is $23^{\rm h}56^{\rm m}$, its
amplitude depends on the magnitude of the shift and its phase depends
on the direction of the shift relative to the baseline direction. It
is zero only in the abscence of a core shift, and therefore cannot be
neglected. The atmospheric phase errors also scale with frequency, and
hence $(\phi_{\rm atm}^{\rm \nu_t}-r\tilde{\phi}_{\rm atm}^{\rm
\nu_r})=0$, but the ionospheric phase noise is difficult to determine;
something which we, unfortunately, did not discover beforehand the
observations presented here. It can probably be determined using
interspersed, wide-band scans at a low frequency, e.g., in the
1.4\,GHz band, where the effect is strong. Similar to
Eq.~\ref{eq:vis4}, the remaining terms describe the difference between
the target frequency phase and the interpolated reference frequency
phase as the target frequency visibility phase plus the position
offset and an interpolation term, $\phi_{\rm int}$:

\begin{equation}
\phi_{\rm \nu_t}-\tilde{\phi}_{\rm \nu_r}=
\phi_{\rm t}+
(\phi_{\rm pos}^{\rm \nu_t}-r\tilde{\phi}_{\rm pos}^{\rm \nu_r})+\phi_{\rm int}.
\label{eq:ffs-phase4}
\end{equation}

The result is the high-frequency visibilities phase-referenced to the
source at low frequency, and so the technique can measure the position
shift of AGN cores with frequency and can prolong coherence.  The
coherence time can in principle be prolonged indefinitely, however, in
our project, unmodelled ionospheric path length changes limited the
coherence to half an hour in the worst case. Thus, in the following
sections, we use fast frequency switching to calibrate the short-term
phase fluctuations and then use self-calibration at the target
frequency with a half-hour solution interval to remove the long-term
phase drifts. The extension from the 30\,s atmospheric coherence time
to a coherent integration time of 30\,min yields a large sensitivity
improvement.\\

A similar observing strategy has been developed by \cite{Kassim1993}
for the VLA, who used scaled phase solutions from 330\,MHz to
calibrate simultaneously observed 74\,MHz data. In this case, the
dominant source of phase errors was ionospheric path length
changes. Using their new calibration technique, they were able to
increase the coherent integration time from $<$\,1\,min to
$>$\,10\,min and to make 74\,MHz images of several radio sources. Fast
frequency switching is also being considered as a standard calibration
mode for ALMA in the future (\citealt{DAddario2003}).\\

\section{Observations}

We observed a pilot project (BR073) on January 5, 2002, to attempt to
measure the diameter of the AGN in M81 (M81*) at 86\,GHz. We
obtained only very weak detections of M81*, but we learnt a number
of important points about designing fast-frequency switching
experiments (\citealt{Middelberg2002}). We incorporated those
considerable improvements in the observing strategy for our next
project (BM175C), which we describe here.

\begin{table*}[htpb!]
\center
\begin{tabular}{lcc|lcc}
\hline
Source   & Duration  & Freq.  & 3C279    &  5   & 43 \\ 
         & (min)     & pair   & 3C273    &  5   & 43 \\ 
(1)      & (2)       & (3)    & 3C273    &  5   & 1  \\ 
\cline{1-4}
OJ287    & 10   & 15-86  & NGC4261  & 25   & 15-43  \\  
OJ287    & 10   & 15-43  & 3C273    &  5   & 15-43  \\  
NGC4261  & 25   & 15-43  & 3C273    &  5   & 15-86  \\  
3C273    &  5   & 15-43  & NGC4261  & 25   & 15-86  \\  
NGC4261  & 25   & 15-86  & 3C279    &  5   & 15-43  \\  
3C273    &  5   & 15-86  & 3C279    &  5   & 15-86  \\  
3C273    &  5   & 15-43  &          &      &        \\  
         &      &        & 3C273    &  10  & 15-43  \\  
OJ287    &  5   & 15$^a$ &          &      & 15-    \\  
OJ287    &  5   & 43$^a$ & NGC4261  & 25   & 15-43  \\  
OJ287    &  5   & 86$^a$ & 3C273    &  5   & 15-43  \\  
         &      &        & 3C273    &  5   & 15-86  \\  
NGC4261  & 25   & 15-43  & NGC4261  & 25   & 15-86  \\  
3C279    &  5   & 15-43  & 3C279    &  5   & 15-86  \\  
3C279    &  5   & 15-86  & 3C279    &  5   & 15-43  \\  
3C273    &  5   & 15-86  &          &      &        \\  
3C273    &  5   & 15-43  & 3C279    &  5   & 15$^a$ \\  
NGC4261  & 25   & 15-43  & 3C279    &  5   & 43$^a$ \\  
3C273    &  5   & 15-43  & 3C279    &  5   & 86$^a$ \\  
3C273    &  5   & 15-86  &          &      &        \\  
NGC4261  & 25   & 15-86  & 3C273    &  5   & 15-43  \\  
3C279    &  5   & 15-86  & 3C273    &  5   & 15-86  \\  
3C279    &  5   & 15-43  & NGC4261  & 25   & 15-86  \\  
         &      &        & 3C273    &  5   & 15-86  \\  
3C273    & 10   & 15-43  & 3C273    &  5   & 15-43  \\  
         &      &        & NGC4261  & 25   & 15-43  \\  
NGC4261  & 25   & 15-43  & 3C279    &  5   & 15-43  \\  
3C273    &  5   & 15-43  & 3C279    &  5   & 15-86  \\  
3C273    &  5   & 15-86  & 3C273    &  5   & 15-43  \\  
NGC4261  & 25   & 15-86  &          &      &        \\  
3C279    &  5   & 15-86  & 3C345    &  5   & 15$^a$ \\  
         &      &        & 3C345    &  5   & 43$^a$ \\  
3C279    &  5   & 15$^a$ & 3C345    &  5   & 86$^a$ \\  
3C279    &  5   & 43$^a$            &      &        \\  
3C279    &  5   & 86$^a$ & 3C345    &  10  & 15-43  \\  
         &      &        & 3C345    &  10  & 15-86  \\  
\hline
\end{tabular}
\caption[Summary of project BM175]{Summary of project BM175 observed on May 5, 2003. The sources
observed are in column (1), the scan duration is in column (2) and the
frequency pairs in GHz are in column (3). An $a$ indicates fringe finder
scans during which no frequency switching was done.}
\label{tab:scans}
\end{table*}

We observed NGC\,4261 as a fast frequency switching target and 3C\,273
and 3C\,279 occasionally for the inter-frequency offset and to test
the technique on strong sources on May 5, 2003 (project BM175C). A
previous run (BM175B) had to be re-observed because of high fringe
rates due to weather and a number of equipment failures. BM175C was
observed in a period of superb weather, using 256\,Mbps to record a
bandwidth of 64\,MHz with 2-bit sampling. We recorded LCP only, the
data were divided into 8 IFs with 8\,MHz bandwidth, each of which were
subdivided into 64 spectral channels 125\,kHz wide. The correlator
integration time was one second to allow monitoring of the phases with
high time resolution. All antennas performed well, except for Fort
Davis, where a receiver problem caused complete loss of 43\,GHz
data. A summary of this observing run is given in
Table~\ref{tab:scans}.\\

Several considerations influenced the experiment design: 

\begin{itemize} 
  
\item{\bf Frequency Choice} The target frequency should be an integer 
multiple of the reference frequency to avoid having to unwrap phase
wraps. For example, if the frequency ratio, $r$, is the non-integer
value of 2.5, and the reference frequency phase wraps from
$359^{\circ}$ to $0^{\circ}$, then the scaled target frequency phase
will jump from $897.5^{\circ}$ ($=177.5^{\circ}$) to $0^{\circ}$,
introducing a phase jump of $177.5^{\circ}$ into the calibration phase
for a $1^{\circ}$ phase change at $\nu_{\rm ref}$. In contrast,
choosing $r=2.0$, when the reference frequency phase wraps from
$359^{\circ}$ to $0^{\circ}$, the scaled target frequency phase
changes from $718^{\circ}$ ($=358^{\circ}$) to $0^{\circ}$,
corresponding to a change of $2^{\circ}$ for a $1^{\circ}$ phase
change at $\nu_{\rm ref}$. Hence, for integer values of $r$, phase
wraps at the reference frequency introduce integer multiples of
$360^{\circ}$ at the target frequency and so have no effect.

We chose a reference frequency of 14.375\,GHz since the third and
sixth harmonics at 43.125\,GHz and 86.25\,GHz lie within the VLBA
receiver bands. For convenience, we will refer to these frequencies as
``15\,GHz'', ``43\,GHz'' and ``86\,GHz'', respectively. In this
document, two consecutive integrations at $\nu_{\rm ref}$ and
$\nu_{\rm t}$ will be called a ``cycle'', each integration of which is
called a ``half-cycle'', and a sequence of cycles on the same source
is called a ``scan''. Long (several minutes), continuous integrations
on a single source at one frequency, e.g. fringe-finder observations,
will also be called ``scans''.

\item{\bf Integration Times} We chose a cycle time of 50\,s, of
which 22\,s were spent at the reference frequency of 15\,GHz and the
remaining 28\,s were spent at the target frequency, either 43\,GHz or
86\,GHz.  An average time of 7\,s per half-cycle was lost in moving
the subreflector between the feed horns, resulting in net integration
times of 15\,s at $\nu_{\rm ref}$ and 21\,s at $\nu_{\rm t}$. The
integration times are a compromise between source brightness, antenna
sensitivity and expected weather conditions. This setup yielded a
$5\,\sigma$ detection limit of 89\,mJy in 15\,s at 15\,GHz for the
VLBA.

\item{\bf Calibrator Scans} Calibrators must be observed frequently to
monitor the phase offset $\Delta\Phi$ between the two frequencies. An
important constraint is that the calibrators must be achromatic, i.e.,
they must not have their own frequency-dependent core shifts. We
included adjacent scans on two different calibrators to measure
$\Delta\Phi$ before and after target source observations. Five of the
calibrator scans were twice as long to provide more time for tests
with strong signals.

\end{itemize}

\section{Data Reduction}

\subsection{Standard Steps}

Data reduction was carried out in AIPS. A calibration table
entry was generated every 4.8\,s to provide high temporal
resolution. The amplitudes were calibrated using $T_{\rm sys}$ and
gain measurements provided by the VLBA's automated calibration
transfer, and amplitude corrections for errors in the sampler
thresholds were performed using autocorrelation data. Phase
corrections for parallactic angles were applied and a simple
bandpass correction was derived at each frequency from one of the
fringe finder scans.\\

The VLBA's pulse calibration system did not deliver usable data to
calibrate the phase offsets between the IFs because the frequencies
were changed too quickly. The pulse calibration system has a default
integration time of 10\,s, plus one second for readout. If any part of
that integration time is during a part of the scan that is flagged by
the online system, then the integration is not accepted. Thus, one
looses the first two integrations because the online system
conservatively flags 10\,s to 11\,s after a frequency change, and
there is an overlap between the flagged time and the second
integration. During the whole experiment, only $\sim10$ useful
pulse-cal measurements were recorded per station at 15\,GHz
(half-cycle time 22\,s), and $\sim100$ at 43\,GHz (half-cycle time
28\,s), compared to the number of half-cycles at these frequencies of
$\sim630$ and $\sim230$, respectively. We therefore used
self-calibration on the same fringe finder scans as were used for
bandpass calibration to correct for instrumental delays and phase
offsets. These offsets were found to be stable over the experiment.\\

\subsection{Ionospheric Correction}

Although the frequencies used in the project are quite high,
ionospheric effects can not be neglected and will prevent a successful
phase transfer if uncorrected. A typical ionospheric delay at a
frequency of $\nu_1=100$\,MHz is 0.1\,$\mu$s (10 turns of phase), the
exact number depending on the time of day. The delay scales with
frequency as $\nu^{-2}$, and so for a delay at 100\,MHz of
0.1\,$\mu$s, the ionospheric delays at 14.375\,GHz, 43.125\,GHz and
86.250\,GHz are 4.84\,ps, 0.54\,ps and 0.13\,ps, respectively,
corresponding to $25^\circ$, $8.4^\circ$ and $4.0^\circ$ of phase. In
contrast, the path length through the troposphere is non-dispersive
and so the linear phase versus frequency scaling law used by fast
frequency switching cannot correct phase changes that are induced by
the ionosphere. Ionospheric phase changes have much longer time scales
than tropospheric changes, and they can be calibrated before
fringe-fitting when the electron content of the ionosphere along the
line of sight is known.  The AIPS task TECOR can use maps of
ionospheric total electron content derived from GPS data to calculate
phase and delay corrections. Unfortunately, the error in these maps
can be quite high, up to 20\,\% when the TEC is as high as a few tens
of TEC units (1\,TECU=$10^{16}\,{\rm electrons\,m^{-2}}$), and up to
50\,\% or higher when the TEC is of the order of a few TEC units. We
have used the TEC maps produced by the Center for Orbit Determination
in Europe
(CODE\footnote{\url{http://www.aiub.unibe.ch/ionosphere.html}}) to
calibrate the effects of the ionosphere. We found that these maps
yielded better results than those from the Jet Propulsion Laboratory
(JPL), i.e., the residual phase errors after scaling were smaller. We
do not know whether this finding is coincidental or the CODE maps
generally give better results.

\subsection{Flagging}

\begin{table*}[ht!]
\center
\begin{tabular}{lcccccccccc}
\hline
\hline
Frequency & BR & FD & HN & KP & LA & MK & NL & OV & PT & SC\\
\hline
15\,GHz   & 6  & 6  & 6  & 7  & 6  & 7  & 7  & 6  & 8  & 7\\
43\,GHz   & 6  & -  & 7  & 7  & 5  & 5  & 6  & 6  & 7  & 6\\
86\,GHz   & 7  & 7  & 7  & 8  & 7  & 8  & 8  & 7  & 9  & 8\\
\hline
\end{tabular}
\caption[Flagging times]{Flagging times in seconds at each station and
frequency applied to the beginning of each half-cycle with QUACK.}
\label{tab:flagging}
\end{table*}

From the start of a new half-cycle 1\,s to 2\,s are required to set up
the electronics and 5\,s to 8\,s are required to position the
subreflector. Data during that time should be flagged. The flags
generated by the VLBA online system turned out to be too conservative,
flagging more data than needed and sometimes leaving only 10\,s of
data per integration at either $\nu_{\rm ref}$ or $\nu_{\rm t}$. We
used amplitude vs. time plots of calibrator scans to estimate the
switching times for each antenna separately. We found that the times
needed to move the subreflector are frequently asymmetric (e.g.,
moving it from $\nu_{\rm t}$ to $\nu_{\rm ref}$ took more time than
moving it back). Furthermore, they vary slowly with time, presumably
due to elevation-dependent gravitational force on the
subreflectors. Compromise flagging times for each antenna and
frequency at the start of each half-cycle were applied using the AIPS
task QUACK (Table~\ref{tab:flagging}).\\

\begin{SCfigure}[][htpb]
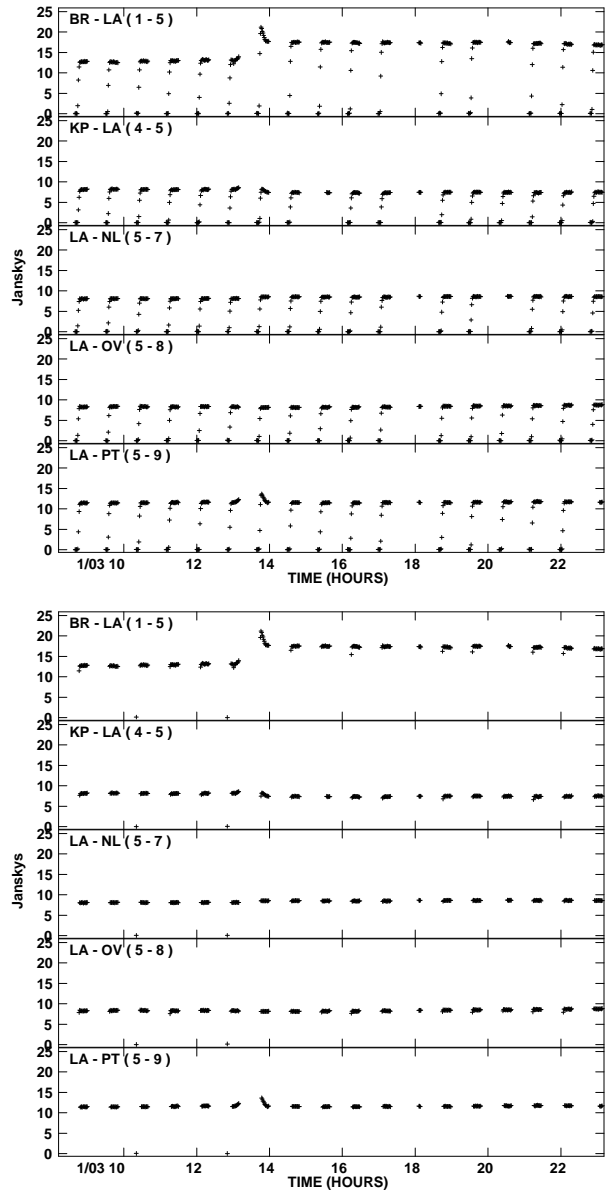

\begin{minipage}[b]{8cm}
\includegraphics[width=8cm, angle=270]{chap6/plots/amplitudes_unflagged.ps}
\includegraphics[width=8cm, angle=270]{chap6/plots/amplitudes_flagged.ps}
\end{minipage}
\caption[Illustration of the flagging scheme (amplitudes)]  {Illustration of the
flagging scheme. The period 03:08-03:13 UT shows 3C\,279 data, the
period 03:13-03:23 UT shows 3C\,273 data. Occasional outliers are due
to varying subreflector rotation times. {\it Top:} 15\,GHz visibility
amplitudes on five baselines to the reference antenna, LA.  No
flagging has been applied. {\it Bottom:} The same visibility
amplitudes after flagging.}
\label{fig:flagging}
\end{SCfigure}

The effect of the flagging is illustrated in Fig.~\ref{fig:flagging}.
Amplitudes need 2\,s to 3\,s longer to reach their final values than
do the phases. Thus, the visibility phases are not very sensitive to
even large errors in subreflector rotational position. The
repeatability of the phases from cycle to cycle shows that
positioning the subreflector along the optical axis is repeatable to
$<5^{\circ}$ of phase at 15\,GHz, which is much less than other
sources of phase error in fast frequency switching.\\

Occasionally, a subreflector does not move at all, resulting in a
half-cycle of noise. These events have been flagged by calculating the
duration of the entries in the list of flags provided by the VLBA and
flagging those cycles that have flagging entries for longer than 30\,s
and therefore exceeded the half-cycle time.

\subsection{Fringe-Fitting and Phase Scaling}

\begin{SCfigure}[][htpb]
 \includegraphics[width=8cm]{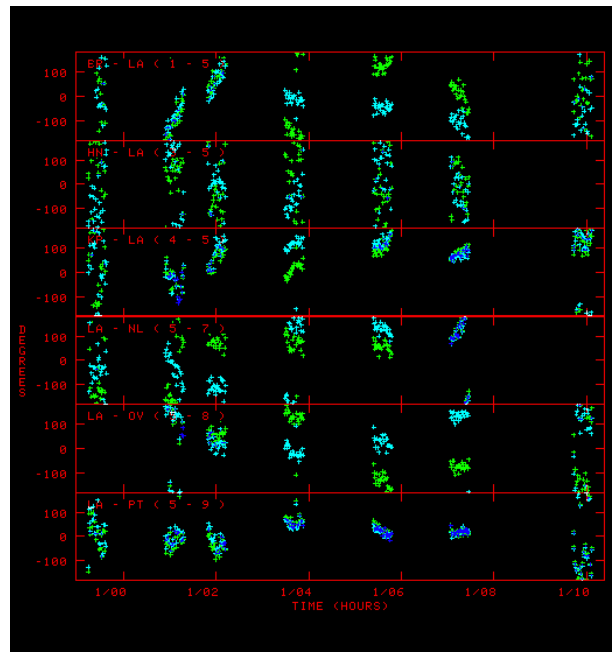}
\caption[The effect of using a model in self-calibration]{The effect of
using a model in self-calibration. NGC\,4261 calibrated 43\,GHz
visibility phases on baselines to Los Alamos. Calibration has been
done with scaled-up phase solutions from fringe-fitting at 15\,GHz
using a point source model (light blue) and a clean component model
(green), showing the effect of source structure at 15\,GHz in the
calibration at 43\,GHz. }
\label{fig:model+nomodel}
\end{SCfigure}

We fringe-fitted the 15\,GHz data using the AIPS task FRING, with an
SNR threshold of 5 and delay and rate search windows of 20\,ns and
50\,mHz, respectively. As NGC\,4261 has an extended, double-sided jet
structure at 15\,GHz, we made a 15\,GHz image that we used as a source
model in a second run of FRING, so that the phase solutions did not
contain structural phase contributions
(Fig.~\ref{fig:model+nomodel}). The solution interval was set to
1\,min, yielding one phase, delay and rate solution per half
cycle. The detection rate was $\sim90\,\%$. The 15\,GHz solution (SN)
table was written to a text file with TBOUT to do the phase scaling
outside AIPS. We have written a Python program (FFSTG, the {\it F}ast
{\it F}requency {\it S}witching {\it T}able {\it G}enerator) that
processes a SN table in the following way.\\

First, a series of timestamps is generated from each pair of
consecutive entries in the input table such that they coincide with
the $\nu_{\rm t}$ half-cycles. Second, from each pair of consecutive
$\nu_{\rm ref}$ phase solutions, a solution is interpolated for the
new timestamps consisting of a phase, a phase rate derived from the
$\nu_{\rm ref}$ phase solutions and the time interval between the
$\nu_{\rm ref}$ scans, and an interpolated delay. Both the phase rate
and the delay do not need to be scaled by $r$ because the rate is
stored in a frequency-independent format (in units of ${\rm
s\,s^{-1}}$) and the delay is non-dispersive. Third, the interpolated
solution is scaled by the frequency ratio, $r$, yielding $\nu_{\rm t}$
phase solutions for the times at which the source was observed at
$\nu_{\rm t}$.

We did not use the phase rates derived by FRING because each of those
was derived from one half-cycle of 22\,s length, whereas the phase
rates interpolated from two consecutive half-cycles as described above
used two half-cycles separated by 50\,s and therefore have much better
SNR. A further advantage of deriving phase rates from pairs of phase
solutions is that each $\nu_{\rm ref}$ phase solution is used both in
the determination of the phase rate to the preceding and the
succeeding $\nu_{\rm ref}$ solution, causing a smoothing of the phases
with time and reducing the effect of outliers. Finally, the
interpolated phase, phase rate and delay solutions are stored in an
output table together with the $\nu_{\rm t}$ frequency ID. FFSTG
provides an interface to Gnuplot to plot and inspect the input and
output phases. One can select regions of interest and check the
results of the scaling.

The table is imported to AIPS using TBIN and can be used to update the
most recent calibration table at the target frequency.\\

\section{Results}

\subsection{43\,GHz}

\begin{SCfigure}[][htpb]
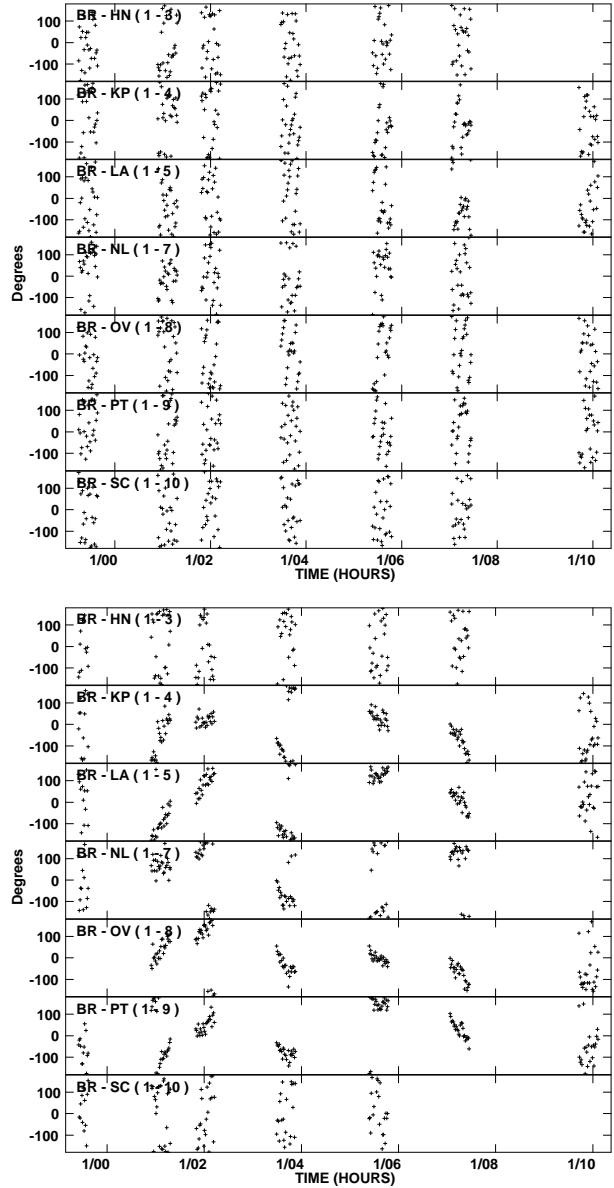

\begin{minipage}[b]{8cm}
 \includegraphics[width=8cm, angle=270]{chap6/plots/4261_43_tecor_BR.ps}
 \includegraphics[width=8cm, angle=270]{chap6/plots/4261_43_tecor+ffs_BR.ps}
\end{minipage}
\caption[Raw and calibrated visibility phases on baselines to
Brewster]{Raw and calibrated visibility phases on baselines to
Brewster. {\it Top:} NGC\,4261 raw 43\,GHz visibility phases on
baselines to Brewster with only delay calibration applied. 
{\it Bottom:} NGC\,4261 calibrated 43\,GHz visibility phases on baselines
to Brewster. Calibration used scaled-up phase solutions from
fringe-fitting with a clean component model at 15\,GHz.}
\label{fig:results_BR}
\end{SCfigure}

NGC\,4261 was detected on most baselines at all times after scaling
the 15\,GHz solutions to 43\,GHz, with correlated flux densities of
30\,mJy (800\,M$\lambda$) to 160\,mJy (30\,M$\lambda$). The 43\,GHz
half-cycle average visibilities on baselines to BR and LA are shown in
Figs.~\ref{fig:results_BR} and \ref{fig:results_LA}. The short-term
fluctuations introduced by the troposphere are almost perfectly
calibrated, but residual phase drifts remain on longer timescales,
especially at the beginning and at the end of the experiment, when the
sun was setting or rising.\\

\subsubsection{Structure Functions}

\begin{figure}[htpb]
\centering
 \includegraphics[width=0.8\linewidth]{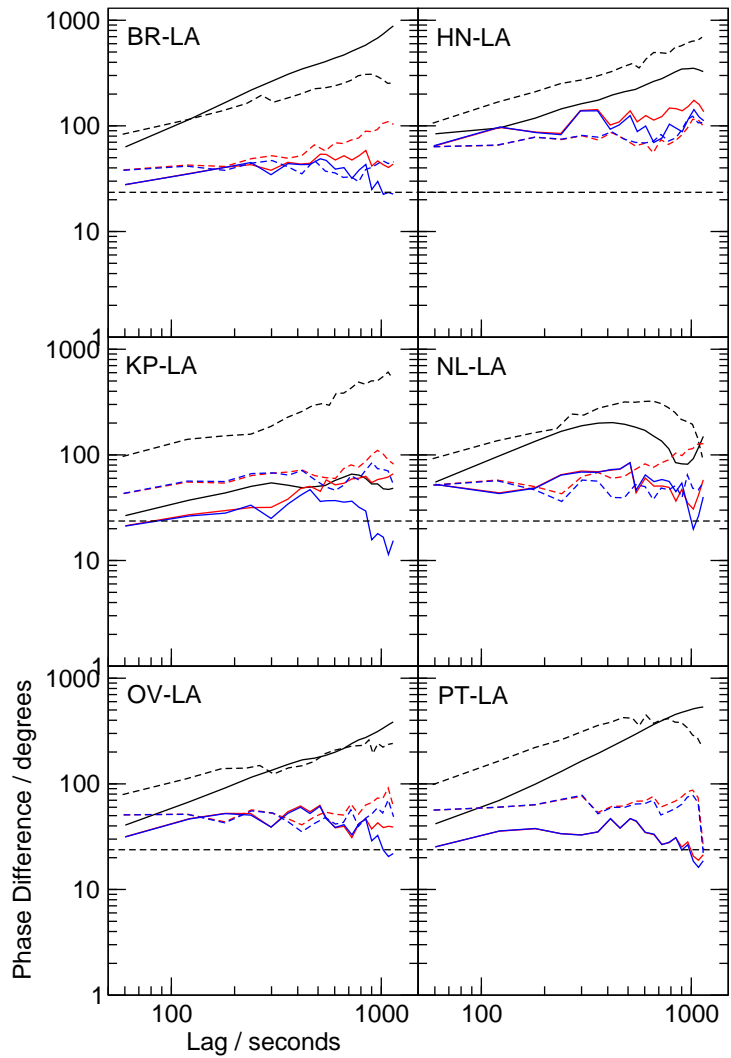}
 \caption[Structure functions of raw and calibrated 43 GHz
 data]{Structure functions from 25\,min of observation of NGC\,4261 at
 43\,GHz (the structure functions have been cut off after 1150\,s due
 to a lack of points at longer lags). Solid black lines show raw data
 observed at night with delay calibration only, solid red lines show
 the same data calibrated with scaled-up phase solutions from
 fringe-fitting at 15\,GHz, and solid blue lines show the same data
 without the residual phase rate. Dashed lines show the same stages of
 calibration, respectively, but from a scan observed during the late
 evening, where ionospheric effects increased the phase noise. The
 horizontal dashed black lines indicate the theoretical phase noise.
 The median residual phase noise of calibrated data, with residual
 phase rates, is $33^{\circ}$ at night and $50^{\circ}$ during sunset,
 and without residual phase rates is $31^{\circ}$ at night and
 $44^{\circ}$ during sunset.}
\label{fig:struct_func_bm175c}
\end{figure}

Structure functions of the 43\,GHz visibility phases from two 25\,min
scans on NGC\,4261 are shown in Fig.~\ref{fig:struct_func_bm175c}.
Solid black lines show raw data observed at night with delay
calibration only, and solid red lines show the same data calibrated
with scaled-up phase solutions from fringe-fitting at 15\,GHz. The
dashed lines show the same stages of calibration from a 25\,min scan
observed during the evening.

The structure functions constructed from the calibrated phase series
show a residual phase noise on the shortest timescales (60\,s) of
$50^{\circ}$ during sunset and $33^{\circ}$ at night (taking the
median of the phase noise of all baselines). The phase noise increases
toward longer timescales, probably due to errors in the ionospheric
models. Evidence for the long-term phase noise being dominated by
ionosphere is as follows.

\begin{SCfigure}[][htpb]
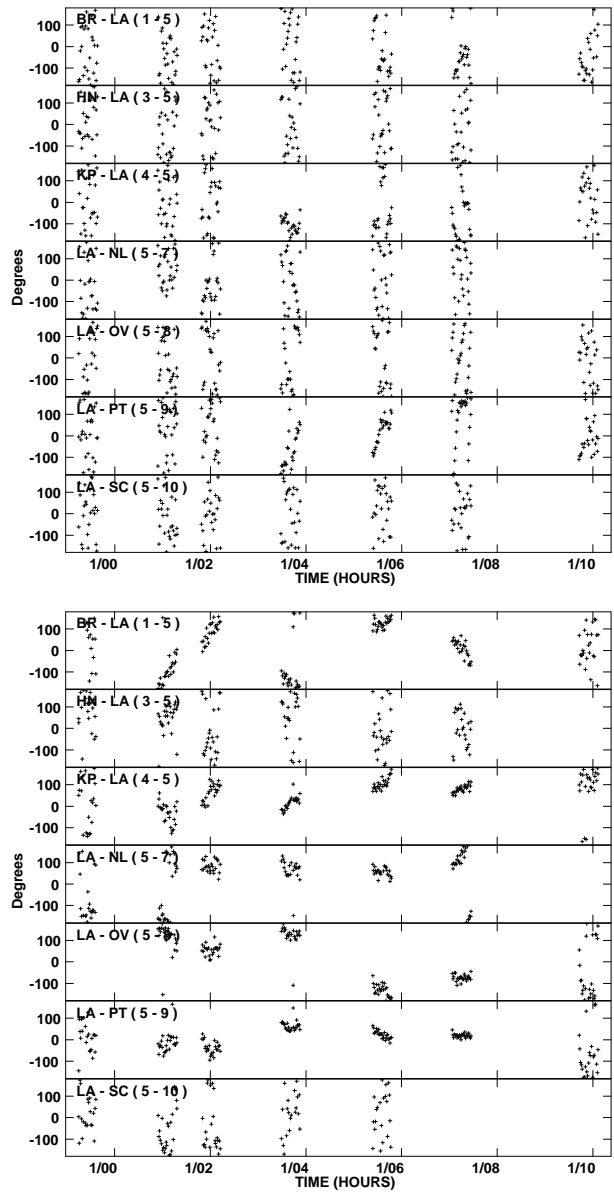

\begin{minipage}[b]{8cm}
 \includegraphics[width=8cm, angle=270]{chap6/plots/4261_43_tecor_LA.ps}
 \includegraphics[width=8cm, angle=270]{chap6/plots/4261_43_tecor+ffs_LA.ps}
\end{minipage}
\caption[Raw and calibrated visibility phases on baselines to Los
Alamos]{Raw and calibrated visibility phases on baselines to Los
Alamos. {\it Top:} NGC\,4261 raw 43\,GHz visibility phases on
baselines to Los Alamos with only delay calibration applied. {\it
Bottom:} NGC\,4261 calibrated 43\,GHz visibility phases on baselines
to Los Alamos. Calibration used scaled-up phase solutions from
fringe-fitting with a clean component model at 15\,GHz.}
\label{fig:results_LA}
\end{SCfigure}

\begin{SCfigure}[][htpb]
 \includegraphics[width=8cm]{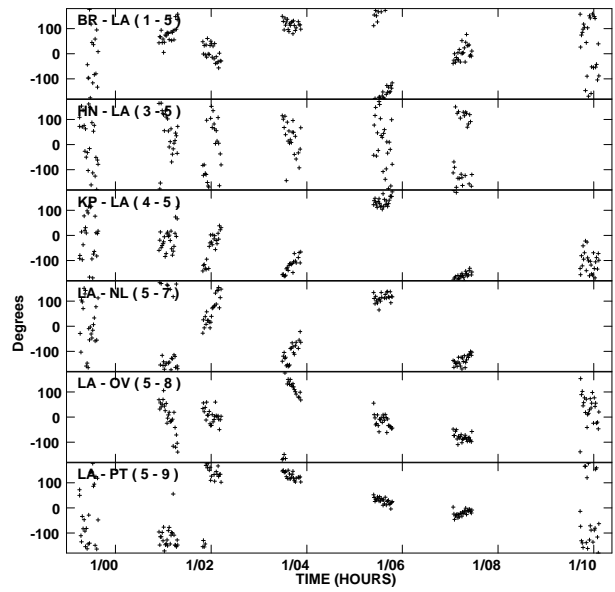}
\caption[The effect of ionospheric corrections]{The effect of
ionospheric corrections. NGC\,4261 calibrated 43\,GHz visibility
phases on baselines to Los Alamos as in Fig.~\ref{fig:results_LA}
(lower panel), but no ionospheric correction has been applied. The
residual phase rates are much higher.}
\label{fig:no_tecor}
\end{SCfigure}

First, applying an ionospheric TEC correction significantly improved
the coherence, as one can see by comparing
Figs.~\ref{fig:results_LA} (lower panel) and
\ref{fig:no_tecor}. Second, the residual phase errors decreased
at night. After sunset at 19:56 local time in Los Alamos (=02:56 UT),
representative for the south-western antennas, the phase errors
rapidly dropped.

We used FRING on the 43\,GHz data calibrated with fast frequency
switching with a solution interval of 30\,min, to self-calibrate the
data and so to remove the residual long-term phase drifts. The
structure functions after applying this correction are shown as blue
lines in Fig.~\ref{fig:struct_func_bm175c}; the long-term phase noise
is reduced as expected, and the phase noise on 60\,s timescale is
slightly lower, with a median of $44^{\circ}$ during sunset and
$31^{\circ}$ at night.\\

\subsubsection{Expected Phase Noise}

The expected phase noise in the visibilities calibrated with fast
frequency switching consists of three parts: (1) thermal phase noise
at the reference frequency scaled by the frequency ratio, (2) thermal
phase noise at the target frequency and (3) tropospheric phase changes
the two integrations. One can estimate the expected rms phase noise
from sensitivity calculations as follows. The $1\,\sigma$ thermal
noise on a baseline is given by

\begin{equation}
\Delta S = \frac{1}{\eta_{\rm s}} \times \frac{SEFD}{\sqrt{2\times\Delta\nu\times\tau}}
\end{equation}

(\citealt{Walker1995}), where $\eta_{\rm s}$ is a scaling factor,
$SEFD$ is the antenna's system equivalent flux density, $\Delta\nu$ is
the bandwidth and $\tau$ the integration time. The efficiency factor
$\eta_{\rm s}$ depends on accurately known effects like losses due to
signal quantisation, clock drifts, and others, but also on
time-variable, unknown effects like antenna pointing errors and
moisture on the feeds and reflecting surfaces. It therefore has to be
determined empirically, and for the VLBA usually is assumed to be 0.5
for 1-bit sampling. The observations presented here were done using
2-bit sampling, increasing the sampling efficiency from 0.637 to
0.881, or by 38\,\%, and $\eta_{\rm s}$ then is 0.69.

According to the VLBA Observational Status
Summary\footnote{\url{http://www.aoc.nrao.edu/vlba/obstatus/obssum/obssum.html}},
the $SEFD$ for a VLBA dish is 550\,Jy at 15\,GHz, 1436\,Jy at 43\,GHz
and 5170\,Jy at 86\,GHz. In our observations, $\tau$ was 15\,s at
15\,GHz and 21\,s at 43\,GHz and 86\,GHz, and $\Delta\nu$ was
64\,MHz. One therefore expects thermal noise levels of 17.8\,mJy,
39.2\,mJy and 141\,mJy at 15\,GHz, 43\,GHz and 86\,GHz,
respectively. At 15\,GHz, NGC\,4261 has a correlated flux density,
$S$, between $60\,{\rm mJy}$ and $300\,{\rm mJy}$, depending on
baseline length, and we do the following calculations assuming either
$80\,{\rm mJy}$, representative for long baselines, or $200\,{\rm
mJy}$, representative for short baselines (the results of which we
give in brackets).\\

\paragraph{(1)} On a single baseline, the expected SNR of a detection
at 15\,GHz is $S/\Delta S=80\,{\rm mJy}/17.8\,{\rm mJy}=4.49$ (11.2)
when averaging over the band. Fringe-fitting, however, uses all
baselines to one particular antenna to derive a phase correction, and
so the SNR is increased by $\sqrt{N}$, where $N$ is the number of
baselines. At any given time, $N\approx8$, so that the SNR of a
detection at 15\,GHz increases to 12.7 (31.7), corresponding to a
phase error of $1\,{\rm rad/SNR}=4.51^{\circ}$ ($1.81^{\circ}$). These
errors are scaled by the frequency ratio to $13.5^{\circ}$
($5.43^{\circ}$) at 43\,GHz and $27.1^{\circ}$ ($10.9^{\circ}$) at
86\,GHz. 

\paragraph{(2)} Assuming that NGC\,4261 has a compact flux density of
100\,mJy at both frequencies, the thermal noise contributions are
$21.4^{\circ}$ and $54.7^{\circ}$, respectively. Adding those in
quadrature to the scaled-up thermal rms phase noise from 15\,GHz
yields $25^{\circ}$ ($22.1^{\circ}$) at 43\,GHz and $56.3^{\circ}$
($55.0^{\circ}$) at 86\,GHz.

\paragraph{(3)} We estimated the tropospheric phase noise within the
switching cycle time using structure functions of 3C\,273. On 3C\,273,
the thermal noise contributions at 15\,GHz and 43\,GHz are
$0.04^{\circ}$ and $0.4^{\circ}$, respectively, so that the visibility
phases are essentially free of thermal noise and any phase changes
during and between the half-cycles are due to changes in the
troposphere. We found the median rms phase noise after
self-calibration with a 30\,min solution interval to remove the
residual long-term phase drift to be $13.3^{\circ}$ at 43\,GHz, or
$26.6^{\circ}$ at 86\,GHz. 

Adding those three constituents in quadrature yields $28.3^{\circ}$
($25.8^{\circ}$) at 43\,GHz and is in good agreement with the measured
rms phase noise of $31^{\circ}$.\\

\subsubsection{Coherence}

\begin{figure}[ht!]
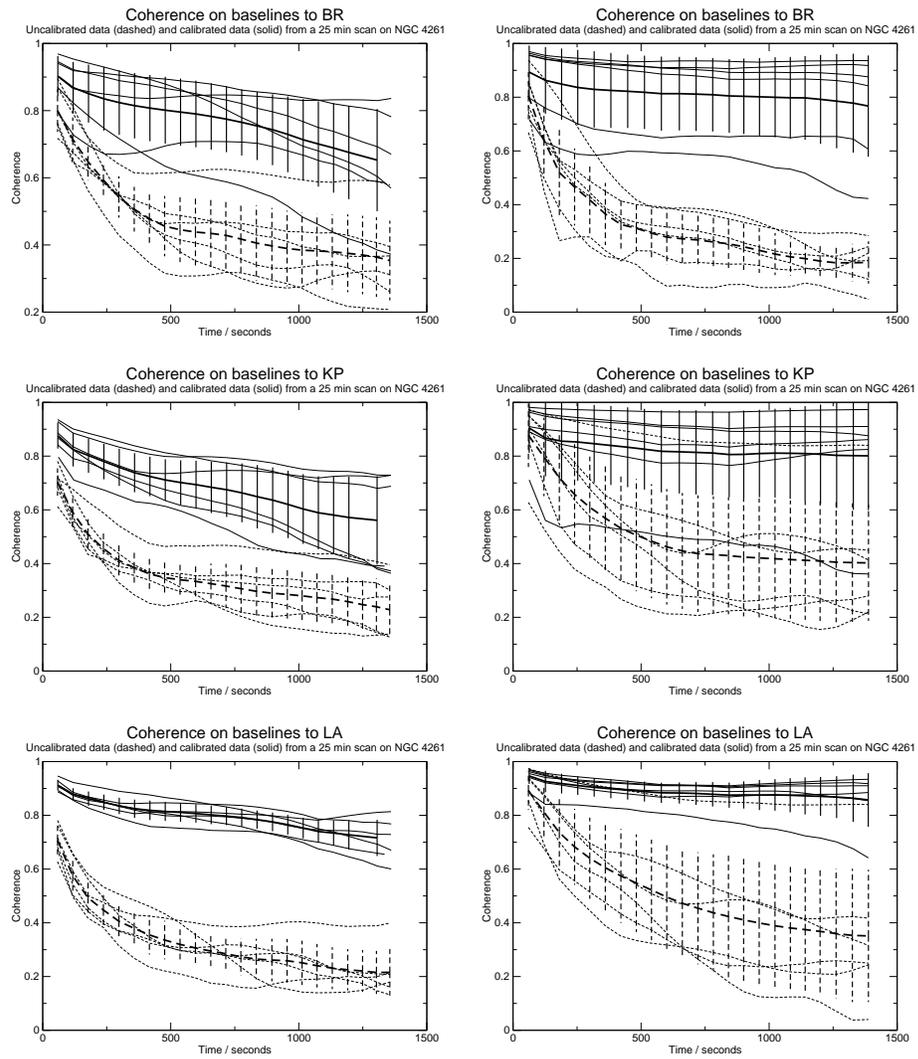

\centering
 \hbox{
 \includegraphics[width=0.48\linewidth]{chap6/plots/coherence_4261_BR_day.eps}
 \hspace*{3mm}           
 \includegraphics[width=0.48\linewidth]{chap6/plots/coherence_4261_BR_night.eps}
 }
 \vspace{3.5mm}          
 \hbox{                  
 \includegraphics[width=0.48\linewidth]{chap6/plots/coherence_4261_KP_day.eps}
 \hspace*{3mm}    
 \includegraphics[width=0.48\linewidth]{chap6/plots/coherence_4261_KP_night.eps}
 }                
 \vspace{3.5mm}   
 \hbox{           
 \includegraphics[width=0.48\linewidth]{chap6/plots/coherence_4261_LA_day.eps}
 \hspace*{3mm}    
 \includegraphics[width=0.48\linewidth]{chap6/plots/coherence_4261_LA_night.eps}
 }
 \caption[Coherence improvement at 43\,GHz]{Coherence improvement at
 43\,GHz. NGC\,4261 was observed in the early evening (centred on
 01:06 UT, left column) and at night (centred on 03:40 UT, right
 column) at 43\,GHz with interleaved 15\,GHz observations for
 calibration. 43\,GHz data with (solid lines) and without (dashed
 lines) phase-calibration from 15\,GHz have been averaged in frequency
 and in time over each 43\,GHz half-cycle. The normalized vector sum has
 been computed on all baselines to a given station over progressively
 larger time intervals and assuming unit amplitude. Bold lines show
 the average over all baselines.}
 \label{fig:coherence}
\end{figure}

Another way of illustrating the performance of the corrections is with
coherence diagrams (Fig.~\ref{fig:coherence}). For the plots shown,
NGC\,4261 43\,GHz data from all baselines to three stations, BR, KP
and LA, have been used. All data were calibrated using fast frequency
switching (no subsequent fringe-fitting), and were averaged over the
band and in time to obtain one visibility per baseline and
half-cycle. Assuming unit amplitude, the vector sum of these
visibilities has been computed over progressively longer time
intervals and the length of the resulting vector has been normalized
to one by dividing by the number of visibilities. All of these sums
have then been averaged for each baseline separately to yield one
coherence measurement per baseline and coherent averaging time
interval. The left panels show data from a 25\,min scan centred on
01:06 UT (early evening) and the right panels show data from a 25\,min
scan centred on 03:40 UT (at night). Bold lines show the average over
all baselines and the error bars show the $1\,\sigma$ scatter of the
baseline average. The coherence achieved at night with fast frequency
switching is significantly better than that achieved during the early
evening.\\

Before making an image from the 43\,GHz visibilities calibrated with
fast frequency switching, the residual phase offsets and phase rates
were removed using FRING with a solution interval of 30\,min so that
one solution per scan was obtained. The resulting dirty image
(Fig.~\ref{fig:NGC4261_43GHz_FFS_raw}) has a peak flux density of
$79\,{\rm mJy\,beam^{-1}}$ and an rms noise of $4.4\,{\rm
mJy\,beam^{-1}}$, yielding a dynamic range of 18:1. After several
cycles of phase self-calibration with a solution interval of 30\,s and
one cycle of amplitude self-calibration with a solution interval of
12\,h, the peak flux density was $95\,{\rm mJy\,beam^{-1}}$ and the
rms noise was $0.78\,{\rm mJy\,beam^{-1}}$, so the dynamic range
improved to 122:1 (Fig.~\ref{fig:NGC4261_43GHz_FFS}). The theoretical
rms noise at 43\,GHz was expected to be $0.46\,{\rm
mJy\,beam^{-1}}$. Unfortunately, the ionospheric phase errors and the
need to run FRING to calibrate them did not allow a core shift
measurement to be made.

The image dynamic range can be estimated using

\begin{equation}
DNR=\frac{\sqrt{M}N}{\phi_{\rm rms}}
\end{equation}
(\citealt{Perley1986}), where $M$ is the number of independent
measurements, $N$ is the number of antennas and $\phi_{\rm rms}$ is
the rms of the phase noise. The number of independent measurements is
given by the number of 43\,GHz half-cycles, corresponding to the basis
on which $\phi_{\rm rms}$ was calculated. With seven 43\,GHz scans, 30
half-cycles per scan, and multiplying by the number of baselines that
gave good data ($N=8$, so 28 baselines), $M=5880$. The average of
$\phi_{\rm rms}$ during the evening and at night was
$(44^{\circ}+31^{\circ})/2=38^{\circ}$, and the expected dynamic range
follows to 16:1. This is in excellent agreement with what we measured
in the first 43\,GHz image (18:1). Note that this agreement is
independent of the half-cycle times. Doubling the cycle time decreases
$\phi_{\rm rms}$ by $\sqrt{2}$, but $M$ is decreased by two at the
same time, and the result remains unchanged.

\subsection{86\,GHz}

\begin{SCfigure}[][htpb]
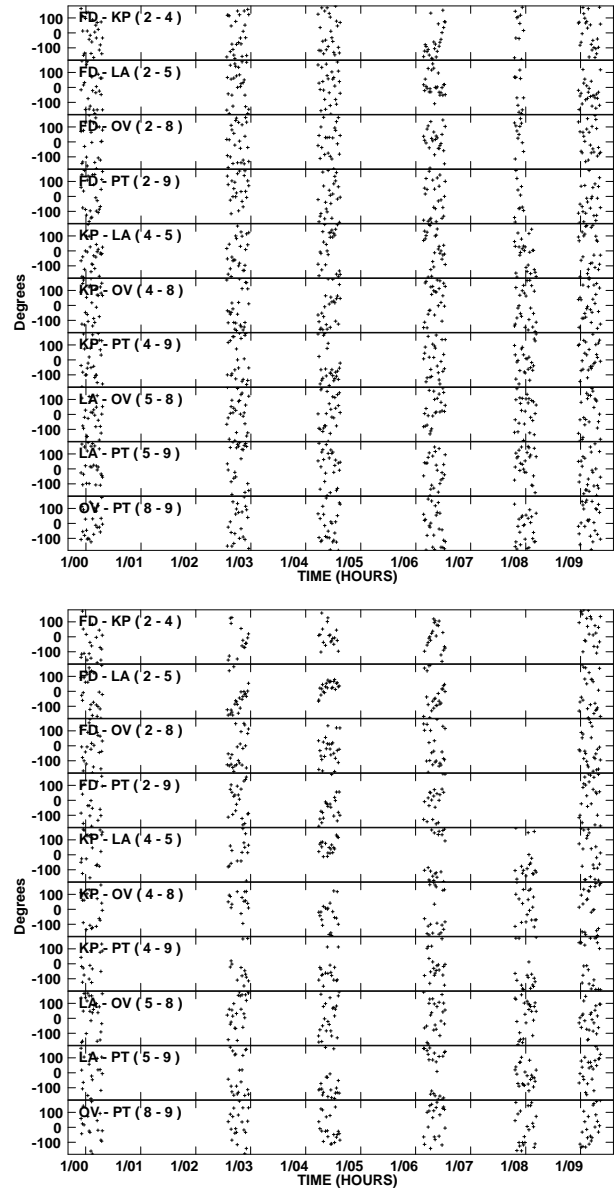

\begin{minipage}[b]{8cm}
 \includegraphics[width=8cm, angle=270]{chap6/plots/4261_86_CL11_tecor.ps}
 \includegraphics[width=8cm, angle=270]{chap6/plots/4261_86_CL17_tecor+ffs.ps}
\end{minipage}
\caption[86\,GHz visibility phases]{86\,GHz visibility phases. {\it
Top:} NGC\,4261 raw 86\,GHz visibility phases on baselines among Fort
Davis, Kitt Peak, Los Alamos, Owens Valley and Pie Town with only
delay calibration applied. {\it Bottom:} 86\,GHz visibility phases on
baselines among Fort Davis, Kitt Peak, Los Alamos, Owens Valley and
Pie Town. Calibration has been done with scaled-up phase solutions
from fringe-fitting at 15\,GHz using a clean component model. Good
detections were made during almost every 25\,min scan observed at
night between 2:00\,h UT and 7:00\,h UT, when the ionosphere was
stable.}
\label{fig:4261_86}
\end{SCfigure}

\begin{figure}
 \centering
 \includegraphics[width=0.9\linewidth]{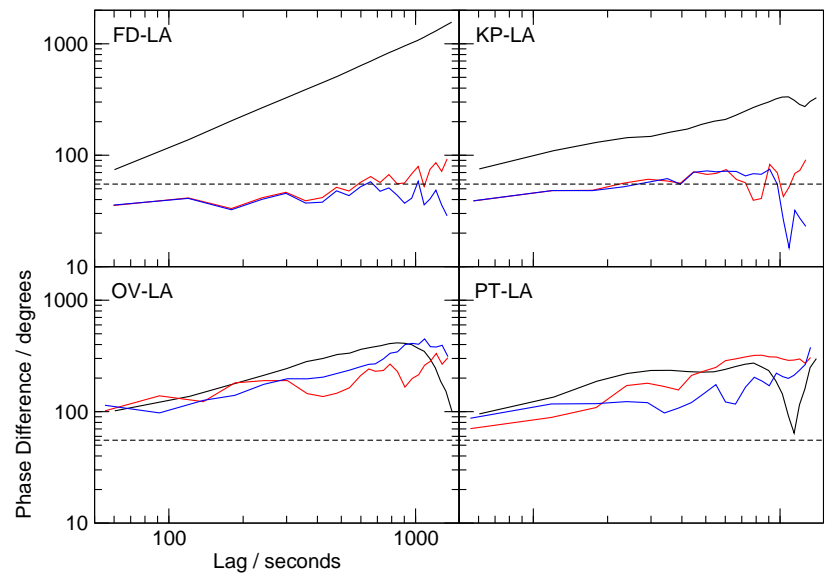}
 \caption[Phase structure functions of a 25\,min scan at 86\,GHz]{Structure
 functions of a 25\,min scan at 86\,GHz. Colours show the same stages of calibration as 
 is Fig.~\ref{fig:struct_func_bm175c}.}
 \label{fig:struct_funct_86GHz_bm175c}
\end{figure}

\begin{figure}[ht!]
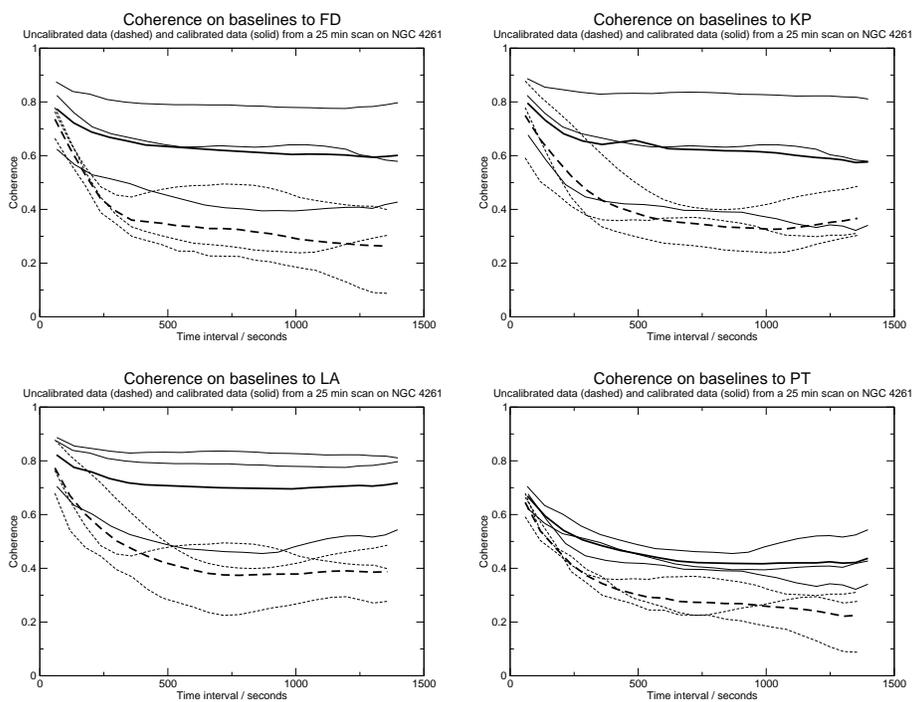

 \hbox{
 \includegraphics[width=0.48\linewidth]{chap6/plots/coherence_4261_FD_86.eps}
 \hspace*{3mm}
 \includegraphics[width=0.48\linewidth]{chap6/plots/coherence_4261_KP_86.eps}
 }
 \vspace{3.5mm}
 \hbox{
 \includegraphics[width=0.48\linewidth]{chap6/plots/coherence_4261_LA_86.eps}
 \hspace*{3mm}
 \includegraphics[width=0.48\linewidth]{chap6/plots/coherence_4261_PT_86.eps}
 }
 \caption[Coherence improvement at 86\,GHz]{The coherence improvement
 at 86\,GHz with fast frequency switching on some of the baselines
 shown in Fig.~\ref{fig:4261_86}. Errors have not been
 computed due to the reduced number of data points.}
 \label{fig:coherence_86}
\end{figure}

\begin{SCfigure}[][htpb]
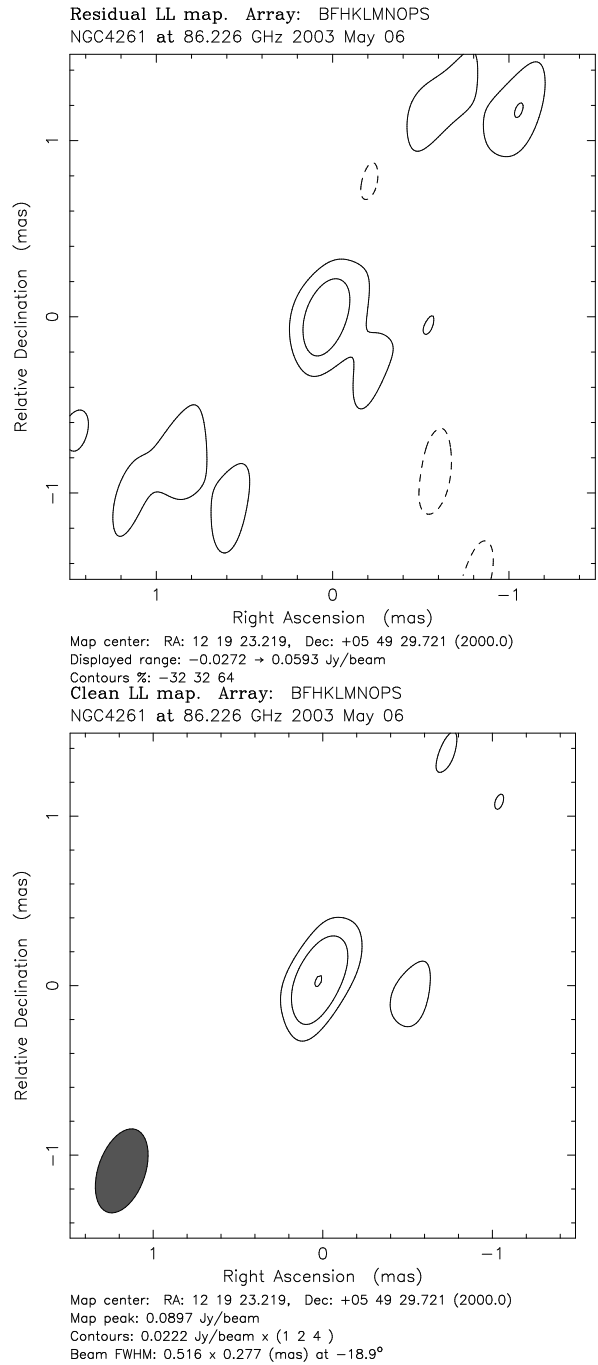

\begin{minipage}[b]{8cm}
 \includegraphics[width=9cm, angle=270]{chap6/plots/NGC4261_86_CL17_raw.ps}
 \includegraphics[width=9cm, angle=270]{chap6/plots/NGC4261_86_CL17_it1.ps}
\end{minipage}
\caption[86\,GHz images of NGC\,4261]{86\,GHz images of NGC\,4261. {\it Top:} Naturally weighted,
 full-resolution image of NGC\,4261 at 86\,GHz, calibrated with
 scaled-up phase solutions from 15\,GHz. FRING has been used to solve
 for one residual phase and rate solution per 25\,min scan before
 exporting the data to  Difmap. No further self-calibration has
 been applied. The image noise is 8.4\,mJy\,beam$^{-1}$ and the
 dynamic range is 7:1. {\it Bottom:} Data and imaging parameters
 as above, but several cycles of
 phase self-calibration have been applied. The image noise
 is 7.4\,mJy\,beam$^{-1}$ and the dynamic range is 12:1.}
\label{fig:NGC4261_86GHz}
\end{SCfigure}

Following the same data reduction path as for the 43\,GHz data, we
obtained good detections of NGC\,4261 at 86\,GHz on baselines among
the four stations FD, KP, LA, and PT. Uncorrected and corrected visibility phases are plotted in
Figs.~\ref{fig:4261_86}, structure functions and coherence plots
from a 25\,min scan are shown in
Figs.~\ref{fig:struct_funct_86GHz_bm175c} and
\ref{fig:coherence_86} and images are shown in
Fig.~\ref{fig:NGC4261_86GHz}.  The peak flux density in the upper panel of 
Fig.~\ref{fig:NGC4261_86GHz} is 59.3\,mJy. This is, we
believe, the first detection of NGC\,4261 with VLBI at 86\,GHz. It is
also probably the weakest continuum object ever detected with VLBI at
this frequency, because it is less than the 85\,mJy in the 86\,GHz
image of $1308+328$ by \cite{Porcas2002}, using conventional
phase-referencing to a calibrator $14.3^{\prime}$ away. With only
delay calibration applied, the median rms phase noise of the best
25\,min scan is $104^{\circ}$, after applying the scaled 15\,GHz phase
solutions is $70^{\circ}$ and after self-calibration with a 30\,min
solution interval is $80^{\circ}$. The increase in rms phase noise
after removal of phase rates is due to the small number of stations in
the sample. The expected thermal phase noise is the quadrature sum of
the scaled-up 15\,GHz noise and the 86\,GHz noise, the former of which
we estimated to be $27.1^{\circ}$ on long and $10.9^{\circ}$ on short
baselines, and the latter, assuming a correlated flux density of
100\,mJy, is $54.7^{\circ}$. The quadrature sum is $61.0^{\circ}$ on
long and $55.7^{\circ}$ on short baselines, and adding the
tropospheric phase errors of $26.6^{\circ}$ is $66.6^{\circ}$ and
$61.7^{\circ}$, in agreement with the measured noise levels. The rms
noise in the final 86\,GHz image is $7.4\,{\rm mJy\,beam^{-1}}$,
compared to an expected thermal noise of $3.7\,{\rm
mJy\,beam^{-1}}$. The dynamic range in the first image was 7:1, and
assuming seven antennas and an average rms phase noise of $64^{\circ}$
gives an expected dynamic range of 9:1. Thus, the results of fast
frequency switching agree very well with the theoretical expectations.

\section{Summary}

\begin{itemize}

\item Fast-frequency switching can be used to calibrate tropospheric
phase fluctuations if the switching cycle time is shorter than the
atmospheric coherence time. The accuracy is limited by tropospheric
phase changes between and during the reference frequency integrations.

\item Insufficient knowledge about the ionosphere's total electron
content (TEC) has prevented calibration of the inter-band phase offset
and hence making a pure phase-referenced image without using selfcal
was not possible. This also prevented the detection of a core
shift. Current global TEC models derived from GPS data have errors
that are too large to sufficiently calibrate the ionospheric component
of the phase changes. A possible solution could be to insert frequent
(every 10\,min) scans at L-band with widely separated IF frequencies
to derive the effective TEC on the line of sight to the target source.

\item The VLBA pulse calibration system might be able to monitor
instrumental phase changes if it is modified to cope with the
unusually short integration times.

\item The instrumental phase offset seen on the LA-PT
baseline was stable to $<$1\,rad over 10\,h, and it seems to be
sufficient to determine the offset a few times throughout the
experiment. Measuring it every 30\,min like we attempted therefore is
not necessary.

\item Truly simultaneous observations at two bands would reduce the
residual phase errors even further because tropospheric phase changes
within the switching cycle time would be entirely removed.

\end{itemize}

The primary use of fast frequency switching is to detect sources that
are too weak for self-calibration at the target frequency within the
atmospheric coherence time, but can be reliably detected at a lower
frequency. The $5\,\sigma$ detection limit of the VLBA at 43\,GHz
within 120\,s, with 64\,MHz bandwidth and 2-bit sampling, and
neglecting coherence loss due to tropospheric phase changes, is
81.9\,mJy (i.e., a thermal noise level of 16.4\,mJy on a single
baseline). The $1\,\sigma$ noise level using fast frequency switching
with half-cycle times of 22\,s at 15\,GHz and 28\,s at 43\,GHz
(yielding net integration times of 15\,s and 21\,s, respectively),
after one cycle is 66.2\,mJy, of which 39.1\,mJy is thermal noise in
the raw 43\,GHz visibility, 53.4\,mJy is thermal noise in the phase
solutions after scaling from 15\,GHz, and neglecting noise due to
tropospheric phase changes. The $1\,\sigma$ noise level after 120\,s
of fast frequency switching (i.e., 2.4 cycles) is 42.7\,mJy. The noise
level of fast frequency switching reaches that of a conventional
120\,s integration at 43\,GHz after 815\,s (13.6\,min),
i.e. 16.4\,mJy. Any longer integration with fast frequency switching
then yields detection thresholds that are out of reach with
conventional methods.\\

Fast frequency switching, however, is limited by the source strength
at the reference frequency, and 100\,mJy at 15\,GHz is, from our
experience, a reasonable minimum required for a successful
observation. However, many sources exist that meet this criterion. The
2\,cm survey by
\cite{Kellermann1998}, for example, detected 130 sources with peak
flux densities of 100\,mJy or more; all of them can be observed at
86\,GHz using fast frequency switching. Another estimate of the number
of new sources made accessible to 86\,GHz VLBI by fast frequency
switching can be made by means of the $\log N-\log S$ relation. We
assume that the sources observable with conventional 86\,GHz VLBI are
distributed homogeneously in the universe, i.e., lowering the
detection limit does not yield the detection of a new population, so
the number of sources above a given flux density limit, $S$, increases
as $S^{-2.5}$. We further make the conservative assumption that these
sources have $\alpha=0$ between 15\,GHz and 86\,GHz, because they are
compact. The current $5\,\sigma$ detection limit of the VLBA at
86\,GHz is 600\,mJy (thermal noise of a 30\,s integration with
128\,MHz bandwidth on a single baseline). In fringe-fitting, all $N$
antennas are used to derive a phase solution, so the noise level is
reduced by $\sqrt{N}$, and with $N=7$, assuming one VLBA station
capable of 86\,GHz observations is missing at any time, the
$5\,\sigma$ detection limit is 230\,mJy. Fast frequency switching
allows one to observe any compact source with $S_{\rm 86\,GHz}\ge
100\,{\rm mJy}$, a factor of 2.3 fainter than with conventional
methods, and one expects eight times more sources brighter than
100\,mJy than there are brighter than 230\,mJy, following the $\log
N-\log S$ relation. A little less conservative estimate would be to
allow for spectral indices as steep as $\alpha=-0.3$, so that sources
with $S_{\rm 15\,GHz}=100\,{\rm mJy}$ have $S_{\rm 86\,GHz}>60\,{\rm
mJy}$. The number of observable sources would then increase by a
factor of 29 of the number accessible to conventional techniques.

The absolute number of presently observable sources is much more
difficult to determine due to selection effects in the various radio
source catalogues. However, we can make a simple estimate using the
recent 86\,GHz VLBI survey by \cite{Lobanov2002}. They aimed at
observing more than 100 sources collected from the literature that had
an expected compact flux density of $S_{\rm 86\,GHz}>300\,{\rm
mJy}$. Applying the fast frequency switching sensitivity estimates
then yields between 800 and 2900 sources observable with the VLBA at
86\,GHz.

Furthermore, the proposed capability of fast frequency switching to
detect core shifts in AGN can supplement jet physics with
observational constraints which are otherwise difficult to obtain.

\section{A Recipe}

This section summaries the experiment design and the data reduction
for those who want to use fast frequency switching for their
observations.

\subsection{Experiment Design}

\begin{itemize}

\item{\bf Frequencies} The choice of target and reference frequencies
needs to be made with careful attention to the following issues. The
calibration of $\nu_{\rm t}$ critically depends on the ability to
self-calibrate the $\nu_{\rm ref}$ data. Observers should check
whether the $\nu_{\rm ref}$ provides sufficient baseline sensitivity
within the available $\nu_{\rm ref}$ integration time to yield
reliable detections. Second, the frequency ratio $r$ needs to be
integer. As this can be true for only an infinitely small bandwidth,
one should select the frequencies such that the {\it centres} of the
$\nu_{\rm t}$ and $\nu_{\rm ref}$ bands have an integer ratio (this
condition also prefers 2-bit sampling over 1-bit sampling because of
the smaller bandwidth at the same recorded bitrate). If the bandwidth
is 64\,MHz and the centre frequencies are $\nu_{\rm ref}=14.375$\,GHz
and $\nu_{\rm t}=43.125$\,GHz or $\nu_{\rm t}=86.250$\,GHz, the
frequency ratios at the lower edge of the bands (where the deviation
from an integer ratio reaches a maximum) are

\begin{displaymath}
r=\frac{{\rm 43.125\,GHz-32\,MHz}}{{\rm 14.375\,GHz}} = 2.9978
\end{displaymath}

\begin{displaymath}
r=\frac{{\rm 86.250\,GHz-32\,MHz}}{{\rm 14.375\,GHz}} = 5.9978.
\end{displaymath}

Thus, for each phase turn at 15\,GHz, one accumulates a phase error of
$0.79^{\circ}$ at 86\,GHz, and after 72 turns the error would be one
radian. However, averaging reduces this error because it has opposite
sign at the other edge of the band. Furthermore, the phase does
normally not wind constantly in one direction throughout the
observation.

\item{\bf Ionospheric corrections} Given the results from the data
  presented in this memo, we think that regular (every 10\,min), short
  (1\,min) scans at L-band with widely separated IFs should be
  observed during the observations to derive the ionospheric
  dispersive delay. This should allow one to separate the ionospheric
  and tropospheric phase errors and to obtain coherence over several
  hours even during dusk and dawn. We have not tested the procedure
  but have estimated the effect on a data set from a different project
  (BM170), where the flux density on the longest baselines was
  50\,mJy. We obtained one phase solution per IF every 30\,s using
  fringe-fitting, and the delays were in good agreement with our
  expectations, although the effect was small due to the small
  bandwidth that was covered by the IFs.

\item{\bf Switching cycles} The switching cycle times depend very much
  on the weather conditions. As they are unknown, one should apply for
  dynamic scheduling so that a certain level of ``good'' weather can
  be assumed. During the observations presented here, the coherence
  times were mostly 5\,min to 10\,min at 43\,GHz, occasionally
  decreasing to 2\,min. We chose a cycle time of 50\,s, short enough
  to follow the phase at $\nu_{\rm ref}$ most of the time.

\item{\bf Calibrator scans} One should insert regular scans on strong,
  nearby calibrators to be able to check the phase transfer from
  $\nu_{\rm ref}$ to $\nu_{\rm t}$ during postprocessing and to
  measure the inter-frequency instrumental offset. The calibrators
  must be strong enough for self-calibration at both $\nu_{\rm t}$ and
  $\nu_{\rm ref}$ within the respective integration times. At least 5
  to 7 of those calibrator observations should be 10\,min to 15\,min
  long to estimate the atmospheric coherence times and to help debug
  the calibration. There is the danger of transferring a possible core
  shift to the target. This can be avoided if at least two, but
  preferably more, calibrators are used to check the calibration.
  
\item{\bf Correlation} The correlator integration time should be set
  to 1\,s to allow good tracking of the visibility phases with time.

\end{itemize}

\subsection{Data Reduction}

\begin{itemize}

\item{\bf Loading data} A sufficiently small interval between the
calibration table entries should be chosen, 5\,s to 8\,s seemed
appropriate to us.

\item{\bf Flagging} Flagging of the data that were recorded during a
frequency change is required. The times that need to be flagged at the
beginning of a half-cycle with QUACK can be estimated using plots of
visibility amplitude with time on calibrator scans.

\item{\bf Standard steps} The following steps do not differ from
standard VLBI data reduction: amplitude calibration using $T_{\rm
sys}$ and gain measurements and calibration of sampler threshold
errors, removal of parallactic angles from the visibility phases,
bandpass calibration and manual calibration of the inter-IF phase
offsets using self-calibration on a strong source.

\item{\bf Ionospheric errors} The ionospheric delay needs to be
calibrated carefully. GPS-based models applied with TECOR were not
accurate enough. If the source has been observed at L-band, the
dispersive delay should be estimated and applied to all
frequencies. We have no experience with this procedure and cannot give
advice on details.

\item{\bf Self-calibration} Self-calibration should be carried out at
$\nu_{\rm ref}$ and will yield one phase, delay and rate solution per
half-cycle if the solution interval exceeds the half-cycle time.

\item{\bf Phase scaling} The solution (SN) table from self-calibration
is written to a text file using TBOUT. The table is read and processed
outside AIPS with FFSTG, the fast frequency switching
table generator, and read into AIPS with TBIN. The scaled-up
phase solutions are applied to the most recent $\nu_{\rm t}$
calibration table. The data can now be averaged in time and frequency
to increase the SNR and for scientific interpretation. Contact Enno
Middelberg (emiddelb@mpifr-bonn.mpg.de) for help installing Python and
using FFSTG.

\end{itemize}

\section{Figures}

\begin{SCfigure}[][htpb]
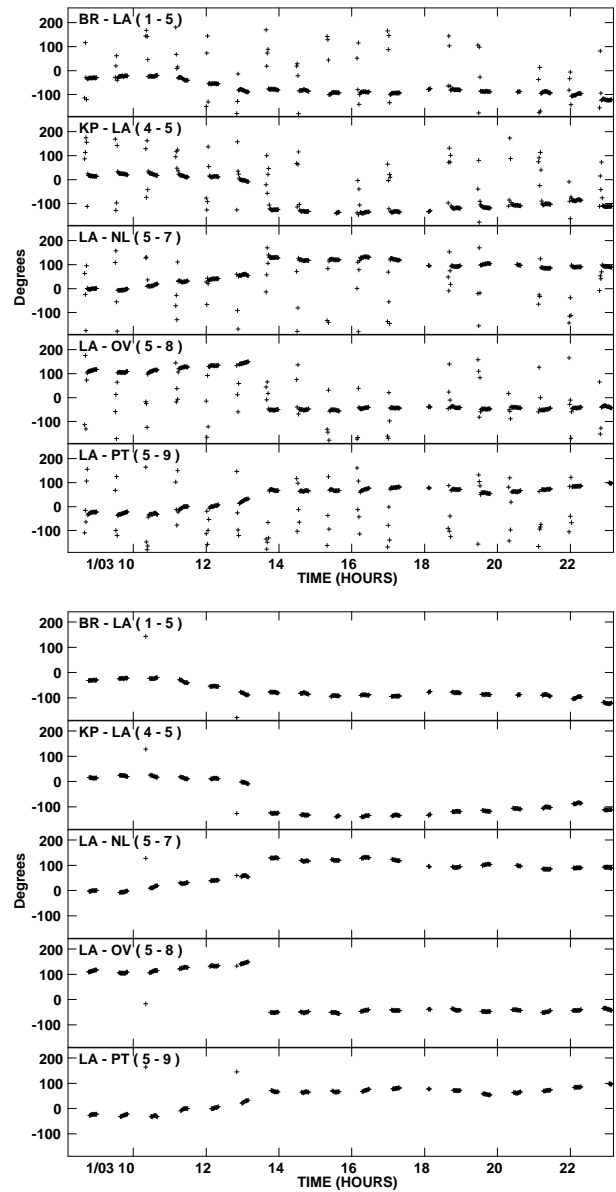

\begin{minipage}[b]{8cm}
 \includegraphics[width=8cm, angle=270]{chap6/plots/phases_unflagged.ps}
 \includegraphics[width=8cm, angle=270]{chap6/plots/phases_flagged.ps}
\end{minipage}
\caption[Illustration of the flagging scheme (phases)]{Phases of the
same visibilities shown in Fig.~\ref{fig:flagging}. {\it Top:} Phases
before flagging. {\it Bottom:} Phases after flagging.}
\label{fig:flagging_phases}
\end{SCfigure}

\begin{SCfigure}[][htpb]
\begin{minipage}[b]{8cm}
 \includegraphics[width=8cm, angle=270]{chap6/plots/4261_43_tecor_BR_short.ps}
 \includegraphics[width=8cm, angle=270]{chap6/plots/4261_43_tecor+ffs_BR_short.ps}
\end{minipage}
\caption[Expanded sections of Fig.~\ref{fig:results_BR}]{Expanded
sections of Fig.~\ref{fig:results_BR}}
\label{fig:results_BR_expanded}
\end{SCfigure}

\begin{SCfigure}[][htpb]
\begin{minipage}[b]{8cm}
 \includegraphics[width=8cm, angle=270]{chap6/plots/4261_43_tecor_LA_short.ps}
 \includegraphics[width=8cm, angle=270]{chap6/plots/4261_43_tecor+ffs_LA_short.ps}
\end{minipage}
\caption[Expanded sections of Fig.~\ref{fig:results_LA}]{Expanded
sections of Fig.~\ref{fig:results_LA}}
\label{fig:results_LA_expanded}
\end{SCfigure}

\begin{figure}[ht!]
\centering
 \includegraphics[width=0.7\linewidth, angle=270]{chap6/plots/NGC4261_43_CL16_no_rates_flagged.ps}
 \caption[NGC\,4261 43\,GHz dirty image]{Naturally weighted,
 full-resolution image of NGC\,4261 at 43\,GHz, calibrated with
 scaled-up phase solutions from 15\,GHz. FRING has been used to solve
 for one residual phase and rate solution per 25\,min scan before
 exporting the data to  Difmap. No further self-calibration has
 been applied. The image noise is 4.4\,mJy\,beam$^{-1}$ and the
 dynamic range is 18:1.}  \label{fig:NGC4261_43GHz_FFS_raw}
\end{figure}

\begin{figure}[ht!]
\centering
 \includegraphics[width=0.7\linewidth, angle=270]{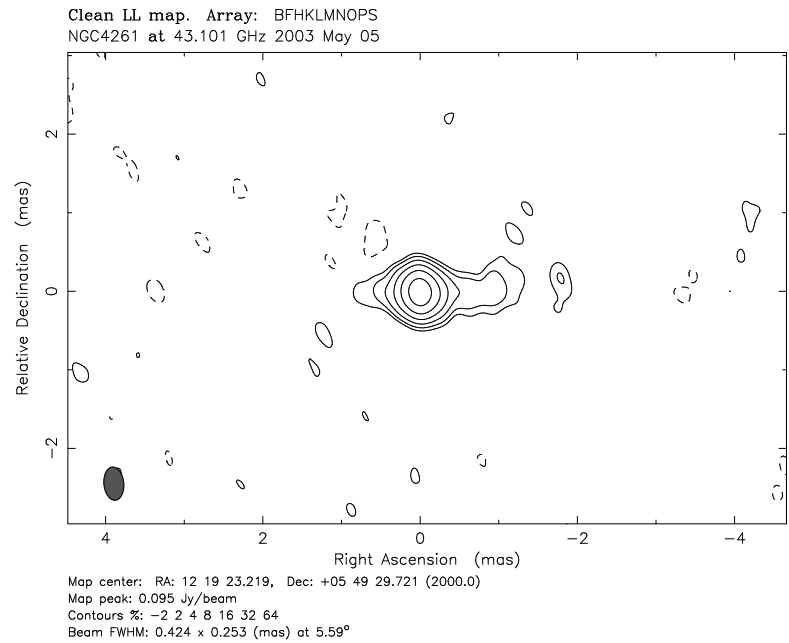}
 \caption[NGC\,4261 43\,GHz clean image]{Data and imaging parameters as
 in Fig.~\ref{fig:NGC4261_43GHz_FFS_raw}, but several cycles of phase
 self-calibration with a solution interval of 30\,s and one cycle of
 amplitude self-calibration with a solution interval of 12\,h have
 been applied. The image noise is 0.78\,mJy\,beam$^{-1}$ and the
 dynamic range is 122:1.}  \label{fig:NGC4261_43GHz_FFS}
\end{figure}

\chapter{Conclusions}

In an attempt to measure the magnetic field strength in the sub-pc
scale environment of AGN through Faraday rotation, we have used the
VLBA at 15.4\,GHz to look for polarized emission in NGC\,3079,
NGC\,1052, NGC\,4261, Hydra~A, Centaurus~A and Cygnus~A.  We selected
these nearby ($D<200\,{\rm Mpc}$) objects for high linear resolution
and good evidence for free-free absorption in a foreground medium. If
the medium is interspersed with a magnetic field, Faraday rotation is
expected to occur. Surprisingly, we did not detect any polarized
emission in five out of six sources, and we found one source
(Cygnus~A) to be polarized at a level of only $(0.44\pm0.3)\,\%$. The
results for one source (Centaurus~A) were inconclusive due to poor
$(u,v)$ coverage.  The non-detection of polarized emission is not
unexpected in regions where a dense free-free absorber is seen, but is
unexpected for regions that apparently are unabsorbed.  We have
supplemented our observations of NGC\,3079 with multi-epoch,
multi-frequency VLBI observations to study spectral properties and
proper motions. From our study, we draw the following conclusions.\\

1. The non-detection of polarized emission is significantly different
from VLBI surveys targeted at highly beamed quasars. We find that our
selection criteria biased the sample towards unbeamed, double-sided
radio jets in which circumnuclear absorbers are seen edge-on.\\

2. We expect that the radio emission from the sources is synchrotron
emission which is intrinsically polarized, and conclude that the
radiation is depolarized in a foreground screen. The rotation measure
in the screen must be variable on scales much smaller than the
synthesized beam size so that differential Faraday rotation causes a
net decrease of polarized emission when averaged across the beam
area. In our sources, such screens are detected through their
free-free absorption only in very small regions across parts of the
radio jets where they are interpreted as free-free absorption in
circumnuclear tori, but their opacity rapidly decreases away from that
region and they are not detected along the jets. This imposes
constraints on the physical parameters in the foreground screen. The
screen must be dense enough to depolarize at the relatively high
frequency of 15.4\,GHz and at the same time, must be transparent to
radio emission along most of the jet. We find that the electron
density in the screen must be of the order of $25\,{\rm cm^{-3}}$ for
magnetic fields of the order of 0.1\,mG and for path lengths of a few
parsecs.\\

3. Considering possible absorbers in front of the jets in which the
Faraday depolarization occurs, we present a model in which ionized
material is accreted in a spherical flow from scales of $>10\,{\rm
pc}$ to sub-pc scales (Bondi accretion). Assuming equipartition
magnetic fields, the model predicts the necessary Faraday rotation
and, at the same time, is transparent to radio emission at
15.4\,GHz. The accretion rates are probably signifcant lower in the
Seyferts NGC\,3079 and NGC\,1052 than in the other four objects, and
the rotation measures predicted by the model are too low. However, we
cannot rule out that the depolarization in these two objects is due to
a combination of the ionized fraction of the interstellar gas,
H\,{\small II} regions and NLR clouds.\\

4. Our observations of NGC\,3079 have yielded the following additional
results. In two radio components, $A$ and $E$, we find good evidence
for a foreground free-free absorber, with column densities of the
order of $10^{22}\,{\rm cm^{-2}}$. The intrinsic (unabsorbed) spectrum
of $E$ is extremely steep, with $\alpha=-4.1$, the best explanation of
which is by a non-power-law distribution of electron energies. The
spatial distribution of peak frequencies found in four components in
NGC\,3079 is suggestive of a foreground absorber with a density
gradient.

We find proper motions of the order of $0.1\,c$ among the three
strongest components, $A$, $B$ and $E$. The lack of absolute positions
for all but the last epoch did not allow us to define a kinematic
centre in NGC\,3079, but we prefer a scenario in which either $B$ is
stationary (possibly equvalent to the case in which the kinematic
centre is not coincident with any of the components) or the components
move on helical trajectories. If $B$ is stationary, component $E$ is
moving towards the AGN, which is unexpected.

In a statistical analysis of VLBI observations of Seyfert galaxies, we
find that the spectral and structural properties of NGC\,3079 are not
uncommon, though the spectra are at the positive end of the
distribution.\\

5. For higher resolution observations of nearby and hence mostly weak
AGN, we have developed the phase-referencing strategy of fast
frequency switching, in which the source is observed continuously
while rapidly switching between the target frequency and a lower
reference frequency. Phase solutions from self-calibration at the low
frequency are multiplied by the frequency ratio and are used to
calibrate the high-frequency visibilities. We describe the strategy
and results of an observation of NGC\,4261. We found that short-term
phase fluctuations at 43\,GHz and 86\,GHz can be calibrated with
residual rms phase errors of $33^\circ$ and $70^\circ$, respectively
using interleaved 15\,GHz scans. Ionospheric phase fluctuations on
timescales of minutes to hours caused residual phase drifts of up to a
turn of phase and did not allow us to fully calibrate the data without
self-calibration. We present the first detection of NGC\,4261 at
86\,GHz which was not possible with conventional VLBI because it is
weak. To our knowledge, this is so far the weakest source detected
with VLBI at that frequency.

\bibliography{refs}

\chapter{Danksagung}
\selectlanguage{german}

Die Danksagung wird oft als das schwierigste Kapitel einer
Doktorarbeit bezeichnet, und angesichts der grossen Anzahl von
Menschen, denen ich zu Dank verpflichtet bin, beinhaltet diese Aussage
wohl einen Funken Wahrheit.\\

Zun\"achst m\"ochte ich meinen Referenten Prof. Ulrich Klein und
Priv. Doz. Walter Huchtmeier danken, die die Arbeit des Nachsehens auf
sich genommen haben. Sie haben mir dar\"uber hinaus in zahlreichen
Gespr\"achen hilfreich zur Seite gestanden und wichtige Impulse
gegeben, die meine Arbeit in die vorliegende Form gebracht haben.\\

Den Direktoren des MPI m\"ochte ich f\"ur die M\"oglichkeiten danken,
die mir durch meine Anstellung am Institut zuteil wurden. Insbesondere
die zahlreichen Dienstreisen und Konferenzen haben meinen Horizont
betr\"achtlich erweitert. Dr. Anton Zensus bin ich dar\"uber hinaus
f\"ur zahlreiche Gespr\"ache dankbar, die den Fortgang meiner Arbeit
und die Planungen f\"ur die Zeit danach betrafen.  Auch Dr. Arno
Witzel hat mir hier oft weiter geholfen.\\

Einer der sch\"onsten Aspekte meiner Zeit am MPI war sicherlich das
Arbeiten in einer so freundlichen Gruppe mit so vielen Menschen, die
ihr Wissen bereitwillig weiter geben, und allen geb\"uhrt mein
aufrichtiger Dank. Insbesondere danken m\"ochte ich Dr. Andrei Lobanov
und Dr. Eduardo Ros f\"ur zahlreiche Diskussionen zur AGN-Physik und
Astrometrie; Dr. Richard Porcas f\"ur geduldige Erkl\"arungen und
interessante Diskussionen zu den zahlreichen Fallstricken in VLBI und
f\"ur das Nachsehen von Teilen dieser Arbeit; Dr. Thomas Krichbaum
f\"ur seine Unterst\"utzung meiner Projekte und Ratschl\"age zu
Proposals sowie ebenfalls f\"ur Korrekturen; Dr. Thomas Beckert f\"ur
zahlreiche Erl\"auterungen theoretischer Zusammenh\"ange; Dr. Walter
Alef f\"ur Erkl\"arungen zum Korrelator und seine nimmerm\"ude
Hilfsbereitschaft in Computerangelegenheiten; Dr. Alex Kraus f\"ur
seine Hilfe in Effelsberg und zur Datenreduktion und Prof. Heino
Falcke f\"ur seine Beitr\"age und Ideen zu meiner Arbeit. Macht weiter
so!\\

Nicht zu untersch\"atzen ist auch der Beitrag der Studenten zu diesem
Arbeitsklima, und daf\"ur bin ich allen dankbar. Hervorheben m\"ochte
ich Tim Huege und Elmar K\"ording, die mir in Dingen der Mathematik
hilfreich zur Seite standen (au"serdem werde ich den t\"aglichen
12~Uhr-Anruf vermissen). Matthias Kadler, Violetta Impellizzeri und
Uwe Bach haben nicht nur eine sehr angenehme B\"uroatmosph\"are
geschaffen, sondern waren immer zu Diskussionen bereit, um die
t\"aglichen Probleme zu l\"osen. Mit Uwe Bach verbindet mich nun auch
ein nahezu neunj\"ahriges Studium, und ohne ihn w\"aren so einige
H\"urden h\"oher und viele Probleme schwieriger gewesen, und daf\"ur
bin ich sehr dankbar. Au"serdem geb\"uhrt allen anderen Mitarbeitern
des MPI, von der Materialstelle \"uber die Verwaltung bis zur
Rechnerabteilung, ein recht herzliches Dankesch\"on.\\

Von au"serhalb des MPI habe ich Dr. Andrew Wilson f\"ur seine
ausf\"uhrlichen Bemerkungen zur Statistik der Seyfert-Galaxien und
Dr. Denise Gabuzda f\"ur ihre Hilfe bei der Polarimtrie zu
danken. Dr. Craig Walker hatte einen erheblichen Anteil an der
Entwicklung des Fast Frequency Switchings, und die Zeit in Socorro
bleibt unverge"slich.\\

F\"ur Dr. Alan Roy fehlen mir die Worte und es w\"are sinnlos, das
durch unzureichende Umschreibungen zu verdecken. Vielen, vielen Dank,
Alan.\\

Meinen Eltern danke ich f\"ur ihre Unterst\"utzung, ihre Liebe und
ihren Optimismus, und Angelika f\"ur alles, was uns verbindet.

\end{document}